\documentclass[a4paper,12pt,headsepline,twoside,bibliography=totoc]{scrartcl}
\newcommand{\titleDocument}{PhD Thesis}
\newcommand{\subjectDocument}{in the Physical Sciences}

\usepackage[hyphens,obeyspaces,spaces]{url}

\usepackage[bibstyle=numeric,citestyle=numeric-comp,sorting=none]{biblatex}
\usepackage{graphicx}
\usepackage[table]{xcolor}
\usepackage{colortbl}

\usepackage{comment}

\usepackage{unitsdef}

\usepackage{graphicx}

\usepackage[right]{eurosym}

\usepackage[utf8]{inputenc}

\usepackage[T1]{fontenc}

\usepackage{eufrak}

\usepackage{lmodern}
\usepackage{fix-cm}

\usepackage{longtable}

\usepackage{geometry}
\usepackage{pgfplots}
\pgfdeclarelayer{foreground}
\pgfsetlayers{background,main,foreground}
\usepackage{xcolor}
\usepackage{rotating}
\usepackage{amsfonts}
\usepackage{mathtools}
\DeclarePairedDelimiter\bra{\langle}{\rvert}
\DeclarePairedDelimiter\ket{\lvert}{\rangle}
\DeclarePairedDelimiter\braket{\langle}{\rangle}
\DeclarePairedDelimiterX\Braket[2]{\langle}{\rangle}{#1 \delimsize\vert #2}

\usepackage{appendix}
\usepackage{titletoc}

\usepackage{dsfont}
\usepackage{slashed}
\newcommand\hatslashed[1]{{\hat{#1}\mathllap{\slashed{#1}}}}

\usepackage[thinlines]{easytable}

\usepackage[super]{nth}
\usepackage{fancybox}

\let\stdsection\section
\renewcommand\section{\clearpage\stdsection}

\usepackage{color}

\usepackage{enumerate}

\usepackage{xspace}

\usepackage{float}

\usepackage{amsmath}
\usepackage{amssymb}
\usepackage{revsymb}
\usepackage[polish,british]{babel}

\numberwithin{equation}{section}

\usepackage{subcaption}

\usepackage[bookmarksnumbered,pdftitle={\titleDocument},hyperfootnotes=false,hidelinks]{hyperref} 
\usepackage{cleveref}

\usepackage{fancyhdr} %Paket laden
\fancypagestyle{mainpart}{ %eigener Seitenstil
\fancyhf{} %alle Kopf- und Fußzeilenfelder bereinigen
\fancyhead[L]{\nouppercase{\leftmark}} %Kopfzeile links
\fancyhead[C]{} %zentrierte Kopfzeile
\fancyhead[R]{\thepage} %Kopfzeile rechts
\fancyfoot[C]{\thepage} %Seitennummer

} %untere Trennlinie

\fancypagestyle{preamb}{ %eigener Seitenstil
\fancyhf{} %alle Kopf- und Fußzeilenfelder bereinigen
\fancyhead[L]{\nouppercase{\leftmark}} %Kopfzeile links
\fancyhead[C]{} %zentrierte Kopfzeile
\fancyhead[R]{\roman{\thepage}} %Kopfzeile rechts
\fancyfoot[C]{\roman{\thepage}} %Seitennummer
}

\usepackage{array}

\usepackage{setspace}

\usepackage{capt-of}

\usepackage{makeidx}

\usepackage{csquotes}

\usepackage{listings}
\makeindex

\usepackage[super]{nth}

\usepackage[german]{nomencl}

\makenomenclature
\DeclareOldFontCommand{\rm}{\normalfont\rmfamily}{\mathrm}

\newcommand{\be}{\begin{equation}}
\newcommand{\ee}{\end{equation}}
\newcommand{\bea}{\begin{eqnarray}}
\newcommand{\eea}{\end{eqnarray}}

\newcommand\ie{\mbox{\textit{i.\,e.}}\xspace}
\newcommand\cf{\mbox{c.\,f.}\xspace}
\newcommand\eg{\mbox{e.\,g.}\xspace}

\newcommand\D{\mathrm{d}}

\newcommand\Ord{\mathcal{O}}

\newcommand{\HRule}{\rule{\linewidth}{0.5mm}}
\newcommand\op{\hat{\mathcal{O}}}
\newcommand\hil{\mathcal{H}}
\newcommand{\var}[1]{\left(\sigma^2_{#1}\right)}
\newcommand{\id}{\mathds{1}}

\newcommand{\hx}{\hat{x}}
\newcommand{\hp}{\hat{p}}
\newcommand{\hX}{\hat{X}}
\newcommand{\hP}{\hat{P}}
\newcommand{\mpar}{\dot{\partial}}
\newcommand{\vel}{\mathcal{V}}
\newcommand{\reals}{{\rm I\!R}}
\newcommand{\Mod}[1]{\ (\mathrm{mod}\ #1)}
\xdefinecolor{myblue}{rgb}{0.2,0.3,0.65}
\definecolor{burgundy}{rgb}{0.5, 0.0, 0.13}
\xdefinecolor{mix}{rgb}{0.15,.37, .40} 

\DeclareMathOperator\arctanh{arctanh}

\makeatletter
\newcommand{\vast}{\bBigg@{4}}
\newcommand{\Vast}{\bBigg@{5}}
\makeatother

\makeatletter
\newcommand*\bigcdot{\mathpalette\bigcdot@{.5}}
\newcommand*\bigcdot@[2]{\mathbin{\vcenter{\hbox{\scalebox{#2}{$\m@th#1\bullet$}}}}}
\makeatother

\def\makeheadline{\vbox to 0pt{\vskip-22.5pt
  \hbox to \hsize{\vbox to 8.5pt{}\the\headline}\vss}\nointerlineskip}

\begin{document}

% Titelseite %
% das Papierformat zuerst
%\documentclass[a4paper, 11pt]{article}

% deutsche Silbentrennung
%\usepackage[ngerman]{babel}

% wegen deutschen Umlauten
%\usepackage[ansinew]{inputenc}

% hier beginnt das Dokument
%\begin{document}

\thispagestyle{empty}

%\begin{figure}[t]
% \includegraphics[width=0.6\textwidth]{abb/fh_koeln_logo}
%\end{figure}

\begin{figure}[t]
\centering
\begin{minipage}{0.5\linewidth}
\noindent\includegraphics[width=.8\linewidth]{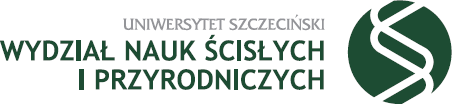}
\end{minipage}
\hfill
\begin{minipage}{0.3\linewidth}
\includegraphics[width=.8\linewidth]{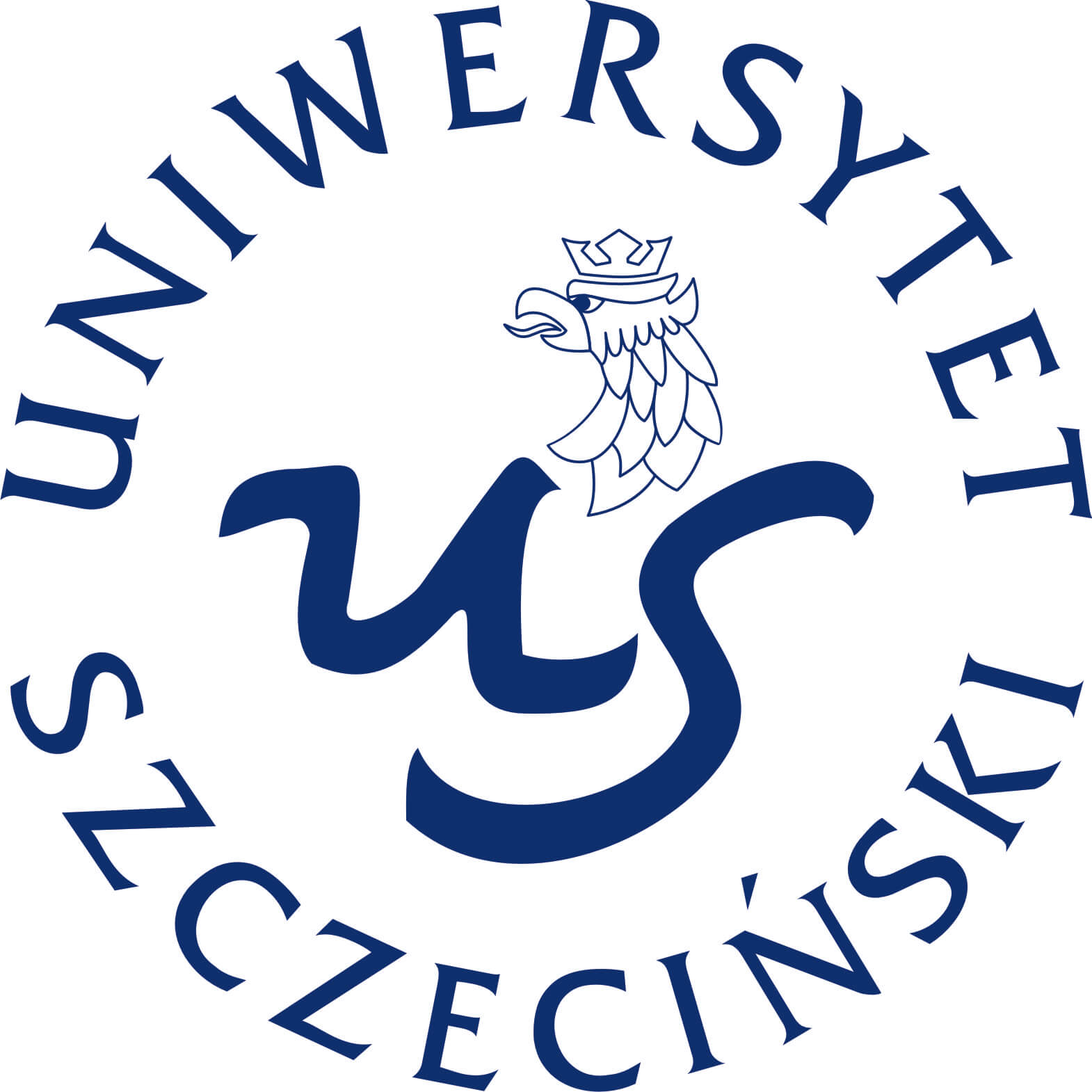}
\end{minipage}
\end{figure}

\begin{verbatim}





\end{verbatim}

\begin{center}
\Large{Uniwersytet Szczeci\'{n}ski}\\
\end{center}

\begin{center}
\Large{Wydzia\l{} Nauk \'{S}cis\l{}ych i Przyrodniczych}
\end{center}
\begin{verbatim}



\end{verbatim}
\begin{center}
\doublespacing
\textbf{\LARGE{\titleDocument}}\\
\singlespacing

{~\subjectDocument~}
\end{center}

\begin{center}
to receive the academic degree of\\ Doctor of Physical Sciences
\end{center}
\begin{verbatim}

\end{verbatim}

\begin{center}
\doublespacing
\HRule\\[0.3cm]
\textbf{\Large{Modified uncertainty relations from classical and quantum gravity}}\\
\HRule
\end{center}
\begin{verbatim}




\end{verbatim}
\begin{flushleft}
\begin{tabular}{llll}
\textbf{Author:} & & Fabian László Konstantin Wagner & \\
\textbf{PESEL:}& & 12019415155 & \\
%\textbf{Version vom:} & & \today &\\
& & \\
\textbf{Supervisor:} & & Prof.  Mariusz P.  D\k{a}browski &\\
\textbf{Auxiliary Supervisor:} & & Dr. Hussain Gohar &\\
& & \\
\textbf{Submission date:} & & \nth{20} of May 2022
\end{tabular}
\end{flushleft}

\newpage

\cleardoublepage

\pagestyle{empty}

%\input{Main_and_Chapters/Declaration}

%\newpage

\cleardoublepage

\pagestyle{mainpart}
\setcounter{page}{1}
\pagenumbering{roman}

\section*{Abstract}
\markright{Abstract}
A good hundred years after the necessity for a quantum theory of gravity was acknowledged by Albert Einstein, the search for it continues to be an ongoing endeavour. Nevertheless, the field still evolves rapidly as manifested by the recent rise of quantum gravity phenomenology supported by an enormous surge in experimental precision. In particular, the minimum length paradigm ingrained in the program of generalized uncertainty principles (GUPs) is steadily growing in importance. 

The present thesis is aimed at establishing a link between modified uncertainty relations, derived from deformed canonical commutators, and curved spaces - specifically, GUPs and nontrivial momentum space as well as the related extended uncertainty principles (EUPs) and curved position space. In that vein, we derive a new kind of EUP relating the radius of geodesic balls, assumed to constrain the wave functions in the underlying Hilbert space,  with the standard deviation of the momentum operator, suitably made compatible with the curved spatial background. This result is gradually generalized to relativistic particles in curved spacetime in accordance with the 3+1 decomposition, thereby relating semiclassical gravity with the EUP. 

The corresponding corrections to the relation in flat space depend on the Ricci scalar of the effective spatial metric, the lapse function and the shift vector, as well as covariant derivatives thereof. The ensuing inequality is evaluated in Rindler, de Sitter and Schwarzschild backgrounds, at lowest approximation leading to identical effects, as well as to rotating geometries like Kerr black holes and their analogues in higher-order theories of gravity. 

In a sense pursuing the inverse route, we find an explicit correspondence between theories yielding a GUP, possibly including a noncommutative geometry, and quantum dynamics set on non-Euclidean momentum space. Quantitatively, the coordinate noncommutativity translates to momentum space curvature in the dual description, allowing for an analogous transfer of constraints from the literature. However, a commutative geometry does not imply trivial dynamics; the corresponding types of GUP lead to a flat momentum space, described in terms of a nontrivial basis, permitting the import of further bounds. 

Finally, we find a formulation of quantum mechanics which proves consistent on the arbitrarily curved cotangent bundle. Along these lines, we show that the harmonic oscillator can, given a suitable choice of operator ordering, not be used as a means to distinguish between curvature in position and momentum space, thereby providing an explicit instantiation of Born reciprocity in the context of curved spaces.

\section*{Acknowledgements}
\markright{Acknowledgements}

Considering the public importance of personal prizes and accomplishments like the awarding of a PhD, one might get the impression that science is done by savvy, independent individuals living monkish lives. Nothing, however, could be further from the truth. 

Where science's positive influence on everybody's life is apparent as in new technologies simplifying the daily grind or, especially in the light of recent events, medicine, there is also the kind of research whose impact on society cannot be as immediate. This thesis clearly belongs to the latter kind. In a way, the work contained in it may be understood as a luxury for society, a minute step towards understanding the universe. Therefore, I am first and foremost indebted to the people whose contribution to society makes this work possible, in this case the citizens of the European Union, in particular, of course, the Polish ones.

Apart from that scientific endeavour is seldom the work of few. We are nothing without the colleagues around us, providing important ideas and feedback. Where biases cloud our own judgement, it is important to know, which ideas have potential and which belong to the dustbin. 

Let me first express my respect and appreciation to my supervisor Prof. Mariusz D\k{a}\-browski, who showed a great amount of openness whenever confronted with my crazy thoughts, yet taught me how to extract the important bits; who did not object when I decided to take a different route than the one he had envisioned for me and thus gave me the freedom to develop my own interest in research.

Much of the inspiration and motivation for my work as well as much of the valuable critique I have been offered, originated in the regular Zoom meetings I had with Luciano Petruzziello, collaborator and friend. Whenever I was stuck, however, I reverted to my dear colleague Samuel Barroso-Bellido, who, in many ways better-versed than me on the mathematical side, often made questions or comments thinking about which lead to the solution of many a hard nut to crack. Let me also thank the other members of the Szczecin cosmology group for everything they have taught me and all the interesting discussions (and pizzas) we have had.

Writing a PhD thesis is not just a techno-scientific effort, though. It is the concentration of all means accessible to a person towards one sole goal. In and of itself this is a very self-centered act, often at the expense of the social environment. 

In this sense, I would like to convey my deepest gratitude and love to my family, without whose support I would have never made it to this point, who always nourished my interest in science in general and physics in particular. 

Mama, you have made my life, as it is, possible - with an effort, which sometimes appears superhuman to me. Laura and me, we are, in so many ways, your creation, your life's work. As long as you are behind me, I shall never feel insecure.

Papa, it took me a long time to understand, how many people could everlastingly strive for their fathers' appreciation, and perhaps even longer to grasp, how lucky I should be about not understanding that. Thank you so much for your continuous support, whatever may come, and always telling me how proud you are irrespective of what I achieve.

Laura, we have always been and will always be a team. In you, I always know my biggest supporter as you do in me. Irrespective of the circumstances, siblings stick together. The same, of course, goes for Maxi, who already shares my love for a good book on a Sunday morning - what I consider one of my best traits.

Without Fritz, my recently deceased grandfather, I may have never turned to science. He was the one, whose popular science journals I was reading, when I would get up too early as a child,  that is every day. He was also the one to get me a subscription to the German version of the Scientific American for every birthday from when he noticed my interest in science onward.

Let's be honest, as probably everybody following this route, I experienced tough moments during the time I spent as a PhD student. Yet, whenever I was feeling at the cusp of despair, you, Niklas, Elena, Robert, Anna, Rike, Dome, Lucas, Flo, Mikel, Luca, Moritz, Paolo, Enrico, Leo, Julian and so many more, were there for me. If I wanted to recount all the ways in which you supported and, in fact, shaped me during the last years, I would probably have to write a text of the length of the thesis itself.

As usual the most important piece comes last - Aim\'ee, ma bien-aim\'ee! If there is anyone on this planet, who embodies fighting fire with fire, it is you. In you I see a passion for life and a willingness to fight for it, I deeply admire. In this way, you are my role model. You were the one to instill confidence in me when I stopped to dare believing, the confidence needed to accomplish this piece of work. Not only did you support me unconditionally, you also had to deal with me being disappointed and unbearable at times, when things did not end up the planned way, and often did so without a word of anger. Whatever may be, I shall be content if I am with you.

\tableofcontents

\newpage

\setcounter{page}{1}
\pagenumbering{arabic}

\begin{refsection}
\nocite{Dabrowski19,Dabrowski20,Petruzziello21,Wagner21a,Wagner21b,Wagner21c}
\defbibnote{myPrenote}{
    The present thesis is based on the papers enumerated on this page. In particular, the sections \ref{sec:3DEUP} -- \ref{sec:relgen}, \ref{sec:modcomorcurspa} and \ref{sec:qmhamilton} contain work published in Refs. \cite{Dabrowski20,Petruzziello21,Wagner21b,Wagner21a,Wagner21c}, respectively, while section \ref{sec:GUPsEUPs} contains some information from Ref. \cite{Dabrowski19}. Parts and ideas extracted from the Refs. \cite{Dabrowski19,Dabrowski20,Petruzziello21,Wagner21b} are combined and extended in section \ref{sec:applications}.
}
\defbibnote{myPostnote}{
Note that Ref. \cite{Wagner21c} has not been published yet. Further publications prepared and coauthored during the duration of the studies but not included in the thesis involve :
}
\printbibliography[
    heading=bibintoc,
    title={Author's Contributions},
    prenote=myPrenote,
    postnote=myPostnote
]
\end{refsection}

\begin{refsection}
\DeclareFieldFormat{labelnumber}{%
  \ifinteger{#1}
    {\number\numexpr#1+6\relax}
    {#1}}
\nocite{Bellido22,Bosso:2022vlz,Wagner:2022rjg,Loll:2022ibq,Bosso:2022ogb,Wagner:2022dkc,Guendelman:2022ruu}
\printbibliography[heading=none]
\end{refsection}

\nocite{Dabrowski19,Dabrowski20,Petruzziello21,Wagner21a,Wagner21b,Wagner21c,Bellido22,Bosso:2022vlz,Wagner:2022rjg,Loll:2022ibq,Bosso:2022ogb,Wagner:2022dkc,Guendelman:2022ruu}

\section{Introduction}\label{sec:intro}

The considerations around which this thesis is centered are situated well within a particularly old field of research. As early as 1916, in his famous paper on gravitational waves Albert Einstein observed that
\begin{displayquote}
    it appears that the quantum theory must modify not only Maxwell's electrodynamics but also the new theory of gravitation. \cite{Einstein16}
\end{displayquote}
At the time the development of such a theory was thought to be a necessary but straightforward step, a sentiment well palpable in Wolfgang Pauli's and Werner Heisenberg's remark from 1929:
\begin{displayquote}
    Let it be mentioned that a quantization of the gravitational field, while appearing to be necessary on physical grounds,[\dots] should be feasible without new difficulties. \cite{Heisenberg29}
\end{displayquote}
Unfortunately, history has proven them wrong.

Now, almost a hundred years later, this problem continues to attract generations over generations of practitioners of fundamental physics. Nevertheless, the end of this endeavour is not even remotely in sight. On the contrary, new results stemming from a plethora of approaches continue to amaze even the most pragmatic researcher. However, the interest of the community is slowly diverging from developing new approaches to finally discussing observables. 

The present thesis is intended to be a step towards understanding some of the peculiarities of quantum gravity in the phenomenological context to finally try and find paths towards experimental falsification of some of its underlying concepts. First, however, we need to understand why we need quantum gravity and why it has become such a challenge to find a suitable candidate.

\subsection{Insurmountable inconsistencies}

The world is quantum. In particular, the fundamental laws underlying the standard model, describing all of matter and its interactions between sizes of $10^{-22}\meter$ (largest possible radius of the electron compatible with observation  \cite{Dehmelt88}) and the millimeter scale (smallest detected gravitational source \cite{Westphal21}), are expressed in the language of quantum field theory. The classical environment we perceive arises from this framework following a highly nontrivial, yet consistent limiting procedure. 

The other pillar of modern physics, Einstein's theory of general relativity, in contrast, is an inherently classical framework, providing an understanding of the very large, \ie astrophysical and cosmological, scales. From the microscopic perspective, however, it tells a tale of incoherence as was famously proven by Hawking and Penrose \cite{Penrose65,Hawking70}. In particular, gravitational collapse inevitably ends up in a singularity, thereby breaking the smooth manifold structure at the very heart of the theory. Similar considerations hold for the beginning of the universe. In other words, general relativity predicts its own demise.

Following the reductionist approach, that proved so successful in the development of the standard model, we should expect that there is an underlying microscopic theory coupled to matter, from which Einstein's equations emerge through coarse graining. Since Niels Bohr's and L\'eon Rosenfeld's study on the measurability of the electromagnetic field \cite{Bohr33}, published in 1933, there has been a successive development of more and more convincing arguments, implying that classical and quantum theories cannot be coupled to each other \cite{deWitt13,Peres01,Marletto17a} consistently (\cf Ref. \cite{Rosenfeld63,Penrose96} for a different view). Hawking, for example, pointed out an explicit paradox leading to loss of information in black hole evaporation \cite{Hawking75}, which would be in stark contradiction with quantum theory, when trusting this semiclassical approach up to high energies \cite{Hawking76}.  No less problematic, it can only be applied to semiclassical states. What, for example, would be the gravitational field of an object in a superposition of distinct locations \cite{Carlesso19}? Apart from that, there is indirect experimental evidence for quantum gravity under the assumption of an interpretation of quantum mechanics involving unitary evolution \cite{Page81}. These and more arguments are explained in more depth in the recent review \cite{Wallace21}. Summarizing it in Richard Feynman's words,
\begin{displayquote}
    it seems clear [\dots] that we're in trouble if we believe in quantum mechanics but don't quantize gravitational theory. \cite{DeWitt57}
\end{displayquote}
Evidently, the need for such a theory was established early on. How did the same Feynman around the same time thus come to the point of expressing his anger and disbelieve in a letter to his wife as
\begin{displayquote}
    Remind me not to come to any more gravity conferences! \cite{Feynman88}
\end{displayquote}

\subsection{Quarrels with quantum gravity}

Quantum gravity is hard. The holy grail of this research program would be a theory, capable of describing everything down to scales characterized by the Planck units, which are constructed from the speed of light $c,$ Newton's constant $G$ and Planck's constant $\hbar,$ representing relativity, gravity and the quantum, respectively, and quantum general relativity together. As a matter of fact, Planck himself immediately realized the importance of the "new" units (George Stoney had invented a similar system before \cite{Stoney81}) when introducing his constant in 1900, stating that they
\begin{displayquote}
    necessarily retain their significance for all times and for all cultures, even alien and non-human ones. \cite{Planck00}
\end{displayquote}
Throughout this thesis, we will mainly encounter measures of distance and energy, rendering it instructive to express them in terms of the Planck length and the Planck mass
\begin{equation}
    l_p=\sqrt{\frac{\hbar G}{c^3}}\approx 1.6*10^{-35}\meter,\hspace{1cm}
    m_p=\sqrt{\frac{\hbar c}{G}}\approx 2.2*10^{-8}\kilogram,
\end{equation}
while the speed of light will hereafter be set equal to one. It was claimed above that the standard model retains its validity down to distances of $10^{-22}\meter.$ To be more precise, this value marks the limits of our experimental precision. In order to get an impression of the minuteness of the Planck length, note that it is situated 13 orders of magnitude below this, as of yet, highest ever achieved accuracy.

For comparison, ordinary quantum field theories, on the one hand, are placed on an \emph{a priori} existing stage. The underlying Minkowskian spacetime manifold and its global Poincar\'e symmetry are indispensable, \eg for the particle-concept and the definition of scattering processes. On the other hand, in general relativity it is the stage itself that becomes the actor. Einstein's field equations \emph{dynamically} relate spacetime to the distribution of matter and energy - to echo Wheeler's popular bon mot
\begin{displayquote}
    Spacetime tells matter how to move; matter tells spacetime how to curve. \cite{Wheeler10}
\end{displayquote}

Granted, the weak field-limit provides a way to deal with linearized gravity on flat spacetime, \ie excitations as gravitational waves, and thus retain the advantages of perturbative quantum field theory \cite{Donoghue94,Donoghue95}. In fact, this is exactly what Heisenberg and Pauli had in mind in the above statement. Yet, it should not come as a surprise that this approach cannot be the last word, but should be understood as an effective theory well below the Planckian regime. 

In particular, perturbative quantum gravity is nonrenormalizable at two loops \cite{Goroff85,vandeVen91}. \emph{Ipso facto}, the theory requires the experimental determination of an infinite amount of coupling constants beforehand to describe processes at the Planck scale, rendering it devoid of any predictive value. In other words, every renormalized amplitude $\mathcal{M},$ measured at some energy $E,$ has to be expanded in a series of the form
\begin{equation}
    \mathcal{M} = \sum_n \mathcal{M}_n\left(\frac{E}{m_p}\right)^{2n}\label{qgamp},
\end{equation}
which clearly needs an infinite amount of input above the Planck scale. Fundamental reasoning about gravity can, therefore, only be of the nonperturbative kind.

Wheeler's statement further touches upon the universality of gravity. According to the equivalence principle \cite{Einstein08,Tino20}, the interaction couples to all kinds of mass and energy, including itself. Additionally the assumption of the weak energy condition \cite{Hawking73,Kontou20}, \ie the nonnegativity of the energy-density perceived by timelike observers, consistently precludes screening. Correspondingly, Einstein's equations in and of themselves are highly nonlinear, and cannot be solved in general. Needless to say, at the Planck scale, it is expected that the concept of spacetime itself loses its meaning, leading to a possibly fractal behaviour \cite{Ambjorn05}. Correspondingly, it is not even known whether the ground state of quantum gravity can be understood as a flat manifold as assumed in the perturbative context - more on this below.

Running the risk of becoming redundant, quantum gravity is hard. Unfortunately, the traditional, perturbative approach fails, necessitating nonperturbative ans\"atze. In situations alike, the theory community usually asks the experimenters for guidance.

\subsection{What about experiments?}

Gravity is weak. In fact, everybody who has succeeded in countering the effect of the whole planet by lifting up a coin with a permanent magnet knows this. To quantify exactly how weak it is, compare the acceleration $a$ induced by the gravitational and the electrostatic fields between a proton and an electron, known through Dirac's large number hypothesis \cite{Dirac37},
\begin{equation}
    \frac{a_{\text{grav}}}{a_{\text{el}}}\sim 10^{-39}.
\end{equation}
The difference in strength between the interactions amounts to a whopping 39 orders of magnitude. In that vein, the first quantum gravitational correction to the amplitude \eqref{qgamp} when evaluated at the energy scale of the Large Hadron Collider at CERN, \ie the most extreme conditions ever devised by humankind, is of the order $10^{-32}$. This is why, the standard model proves so precise at microscopic scales in the first place. Unfortunately, this is also why experimental progress has been limited thus far. Worse even, Freeman Dyson gave a coherent argument that a detector capable of finding gravitons, the hypothetical mediators of gravity in the weak-field limit, practically has to be so dense as to collapse to a black hole \cite{Dyson04,Rothman06}. Similarly, in accordance with Penrose's cosmic censorship conjecture \cite{Penrose69} it is believed that regions of exceedingly strong curvature like black holes, the natural arena of quantum gravity, are generally veiled by horizons.

\subsection{\emph{Quo vadis} quantum gravity?}

\begin{figure}[t!]
    \centering
    \includegraphics[scale=.9]{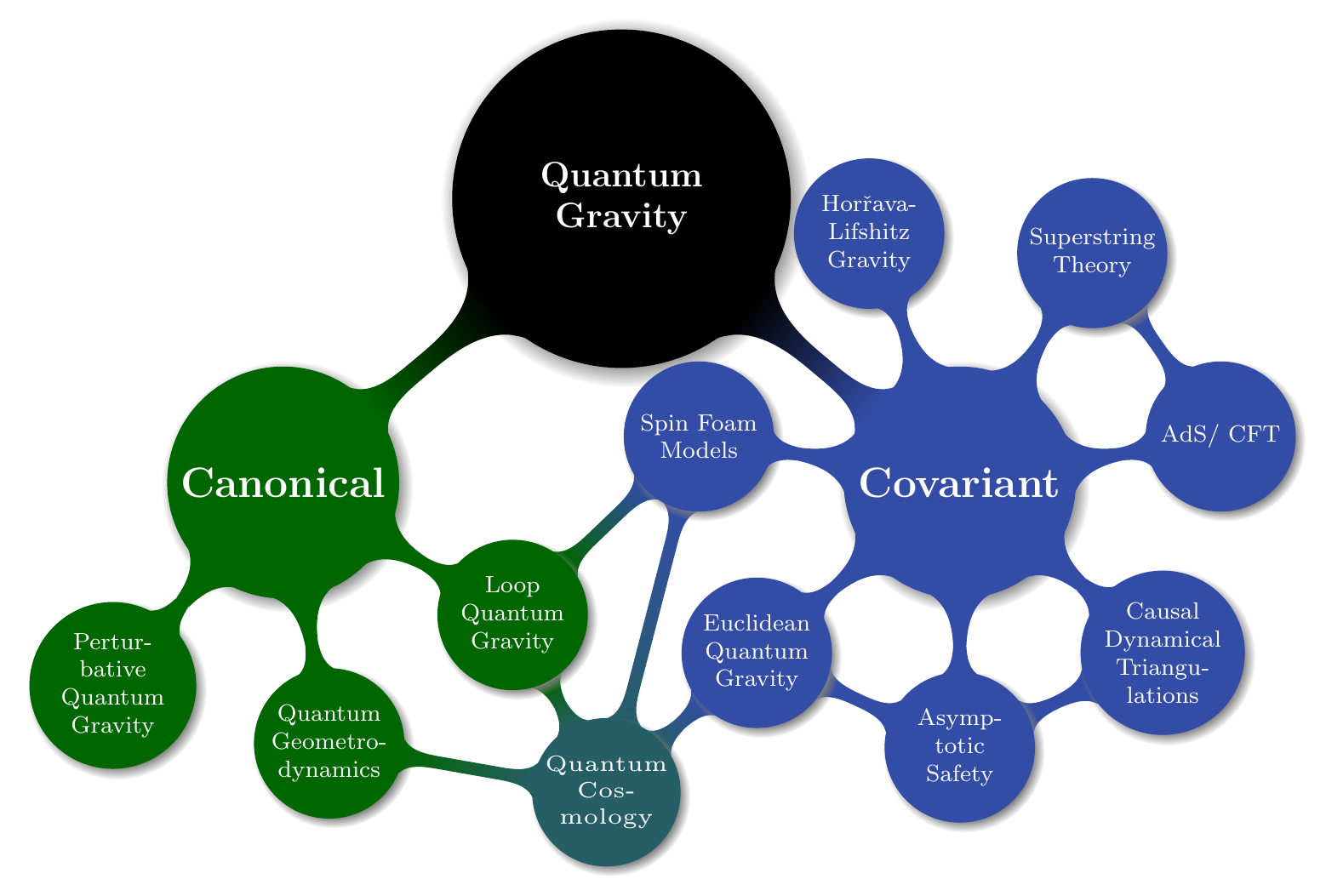}
    \caption{Some of the different approaches to quantum gravity and their main connections.}
    \label{fig:mindmap}
\end{figure}

Where there is little headway through observation, the opportunities for creative reasoning blossom. In the gravitational case, this has lead to myriads of candidate completions in the ultraviolet, some of which and their interrelations are displayed in Fig. \ref{fig:mindmap}. All of these approaches differ in their underlying assumptions. Some, like quantum geometrodynamics \cite{Wheeler57,DeWitt67}, loop quantum gravity \cite{Ashtekar86,Rovelli89,Thiemann96} and the related spin foam models \cite{Baez97} focus on diffeomorphism invariance, leading to the problem of time \cite{Isham92,Anderson12} and possibly nonunitary evolution. Others are manifestly unitary but only defined perturbatively like superstring theory \cite{Green84}, or explicitly break diffeomorphism invariance like Ho\v{r}ava-Lifshitz gravity \cite{Horava09}. Holography, \ie the AdS/CFT correspondence \cite{Maldacena97,Witten98}, causal dynamical triangulations \cite{Loll00}, asymptotic safety \cite{Weinberg79,Reuter96} and through this principle Euclidean quantum gravity \cite{Gibbons76,Hawking78,Hartle83}, on the other hand, are supposed to provide the means to satisfying both properties at the same time. 

To put it in a nutshell, there is a wide variety of approaches, some heavily interrelated and depending on each other, others so distinct that it is hard to even find a common language, not to say shared objectives. This issue as well of some of the discontent resulting from the competition between the different communities can, for example, be inferred from a recently published series of interviews with some of their representatives \cite{Armas21}. How, then, can the field as a whole find a way out of this dilemma?

\subsection{A new philosophy: quantum gravity phenomenology}

These times, the focus of parts of the community is starting to shift from futile debates about the correct approach to finding commonalities, thereby obtaining comparably robust predictions of the concept of quantum gravity itself. A first example of this overarching endeavour may be seen in the reduction of the spectral dimension of spacetime at the Planck scale, displaying its fractal behaviour, which has been investigated from manifold perspectives \cite{Lauscher05,Connes06,Modesto08,Modesto09,Carlip09,Horava09b,Padmanabhan15} after its discovery in causal dynamical triangulations \cite{Ambjorn05}.

According to this new school of thought, developed since the late 1990s \cite{Amelino-Camelia99}, Planck scale effects need not be out of reach of current sensitivity when exploiting natural mechanisms of amplification. This observation points towards the essence of the program of quantum gravity phenomenology \cite{Amelino-Camelia08,Addazi21}, which itself has been boosted by strong recent progress on the experimental side \cite{Amelino-Camelia99}.

On the one hand, the resulting predictions may stem from perturbative quantum gravity. Recently, for example, there have been a number of proposed experiments \cite{Bose17,Marletto17b} intended to finally settle the debate about the need for a quantization of gravity, mentioned above. In that vein, the authors want to answer the question whether gravity can mediate entanglement between two coherent macroscopic objects, which would amount to an indirect observation of a quantum mechanical mediator, \ie gravitons. Evidently, in this case, the effect is amplified by the number of involved constituents.

On the other hand, the extreme conditions caused by collisions involving very-high-energy gamma rays \cite{CTA17} of natural origin as well as the long travel times of gamma ray bursts from very distant sources \cite{Amelino-Camelia97} serve as probes of nonperturbative effects, among which we find the very subject of the present thesis, an idea, which has been inherent to the discourse on quantum gravity since its very beginnings.

\subsection{Quantum spacetime and the minimum length-paradigm}\label{subsec:minlen}

Can we divide space into ever smaller parts? This question has daunted philosophers and mathematicians alike since the conception of Zeno's paradox in antiquity \cite{Huggett19,Hagar14}. Sure enough, it is possible to imagine a continuous line. However, as David Hilbert put it
\begin{displayquote}
{[a]} homogeneous continuum which admits of the sort of divisibility needed to realize the infinitely small is nowhere to be found in reality. \cite{Hilbert84}
\end{displayquote}
In other words, experience can show space to be discrete; yet, its continuity can never leave the realm of the metaphysical. In fact, this mere idea has troubling consequences for fundamental physics, which where echoed in Feynman's question:
\begin{displayquote}
Why should it take an infinite amount of logic to figure out what one tiny piece of spacetime is going to do? \cite{Feynman65}
\end{displayquote}

To be precise, the mathematical concept of spacetime manifold, or space for that matter, embraces a notion of infinity. In that sense, from Newtonian mechanics to the quantum field theories in the standard model, every physical theory has been staged on a union of an infinite number of zero-dimensional points. The infinite, however, \emph{a priori} exceeds our sensitivity.

Since the first steps towards a quantum theory of gravity \cite{Bronstein30,Bronstein36} were being taken, it has been clear that such an endeavour would question some of the most strongly held principles underlying physics as it was known at the time. Among those may as well be the continuous accessibility of spacetime itself. In the words of the father of quantum gravity, Matvei Bronstein,
\begin{displayquote}
the possibilities of measurement are even more restricted than those due to the quantum-mechanical commutation relations. Without a deep revision of classical notions it seems hardly possible to extend the quantum theory of gravity also to this domain. \cite{Bronstein36}
\end{displayquote}
By pure reasoning on the concepts behind quantum mechanics and general relativity, he had thus arrived at the conclusion that the fundamental accuracy of measurements of positions and momenta proved even more constrained than in quantum mechanics alone. Effectively, he had found a \emph{minimum length}.

Even earlier, Werner Heisenberg had been trying to deform the canonical commutation relations underlying textbook quantum mechanics \cite{Dirac30,Messiah99} to allow for noncommuting coordinates in order to cure divergences in the self-energy of the electron \cite{Heisenberg30,Pauli85}. Clearly, this assumption implied a nonvanishing uncertainty relation for position measurements -- a minimum length in the form of what is nowadays called a generalized uncertainty principle (GUP). Yet, according to his understanding, such a limitation could not be implemented in a relativistically invariant way due to Lorentz contractions of the scale. Shortly after, similar considerations lead Gleb Wataghin to inventing nonlocal field theories which, though Lorentz covariant, allowed for acausal behaviour \cite{Wataghin37}.

These difficulties were remedied by Hartland Snyder \cite{Snyder46,Snyder47,Yang47}. inventing the first Lorentz invariant theory of noncommutative geometry. Yet, even when it had been put on much firmer conceptual ground by the inclusion of gravity into the Heisenberg microscope gedankenexperiment by Alden Mead \cite{Mead64,Mead66}, the idea did not catch on immediately. 

All of which goes to say that the intuition behind the minimum length concept had been present a long time before its rediscovery and subsequent popularization through a series of results in string theory \cite{Amati87,Gross87a,Gross87b,Amati88,Konishi89} reflecting its emergence in string scattering processes at very high energies. Since then, there has been evidence, connecting it to the low-energy regimes of quantum group theory \cite{Maggiore93c}, noncommutative geometry \cite{Battisti08c,Pramanik13}, loop quantum gravity \cite{Rovelli94,Hossain10,Majumder12a,Girelli12,Gorji15}, Ho\v{r}ava-Lifshitz gravity \cite{Myung09b,Myung09c,Eune10}, causal dynamical triangulations \cite{Coumbe15} and supersymmetry breaking \cite{Faizal16}. 

We stress, however, that it is not necessary to revert to specific approaches to lend support to the GUP. Instead, it suffices to consider combined insights from general relativity and quantum mechanics \cite{Mead64,Mead66,Padmanabhan87,Ng93,Maggiore93a,Amelino-Camelia94,Garay94,Adler99a,Scardigli99,Capozziello99,Camacho02,Calmet04,Ghosh10,Casadio13,Casadio15,El-Nabulsi20a} or invoke effective field theory \cite{Faizal17,Nenmeli21} to motivate it in a theory agnostic fashion. Thorough collections of all sorts of motivations can be found in the reviews \cite{Hossenfelder12,Tawfik14,Tawfik15c}.

\begin{figure}[ht!]
    \centering
    \includegraphics[width=\linewidth]{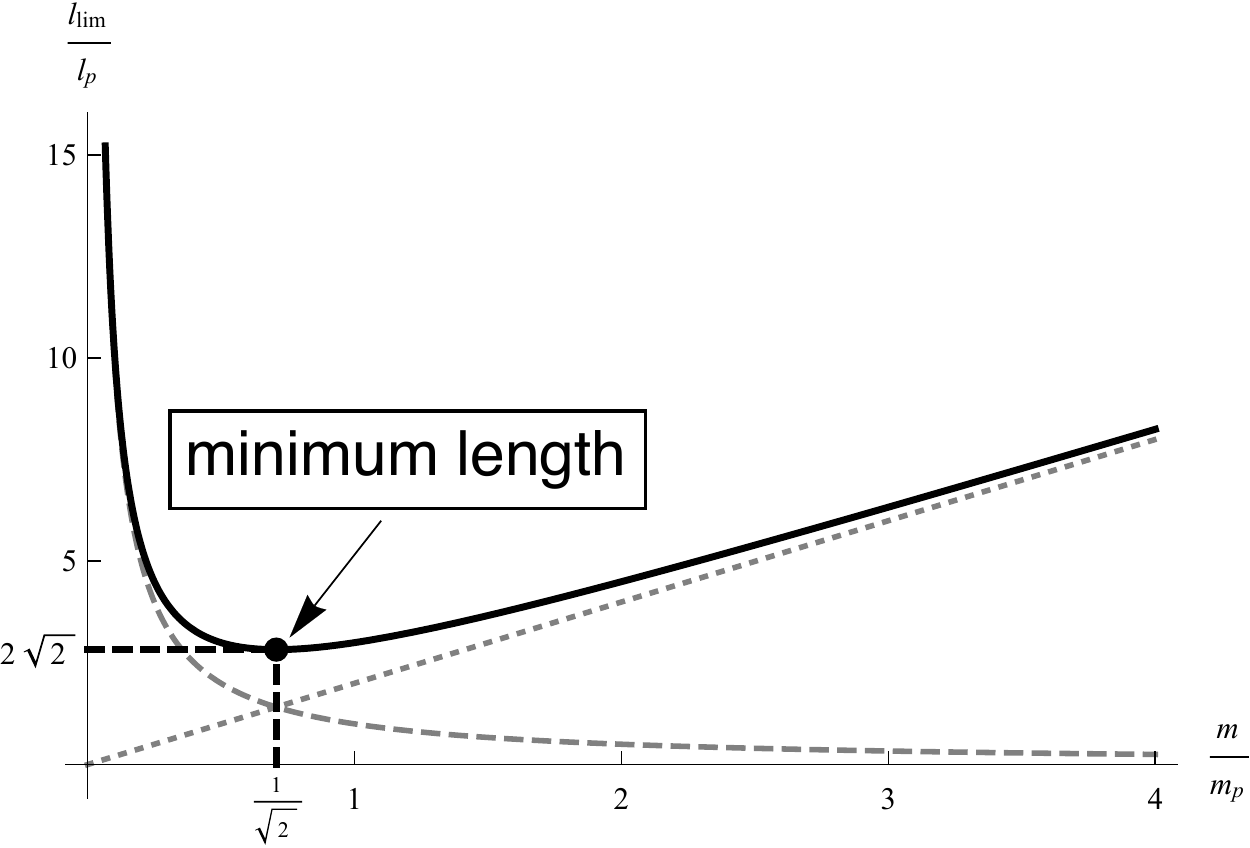}
    \caption{The reduced Compton wavelength $\lambdabar_C$ (dashed) and the Schwarzschild radius $r_S$ (dotted) as functions of the object's mass. The simple superposition of both, $l_{\text{lim}}$, understood as possible interpolation is displayed as black line and its minimum marked by a point.}
    \label{fig:minlength}
\end{figure}

For the purpose of this thesis, we will contend ourselves with a simple argument of scales put forward for example in \cite{Maggiore93a,Susskind04}: Both relativistic quantum mechanics and general relativity predict independent limits to the localizability of free particles of mass $m$ (we resort to a somewhat sloppy usage of the concepts of particle and mass here), the reduced Compton wavelength $\lambdabar_C=\hbar/m$ and the Schwarzschild radius $r_S=2Gm,$ respectively. Clearly, those length scales are inversely proportional to each other, thus governing different regimes. While the reduced Compton wavelength is dominant for particles of small mass, the limiting size of every day-objects is governed by their Schwarzschild radius. This implies that there has to be minimum in between, where both limits exactly equal each other. As expected, this occurs around the Planck scale, \ie $l_{\text{min}}\sim l_p.$ As an example for a generalized limiting length, the sum of the Schwarzschild radius and the reduced Compton length $l_{\text{lim}}=r_S+\lambdabar_C$ is compared to both isolated quantities in Fig. \ref{fig:minlength}. Thus, taking into account both general relativity and quantum mechanics, it is impossible to resolve distances smaller than this \emph{minimum length}. Evidently, this kind of argument cannot in and of itself fix its exact value, which presumably will be predicted from a fully-fledged theory of quantum gravity. Indeed, it should rather be understood as an order-of-magnitude estimate.

Not knowing its exact value, does not prevent theorists from trying to formalize the consequences of the idea in framework such as the GUP, Lorentz invariance violation (LIV) and deformed (or doubly) special relativity (DSR).

\subsection{A world of acronyms: GUP, LIV and DSR}\label{subsec:GUPLIVDSR}

In which way the minimum length is implemented is open for debate. The first step -- changing Einstein's famous dispersion relation $p^2=m^2,$ with the magnitude of the relativistic four-momentum $p$ and the rest mass $m$ -- is relatively uncontroversial. However, a modified dispersion relation alone implies a breaking of Lorentz invariance. Therefore, the resulting theories collected in Alan Kosteleck\'y's standard model extension (SME \cite{Colladay98}) are by far the most radical and, therefore, most constrained ones \cite{Kostelecky11,Addazi21}. 

On the other hand, the year 2000 marked a breakthrough when Giovanni Amelino-Camelia found a deformation of Einstein's special relativity which allowed for the inclusion of an invariant length scale \cite{Amelino-Camelia00}. In that vein, the Lorentz transformations themselves, \ie the algebra of Lorentz generators, are deformed such that they comply with the modified dispersion relation \cite{Magueijo01}, yielding the name DSR \cite{Amelino-Camelia02a,Amelino-Camelia02a,Amelino-Camelia02c,Kowalski-Glikman02a}. Interestingly, Snyder's model \cite{Snyder46,Snyder47} encountered above is just an instance of exactly this theory \cite{Kowalski-Glikman02b}, which may, in fact, also be directly derived from the solvable toy model of quantum gravity in three dimensions \cite{Freidel03}.

Irrespective of relativistic completion, the essence of the minimum length is formulated in terms of GUPs in nonrelativistic quantum mechanics \cite{Kempf94,Kempf96b}. In fact, Lorentz invariance violating as well as deforming theories immediately imply a modified dispersion relation and correspondingly, a GUP \cite{Hossenfelder05}. 

A short argument shows that the minimum length may not be the only fundamental impediment to measurements induced by gravity.  As the observable universe has an apparent horizon, there is a maximum conceivable length for causal connection of structures, characterized by the radius of the cosmological horizon $r_H.$ This can also be understood as a maximal wavelength and, therefore, with Louis de Broglie \cite{deBroglie24}, a \emph{minimal momentum}. Note, however, that this assumption does not require any input from quantum gravity and should therefore be derivable from semiclassical considerations. 

By analogy with the GUP, systems modified by the maximal wavelength are described by quantum mechanical theories exhibiting so-called extended uncertainty principles (EUPs) \cite{Mignemi09,Li08,Bolen04,Ghosh09,CostaFilho16}. Correspondingly, the synthesis of both approaches leads to generalized extended uncertainty principles (GEUPs) \cite{Bambi08a,Wang09,Stetsko12}.

Modifications to Heisenberg's relation shall not be the only subject of the present thesis, however. We would rather like to build a bridge between these and theories of quantum mechanics on curved background manifolds, in particular between curved momentum space and the GUP.

\subsection{Curved momentum space}\label{subsec:introcurvmom}

That geometry may depend not only on the position but also on the direction of motion is a very old idea, which is only gradually attracting attention within physics. The first record of it in mathematics dates back to Bernhard Riemann's habilitation dissertation \cite{Riemann68}. Later, it was mainly developed by Paul Finsler \cite{Finsler18} and \'Ellie Cartan \cite{Cartan34}. An overview of this topic, these days subsumed under the terms Lagrangian and Hamiltonian geometry, can be found in Refs. \cite{Miron01,Miron12}.

From the physical point of view, it was Max Born \cite{Born38,Born49} who observed that quantum mechanics, as well as Hamilton mechanics for that matter, is invariant under the exchange of positions $\hat{x}$ and momenta $\hat{p}$
\begin{equation}
    \hx\rightarrow \hp,\hspace{1cm}\hp\rightarrow -x.
\end{equation}
Owing to the term reciprocal lattice in the theory of condensed matter, these days this duality carries the name "Born reciprocity". According to his reasoning, it is the theory of general relativity which breaks this symmetry by curving space alone. Therefore, a successful unification of gravity and quantum mechanics had to involve curving momentum space, thereby restoring the duality. Note that this would also involve an invariance of the dynamics under diffeomorphims in momentum space. However, there have been recent claims that this symmetry is not realized in nature when considering active transformations \cite{Amelino-Camelia19,Gubitosi21}.

Born's line of reasoning was further developed mainly by Yuri Gol'fand \cite{Golfand59,Golfand62,Golfand63,Golfand71} and Igor Tamm \cite{Tamm65,Tamm72}. From the mathematical side the said endeavour lead to the theory of quantum groups \cite{Drinfeld88,Majid88,Majid90,Majid91,Freidel05}. Furthermore, the canonical quantization of theories on curved momentum space was treated in Refs. \cite{Batalin89a,Batalin89b,Bars10}. These efforts culminated in their recent application to quantum gravity phenomenology \cite{Amelino-Camelia11,Carmona19,Relancio20a} on the one hand. On the other hand, they paved the way for the construction of Born geometry \cite{Freidel13,Freidel14,Freidel15,Freidel17,Freidel18}, which captures all mathematical structures behind Hamiltonian mechanics (symplectic), quantum theory (complex) and general relativity (metric) at once.

How, then, are these ideas connected to the GUP? As all required concepts have been introduced, it is time to describe the aim of the present work.

\subsection{This thesis - from curved manifolds to modified uncertainty relations}

\begin{figure}[ht!]
    \centering
\includegraphics[scale=.74]{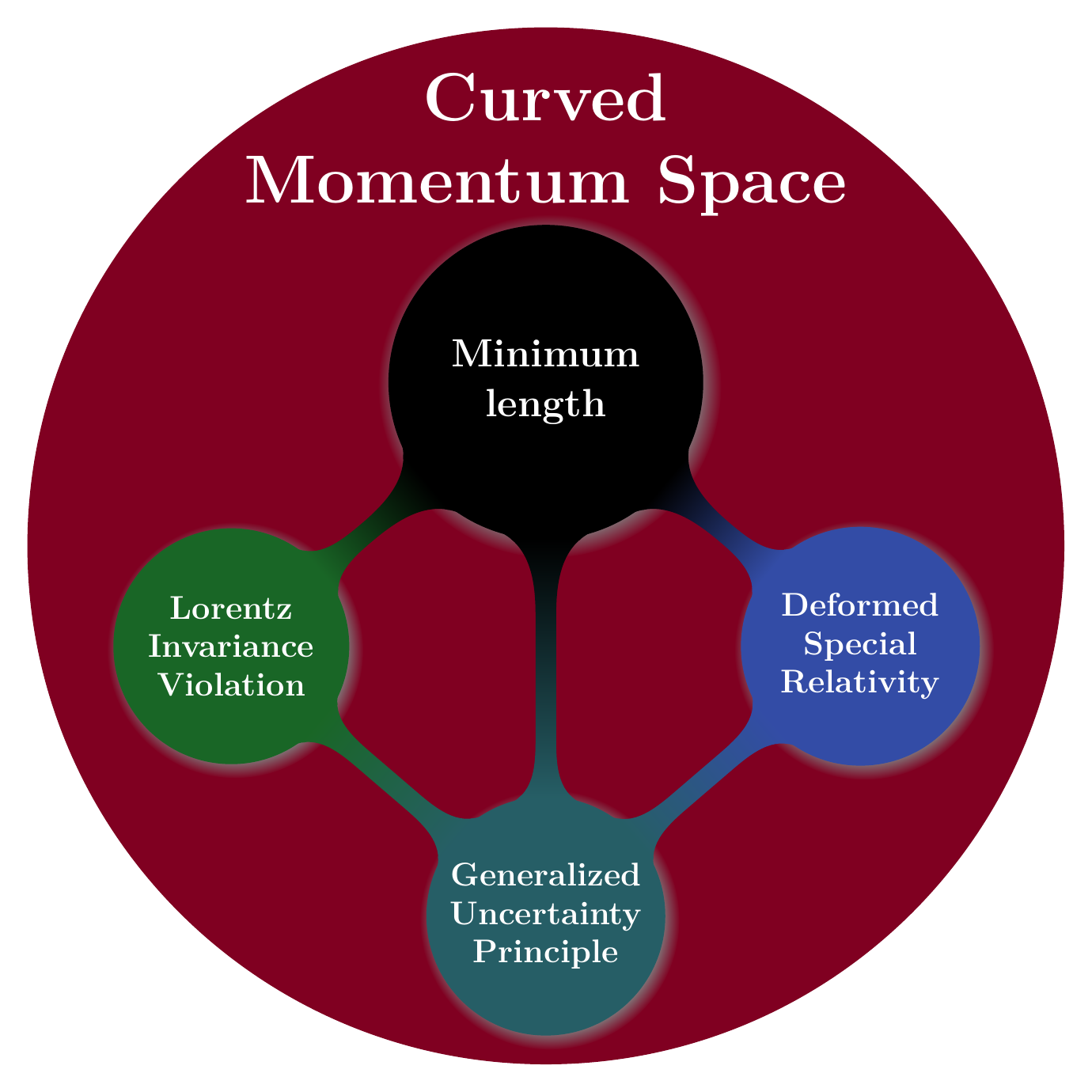}
    \caption{Curved momentum space as the connecting principle of the different expressions of the minimum length-concept.}
    \label{fig:mindmap2}
\end{figure}

It had been known early on that DSR could be expressed as a theory of a de Sitter-shaped momentum space \cite{Kowalski-Glikman02c,Kowalski-Glikman03}. Similarly, Finsler geometry (dual to curved momentum space) is often understood as a typical culprit of LIV.  The present thesis is intended to create an analogous link between theories of nontrivial momentum space and GUP-deformed quantum mechanics. We want to answer the question: can we establish curved momentum space as the overarching principle connecting all areas of quantum gravity phenomenology related to the minimum length (\cf Fig. \ref{fig:mindmap2})? 

In that vein, we will first introduce quantum mechanics on curved backgrounds in section \ref{sec:measures} in accordance with the approach invented by Bryce DeWitt \cite{DeWitt52}. Furthermore, the theory of GUPs and EUPs and the underlying deformations to quantum mechanics are reviewed in section \ref{sec:GUPsEUPs}. Along these lines, we also explain how practitioners usually extract phenomenology out of the said models. Correspondingly, we provide up-to-date constraints on the corresponding parameters, \ie the minimum length, and display a the typical example of this phenomenological reasoning, which the author presented during his studies in the context of a simple model of the deuteron. Note that the said section as well as parts of this very introduction may be understood as a fairly comprehensive review of the literature on GUPs and EUPs that came up during the PhD studies and therefore has, in the author's view, developed its own encyclopaedic value as a matter of convenience for other researchers in the field. This explains the extraordinarily large number of references appearing in both.

In section \ref{sec:3DEUP} we show that an extended uncertainty relation can be derived from a curved three-dimensional background space alone without prior deformation of the canonical commutation relations. The reasoning behind this result includes bounding the studied Hilbert space to a compact domain, specifically to a geodesic ball. Correspondingly, the ball's radius serves as a measure of position uncertainty. Then, the standard deviation of the momentum operator develops a lower bound, which is dependent on exactly that radius, yielding the desired uncertainty relation. This idea is generalized to nonrelativistic particles in curved spacetime in section \ref{sec:4DEUP}, thereby relating the result directly to gravity. Section \ref{sec:relgen} is intended to further extend the obtained inequality to relativistic quantum probes. Thus, we obtain a relativistic EUP induced by semiclassical gravity, exactly as mentioned above. Note that, by the Born reciprocal property of quantum mechanics, these findings can be immediately translated into the language of curved momentum space and GUPs. As the relation itself is quite involved, we apply it to several important spacetimes -- accelerated observers, the cosmological horizon and rotating universes and massive bodies or black holes in general relativity as well as quadratic and infinite-derivative gravity -- in section \ref{sec:applications}.

This link is understood as motivation for further investigation not only of the uncertainty relations but of the underlying algebra of observables. Hence, in section \ref{sec:modcomorcurspa} we find a direct map from theories involving GUP-deformed commutation relations to quantum mechanics on nontrivial momentum space -- the most important result of the present thesis. Accordingly,  to second order in the Planck length the curvature in the momentum space underlying the dual description is proportional to the noncommutativity of the coordinates in the original one. Therefore, we use bounds on the latter to constrain the former.  This does not imply, however, that a commutative set of coordinates leads to a trivial result. On the contrary,  as long as the original theory predicts a GUP, the basis in momentum space deviates from the canonical one in a nonlinear way,  making it possible to compare to existing data on the minimum length.  As an aside, the given map defines conjugate variables for any general GUP, enabling new kinds of analysis like the path integral approach. 

However, along these lines we could only make use of Cartesian coordinates. There was just no formalism describing quantum mechanics on backgrounds which are described by a position- as well as momentum-dependent metric as it may appear after a general coordinate transformation (such as going to spherical coordinates). However, the results of the preceding sections imply that exactly this setting indicates the realm of all kinds of GEUP-physics -- quantum mechanics on the curved cotangent bundle. The first step into the direction of such a theory is taken in section \ref{sec:qmhamilton} by direct generalization of DeWitt's approach. In particular, we promote the Hilbert space measure to an operator. It is merged with the wave function to construct wave densities by analogy with the geometric approach to quantization \cite{Woodhouse97}. After finding the position and momentum representations of the important operators in quantum mechanics -- the position and momentum operators as well as the geodesic distance and the Hamiltonian of a single particle -- with an arbitrary kind of metric and showing the consistency of the formalism, we apply it to central potentials described in a Riemann normal coordinate-like expansion. In particular, we deal with the hydrogenic atom and the isotropic harmonic oscillator. Choosing a suitable operator ordering, the latter in fact becomes an instance of exact Born reciprocity on the curved cotangent bundle. 

\subsection{Conventions and notation}

Indices from the Greek alphabet represent the $d+1$ spacetime dimensions, while Latin letters indicate the $d$ spatial ones. Furthermore, we commit to the mostly positive signature of the spacetime metric $(-,+,+,+)$ and implicitly express the calculations in units, in which $c=1,$ while $G$ and $\hbar$ are retained. 

The Christoffel symbols of the Levi-Civita connection, derived from the metric $g_{\mu\nu}$ and its inverse $g^{\mu\nu},$ read
\begin{equation}
    \Gamma^\lambda_{\mu\nu}\equiv g^{\lambda\sigma}\left(\partial_\mu g_{\sigma\nu}+\partial_{\nu}g_{\mu\sigma}-\partial_\sigma g_{\mu\nu}\right).
\end{equation}
To express the Riemann curvature tensor, we adopt the convention
\begin{equation}
    R_{\rho\mu\nu}^{~~~\lambda}\equiv\partial_\rho\Gamma_{\mu\nu^\lambda}-\partial_\mu\Gamma_{\rho\nu^\lambda}+\Gamma_{\rho\sigma}^\lambda\Gamma_{\mu\nu}^\sigma-\Gamma_{\mu\sigma}^\lambda\Gamma_{\rho\nu}^\sigma.
\end{equation}
Moreover the Ricci tensor and scalar are defined as
\begin{align}
    R_{\mu\nu}&\equiv R^\lambda_{~\mu\lambda\nu},\\
    R&\equiv g^{\mu\nu}R_{\mu\nu}.
\end{align}
Symmetrization and antisymmetrization of two tensorial indices, say of a tensor $T_{\mu\nu}$ are defined as
\begin{align}
    T_{(\mu\nu)}\equiv&\frac{1}{2}\left(T_{\mu\nu}+T_{\nu\mu}\right),\\
    T_{[\mu\nu]}\equiv&\frac{1}{2}\left(T_{\mu\nu}-T_{\nu\mu}\right).
\end{align}
If a tensor is symmetrized over two indices which are, notation-wise not next to each other, vertical bars indicate the indices, which are not (anti-)symmetrized over, for example for a tensor $T_{\mu\nu\rho}$
\begin{equation}
    T_{(\mu|\nu|\rho)}=\frac{1}{2}\left(T_{\mu\nu\rho}+T_{\rho\nu\mu}\right).
\end{equation}
Finally, hats on quantities describe quantum mechanical operators.

\section{Quantum mechanics on curved manifolds}\label{sec:measures}

The formulation of nonrelativistic single-particle quantum mechanics championed in the average undergraduate course reveals several deficits under consideration of deviations from the standard description of flat space in Cartesian coordinates. In principle, the usage of curvilinear, \eg spherical, coordinates suffices to create enormous problems, which in this framework can only be solved in an ad-hoc fashion. However, following an approach invented by Bryce DeWitt \cite{DeWitt52}, it is possible to provide a consistent description of quantum mechanics in general coordinates and on curved backgrounds.

\subsection{Hilbert space}\label{subsec:hilbert}

At the kinematical level the most basic ingredient for any quantum theory, be it a single-particle or field description, is constituted by the space of allowed states, the Hilbert space $\hil.$ Thus, a general state describing a physical system, say $\ket{\psi}$ has to be an element of $\hil.$ There have been a great many studies written highlighting the complexity those possibly infinite-dimensional spaces can accommodate for. For the purpose of this section,  though, we restrict ourselves to the relatively simple examples appearing in the context of single particles in nonrelativistic quantum mechanics. In particular, we are describing them in their position and momentum space representations even though the gained conclusions hold for generic continuous observables. 

As measurable quantities are generally represented by self-adjoint operators, their eigenstates, if properly normalized, furnish an orthonormal basis in terms of which the entirety of states in $\hil$ can be expressed. Exactly how this is accomplished for the eigenstates of the position operator $\ket{x}$ is summarised in the corresponding measure of the scalar product, say $\D\mu.$ In ordinary quantum mechanics on flat space given in the position representation in terms of Cartesian coordinates $x^a,$ the framework usually taught in undergraduate courses, the measure trivially reads $\D\mu=\D^d x,$  where $d$ stands for the number of dimensions. Yet, not only is this generally not the case, the measure in fact constitutes a defining feature of the usually considered representation of the single-particle Hilbert space. In that vein, the latter is usually chosen to be the space of square-integrable functions on a given domain $\mathcal{D}$ with scalar product measure $\D^d \mu,$ in short $\hil=L^2(\mathcal{D},\D\mu).$ In the aforementioned case, those ingredients are $\mathcal{D}=\reals^d$ and $\D\mu=\D^d x.$

Given a Hilbert space $\hil$ and two normalized states contained in it $\ket{\psi}\in\hil$ and its dual $\bra{\phi}\in\hil^*$ the scalar product satisfies
\begin{equation}
    \Braket*{\phi}{\psi}=\Braket*{\psi}{\phi}^*,\label{Hscalarprod}
\end{equation}
where the superscript $*$ implies complex conjugation. Clearly, there is no measure entering here. Thus, the measure reflects the representation that is suited to tackle the specific system at hand, \ie the relevant observables.

Assume as given a self-adjoint operator acting on the states of $\hil$ with continuous spectrum, denoted $\hat{O},$ its eigenstates $\ket{O}$ and eigenvalues $O.$ Then, those eigenstates furnish an orthonormal basis, \ie
\begin{align}
    \id =&\int\D\mu(O)\ket{O}\bra{O},\label{normality}\\
    \ket{O}=&\int\D O'\delta (O'-O)\ket{O'},\label{orthogonality}
\end{align}
with Dirac's delta distribution $\delta(x).$ Note, that this is where the measure appears first, meaning that it is conditional on the observable which is chosen to represent the system. According to Eq. \eqref{normality}, every state $\ket{\psi}\in\hil$ can be expanded as
\begin{align}
    \ket{\psi}=&\int\D\mu\Braket*{O}{\psi}\ket{O}\\
    \equiv &\int\D\mu\psi(O)\ket{O},
\end{align}
where the last equality defines the wave function $\psi(O).$ Thus, the scalar product \eqref{Hscalarprod} can be represented as
\begin{equation}
    \Braket*{\phi}{\psi}=\int\D\mu\phi^*\psi.
\end{equation}
Apart from it being self-adjoint and having a continuous spectrum, no assumption has been made about the observable in question. The best-known examples for this description are clearly the position and momentum operators denoted $\hat{x}^i$ and $\hat{p}_i$ respectively. Those are explicitly dealt with in the following subsection.

\subsection{Complementary observables - position and momentum}\label{subsec:complobs}

In quantum mechanics, observables which obey a non-abelian algebra, \ie which are noncommuting, yield a complementary description of the treated system. For example, we can deal with a problem either in the momentum or the position space representation. In the present section we choose to use the position basis, even though the results also hold in the momentum basis. Furthermore, we assume that there is a (possibly trivial) Riemannian background metric $g_{ij}(x),$ which can be used to raise and lower indices and contract expressions as usual.

In $d$ spatial dimensions position and momentum are represented by vector operators, thus requiring $d$-dimensional measures. The measure appearing in the definition of the position operator is proportional to $\D^dx$ such that it can be expressed as
\begin{equation}
    \D\mu=\mu(x)\D^d x,
\end{equation}
where we introduced the real positive function $\mu(x).$ The representation for the position operator $\hat{x}^i$ follows suit:
\begin{equation}
    \hat{x}^i=\int\D^dx\mu(x)x^i\ket{x}\bra{x},\label{genposop}
\end{equation}
with its eigenvalues $x^i$ and eigenstates $\ket{x}$ which satisfy
\begin{align}
    \id=&\int\D^dx\mu(x)\ket{x}\bra{x},\label{genposid}\\
    \Braket*{x'}{x}=&\frac{\delta^d(x-x')}{\mu(x)},\label{genpostransamp}
\end{align}
in accordance with Eqs. \eqref{normality} and \eqref{orthogonality}. 

The canonical momentum operator $\hat{\pi}_i$, on the other hand, being a complementary observable to the position on an infinite dimensional Hilbert space, has to be represented by a derivative operator.  As observables yield real eigenvalues, they have to be symmetric. Equally so, the momentum operator should satisfy
\begin{equation}
    \Braket*{\phi}{\hat{\pi}_i\psi}=\Braket*{\hat{\pi}_i\phi}{\psi}.
\end{equation}
Furthermore, the operators classically spanning phase space should obey the Heisenberg algebra well-known from textbook quantum mechanics \cite{Dirac30,Messiah99}
\begin{align}
    [\hat{x}^i,\hat{x}^j]=0,\hspace{1cm}[\hat{\pi}_i,\hat{\pi}_j]=0,\hspace{1cm}[\hat{x}^i,\hat{\pi}_j]=i\hbar\delta^i_j.\label{HeisAlg}
\end{align}
Up to a position-dependent one-form, which will be neglected throughout this thesis, the only operator satisfying those requirements acts on wave functions as
\begin{equation}
    \hat{\pi}_i\psi=-i\hbar \frac{1}{\sqrt{\mu}}\partial_i\left(\sqrt{\mu}\psi\right)\psi.\label{csmomop}
\end{equation}
The context within which it arises is provided in the subsequent section.

\subsection{Curved spaces and curvilinear coordinates}\label{subsec:curved}

Coordinate changes are represented by unitary transformations on Hilbert space. In other words, the scalar product of two states $\Braket*{\phi}{\psi}$ is independent of the applied system of coordinates. To respect this invariance without further complication, every integration done in this context has to contain the volume form derived from the background metric $g_{ij}(x).$ Correspondingly, the measure equals
\begin{equation}
    \D\mu=\sqrt{g(x)}\D^dx,\label{curvedspacemeas}
\end{equation}
where $g=\det g_{ij}.$ Hence, nontrivial measures appear naturally when considering systems beyond flat space and Cartesian coordinates (which imply $g=1$). The position operator can then be defined in accordance with Eq. \eqref{genposop} 
\begin{equation}
    \hat{x}^i=\int\D^dx\sqrt{g}x^i\ket{x}\bra{x}.\label{curvposop}
\end{equation}
As stated before (\cf Eqs. \eqref{genposid} and \eqref{genpostransamp}), its eigenvalues $x^i$ and eigenstates $\ket{x}$ obey the relations
\begin{align}
    \id=&\int\D^dx\sqrt{g}\ket{x}\bra{x},\label{curvposid}\\
    \Braket*{x'}{x}=&\frac{\delta^d(x-x')}{\sqrt{g}}.\label{curvpostransamp}
\end{align}

Then, the momentum operator acts on wave functions like a covariant derivative on scalar densities of weight $1/2$ (\cf section \ref{sec:qmhamilton}), \ie \cite{DeWitt52}
\begin{equation}
    \hat{\pi}_i\psi=-i\hbar g^{-1/4}(x)\partial_i\left[g^{1/4}(x)\psi\right]=-i\hbar \left[\partial_i+\frac{1}{2}\Gamma^j_{ij}(x)\right]\psi=-i\hbar\nabla_i\psi,\label{curvedspacemom}
\end{equation}
where $\Gamma^k_{ij}$ denotes the Christoffel symbol and $\nabla_i$ the covariant derivative. Alternatively, if we had started in curved momentum space, we could have performed essentially the same derivation, yielding
\begin{equation}
    \hat{x}^i\tilde{\psi}=i\hbar g^{-1/4}(p)\dot{\partial}^i\left[g^{1/4}(p)\tilde{\psi}\right]=i\hbar \left[\dot{\partial}_i+\frac{1}{2}C^{ij}_{j}\left(p\right)\right]\tilde{\psi}=i\hbar\dot{\nabla}^i\tilde{\psi},\label{curvedmomspaceposop}
\end{equation}
with the connection coefficients $C^{ij}_k$ and the covariant derivative $\dot{\nabla}^i$ in momentum space. However, within the present approach it is unclear, how to describe quantum mechanics on a curved position space in the momentum space representation and vice versa. This will be dealt with in section \ref{sec:qmhamilton}.

Not only the momentum operator, but also its square will be modified on curved backgrounds. How exactly this is done is explained in the next subsection.

\subsection{The free particle}

As the background enters with a nontrivial scalar product measure, it should be expected that the square of the momentum operator, which enters the free-particle Hamiltonian, needs to be modified in similar way. Classically, this function is expressed as
\begin{equation}
    H_{fp}=\frac{1}{2m}g^{ij}(x)\pi_i\pi_j,
\end{equation}
with the mass of the particle $m.$ Clearly, when quantized, this expression, depending on both positions and momenta, harbours an ordering ambiguity. This kind of problem will reappear more prominently in section \ref{subsubsec:opord}. Fortunately though, there is theoretical guidance in this case. Naturally, being the square of the momentum, the representation of this operator should contain two derivatives. Furthermore, it should be symmetric with respect to the measure $\D\mu,$ which implies that it is a scalar quantity. Then, it can only be represented by the Laplace-Beltrami operator
\begin{equation}
    \hat{\pi}^2\psi=-\frac{\hbar^2}{\sqrt{g(x)}}\partial_i\left[\sqrt{g(x)}g^{ij}(x)\partial_j\psi\right]\equiv -\hbar^2\Delta\psi.\label{Laplace-Beltrami}
\end{equation}
Yet, this only accounts for the canonical momentum. The physical one may differ as can be gathered from section \ref{subsubsec:physmomop}. Analogously, the position operator on nontrivial momentum space reads
\begin{equation}
    \hat{x}^2\tilde{\psi}=-\frac{\hbar^2}{\sqrt{g(p)}}\dot{\partial}^i\left[\sqrt{g(p)}g_{ij}(p)\dot{\partial}^j\tilde{\psi}\right]\equiv -\hbar^2\dot{\Delta}\tilde{\psi}.\label{xsqmomrep}
\end{equation}
However, the combination of quantum mechanics and differential geometry harbours some more complications, that need to be dealt with.

\subsection{Vector operators and geometric calculus}\label{subsec:vecop}

Having defined the free-particle Hamiltonian and the momentum operator, all appears to be set to start tackling problems. Yet, the definition of the momentum operator given in Eq. \eqref{curvedspacemom}, though formally correct, conceals a subtlety: Its expectation value, being an integral over a vector, is mathematically not well defined. In particular, we could describe the momentum in two distinct coordinate systems $x^i$ and $y^j,$ expressing the components of a general one-form $\omega_i$ of the former in terms of the latter as
\begin{equation}
    \omega_i=\frac{\partial y^j}{\partial x^i}\omega_j. 
\end{equation}
Then, the expectation value of the conjugated momentum operator in the coordinate system $x^i$ with respect to a general state $\ket{\psi}$ would read
\begin{equation}
    \braket*{\hat{\pi}_i}=\braket*{\frac{\partial y^j}{\partial x^i}\hat{\pi}_j}=\int\D^dx\sqrt{g}\psi^*\left(-i\frac{\partial y^j}{\partial x^i}\nabla_j\right)\psi\neq \frac{\partial y^j}{\partial x^i}\int \D^dx\sqrt{g}\psi^*(-i\nabla_i)\psi,
\end{equation}
where the transformation matrix $\partial y^a/\partial x^i,$ being position dependent, cannot be taken out of the integral. Thus, the expectation value is not diffeomorphism invariant, or
\begin{equation}
    \braket*{\hat{\pi}_i}\neq \frac{\partial y^j}{\partial x^i}\braket*{\hat{\pi}_j}.
\end{equation}
This problem can be circumvented with the help of geometric calculus \cite{Hestenes84}. Expressed in this language, one-forms are expanded in terms of basis vectors $\gamma^i(x),$ which satisfy the generalized Clifford algebra
\begin{equation}
    \{\gamma^i,\gamma^j\}=2g^{ij},\label{Clifford}
\end{equation}
where the curly brackets stand for the anticommutator. These basis vectors can be made independent of the position using the tetrad formalism \cite{Einstein28}. Define the \emph{vielbein} $e^i_a$ such that
\begin{equation}
    g_{ij}\equiv e_i^ae_j^b\delta_{ab}.\label{vielbein1}
\end{equation}
Further denoting its inverse as $e^i_a\equiv(e^a_i)^{-1}$ yields
\begin{equation}
    g^{ij}\equiv e^i_ae^j_b\delta^{ab}.\label{vielbein2}
\end{equation}
Then, according to Eq. \eqref{Clifford}, one can choose a basis such that $\gamma^a=e^a_i\gamma^i\neq\gamma^a(x)$ and
\begin{equation}
    \{\gamma^a,\gamma^b\}=2\delta^{ab}.\label{normalClifford}
\end{equation}
Applying all of this machinery, a one-form $\omega$ can be expressed as $\omega=\gamma^i(x)\omega_i=\gamma^ae_a^i\omega_i.$ Thus, using slash notation familiar from the Dirac equation, we can define the momentum operator \cite{Pavsic01}
\begin{equation}
    \hatslashed{p}\psi\equiv\gamma^i\hat{p}_i\psi=-i\hbar\gamma^i (x)\nabla_i \psi\label{newmomop}
\end{equation}
whose expectation value reads
\begin{equation}
    \braket*{\hatslashed{p}}=\int\D^4x\sqrt{g}\psi^*\left(-i\hbar\gamma^ae_a^i\nabla_i\psi\right)=\gamma^a\int\D^4x\sqrt{g}\psi^*\left(-i\hbar e_a^i\nabla_i\psi\right)=\gamma^a\braket*{e^i_a\hat{p}_i}.
\end{equation}
Here we could take the basis vector out of the integral because, as alluded to above, it is independent of the positions. Thus, it suffices to add in the \emph{vielbein}, \ie describe the system in a local Euclidean frame, to turn the expectation value of the momentum operator into a well-defined object. In ref. \cite{Pavsic01} the operator $\hatslashed{p}$ is shown to be self-adjoint on curved spaces and to generate translations.  Furthermore, it is proven that its square is proportional to the Laplace-Beltrami operator
\begin{equation}
    \hatslashed{p}^2\psi=\hat{p}^2\psi,\label{squaredmom}
\end{equation}
thereby claiming the correct relation to the free particle Hamiltonian. Thus, it fulfils all the requirements expected from the position space representation of the momentum operator in curved space.

Having, thus,  introduced all relevant operators in the curved context, the stage is set to perform calculations within the setting of quantum mechanics on general manifolds. However, this framework is just one side of the considerations making up the present thesis.  The other perspective is the matter of the subsequent section.

\section{Generalized and extended uncertainty relations}\label{sec:GUPsEUPs}

As mentioned in sections \ref{subsec:minlen} and \ref{subsec:GUPLIVDSR}, the minimal length and minimal momentum concepts can be introduced into quantum mechanics by virtue of GUPs and EUPs, respectively. The present section is intended to provide a more profound look into the theory underlying these ideas. Furthermore, we show how phenomenological predictions can be made under the assumption of GUP- and EUP-like deformations and display current constraints on the relevant parameters. Correspondingly, this should be understood as a short review of the field. As an example, we derive the GUP-corrections to the radius of the deuteron.

\subsection{Theory}\label{subsec:theo}

In quantum mechanics fundamental measurement uncertainties of complementary observables are generally linked. This was first formulated by Heisenberg through his celebrated uncertainty relation for positions and momenta \cite{Heisenberg27}
\begin{equation}
    \Delta x\Delta p\sim \hbar,\label{origheisunc}
\end{equation}
where $\Delta x$ and $\Delta p$ stand for the corresponding measures of uncertainty, respectively. An instance of this behaviour was derived by Robertson \cite{Robertson29} and further strengthened by Schrödinger \cite{Schroedinger30}, making use of standard deviations (defined here with respect to a general operator $\hat{O}$)
\begin{equation}
    \sigma_O\equiv\sqrt{\braket*{\hat{O}^2}-\braket*{\hat{O}}^2}
\end{equation}
as measures of uncertainty. Given two symmetric operators $\hat{A}$ and $\hat{B},$  Robertson proved that
\begin{equation}
    \sigma_A\sigma_B\geq \frac{1}{2}\left|\braket*{\left[\hat{A},\hat{B}\right]}\right|\label{Robertson}.
\end{equation}
For the sake of simplicity, we first consider a particle in one dimension to explain how the minimum length and similar impediments to measurement arise. Then, we generalize the approach to $d$ dimensions to be able to compare it to the real world. 

\subsubsection{Emergence of limiting scales in one dimension}\label{subsubsec:GUPEUP1D}

In this simple case, the canonical commutation relations imply that
\begin{equation}
    \sigma_{x}\sigma_{p}\geq\hbar/2.\label{HUP}
\end{equation} 
Under the assumption that they are modified as
\begin{equation}
    \left[\hat{x},\hat{p}\right]=i\hbar f\left(\hat{x},\hat{p}\right),
\end{equation}
where $f$ denotes a general nonsingular function of the position and momentum operators, the uncertainty relation is altered according to Eq. \eqref{Robertson}. This results in the inequality
\begin{equation}
    \sigma_{x}\sigma_{p}\geq\frac{1}{2}\left|\braket*{f}\right|.\label{formalGEUP}
\end{equation}
 In the literature the term GUP implies a momentum-dependent function $f$ \cite{Maggiore93a,Scardigli99,Maggiore93c,Maggiore93b}, while the EUP \cite{Park07,Zhu08,Mignemi09} and the GEUP \cite{Bambi08a,Izadparast20} require a dependence on positions and both positions and momenta, respectively.
 
For GUPs the inequality \eqref{formalGEUP} can be rewritten as
\begin{equation}
    \sigma_{x}\geq\frac{1}{2}\frac{\left|\braket*{f}\right|\left(\braket{\hat{p}},\braket{\hat{p}^2}\right)}{\sigma_{p}}=\frac{1}{2}\frac{\left|\braket*{f}\right|\left(\braket{\hat{p}},\sigma_p\right)}{\sigma_{p}}.
\end{equation}
If the right-hand side of this inequality has a minimum greater than zero, this implies that the corresponding quantum theory contains a minimum length. A different choice of $f$ in this context may lead to a maximal momentum. By analogy, EUPs imply relations of the form
\begin{equation}
    \sigma_p\geq\frac{1}{2}\frac{\left|\braket*{f}\right|\left(\braket{\hat{x}},\sigma_x\right)}{\sigma_x},
\end{equation}
which may yield a minimal momentum or a maximum length. Finally, GEUPs are supposed to combine both effects.

Phenomenologically, the function $f$ can be expanded in terms of the involved fundamental length scales.  Following the motivation in section \ref{subsec:minlen}, the ratio of the Schwarzschild radius and the reduced Compton wave length is quadratic in the mass. For this and other theoretical reasons, the most often applied approach to the GUP in the literature contains quadratic corrections to the Heisenberg algebra \cite{Maggiore93c,Kempf94,Kempf96a,Benczik02,Mignemi09,Ghosh09,Das12,CostaFilho16}
\begin{equation}
    \left[\hat{x},\hat{p}\right]=i\hbar\left[1+\beta\left(\frac{ l_p\hat{p}}{\hbar}\right)^2\right],\label{GUPalg1D}
\end{equation}
where $\beta$ denotes a dimensionless parameter. Depending on the sign of $\beta,$ this algebra implies a minimum length $\sigma_x\geq 2\sqrt{\beta} l_p$ or a maximal momentum $\sigma_p\leq m_p/\sqrt{-\beta}.$ Hence, the newly introduced parameter determines the size of the minimum length or maximum momentum in units of the Planck length or Planck mass, respectively, and is expected to be of order one to be in accordance with section \ref{subsec:minlen}.

Note that approaches yielding linear corrections to the uncertainty relations also exist \cite{Bambi08b,Nozari12b,Bosso18a,Bosso18b,Vagenas18}. Furthermore, a couple of nonperturbative relations have been proposed \cite{Maggiore93c,Pedram11c,Pedram10b,Pedram11d,Shababi17,Perivolaropoulos17,Chung18,Chung19a,Chung19d,Skara19,Hamil20a,Petruzziello20a,Farahani20,Fadel21}. As linear relations are not motivated as well as the quadratic ones, and nonperturbative completions will not change the observational outcome in testable regimes, we only deal with quadratic modifications to the Heisenberg algebra throughout this thesis.

Similarly to the GUP, the EUP is supposed to contain at most second-order corrections, yielding
\begin{equation}
    \left[\hat{x},\hat{p}\right]=i\hbar\left[1+\alpha\left(\frac{ \hat{x}}{r_H}\right)^2\right],\label{EUPalg1D}
\end{equation}
where the parameter $\alpha$ should in principle be determinable by semiclassical gravity. Again, depending on its sign, this leads to the appearance of a minimal momentum $\sigma_p\geq 2\sqrt{\alpha}\hbar/r_H$ or a maximum length $\sigma_x\geq r_H/\sqrt{-\alpha}.$ Clearly, the treatment of generalized and extended uncertainty relations is entirely analogous. 

As they are understood as corrections in a Taylor expansion, both approaches can be combined by simple superposition
\begin{equation}
    \left[\hat{x},\hat{p}\right]=i\hbar\left[1+\alpha\left(\frac{ \hat{x}}{r_H}\right)^2+\beta\left(\frac{ l_p\hat{p}}{\hbar}\right)^2\right].\label{GEUPalg1D}
\end{equation}
This leads to a modification of the uncertainty relation, reading
\begin{align}
    \sigma_x\sigma_p\geq& \frac{\hbar}{2}\left(1+\alpha\frac{\sigma_x^2-\braket{\hat{x}}^2}{r_H^2}+\beta l_p^2\frac{\sigma_p^2-\braket{\hat{p}}^2}{\hbar^2}\right)\\
    \geq&\frac{\hbar}{2}\left[1+\alpha\left(\frac{\sigma_x}{r_H}\right)^2+\beta\left(l_p\frac{\sigma_p}{\hbar}\right)^2\right],
\end{align}
where the last equality holds for positive $\alpha$ and $\beta.$ Such a synthesis of the two approaches allows for the restoration of the symmetry between position and momentum uncertainties $\beta l_p^2\sigma_p^2/\hbar^2\leftrightarrow\alpha\sigma_p^2/r_H^2$ \cite{Bambi08a,Mignemi15}. Furthermore, it shows a dual behaviour relating small and large momenta and distances according to the transformations $\sigma_x\leftrightarrow r_H^2/\alpha\sigma_x$ and $\sigma_p\leftrightarrow\hbar^2/\beta l_p^2\sigma_p$ in the pure EUP and GUP sectors, respectively. This is reminiscent of T-Duality in string theory \cite{Sathiapalan86}.

To put it in a nutshell, modifying the Heisenberg algebra according to Eqs. \eqref{GUPalg1D} and \eqref{EUPalg1D} incorporates minimum and maximum length effects as well as T-duality-like symmetries into quantum mechanics. Those effects can be combined by superposition of the corrections as in Eq. \eqref{GEUPalg1D}.

\subsubsection{Modified commutators in $d$ dimensions}\label{subsubsec:GUPEUPdD}

Evidently, the world we live in is not one-dimensional. Indubitably, a real theory of modified uncertainty relations should comply with this fact. The present section is aimed at exactly this generalization. Note that the background is assumed to be flat and described in terms of Cartesian coordinates throughout this section as it is usually done in the literature.

When generalized to $d$ dimensions, modifications of the Heisenberg algebra are of the form
\begin{subequations}
\label{GenAlgdD}
\begin{align}
    \left[\hat{x}^a,\hat{x}^b\right]=&i\hbar\theta^{ab}\left(\hat{x},\hat{p}\right),\\
    \left[\hat{p}_a,\hat{p}_b\right]=&i\hbar\tilde{\theta}_{ab}\left(\hat{x},\hat{p}\right),\\
    \left[\hat{x}^a,\hat{p}_b\right]=&i\hbar f_b^a\left(\hat{x},\hat{p}\right),
\end{align}
\end{subequations}
where the functions on the right-hand sides are related through the Jacobi identities
\begin{align}
    2\left[f^a_{[b},p_{c]}\right]=&\left[x^a,\tilde{\theta}_{bc}\right]\\
    2\left[f_a^{[b},x^{c]}\right]=&\left[p_a,\theta^{bc}\right].\label{jacobispacenoncom}
\end{align}
For completeness, it was noticed fairly recently  \cite{Fadel21} that there are more Jacobi identities to be satisfied, which may become nontrivial when considering spin:
\begin{equation}
	\left[\theta^{[ab},\hx^{c]}\right]=\left[\tilde{\theta}_{[ab},\hp_{c]}\right]=0.
\end{equation} 
While this is an interesting fact, it will not be considered further throughout this thesis. 

As the treatment of the EUP is analogous, we will only deal with GUPs for the remainder of this section. The phenomenological version of the algebra then reads \cite{Kempf96a,Benczik02,Sprenger12}
\begin{subequations}
\label{GUPalgdD}
\begin{align}
    \left[\hat{x}^a,\hat{x}^b\right]\simeq&\frac{il_p^2}{\hbar}\left(2\beta-\beta'\right)\hat{J}^{ba},\\
    \left[\hat{p}_a,\hat{p}_b\right]=&0,\\
    \left[\hat{x}^a,\hat{p}_b\right]=&i\hbar\left\{\delta^a_b\left[1+\beta\left(\frac{l_p\hat{p}}{\hbar}\right)^2\right]+\beta'\left(\frac{l_p}{\hbar}\right)^2\hat{p}_a\hat{p}^b\right\},
\end{align}
\end{subequations}
where we purposefully neglected higher-order contributions in $l_p^2\hat{p}^2/\hbar^2$ and introduced the deformed angular momentum operator $\hat{J}^{ab}=2\hat{x}^{[a}\hat{p}^{b]}$ and the squared momentum operator $\hat{p}^2=\hat{p}_a\hat{p}^a.$ In momentum space the observables obeying the said commutation relations can be represented as \cite{Kempf96a,Tawfik14}
\begin{subequations}
\label{GUPrepdD}
\begin{align}
    \hat{p}_a\ket{p}=&p_a\ket{p},\\
    \hat{x}^a\ket{p}=&i\hbar\left[\delta^a_{b}+\left(\frac{l_p}{\hbar}\right)^2\left(\beta p^2\delta^a_b+\beta'p_bp^a\right)\right]\dot{\partial}^b\ket{p}.
\end{align}
\end{subequations}
However, the position operator can only be symmetric, \ie obey $\Braket{\psi}{\hat{x}^i\phi}=\Braket{\hat{x}^i\psi}{\phi}$ if the volume measure in momentum space $\D\mu=\mu(p)\D^dp$ is nontrivial. This may be understood as an indication of curvature. However, the appearance of nontrivial measures is, in general, dependent on the ordering \cite{Bosso20c,Bosso21}. Therefore, this evidence should be understood as circumstantial. A much more thorough argument will be given in section \ref{sec:modcomorcurspa}.

Thus, given a Hamiltonian and knowing the volume measure, it is possible to obtain quantum-gravity-induced corrections to quantum mechanical problems. This approach simplifies dramatically for the choice of parameters $\beta= \beta'/2$.

\subsubsection*{Aside on the case $\beta=\beta'/2$}

This clearly constitutes a distinguished point in parameter space inasmuch as it allows for commutative coordinates and, therefore, for a description in the position representation. For reasons which will become apparent in section \ref{subsec:applquadGUP}, it also marks a most distinguished case from the point of view of curved spaces. The corresponding generalized Heisenberg algebra becomes
\begin{align}
    \left[\hat{x}^a,\hat{x}^b\right]=&0,\\
    \left[\hat{p}_a,\hat{p}_b\right]=&0,\\
    \left[\hat{x}^a,\hat{p}_b\right]=&i\hbar\left[\delta^a_b+\beta\left(\frac{l_p}{\hbar}\right)^2\left(\hat{p}^2\delta^a_b+2\hat{p}_b\hat{p}^a\right)\right].\label{GUPalgdDsimp}
\end{align}
Similarly to the one-dimensional case, this results in an uncertainty relation \cite{Hossenfelder12}
\begin{equation}
    \sigma_{x^a}\sigma_{p_a}\geq\frac{\hbar}{2}\left[1+3\beta\left(\frac{l_p\sigma_{p_a}}{\hbar}\right)^2\right],
\end{equation}
which immediately implies a minimum length or a maximum momentum in accordance with the discussion in one dimension.

In contrast to the general approach, here the resulting quantum kinematics can be simplified, applying the change of variables \cite{Brau06,Bosso18b}
\begin{equation}
    \hat{x}^a\rightarrow\hat{X}^a=\hat{x}^a,\hspace{1cm}\hat{p}_a\rightarrow \hat{P}_a\simeq p_a\left[1-\beta\left(\frac{l_p}{\hbar}\right)^2\hat{p}^2\right].\label{canontrans}
\end{equation}
The resulting phase space coordinates obey the canonical commutation relations \eqref{HeisAlg} up to corrections at higher order.  Anticipating a central result in the present thesis, a similar transformation is be the main content of section \ref{sec:modcomorcurspa}. Note that a change in the algebra implies an altered symplectic form \cite{Arnold13}, implying that this transformation is not canonical. Written in terms of the new coordinates, the Hamiltonian governing the dynamics of a particle subject to a potential $V(\hx)$ reads
\begin{equation}
    \hat{H}=\frac{\hat{p}^2}{2m}+V\left(\hat{x}\right)\simeq\frac{\hat{P}^2}{2m}\left[1+2\beta\left(\frac{l_p\hat{P}}{\hbar}\right)^2\right]+V\left(\hat{x}\right).
\end{equation}
Therefore, this kind of GUP is often applied in position space with which the Hamiltonian reads
\begin{equation}
    \hat{H}\ket{x}=\left[-\frac{\hbar^2\Delta}{2m}\left(1-2\beta l_p^2\Delta\right)+V(x)\right]\ket{x},\label{guphampos}
\end{equation}
with the Laplacian in flat space $\Delta.$ The sole modification appearing in this approach lies in the kinetic term of the Hamiltonian, making it comparably simple to derive predictions. 
%Furthermore, the canonical transformation \eqref{canontrans} implies that there will be a change in the volume measure in momentum space according to the Jacobian determinant
%\begin{equation}
%    \det{\left(\frac{\partial p_i}{\partial P_j}\right)}\simeq 1+3\beta\left(\frac{l_p\hat{P}}{\hbar}\right)^2
%\end{equation}
%where the expansion of is justified as long as only such states are considered which fall off sufficiently fast below the Planck scale.

\subsubsection*{Drawbacks}

Before we deal with the said predictions, however, let us stress a number of subtleties of this particular instantiation of the minimum length concept, many of which are carefully reviewed in Ref.  \cite{Hossenfelder12}. For example, it suffers an inverse soccer problem, rooted in the fact that the corrections to the dynamical variables of the center of mass in multiparticle states are inversely proportional to the number of constituents of the system \cite{Amelino-Camelia13}. This begs the question what a fundamental constituent is supposed to be. Clearly, this problem is related to the strictly nonrelativistic particle paradigm underlying typical applications. Furthermore, the deformed commutator can only yield either a trivial or a divergent classical limit \cite{Casadio20}, implying that it is a purely quantum mechanical effect \cite{Chashchina19}. This just closely saves it from violating Gromov's non-squeezing theorem \cite{Gromov85}, a hallmark of symplectic geometry which may be understood as classical analogue of Heisenberg's uncertainty principle \cite{deGosson09}. In that vein, the GUP may also challenge the second law of thermodynamics, which is closely related to Heisenberg's relation \cite{Hanggi13}. Moreover, its synthesis with the principle of gauge invariance is not thoroughly understood \cite{Chang16} and its relativistic extensions lead to deformations \cite{Ali09} or straight violations \cite{Lambiase17} of Lorentz invariance (\cf section \ref{subsec:GUPLIVDSR}). Of course, this might be seen as a feature rather than a problem. Last but not least, as was alluded to above, the minimum length may be derived from high-energy string scattering amplitudes \cite{Amati87,Amati88}. However, its value differs from the one inferred from D-branes \cite{Shenker95,Douglas96}, making the GUP probe-dependent in string theory.

In spite of these drawbacks, there has been great interest in the community to apply the idea to an immense amount of physical systems.

\subsection{Phenomenology}\label{subsec:pheno}

As modifications to the canonical commutators \eqref{HeisAlg} yield a change on the kinematical level, \ie to the understanding of spacetime and inertia in and of themselves, their effects are expectedly ubiquitous \cite{Das08}. Therefore, it is possible to derive corrections induced by generalized and extended uncertainty relations to virtually every quantum mechanical, and thus physical observable.  In particular, there are two mainly advocated routes towards the investigation of consequences. 

\subsubsection{Deformed Poisson brackets}\label{subsubsec:defpoisson}

First, a large part of the community concentrates on effects on classical problems, basing their reasoning on the correspondence between quantum commutators and classical Poisson brackets \cite{Benczik02}
\begin{equation}
    \frac{1}{i\hbar}[\hat{x}^a,\hat{p}_b]\longleftrightarrow \left\{x^a,p_b\right\}_{\text{PB}}.
\end{equation}
On those grounds, investigations have been carried out in the realm of statistical mechanics \cite{Rama01,Chang01,Nozari06b,Vakili12,Elmashad12,Ali14a,Shababi20}, especially in relation to white dwarfs and neutron stars \cite{Wang10,Ali13b,Moussa15,Mathew17,Bensalem19,Chung19e,Roushan20,Abac20,Viaggiu20,Abac21,Belfaqih21,Moradpour21}, including an ongoing debate about the possible disappearance of the Chandrasekhar limit \cite{Rashidi15,Ong18c,Ong18d,Mathew20}. Furthermore, this approach has been applied to orbits and the equivalence principle \cite{Nozari06f,Casadio09,Tkachuk12,Ghosh13,Ahmadi14,Vagenas17,Blanchette21,Chatterjee21}, cosmology \cite{Bina07,Vakili08a,Vakili08b,Kim08a,Majumder11d,Ali14b,Paliathanasis15,Giacomini20,Paliathanasis21}, the early universe \cite{Nozari06c,Chemissany11,Majumder12c,Tawfik12b,Ali15}, gravitational waves \cite{Khodadi18b,Bosso18a,Moussa21} and electrodynamics \cite{Moayedi13a,Moayedi13b,Moayedi13c}.

However, there are a number of caveats to this kind of approach. On the one hand, as was argued above, the classical limit of the deformed commutator for GUPs, \eg in Eq. \eqref{GUPalg1D}, harbours some subtleties, mainly due to the appearance of $\hbar$ in the denominator of the correction \cite{Casadio20}. The authors showed that, when considering states suitable for the classical limit, the deformation either disappears or diverges rendering the limiting procedure meaningless (see also Ref. \cite{Chashchina19} for similar considerations). The EUP, on the other hand, is supposed to be a semiclassical effect. Therefore, it ought to be possible to include it into classical mechanics just by assuming a curved background. In short, the relevance of this program has been called into question recently. 

\subsubsection{Modification within quantum mechanics}\label{subsubsec:quantpheno}

Secondly, the modified commutation relations can be applied to inherently quantum mechanical systems. Then, the issues mentioned in the previous section can be circumnavigated by not leaving the quantum realm in the first place. Correspondingly, the corrections to various problems within quantum mechanics, many known from undergraduate textbooks, have been computed \cite{Pikovski11,Pedram11a,Nozari05c,Nozari05a,Bouaziz08,Nozari10,Pedram10a,Ghosh11,Vahedi12,Pedram12d,Pedram12a,Pedram12b,Pedram12e,Hassanabadi12,Blado13,Bouaziz13,Ching13,Hassanabadi14,Das14,Rossi16,Dey15,Masood16,Shababi16,Wang16,Bosso16a,Bhat17,Bouaziz17,Bosso17,Villalpando18,Dey18,Vagenas19,Park20}. To provide an example, the next-to-leading order contribution to the radius of the deuteron is determined in the subsequent subsection. Furthermore, the algebra \eqref{GUPalg1D} has been applied to minisuperspace models in quantum cosmology implying a minimum size of the universe but,  interestingly,  not necessarily singularity resolution \cite{Vakili07,Battisti07a,Battisti07b,Battisti07c,Battisti08a,Battisti08b,Majumder11b,Majumder11a,Jalalzadeh14,Pedram15,Faizal15,Garattini15,Bosso19,Gusson20}. 

However, the applications of the GUP in particular have not been restricted solely to the nonrelativistic case. Apart from the invention of a relativistic version of it \cite{Todorinov18,Todorinov21}, its effect on quantum field theory \cite{Kempf96b,Ali10} has been quantified. It has further inspired modifications of the Klein-Gordon \cite{Adler99b,Matsuo05,Nozari05b,Chargui10a,Husain12} and Dirac \cite{Nozari05d,Quesne06,Shibusa07,Hossain08,Chargui10b,Pedram11b,Moayedi11,Pedram12c,Benzair12,Hassanabadi13a,Hassanabadi13b,Menculini13,Pedram14} equations leading to fully fledged gauge theories with minimum length \cite{Camacho03,Kober10,Kober11a,Kober11b,Faizal14a,Bosso20a,Bosso20b}. On this base, it was possible to compute the corrections to the thermodynamics of various types of black holes \cite{Liu03,Liu04,Liu05,Yoon07,Kim07a,Kim07b,Kim07d,Kim07e,Farmany09,Li09,Bina10,Majumder12b,Chen13a,Feng15,Sakalli16,Dehghani15,Ovgun15,Anacleto15a,Anacleto15b,Anacleto15c,Anacleto15d,Bargueno15,Li16,Ovgun17,Gecim17a,Gecim17b,Maluf18b,Kanzi19}, FLRW  and de Sitter spacetimes \cite{Kim06b,Kim10} and Randall-Sundrum models \cite{Kim06a,Nouicer07b}. Furthermore, quantum gravity contributions to the Unruh  \cite{Majhi13,Scardigli18} and Casimir effects \cite{Frassino11,Blasone19a,Blasone19b,Jusufi20b,Carvalho21,Samart21}, the covariant entropy bound in quantum field theory \cite{Kim08b} and the Cardy-Verlinde formula have been obtained \cite{Setare04,Setare05}. 

In particular, horizon thermodynamics can be derived solely from the uncertainty principle itself \cite{Medved04,Cavaglia04,Scardigli06,Park07,Nouicer07a}, an approach which was also applied in ref. \cite{Dabrowski19} cowritten by the present author. Consider an uncertainty relation of the form
\begin{equation}
    \Delta p \simeq \Delta p\left(\Delta x\right),
\end{equation}
where the types of measure of position and momentum uncertainty $\Delta p$ and $\Delta x,$ respectively, are left open for the moment Evidently, the unperturbed uncertainty principle should be of the form
\begin{equation}
    \Delta p \simeq \frac{B\hbar}{\Delta x},\label{genuncunp}
\end{equation}
with a numerical constant $B,$ which depends on the particular kind of relation at hand and equals $1/2$ for Robertson's approach \eqref{HUP}. The position uncertainty basically equals the characteristic scale describing the horizon denoted $l_H,$ \eg the Schwarzschild radius in the Schwarzschild geometry or the inverse acceleration in Rindler space. Thus, set $\Delta x=l_H.$ Furthermore, as it is of black body type \cite{Hawking75}, all information contained of the black hole radiation consists of its characteristic wave length, which is inversely proportional to its temperature. This wavelength, in turn, is inversely proportional to the momentum uncertainty, implying
\begin{equation}
    T_{H,\text{corr}}=C \Delta p \left(\Delta x\right)=C \Delta p \left(l_H\right),
\end{equation}
with the constant of proportionality $C.$ Plugging in the unperturbed relation \eqref{genuncunp} and comparing to the general result \cite{Hawking75,Unruh76,Gibbons77}
\begin{equation}
    T_H=\frac{\hbar}{4\pi l_H},
\end{equation}
the constant can be determined, yielding $C=B/4\pi.$ Thus, the corrected Hawking temperature can be expressed as
\begin{equation}
    T_{H,\text{cor}}=\frac{B}{4\pi}\Delta p\left(l_H\right).
\end{equation}
It was found early on that this modification, when trusted up to Planckian energies, inevitably leads to the creation of black hole remnants \cite{Adler01,Cavaglia03,Scardigli03,Maggiore93d,Chen02,Custodio03,Chen03,Chen04,Maziashvili05,Nozari07,Nozari08,Jizba09,Banerjee10,Nozari12a,Ali12a,Xiang13,Chen13b,Chen13c,Chen14,Dutta14,Alonso-Serrano18a,Alonso-Serrano18b,Ong18b,Li18,Alonso-Serrano20b}, thereby solving the information paradox \cite{Hawking76}. Furthermore, knowing the temperature, it is possible to derive the corrections to the entropy of the horizon in the usual way \cite{Hawking75}, which has been done in many contexts \cite{Nozari05e,Zhao06a,Zhao06b,Ko06,Nozari06h,Nozari06g,Nozari06e,Nozari06d,Myung06,Nouicer07c,Xiang09,Myung09c,Dehghani09,Said11,Majumder11e,Zhao12,Tawfik13,Gangopadhyay13,Faizal14c,Tawfik15a,Tawfik15b,Gangopadhyay15,Hammad15,Kanazawa19,Hassanabadi19}. In the case of the GUP, the resulting contributions are usually logarithmic, yielding an entropy $S$ of the form
\begin{equation}
    S_{\text{BH},\text{cor}}\simeq\frac{A_H}{4l_p^2}+C_S\log\left(\frac{A_H}{l_p^2}\right),
\end{equation}
where the constant $C_S$ depends on the model parameters. As this complies with results from fundamental approaches such as loop quantum gravity \cite{Meissner04} and string theory \cite{Kraus05}, comparison allows for fixing of the parameters \cite{Nozari06a,Kim07c,Scardigli16,Contreras18,Xin-Dong21}. The same can be done for EUPs and semiclassical gravity \cite{Bolen04}.

In another instance of this reversed logic, the adjustment of the entropy has been used to obtain corrections to black hole metrics \cite{Ali12b,Carr11,Sabri12,Nicolini12,Isi13,Carr14,Carr15,Maluf18a,Buoninfante20a,Anacleto21,Carr20}, Newtonian gravity \cite{Setare10,Nozari11,Ali13a} and the Friedmann \cite{Zhu08,Majumder11c,Ali13c,Awad14a,Awad14b,Khodadi16,Atazadeh16,Salah16,Das21a} and Einstein \cite{Majumder13,Alonso-Serrano20a} equations in scenarios of emergent gravity.

Thus, there are manifold avenues towards the determination of consequences of modified uncertainty relations. However, the real phenomenology lies in quantitative comparison to experimental data. 

\subsubsection{Constraints}\label{subsubsec:phenoconst}

In science, theoretical predictions need to stand the challenge of observation. Within the treated subjects, however, this endeavour is  mostly pursued in relation to GUPs because, as was alluded to above, EUPs ought to be derivable from semiclassical physics.

\begin{figure}[h!]
    \centering
    \includegraphics[width=.94\linewidth]{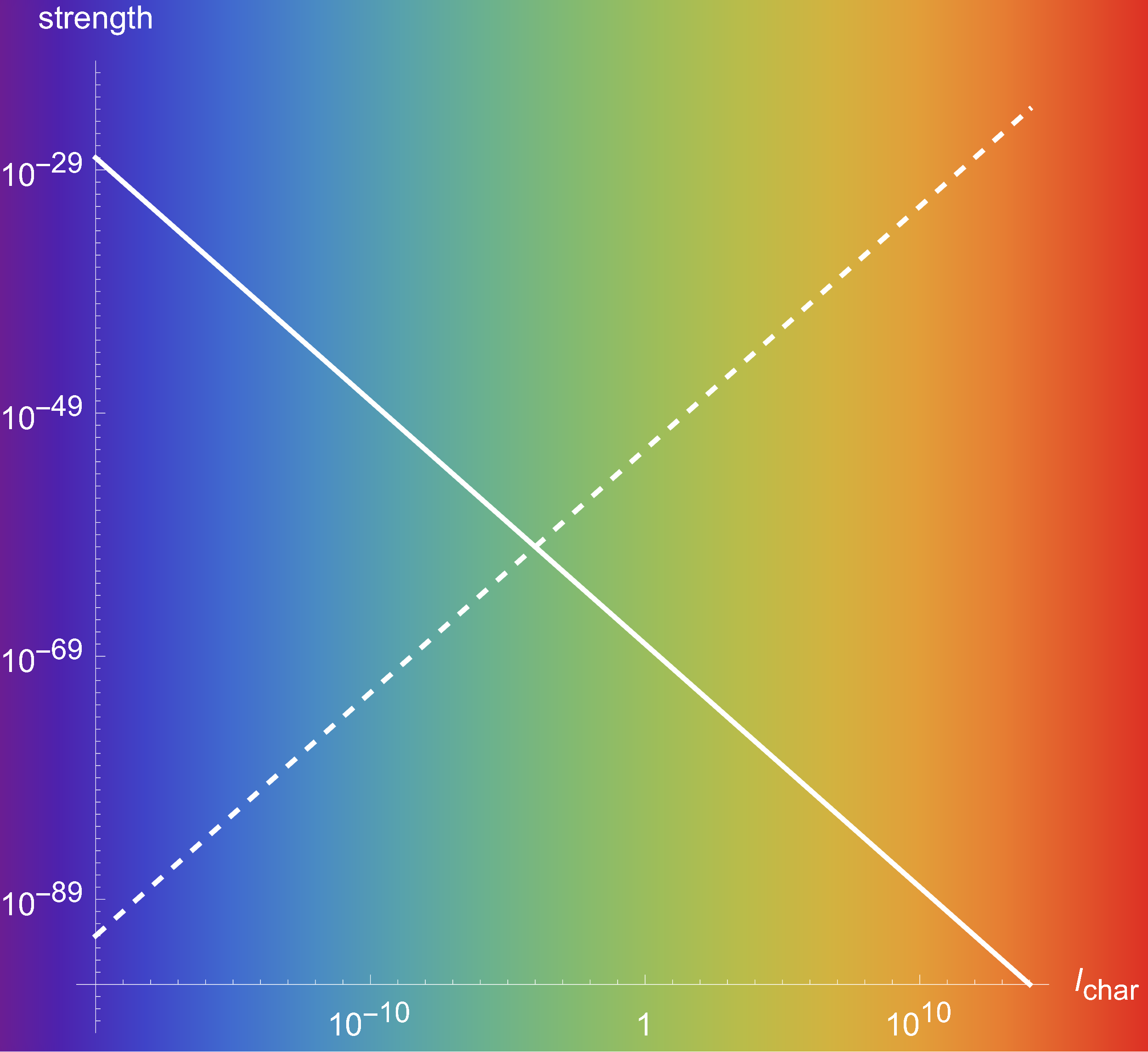}
    \caption{Estimate of magnitude of effects induced by the generalized (connected) and extended (dashed) uncertainty principles at different length scales. The background colour indicates the energy scale, from the ultraviolet to the infrared.}
    \label{fig:GUPEUPstrength}
\end{figure}

Generally, it should be expected that quadratic adjustments derived from the algebra \eqref{GUPalgdDsimp} lead to corrections as of an expansion in $\beta l_p^2/l^2_{\text{char}}$ and $\alpha l^2_{char}/r_{H}^2,$ respectively, where $l_{\text{char}}$ denotes a characteristic length scale of the unperturbed problem. To gain an intuition, these factors, evaluated at distinct characteristic lengths, are displayed in table \ref{tab:factorest} and Fig. \ref{fig:GUPEUPstrength}. Evidently, those contributions are tiny in comparison to order-one processes, which makes it hard to observe them. Therefore, it is useful to look for amplifiers as argued in Refs. \cite{Amelino-Camelia02a,Amelino-Camelia08} in the context of quantum gravity phenomenology in general.

\begin{table}[h!]
    \centering
\begin{tabular}{ c||c c} 
 scale  &  $l_p^2/l^2_{\text{char}}$ & $l^2_{char}/r_{H}^2$\\
 \hline
 solar system & $10^{-96}$ & $10^{-24}$\\ 
 sun & $10^{-84}$ &$10^{-36}$ \\ 
 earth & $10^{-80}$ & $10^{-40}$\\
 human & $10^{-68}$ & $10^{-52}$\\ 
 cell & $10^{-60}$ & $10^{-60}$\\ 
 Buckminsterfullerene & $10^{-50}$ &$10^{-70}$ \\ 
 atom & $10^{-46}$ & $10^{-74}$\\
 proton & $10^{-38}$ & $10^{-82}$ \\
 weak interaction & $10^{-54}$ & $10^{-86}$ \\
 LHC & $10^{-30}$ & $10^{-90}$ 
\end{tabular}
\caption{Estimate of magnitude of effects induced by the generalized and EUPs at different characteristic length scales.\label{tab:factorest}}
\end{table}

Independently of those considerations, every quantitative analysis allows for constraints on the parameters of the GUP \cite{Das08,Das09}. This has been done in the context of lab-based experiments \cite{Ali11b,Bouaziz10,Das11,AntonacciOakes13,Bouaziz13,Jalalzadeh13,Bawaj14,Gao16,Gao17,Khodadi18a,Bushev19,Luciano19,Addazi20,Girdhar20,Jusufi20a,Twagirayezu20} and observations on astrophysical \cite{Wang11,Majumder11f,FaragAli15,Kumar19,Neves19,Tamburini21} and cosmological scales \cite{Tawfik12a,Jalalzadeh13,Marin13,Marin14,Feng16,Giardino20,Das21b}. Furthermore, bounds from Kosteleck\'y 's SME,  an alternative, Lorentz-violating approach to quantum gravity phenomenology mentioned in the introduction, were imported \cite{Lambiase17}. A further constraint was found following the assumption that the parameter $\beta$ in Eq. \eqref{GUPalg1D} is stochastic, leading to quantum gravitational decoherence \cite{Petruzziello20b}. Conveniently, there is a recent collection of bounds \cite{Scardigli19} summarizing most of these contributions. A surely nonexhaustive collection is gathered in tables \ref{tab:labgup}, \ref{tab:gravigup} and \ref{tab:cosmogup} for tabletop experiments, gravitational experiments and observations and cosmological observations, respectively. Clearly, tabletop experiments yield the most precise results because they are related to smaller length scales, while cosmological experiments prove basically useless. Note that the most stringent bounds stem from macroscopic harmonic oscillators comprised of many particles whose number acts as amplifier. Have in mind, though, that this is an area whose validity is controversial due to the inverse soccer ball problem plaguing GUP-induced effects on many bodies \cite{Amelino-Camelia13} (for the soccer ball problem in general consult Refs. \cite{Hossenfelder07,Amelino-Camelia11}). The strongest upper bound that is widely accepted, derived from corrections to the transition amplitudes between stationary states of the hydrogen atom, thus assumes the value $\beta <10^{26}.$

\begin{table}[]
    \centering
\begin{tabular}{ c c || c} 
 experiment & ref.  & upper bound on $\beta$ \\
 \hline
 harmonic oscillators & \cite{Bushev19,Bawaj14} & $10^7$\\ 
 hydrogen state transitions & \cite{Bouaziz10,AntonacciOakes13,Brau99} & $10^{26}$\\
 quantum noise & \cite{Girdhar20} & $10^{28}$\\
 scanning tunnelling microscope & \cite{Pikovski11,Das08} & $10^{33}$\\
 $\mu$ anomalous magnetic moment & \cite{Das11,Das08} & $10^{33}$\\
 lamb shift & \cite{Das08,Ali11b} & $10^{36}$\\ 
 $^{87}$Rb interferometry & \cite{Gao16} & $10^{39}$\\
 Kratzer potential & \cite{Bouaziz13} & $10^{46}$\\
 stimulated emission    &   \cite{Twagirayezu20}    &   $10^{46}$\\
 Landau levels & \cite{Ali11b,Das08,Das09} & $10^{50}$
\end{tabular}
\caption{Upper bounds on the parameter $\beta$ characterizing the quadratic GUP with commutative coordinates \eqref{GUPalgdDsimp} by tabletop experiments not related to gravity. \label{tab:labgup}}
\end{table}

\begin{table}[]
    \centering
\begin{tabular}{ c c || c} 
 experiment & ref.  & upper bound on $\beta$ \\
 \hline
 equivalence principle & \cite{Ghosh13,Gao17,Luciano19} & $10^{20}$\\ 
 gravitational bar detectors & \cite{Marin13,Marin14} & $10^{33}$\\ 
 perihelion precession (solar system) & \cite{Scardigli14,FaragAli15} &$10^{69}$\\
 perihelion precession (pulsars) & \cite{Scardigli14} &$10^{71}$\\
 gravitational redshift & \cite{FaragAli15} & $10^{76}$\\
 black hole quasi normal modes & \cite{Jusufi20a} & $10^{77}$\\
 light deflection & \cite{Scardigli14,FaragAli15} & $10^{78}$\\
 time delay of light & \cite{FaragAli15} & $10^{81}$\\
 black hole shadow & \cite{Neves19,Jusufi20a,Tamburini21} & $10^{90}$
\end{tabular}
\caption{Upper bounds on the parameter $\beta$ characterizing the quadratic GUP with commutative coordinates \eqref{GUPalgdDsimp} by gravitational experiments and observations. \label{tab:gravigup}}
\end{table}

\begin{table}[]
    \centering
\begin{tabular}{ c c || c} 
 experiment & ref.  & upper bound on $\beta$ \\
 \hline
 gravitational waves & \cite{Das21b,Feng16} & $10^{36}$\\
 cosmology (all data) & \cite{Giardino20} & $10^{59}$\\
 cosmology (late-time) & \cite{Giardino20,Kouwn18} & $10^{81}$
\end{tabular}
\caption{Upper bounds on the parameter $\beta$ characterizing the quadratic GUP with commutative coordinates\\
 \eqref{GUPalgdDsimp} on cosmological scales. \label{tab:cosmogup}}
\end{table}

Most of the derivations behind those constraints follow a similar pattern. To gain an intuition into those approaches, an example of how this is done is provided in the subsequent subsection.

\subsubsection{Example: Radius of the deuteron}\label{subsubsec:exdeut}

The derivation presented in this section was performed by the author to provide an instance of the phenomenology of GUPs.% and in its execution bears resemblance to ref. \cite{Faruque14}. 
The main idea behind this calculation lies in the application of the model governed by the algebra \eqref{GUPalgdDsimp} to constrain the parameter $\beta$ (\cf Eq. \eqref{GUPalgdDsimp}) by comparing resulting corrections to measurements of the radius of the deuteron. Thus, it belongs to the category of tabletop experiments.

As mentioned above, the three-dimensional algebra \eqref{GUPalgdDsimp} implies a commutative space, which leads to a considerable simplification of the problem. Furthermore, the radial potential is approximated by a square well to keep the calculations tractable
\begin{equation}
V(\hat{r})=-V_0\Theta(r_0-\hat{r}),
\end{equation}
with the radial position operator $\hat{r},$ the deuteron radius $r_0$ and the Heaviside-function $\Theta.$ According to Eq. \eqref{guphampos}, the time-independent Schrödinger equation then becomes
\begin{equation}
\left[-\frac{\hbar^2\Delta}{2\mu}\left(1+2\beta l_p^2\Delta\right)-V_0\Theta(r_0-r)\right]\psi=E\psi,\label{deuteronschroed}
\end{equation}
with the position space wave function $\psi(x)$ and the reduced mass $\mu$ introduced to effectively turn the initial two-body problem into a one-body problem. As the potential is basically constant, this differential equation can be understood as an eigenvalue equation for a function of the Laplacian. Therefore, we might as well determine the eigenstates of the Laplacian
\begin{equation}
    \left(\Delta+\lambda\right)\psi=0,\label{deutlapeig}
\end{equation}
with the boundary condition that the wave function be normalizable, \ie nondivergent, and approaching zero fast enough at large distances from the origin. Adding this assumption to the problem, we obtain an equation providing a value for $\lambda$
\begin{equation}
    \frac{\hbar^2\lambda}{2\mu}\left(1-2\beta l_p^2\lambda\right)=E+V_0\Theta(r_0-r).
\end{equation}
This equation has two solutions of which only one has a well-defined limit when $\beta\rightarrow 0,$ yielding
\begin{align}
    \lambda &=\left(4l_p^2\beta\right)^{-1}\left(1-\sqrt{1-16\mu\beta l_p^2\frac{E+\Theta (r_0-r)V_0}{\hbar^2}}\right)\\
            &\simeq 2\mu\frac{E+\Theta (r_0-r)V_0}{\hbar^2}\left[1+4\mu\beta l_p^2\frac{E+\Theta (r_0-r)V_0}{\hbar^2}\right],
\end{align}
where we expanded in $\beta l_p^2 E$ to get a grasp on the leading contribution. As the deuteron constitutes a bound state, the energy $E$ is negative and satisfies $E+V_0>0.$ Thus, the eigenvalue $\lambda$ is positive in the interior ($r\geq r_0$) and negative in the exterior of it, a fact that can be expressed as
\begin{align}
    \lambda_{in}=&\left(4l_p^2\beta\right)^{-1}\left(1-\sqrt{1-16\mu\beta l_p^2\frac{V_0-|E|}{\hbar^2}}\right)=|\lambda_{in}|,\\
    \lambda_{ex}=&\left(4l_p^2\beta\right)^{-1}\left(1-\sqrt{1+16\mu\beta l_p^2\frac{|E|}{\hbar^2}}\right)=-|\lambda_{ex}|.
\end{align}
Correspondingly, the differential equation \eqref{deutlapeig} branches off into the two problems
\begin{align}
    \left(\Delta+|\lambda_{in}|\right)\psi|_{r\leq r_0}=&0,\\
    \left(\Delta-|\lambda_{ex}|\right)\psi|_{r > r_0}=&0.
\end{align}
For simplicity, the only configuration considered in this section is the s-wave, \ie ground, state. Thus, the wave function features no angular dependence and we can write the problem in terms of the scalar $u=\psi/r$ yielding
\begin{align}
    \left(\partial_r^2+|\lambda_{in}|\right)u|_{r\leq r_0}=&0,\\
    \left(\partial_r^2-|\lambda_{ex}|\right)u|_{r> r_0}=&0,
\end{align}
which, being a harmonic and an anharmonic oscillator, clearly allow the solutions
\begin{equation}
    u=
    \begin{cases}
    A\sin{\left(\sqrt{|\lambda_{in}|}r\right)}+C\cos{\left(\sqrt{|\lambda_{in}|}r\right)}&r\leq r_0\\
    Be^{-\sqrt{|\lambda_{ex}|}r}+De^{\sqrt{|\lambda_{ex}|}r}&r> r_0.
    \end{cases}
\end{equation}
The boundary condition that $\psi$ be normalizable immediately implies that $C=D=0.$ Furthermore, we have to impose continuity and differentiability at $r=r_0$ leading to the conditions
\begin{align}
A\sin{\left(\sqrt{|\lambda_{in}|} r_0\right)}           				&=Be^{-\sqrt{|\lambda_{ex}|} r_0},\\
A\left[\sqrt{\lambda_{in}} \cos\left(\sqrt{|\lambda_{in}|} r_0\right)-\sin{\left(\sqrt{|\lambda_{in}|} r_0\right)}r_0^{-1}\right]	&=-Be^{-\sqrt{|\lambda_{ex}|} r_0}\left[\sqrt{|\lambda_{ex}|}+r_0^{-1}\right].
\end{align}
Those can be used to determine the radius of the deuteron as the first positive solution to the equation
\begin{equation}
\tan\left(r_0\sqrt{|\lambda_{in}|}\right)=-\sqrt{\left|\frac{\lambda_{in}}{\lambda_{ex}}\right|}.
\end{equation}
As the multivaluedness of the tangent ($\tan(x)=\tan(x+\pi)$) cannot be properly represented by its inverse function, it is necessary to add a term multiplying $\pi$ to obtain
\begin{equation}
r_0=\frac{\pi-\arctan\sqrt{\left|\frac{\lambda_{in}}{\lambda_{ex}}\right|}}{\sqrt{|\lambda_{in}|}}\simeq r_0^{(0)}+r_0^{(1)},
\end{equation}
where $r_0^{(0)}$ and $r_0^{(1)}$ denote the unperturbed radius of the deuteron and the correction induced by the minimum length at first order in $\beta l_p^2\sqrt{\mu |E|}$ and $\beta l_p^2\sqrt{\mu (V_0-|E|)}$, respectively. Explicitly, we obtain
\begin{align}
r_0^{(0)}	&=\hbar\frac{\pi-\arctan{\sqrt{\frac{V_0}{|E|}-1}}}{\sqrt{2\mu(V_0-|E|)}},\\
r_0^{(1)}			&=\frac{\sqrt{2\mu}}{\hbar}\left[\sqrt{|E|}+\sqrt{V_0-|E|}\left(\pi-\arctan{\sqrt{\frac{V_0}{|E|}-1}}\right)\right]\beta l_p^2.
\end{align}
As $V_0/|E|\gg 1$, the first term in $r_0^{(1)}$ can be neglected and $\arctan\sqrt{V_0/|E|-1}\simeq \pi/2.$ Thus, it is possible to roughly estimate the relative effect as
\begin{equation}
\frac{r_0^{(1)}}{r_0^{(0)}}	\simeq \left(\frac{\pi}{2}\right)^2\beta\left(\frac{l_p}{r_0^{(0)}}\right)^2.
\end{equation}
Taking into account the relative precision to which $r_0$ is known ($\sim 10^{-4}$), this result can be used to constrain the GUP parameter $\beta\lesssim 10^{37},$ leading to a minimum length smaller than one hundredth of the deuteron radius.

Clearly, a truly precise calculation of this effect requires a more accurate description of the nucleon-nucleon interaction by a different potential. However, this approach suffices to estimate its magnitude. Note that, over all, this confirms the expectation that effects of the quadratic GUP are of the form $\beta l_p^2/l^2_{\text{char}}.$ In this case, the characteristic length scale is the unperturbed radius of the deuteron $r_0.$ In principle, almost all phenomenological calculations referring to the GUP are of this form.

\section{GUPs and EUPs from the curvature of Riemannian backgrounds}
\label{sec:3DEUP}

According to the reasoning laid out in the preceding section, the main motivations for GUPs and EUPs are the inclusion of a minimal length implied by quantum gravity and a maximum length derived from the curvature of spacetime, respectively. Yet, the connection between fundamental limits to observability and the theory of modified commutation relations is by no means unique. This begs the question whether similar kinds of results could be derived starting at different, less ad hoc assumptions.

In that vein, we base the considerations of this section on the assumption that the three dimensional background manifold describing position or momentum space be curved, while the algebra of observables stays canonical (\cf Eq. \eqref{HeisAlg}). Correspondingly, in this setting there are no modifications to Heisenberg's uncertainty principle \eqref{HUP} via the Robertson relation \eqref{Robertson}. Yet, this inequality is not the only way an uncertainty relation in quantum mechanics can be formulated \cite{Maassen88,Coles17,Maccone14}. The rather vague motivations behind GUPs and EUPs, though,  cannot be deployed as a means of distinguishing between those different approaches. On the contrary, there are a number of alternative approaches towards a minimum length by superposition of geometries \cite{Lake18,Lake19a,Lake19b,Lake20} or direct inclusion into differential geometry \cite{Padmanabhan96,Kothawala13,Kothawala14,Padmanabhan15}, and idea which actually dates back to work of Arthur March in the 30s \cite{March37}. For the following considerations we rely on a recently found alternative, which has the advantages of being rather operational and easily generalizable to curved manifolds \cite{Schuermann09,Schuermann18,Dabrowski19,Dabrowski20,Petruzziello21}. The main idea behind this relation consists in confining the wave function to a compact domain. A covariant invariant measure of the size of this region can then be interpreted as the corresponding uncertainty. As a result, it is possible to find the global minimum of the standard deviation of the complementary observable as a function of that very measure of uncertainty. The corresponding inequality yields the sought-after uncertainty relation.

In this section we show how this can be made precise by finding a definition of the standard deviation of the momentum operator appropriate for curved position space, explaining how to treat the compact domain and posing the exact problem in subsection \ref{subsec:uncrel}. Subsection \ref{subsec:explicitsol} is devoted to solving this problem first in flat and subsequently in slightly curved space, thereby assuming small position uncertainties. In this manner, we provide perturbative corrections to the uncertainty relation to fourth order. In subsection \ref{subsec:AGEUP} we generalize the result to modified commutation relations, according to the paradigm introduced in the preceding section.

Being extracted mainly from the publication \cite{Dabrowski20} with an appendix from \cite{Petruzziello21}, both coauthored by the present author with EUPs in mind, the treatment given in the present chapter is based on curved position space. Have in mind, though, that Born reciprocity \cite{Born38} implies that an analogous effect would be expected for curved momentum space, obtaining a GUP instead. In fact, it turns out that the specific case investigated in section \ref{subsec:explicitsol} is, in its entirety, translatable to the picture of curved momentum space.

\subsection{Uncertainty relation}\label{subsec:uncrel}

The notion of uncertainty relation harbours many more subtleties in curved space than in the ordinary flat case. However, the manifestly covariant invariant approach introduced in ref. \cite{Schuermann18}, though unusual, proofs very effective in this context. This subsection is devoted to explaining exactly what constitutes the mathematical basis for the deduction of the EUP in the subsequent subsection. 

\subsubsection{Standard deviation of the momentum operator}\label{subsubsec:momunc}

In accordance with section \ref{subsec:curved}, we assume that the three-dimensional background manifold of the treated system be characterized by the length element
\begin{equation}
    \D s^2=g_{ij}(x)\D x^i\D x^j,
\end{equation}
featuring the position-dependent metric $g_{ij}(x).$ Correspondingly, the Hilbert space measure is described by Eq. \eqref{curvedspacemeas}. Further asserting the canonical commutation relations \eqref{HeisAlg} to be satisfied, the position space representation of the momentum operator is provided by Eq. \eqref{newmomop}.

Those are all the ingredients required to define the standard deviation of the momentum operator in curved space as
\begin{equation}
    \sigma_\pi\equiv\sqrt{\braket*{\hatslashed{\pi}^2}-\braket*{\hatslashed{\pi}}^2}=\sqrt{\braket*{\hat{\pi}^2}-\gamma^{a}\gamma^{b}\braket*{e_a^i\hat{\pi}_i}\braket*{e_b^j\hat{\pi}_j}}=\sqrt{\braket*{\hat{\pi}^2}-\delta^{ab}\braket*{e_a^i\hat{\pi}_i}\braket*{e_b^j\hat{\pi}_j}}\label{defmomstddev}
\end{equation}
where we used Eq. \eqref{normalClifford} and the symmetry under the exchange of indices $i\leftrightarrow j$ for the third equality. Thus, we have to calculate the standard deviation in a local Euclidean frame. In flat space, for example, this forces us to use Cartesian coordinates. 

Note that to be fully consistent we should denote this quantity as $\sigma_\slashed{\pi}.$ However, we decided to use the notation $\sigma_\pi$ to make its meaning more apparent.

\subsubsection*{Aside on Born reciprocity}

Assuming Born reciprocity, we should be able to follow the exact same path outlined here to define a standard deviation of the position operator in curved space. However, there is a problem with this logic: The definition outlined above hinges on the fact that the momentum operator transforms as a vector under spatial diffeomorphisms. The position operator, though, most definitely does not. Understood as generator of translations in momentum space, it obeys the correct transformation law under momentum space diffeomorphisms. Yet, we cannot necessarily say the same about the world we experience \cite{Amelino-Camelia19,Gubitosi21}. 

Sure enough, we can define the squared position operator on a general curved space invoking the geodesic distance from the origin $\Ord$ to the point in question $p$
\begin{equation}
    \sigma (\Ord,p)=\int_\Ord^p\D s,
\end{equation}
leading to a representation proportional to the Laplace-Beltrami operator in momentum space (\cf Eq. \eqref{Laplace-Beltrami})
\begin{equation}
    \hat{\sigma}^2(\Ord,p)\tilde{\psi}(p)=\frac{1}{\sqrt{g}}\dot{\partial}^i\left(\sqrt{g}g_{ij}\dot{\partial}^j\tilde{\psi}\right).
\end{equation}
Even so, it is unclear how to provide an operator $\hat{x}$ by analogy with Eq. \eqref{newmomop} in a mathematically meaningful way because it does not transform as a vector with respect to diffeomorphisms in position space.

Thus, in general we cannot simply take the equivalent road starting at curved momentum space in the derivation of the uncertainty relation. On the contrary, we are witnessing another instance of the breaking of Born reciprocity. Interestingly, though, this problem can be circumvented in the case of perturbations around flat space treated in section \ref{subsubsec:perturb} because, as will be shown below, this only involves the evaluation of expectation values on a flat background.

In short, we generally cannot define the uncertainty of the position operator in the standard way. How then do we accomplish this goal consistently?

\subsubsection{Position uncertainty as size of a compact domain}\label{subsubsec:posunc}

\begin{figure}[h!]
\begin{minipage}{.48\textwidth}
    \centering
    \includegraphics[width=\linewidth]{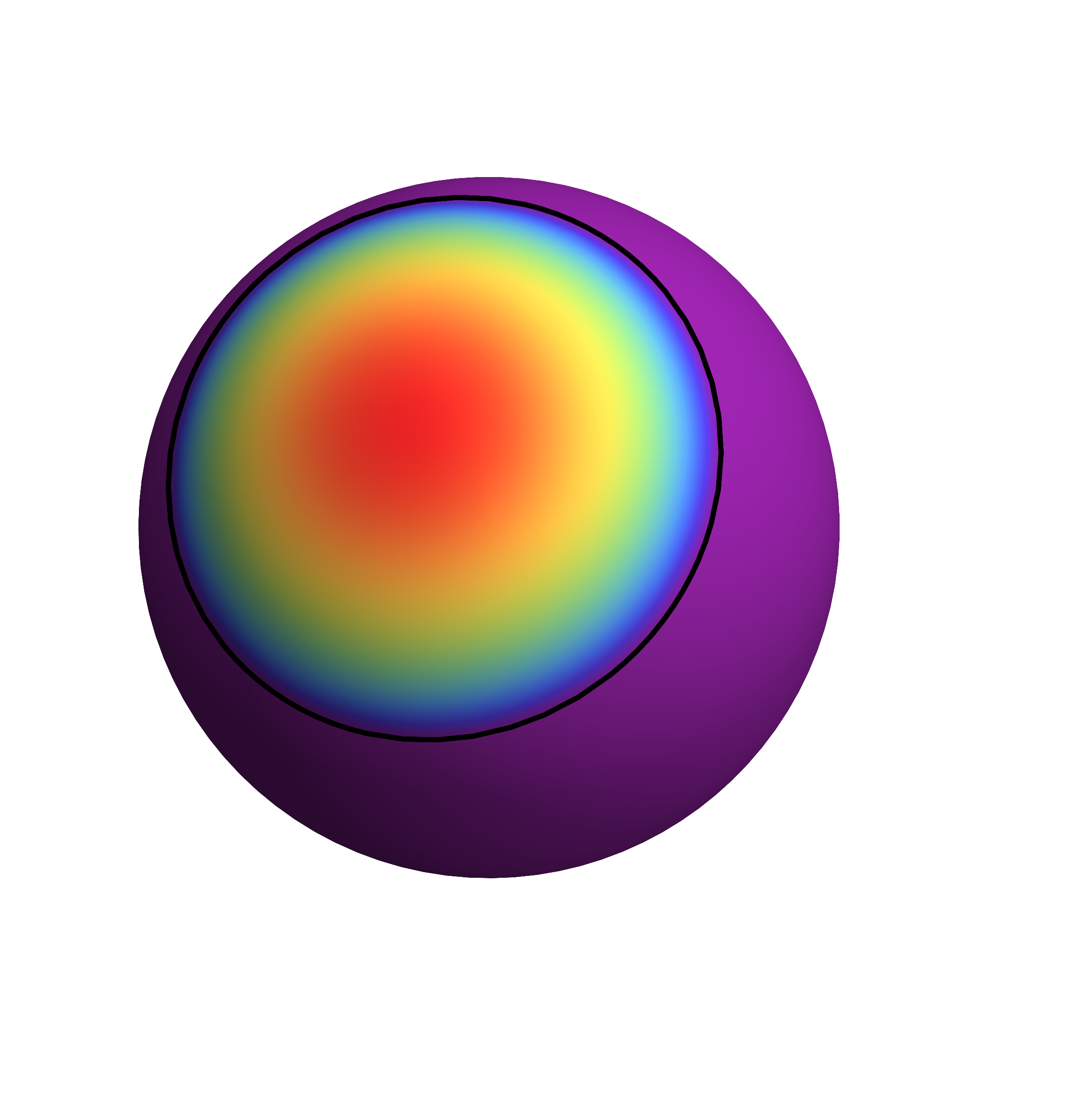}
\end{minipage}
\begin{minipage}{.48\textwidth}
    \centering
    \includegraphics[width=\linewidth]{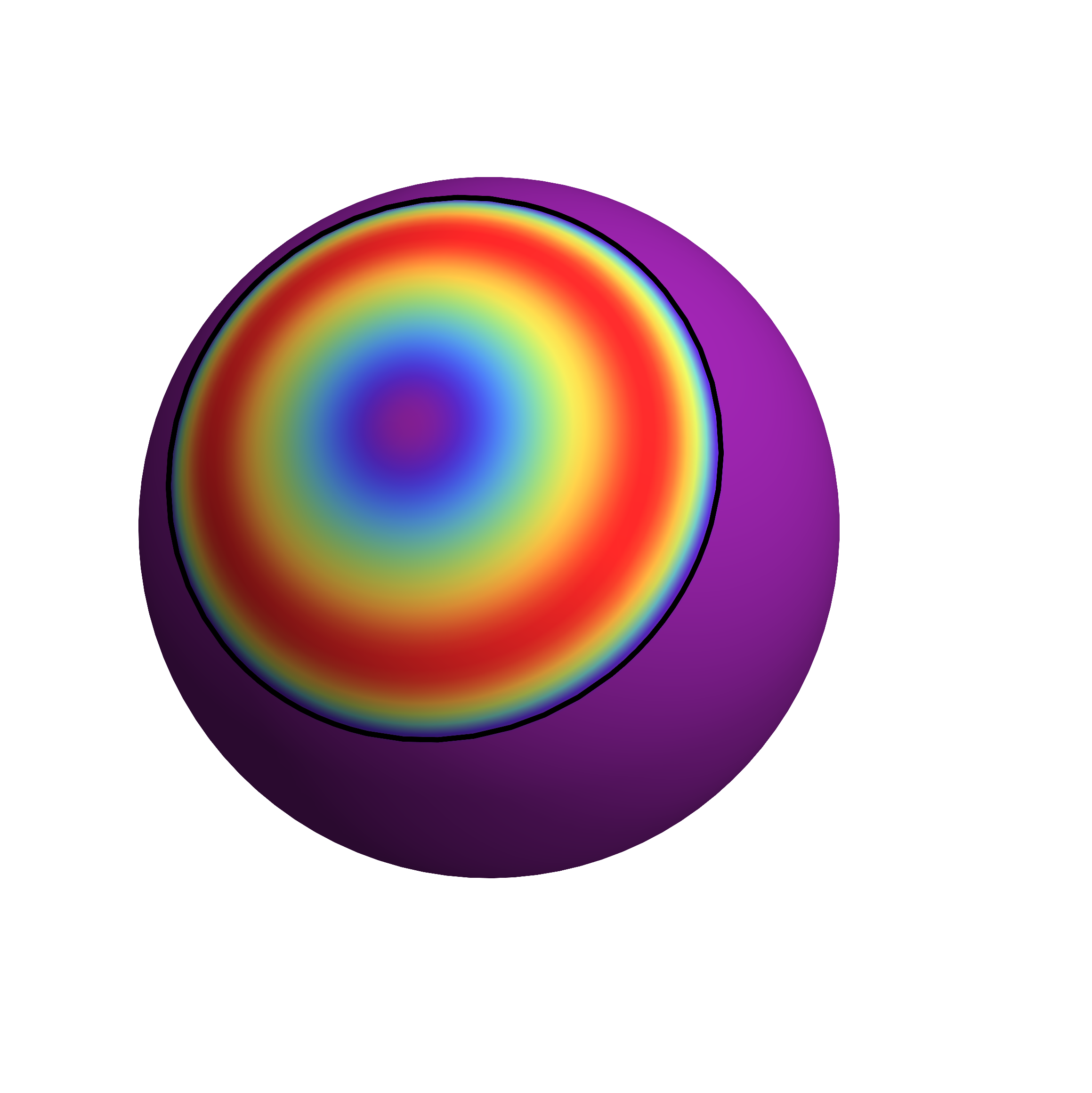}
\end{minipage}
\captionof{figure}{Schematic visualisation of the squared absolute value of two wave functions (eigenfunctions of the Laplacian) colour coded from violet (vanishing) to red on a curved geometry, here as the surface of a sphere embedded in three-dimensional space, and confined to a disk, whose boundary is displayed in black.}
    \label{fig:WaveOnCurvedBackground}
\end{figure}

In this section we take a more operational route towards constructing an instance of position uncertainty. Restricting the support of allowed wave functions in the considered Hilbert space to a compact domain $\mathcal{D}$, \ie choosing it to be given as $\hil =L^2(\mathcal{D},\sqrt{g}\D^3x),$ we clearly localize the system within a controllably sized setting. This can be achieved by imposing Dirichlet boundary conditions. Accordingly, all $\psi\in\hil$ have to satisfy $\psi|_{\partial \mathcal{D}}=0,$ \ie vanish at the boundary and outside of it (see Fig. \ref{fig:WaveOnCurvedBackground} for a visualisation). 

Any diffeomorphism invariant scale characterising the domain's extent would thus yield a measure of position uncertainty. For example, we might use a function of the its volume
\begin{equation}
    V=\int_\mathcal{D} \D^3 x\sqrt{g}.\label{volume}
\end{equation}
In particular,  to provide a scale with dimensions of length $\rho,$ we may choose
\begin{equation}
    \rho \propto\sqrt[3]{V}.\label{genposunc}
\end{equation}
In principle, this approach can be applied to any kind of domain. For reasons of simplicity, however, we will choose to work with geodesic balls, and define the position uncertainty as their radius. This information suffices to specify the Hilbert space which is about to be explored. Thus, we are all set to pose the problem whose solution yields the uncertainty relation.

\subsubsection{Eigenvalue problem}\label{subsubsec:evprob}

The investigated quantum theory is defined within a compact domain on a curved nonsingular background manifold. Therefore, it can be shown \cite{Hilbert12} that the Laplace-Beltrami operator, basically representing the squared momentum operator (\cf Eq. \eqref{squaredmom}), is Hermitian and possesses a discrete spectrum. Thus, its eigenvectors $\psi$ furnish an orthonormal basis of the Hilbert space $\hil.$ Evidently, they have to be solutions to the eigenvalue problem 
\begin{subequations}
\label{evp}
\begin{align}
\Delta \psi + \lambda \psi	&=0~\text{within~}D,\label{evp1}\\
\psi									&=0~\text{on~}\partial D.\label{evp2}
\end{align}
\end{subequations}
In three dimensions, these eigenstates are characterized by three quantum numbers, here represented by the symbol $n$ to avoid index cluttering. A general state $\Psi$ can then be expressed as a linear combination of the eigenstates
\begin{equation}
    \Psi=\sum_{n}a_{n}\psi_{n},\label{genstaten}
\end{equation}
with the coefficients $a_{n},$ satisfying 
\begin{equation}
    \sum_{n}|a_{n}|^2=1.\label{coeflincomb}
\end{equation}
Therefore, we can express the standard deviation of the momentum operator as
\begin{equation}
    \sigma_\pi\left(\Psi\right)=\sqrt{\sum_{n}|a_{n}|^2\hbar^2\lambda_{n}-\left(\sum_{n,n'}a^*_{n}a_{n'}\Braket*{\psi_{n}}{\hatslashed{\pi}_i\psi_{n'}}\right)^2}.\label{stddevmomgenp}
\end{equation}

In general, the real and the imaginary part of the eigenvalue problem \eqref{evp1} are collinear. This implies that the phase of its solutions $\psi_n$ can be removed by rotating the coordinate system. As the locally Euclidean frame is invariant under rotations, we can calculate the expectation value of the momentum operator in any of those related by a rotation. Thus, we can take the eigenfunctions of the Laplacian to be real.  However, the expectation value of the momentum operator with respect to any real wave function $\psi: {\rm I\!R}^3\rightarrow {\rm I\!R}$ vanishes as can be readily verified by
\begin{equation}
    \Braket*{\psi}{\hatslashed{\pi}\psi}=\int\D\mu \psi\hatslashed{\pi}\psi=-\int\D\mu \hatslashed{\pi}(\psi)\psi=-\Braket*{\psi}{\hatslashed{\pi}\psi}=0,\label{vanmomrealwave}
\end{equation}
where we used the Hermiticity of $\hatslashed{\pi}$ and the boundary condition \eqref{evp2}. Hence, the momentum operator has vanishing expectation value if the system is in an eigenstate of the Laplace-Beltrami operator
\begin{equation}
    \Braket*{\psi_{n}}{\hatslashed{\pi}\psi_{n}}=0.\label{vanishingmom}
\end{equation}
Thence, we can rewrite Eq. \eqref{stddevmomgenp} as
\begin{equation}
    \sigma_\pi\left(\Psi\right)=\sqrt{\sum_{n}|a_{n}|^2\hbar^2\lambda_{n}-\left(\sum_{n\neq n'}a^*_{n}a_{n'}\Braket*{\psi_{n}}{\hatslashed{\pi}_i\psi_{n'}}\right)^2}.\label{stddevmomgen}
\end{equation}
Then the precise task to fulfil consists in finding the state $\Psi_0$ yielding the global minimum of this expression such that
\begin{equation}
    \sigma_\pi\left(\Psi\right)\geq\sigma_\pi\left(\Psi_0\right)\equiv\bar{\sigma}_\pi(\rho)\geq 0,\label{uncgen}
\end{equation}
where $\rho,$ recall, denotes the measure of position uncertainty. Multiplication by $\rho$ leads to the uncertainty relation in its usual form, \ie as product of uncertainties
\begin{equation}
    \sigma_\pi\rho\geq\bar{\sigma}_\pi(\rho)\rho.
\end{equation}

In the subsequent sections the search for the state saturating this inequality turns out to be rather simple. In fact, it is given by the ground state of the Laplace-Beltrami operator $\psi_1,$ implying the relation
\begin{equation}
    \bar{\sigma}_\pi(\rho)=\hbar\sqrt{\lambda_1}.
\end{equation}
Whether this is a general feature of the given problem, is conceivable but a proof thereof beyond the scope of this thesis.

\subsection{Explicit solution}\label{subsec:explicitsol}

Having dealt with the theoretical subtleties, we can now derive the uncertainty relation for a general curved background. In order to obtain explicit results, we first have to choose a specific domain, the geodesic ball.  The solution is first obtained analytically for the case of flat space, in a domain-independent fashion and explicitly within geodesic balls, to be further generalized to small perturbations around flat space, \ie an expansion in small position uncertainties to fourth order.

\subsubsection{Geodesic balls}\label{subsubsec:geoball}

Regarding the domain, we restrict ourselves to the rather simple example of a geodesic ball of radius $\rho,$ denoted $B_{\rho}.$ This region is defined by its boundary, every point $p\in \partial B_{\rho}$ on which has a constant geodesic distance 
\begin{equation}
    \sigma(p_0,p)=\int_{p_0}^p\D s=\rho
\end{equation}
to its center $p_0.$ Defined in terms of the geodesic distance, the radius is manifestly diffeomorphism invariant, and may thus serve as measure of position uncertainty. Furthermore, in comparison to other domains, geodesic balls can be scaled up unambiguously in curved as well as in flat space as long as geodesics don't cross. Note that the radius of geodesic balls is not directly proportional to the third root of its volume as in Eq. \eqref{genposunc}. However, for small balls, an assumption made below, the proportionality holds approximately.

\begin{figure}[h!]
\centering
\includegraphics[width=\linewidth]{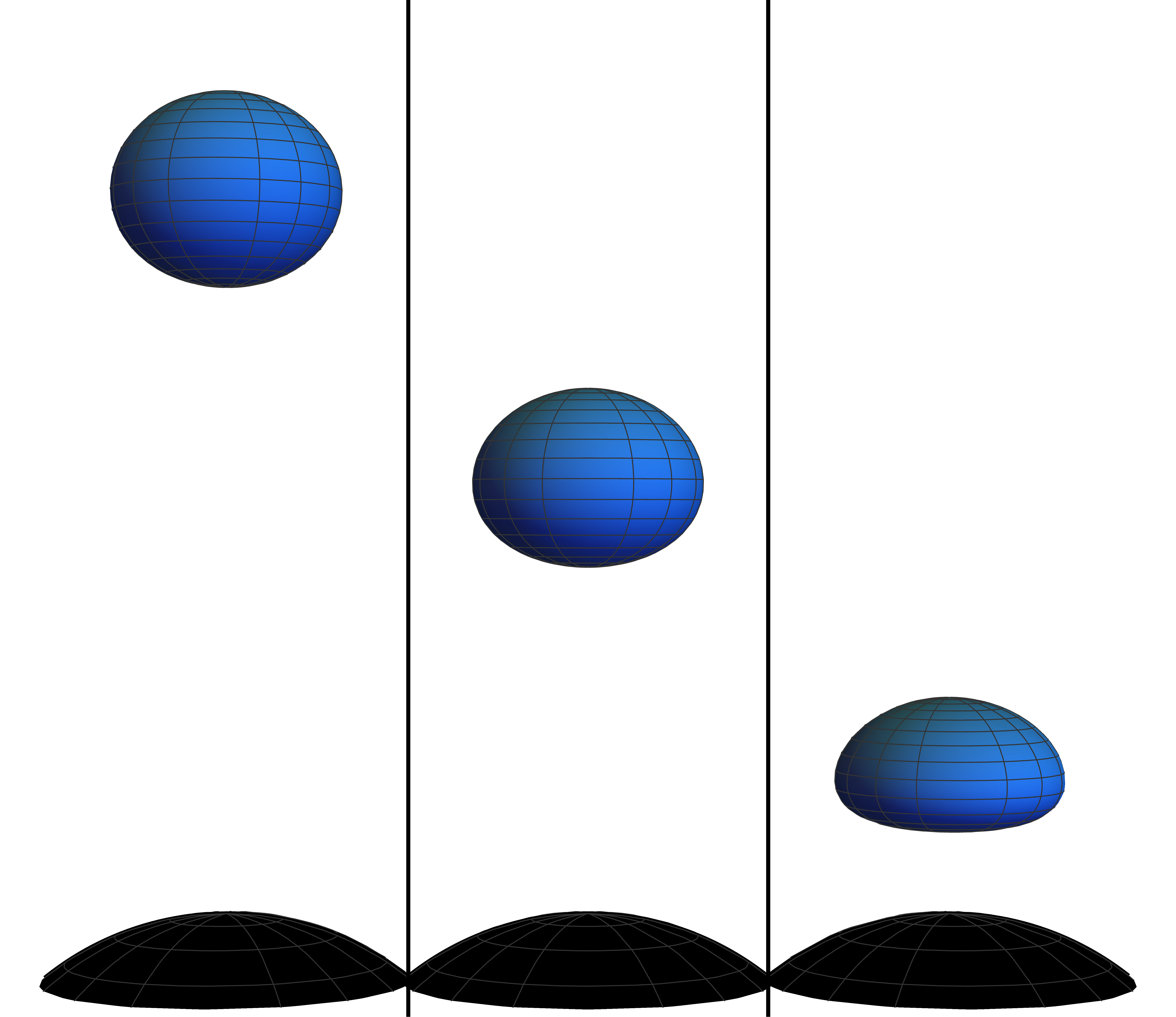}
\caption{Three geodesic balls with equal geodesic radius $\rho=.4R_S$ but different distances from the center of symmetry $r_0=3.5R_S$ (left), $r_0=2.5R_S$ (mid) and $r_0=1.5R_S$ (right) on a spatial section of the Schwarzschild static patch, characterized by the Schwarzschild radius $R_S,$ and described in terms of Schwarzschild coordinates. Surfaces of geodesic balls are coloured blue, while black hole horizons are indicated in black. \label{fig:balls}}
\end{figure}

Counterintuitively, the shape of geodesic balls depends on the chosen set of coordinates and the background geometry, and does not need to even closely resemble ordinary balls in flat space described by Cartesian coordinates. All that can be said about their appearance before choosing a coordinate system consists in the fact that their surface is topologically a sphere. An example of said variability is provided in Fig. \ref{fig:balls}. In this illustration the background consists of constant time slices of the Schwarzschild static patch described by Schwarzschild coordinates. In order to illuminate the distortion, three geodesic balls of equal geodesic radius $\rho$ but different coordinate distances $r_0$ between their center ($p_0$) and the spatial center of symmetry (origin) are compared. The corresponding calculations were done numerically thereby not invoking the small ball approximation made below. Clearly, the distortion of the geodesic balls increases with increasing curvature, \ie decreasing coordinate distance from the center of symmetry.

Fixing $p_0,$ the biscalar $\sigma(x_0,x)$ becomes a scalar field $\sigma(x),$ which measures the distance along the shortest geodesic connecting the center of the geodesic ball $p_0$ and a general point $p$ (here $x^i_0$ and $x^i$ describe the coordinate positions of the points $p_0$ and $p,$ respectively). Hence, it has to satisfy the differential equation
\begin{align}
g^{ij}\partial_i\sigma\partial_j\sigma=1,
 \label{def_geo_dist} 
\end{align}
and the boundary condition $\sigma(x_0)=0.$ As it consistently describes surface normals of the spheres limiting the geodesic balls, we interpret this scalar as radial coordinate. The boundary of the domain $\partial B_\rho$ can then be conveniently defined by the relation $\sigma=\rho.$ 

The coordinates, whose exterior derivatives are normal to $\D\sigma,$ can be understood as surface parameters of geodesic spheres of radius $\sigma.$ In this thesis we denote these as $\chi$ and $\gamma$ to construct the geodesic coordinate system $\sigma^i=(\sigma,\chi,\gamma).$ Imposing orthogonality, the "angular" coordinates have to satisfy
\begin{align}
g^{ij}\partial_i\sigma\partial_j\chi=g^{ij}\partial_i\sigma\partial_j\gamma=0\label{ortho}.
\end{align}
From a geometric point of view, these coordinates describe the background in such a way that the geodesic balls are not distorted giving them the appearance of ordinary balls. This might be quite unnatural for arbitrary metrics, but it simplifies the solution to the eigenvalue problem \eqref{evp} enormously.

\subsubsection*{Interlude - Geodesic coordinates and gravitational waves}

The geodesic coordinates satisfying Eqs. \eqref{def_geo_dist} and \eqref{ortho} are reminiscent of the set which is usually employed in the description of gravitational waves \cite{Schutz85}. In fact, both coordinate systems show striking similarities. In this vein, the coordinate describing the geodesic direction of propagation of the gravitational waves is analogous to the geodesic distance coordinate employed throughout this thesis. The other two coordinates are normal to the geodesic motion, \ie they extend along hypersurfaces of constant geodesic distance, and can thus be considered analogous to the "angular"  subset of the geodesic coordinates. 

However, the representation which is normally used to describe gravitational waves is given in the so-called transverse-traceless (TT) gauge, which provides four conditions in the 4-dimensional Lorentzian system, while in our 3-dimensional approach there are only three conditions. Therefore, the metric is not traceless in the geodesic coordinates defined above - it is merely expressed in a transverse gauge.

\subsubsection{Flat space}\label{subsubsec:flat}

Euclidean space has two simplifying advantages over other Riemannian manifolds. First, any kind of domain can be easily scaled up without ambiguities. Secondly, the local Euclidean frame is, in fact, global, described by Cartesian coordinates. Therefore, we are able to provide a result for general domains, following a simple conformal argument. Afterwards, we explicitly find the state saturating the uncertainty relation \eqref{uncgen} for the example of a geodesic ball. 

\subsubsection*{General domain}

Assume the measure of position uncertainty to be of the form \eqref{genposunc}. Increasing the volume of a general extended object in flat space by a constant factor $a^d,$ \ie transforming $\rho\rightarrow \tilde{\rho}=a\rho,$ is then equivalent to a conformal transformation of the metric $\delta_{ij}\rightarrow a^2\delta_{ij}.$ Correspondingly, the Laplacian transforms as $\Delta\rightarrow\Delta/a^2,$ and, therefore, the $\text{n}^{\rm{th}}$ eigenvalue of the transformed Laplacian, denoted $\tilde{\lambda}_n$, satisfies
\begin{equation}
    \left(\frac{\Delta}{a^2}+\tilde{\lambda}_n\right)\tilde{\psi}_n=0.
\end{equation}
Evidently, the eigenvalues of the Laplacian transform accordingly, \ie $\tilde{\lambda}_n=\lambda_n/a^2.$ Hence, we immediately see that
\begin{equation}
    \frac{\lambda_n}{\tilde{\lambda}_n}=\left(\frac{\tilde{\rho}}{\rho}\right)^2.
\end{equation}
As $C_n=\tilde{\lambda}_n\tilde{\rho}^2$ is just a dimensionless parameter independent of the scale $a,$ the entire dependence of the eigenvalues of the Laplacian on it has to be summarized in $\rho^{-2}.$ Thus, we obtain
\begin{equation}
    \lambda_n (a)=\frac{C_n}{\rho^2(a)},
\end{equation}
where the exact value of $C_n$ depends on the shape of the domain and the exact form of the position uncertainty $\rho$. Clearly, we could have also arrived at this result by pure dimensional analysis

If the state saturating the uncertainty relation is an eigenvector of the Laplacian, represented as $n=1$, as is shown explicitly below for geodesic balls, the uncertainty relation in flat space reads
\begin{equation}
    \sigma_\pi\rho\geq\hbar C_1,
\end{equation}
which shows the same dependence as Heisenberg's celebrated inequality \eqref{origheisunc}. The value of $C_1$ is determined in the subsequent section for the specific choice of domain being a geodesic ball.

\subsubsection*{Geodesic ball}

Rewritten in terms of spherical coordinates $\sigma^i=(\sigma,\chi,\gamma)$ and the explicit quantum numbers in three dimensions $n,l$ and $m,$ the eigenvalue problem \eqref{evp} becomes
\begin{align}
   \left[ \partial_{\sigma}^2+\frac{2}{\sigma}\partial_{\sigma}+\frac{1}{\sigma^2}\left(\partial_{\chi}^2+\cot\chi\partial{\chi}+\sec^2\chi\partial_{\gamma}^2\right)+\lambda^{(0)}_{nlm}\right]\psi^{(0)}_{nlm}=&0,\\
   \left.\psi^{(0)}_{nlm}\right|_{\sigma =\rho}=&0,
\end{align}
where the superscript $(0)$ stands for the zeroth order of the perturbative expansion we perform below. This problem can be solved by separation of variables, yielding the result
\begin{align}
\psi^{(0)}_{nlm}	&=\sqrt{\frac{2}{\rho^3j^2_{l+1}(j_{l,n})}}j_{l}\left(j_{l,n}\frac{\sigma}{\rho}\right)Y^l_m(\chi,\gamma)\label{flatsol}\\
\lambda^{(0)}_{nlm}		&=\left(\frac{j_{l,n}}{\rho}\right)^2\label{flateig}
\end{align}
with the spherical harmonics $Y_m^l,$ the spherical Bessel function of first kind $j_l(x)$ and the $\text{n}^{\text{th}}$ zero of the spherical Bessel function of first kind $j_{l,n}.$ In particular, as is shown below, the state saturating the uncertainty relation is the ground state of the Laplacian, which reads
\begin{equation}
    \psi^{(0)}_{100}=\frac{1}{\sqrt{2\pi\rho}}\frac{\sin\left(\pi\frac{\sigma}{\rho}\right)}{\sigma},\label{unperteig100}
\end{equation}
and corresponds to the eigenvalue $\lambda_{100}^{(0)}=\pi^2/\rho^2.$ This constitutes a rather intuitive result because, being the ground state, it is the only distinguished state in the system. Let us have a closer look at this problem, though.

According to Eq. \eqref{genstaten}, every general state $\ket{\Psi}\in\hil^{(0)}$ can be expressed in terms of the basis \eqref{flatsol}. From there we can derive the standard deviation of the momentum operator \eqref{stddevmomgen}, whose global minimum we want to find. Thus, the momentum uncertainty increases strongly with greater quantum numbers $n$ and $l$ and can be decreased by contributions to the expectation value of the momentum operator.
However, by virtue of Eq. \eqref{vanishingmom} such terms can only stem from relative phases in linear combinations of basis states. In short, we have to concentrate on complex linear combinations. 

Since, as can be checked explicitly, transition amplitudes featuring the momentum operator follow the proportionality
\begin{equation}
\Braket*{\psi^{(0)}_{nlm}}{\hatslashed{\pi}             \psi^{(0)}_{n'l'm'}}\propto\delta_{l,l'\pm 1},
\end{equation}
a linear combination of $N$ eigenstates of the Laplacian just contributes $(N-1)$ terms to the uncertainty relation weighted by coefficients satisfying Eq. \eqref{coeflincomb}. Hence, such a combination of any number of basis states in Eq. \eqref{genstaten} shows the same behaviour as a linear combination of just two of them. Therefore, no further decrease of the momentum uncertainty can be achieved by combining more than two states, which is why we only deal with this case. Up to a global phase, such a state generally reads
\begin{equation}
    \Phi=\sqrt{a} \psi^{(0)}_{nlm}+e^{i\phi}\sqrt{1-a}\psi^{(0)}_{n'l'm'},\label{lincomb2}
\end{equation}
with the real coefficient $a\in (0,1)$ and the relative phase $\phi\in [0,2\pi).$
For the above state, we obtain the momentum uncertainty 
\begin{align}
    \sigma_\pi(\Phi)  =&\Big[\hbar^2\frac{aj_{l,n}^2+(1-a)j_{l',n'}^2}{\rho^2}-4a(1-a)\text{Re}\left(e^{i\phi}\Braket*{\psi^{(0)}_{nlm}}{\hatslashed{\pi} \psi^{(0)}_{n'l'm'}}\right)^2\Big]\\
    \geq & \Big[\hbar^2\frac{aj_{l,n}^2+(1-a)j_{l',n'}^2}{\rho^2}-4a(1-a)\text{MaxRe}\Braket*{\psi^{(0)}_{nlm}}{\hatslashed{\pi} \psi^{(0)}_{n'l'm'}}^2\Big],
\end{align}
where MaxRe stands for a choice of $\phi$ such that the real part is maximal. This quantity is evaluated in detail in appendix \ref{app_maxre}, from which we can extract the inequality \eqref{res_app_maxre}
\begin{align}
    \text{MaxRe}\left[e^{i\phi}\Braket*{\psi_{n'l'm'}}{\hatslashed{\pi}\psi_{nlm}}\right]^2\leq & \left(\frac{\hbar}{\rho}\right)^2\frac{\lambda_{nl}\lambda_{n'l'}}{\left(\lambda_{n'l'}-\lambda_{nl}\right)^2}\nonumber\\
    &\times \Big\{\delta^{l'+1}_{l}\left(\delta^{m+1}_{m'}+\delta^{m-1}_{m'}+\delta^m_{m'}\right)\nonumber\\
    &+\delta^{l'}_{l+1}\left(\delta^{m}_{m'+1}+\delta^{m}_{m'-1}+\delta^m_{m'}\right)\Big\}.\label{maxrefromapp}
\end{align}
The quantity on the right-hand side is nonvanishing only for $\Delta l=|l'-l|=1$ and $\Delta m=|m'-m|=0,1,$ but its magnitude is independent of $\Delta m$. Thus, we can estimate
\begin{equation}
    \text{MaxRe}\left[e^{i\phi}\Braket*{\psi_{n',l',m'}}{\hatslashed{\pi}\psi_{nlm}}\right]^2\leq \left(\frac{\hbar}{\rho}\frac{j_{l',n'}j_{l,n}}{j_{l',n'}^2-j_{l,n}^2}\right)^2,
\end{equation}
where  $\Delta l=1$ and $\Delta m=0,1$ are understood from this point onwards.

In order for the ground state of the Laplacian not to be the one of smallest uncertainty, there should be a state $\Phi$ obeying the inequality $\sigma_\pi(\psi_{100})>\sigma_\pi(\Phi)$ which can be recast as
\begin{equation}
    \left(\frac{j_{l',n'}j_{l,n}}{j_{l',n'}^2-j_{l,n}^2}\right)^2\geq\left(\frac{\rho}{\hbar}\text{MaxRe}\Braket*{\psi^{(0)}_{n',l',m'}}{\hatslashed{\pi}\psi^{(0)}_{nlm}}\right)^2>C(a),\label{ineqstart}
\end{equation}
where we introduced the function
\begin{equation}
    C(a)=\frac{j_{l',n'}^2+\frac{a}{1-a}j_{l,n}^2-\frac{\pi^2}{1-a}}{4a}.
\end{equation}
As for the allowed $n,l$ we have $\pi\leq j_{n,l},$ this function diverges positively for $a\rightarrow 0$ and $a\rightarrow 1$ and is continuous in between. Hence, it has to reach a minimum at
\begin{equation}
    a_{min}=\frac{j_{l',n'}^2-\pi^2-\sqrt{j_{l,n}^2j_{l'n'}^2-\pi^2j_{l'n'}^2-j_{l,n}^2\pi^2+\pi^4}}{j_{l',n'}^2-j_{n,l}^2}.
\end{equation}
Observe that $j_{l',n'}=\pi,$ \ie $(n,l)=(1,0),$ immediately implies that $a_{min}=1$ while $j_{l,n}=\pi,$ meaning $(n',l')=(1,0),$ would lead to $a_{min}=0.$ In other words, linearly combining the ground state of the Laplacian with any distinct eigenstate cannot lead to a decrease in the uncertainty. For a general $0\leq a_{min}\leq 1,$ we deduce that
\begin{equation}
    C(a)\geq C(a_{min})=\frac{1}{4}\left(j_{l,n}^2+j_{n',l'}^2-2\pi^2+2\sqrt{\left(\pi^2-j_{l,n}^2\right)\left(\pi^2-j_{l',n'}^2\right)}\right).
\end{equation}
Thus, the transition amplitude of any linear combination of two basis states whose uncertainty is smaller than the one of the ground state has to satisfy the inequality
\begin{equation}
    \left(\frac{j_{l',n'}j_{l,n}}{j_{l',n'}^2-j_{l,n}^2}\right)^2>C(a_{min}),\label{ineq}
\end{equation}
which is independent of the parameters $a$ and $\phi.$ As will be argued below, this assumption leads to a contradiction.

The inequality \eqref{ineq} is invariant under the exchange $j_{l,n}\leftrightarrow j_{l',n'}.$ Therefore, assuming that $j_{l',n'}>j_{l,n}$ without loss of generality, it can be weakened and rearranged to read
\begin{equation}
    \frac{1}{\left(1-\frac{j_{l,n}^2}{j_{l',n'}^2}\right)^2}>-\pi^2+j_{l,n}^2.
\end{equation}
%At this point we have to distinguish the cases $j_{l',n'}\gg j_{l,n}$ and $j_{l',n'}\gtrsim j_{n,l}.$ In the former case this inequality becomes asymptotically
%\begin{equation}
%    1>-\pi^2+j_{n,l}^2,
%\end{equation}
%thus being trivially satisfied as long as $(n,l)\neq (1,0),$ a case that has been ruled out above. 
The left-hand side of this inequality increases as $j_{l',n'}$ and $j_{l,n}$ approach each other. Further recall, that Eq. \eqref{maxrefromapp} implies that we have to consider transition amplitudes of states satisfying either $l=l'+1$ or $l'=l+1.$ In both cases we want to minimize the quantity $|j_{l,n}-j_{l',n'}|$ subject to the constraint that $j_{l',n'}>j_{l,n}.$ It is a well-known mathematical fact \cite{Watson44} that the zeros of the Bessel functions $J_\nu$ and $J_{\nu+1}$ (here we denote the $\text{n}^\text{th}$ zero as $J_{\nu,l}$) are interlaced as $J_{\nu ,n}<J_{\nu +1,n}<J_{\nu ,n+1}.$ This carries over to the spherical Bessel functions $j_l\propto J_{l+1/2}.$ Therefore, the states maximising the left-hand side of the inequality \eqref{ineq} are characterized by the relations $l=l'+1,$ $n'=n+1$ and $l'=l+1,$ $n'=n$ leading to the stronger inequalities
\begin{align}
    \left(1-\frac{j_{l,n}^2}{j_{l+1,n}^2}\right)^2\left(j_{l,n}^2-\pi^2\right)< &1,\\
    \left(1-\frac{j_{l,n}^2}{j_{l-1,n+1}^2}\right)^2\left(j_{l,n}^2-\pi^2\right)< &1,\label{tempineq}
\end{align}
respectively. The zeroes of spherical Bessel functions do \emph{not} obey these inequalities unless $(l,n)=(1,0),$ a case that has been ruled out above. It is instructive to apply McMahon's expansion for large $n$ (\cf Ref. \cite{Watson44}) $j_{l,n}\simeq n+l/2,$ from which we can derive that
\begin{align}
    \left(1-\frac{j_{l,n}^2}{j_{l+1,n}^2}\right)^2> &\frac{\pi^2}{j_{l+1,n}^2},\\
    \left(1-\frac{j_{l,n}^2}{j_{l-1,n+1}^2}\right)^2> &\frac{j_{l+1,n}^2\pi^2}{j_{l-1,n+1}^4}.
\end{align}
Plugging those relations into the inequality \eqref{tempineq}, thereby further weakening it, we obtain
\begin{align}
    \pi^2\frac{j_{l,n}^2-\pi^2}{j_{l+1,n}^2}&<1,\\
    \pi^2\frac{j_{l+1,n}^2}{j_{l-1,n+1}^2}\frac{j_{l,n}^2-\pi^2}{j_{l-1,n+1}^2}&<1.
\end{align}
The left-hand sides of both inequalities are monotonically increasing with $n$ and $l.$ Excluding linear combinations containing the ground state, we can then estimate that
\begin{align}
    4.9\simeq &\pi^2\frac{j_{0,2}^2-\pi^2}{j_{1,2}^2}<\pi^2\frac{j_{l,n}^2-\pi^2}{j_{l+1,n}^2}<1,\\
    2.2\simeq &\pi^2\frac{j_{1,2}^2}{j_{0,3}^2}\frac{j_{1,2}^2-\pi^2}{j_{0,3}^2}<\pi^2\frac{j_{l+1,n}^2}{j_{l-1,n+1}^2}\frac{j_{l,n}^2-\pi^2}{j_{l-1,n+1}^2}<1.
\end{align}
Thus, by weakening the statement \eqref{ineqstart}, we obtain a contradiction. This, in turn, implies that said inequality is not obeyed by any linear combination of eigenstates of the Laplacian.

\begin{figure}[h!]
    \centering
    \includegraphics[width=\linewidth]{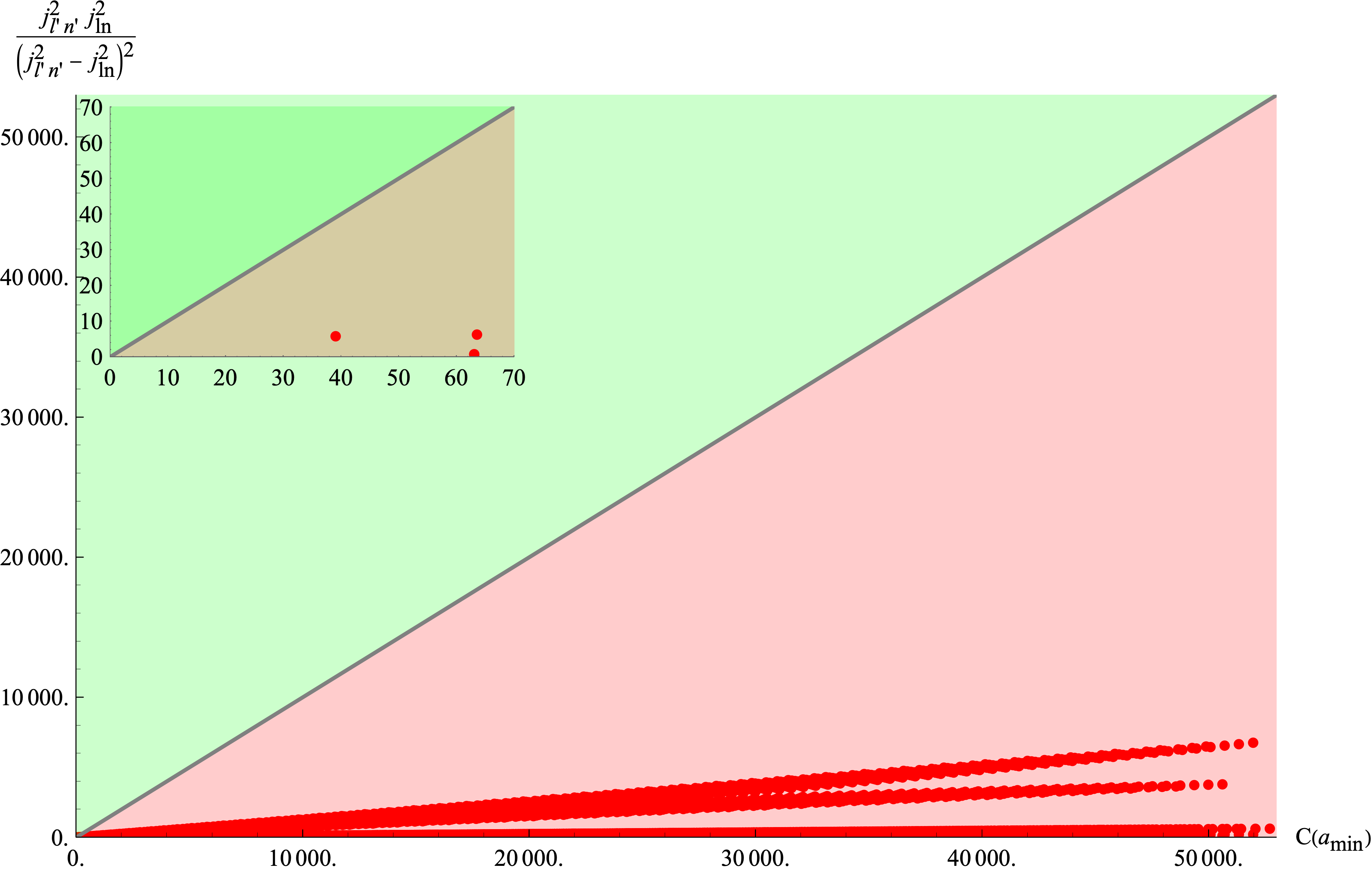}
    \caption{Visualisation of the left side and the right-hand sides of inequality \eqref{ineq} for linear combinations of all eigenstates of the Laplacian with $n\leq 50,$ $n'\leq 51$. Numerical results are symbolized by red dots. The inequality is satisfied in the green area, while it does not hold in the red area. The plot set in the upper left corner depicts the behaviour closer to the origin, \ie for smaller $n,n'$.}
    \label{fig:ineq}
\end{figure}

This result was clearly premised on the accuracy of McMahon's expansion,\ie it applies to large $n,n'.$ In order to shed more light on the problem and provide further evidence for the correctness of our claim, the right and left-hand sides of the inequality \eqref{ineq} are displayed in Fig. \ref{fig:ineq} for all allowed $(n,l,n',l')$ with $n\leq 50$ and $n'\leq 51$ where the green colour points at the area, in which it is satisfied. Obviously, none of the investigated states obeys the inequality \eqref{ineq} and the difference to the line separating the two generally increases with increasing $n,l,n',l'.$ There is no reason to expect this to change for higher values of $n,l,n',l'.$ 

Therefore, we conclude that the ground state $\psi^{(0)}_{100}$ indeed saturates the uncertainty relation, \ie
\begin{equation}
    \Psi_0=\psi_{100}^{(0)}.\label{lowestunc}
\end{equation}
As perturbations (see section \ref{subsubsec:perturb}) are small by definition, this insight immediately carries over to the slightly curved case. Furthermore, we can infer from this result, that $C_1=\pi$ for geodesic balls yielding the flat-space inequality
\begin{equation}
    \sigma_\pi\rho\geq\pi\hbar.\label{flatres}
\end{equation}
This resembles but does not equal Heisenberg's relation because it describes a different setup which is nonlinearly related to the original one.

It was generalized to manifolds of constant curvature $K$ in ref. \cite{Schuermann18}, culminating in the inequality
\begin{equation}
\label{Kgeom}
\sigma_\pi \rho\geq \pi\hbar \sqrt{1-\frac{K}{\pi^2}\rho^2}.
\end{equation}
However, in the course of said derivation, the author neglects complex linear combinations of basis vectors, thus implicitly considering only real wave functions. Then, Eq. \eqref{vanmomrealwave} immediately implies Eq. \eqref{lowestunc}, simplifying the problem. Although an analogous argument to the one made in this section can probably also be applied to spaces of constant curvature and the result \eqref{Kgeom} definitely holds in the slightly curved case as we verify in section \ref{subsubsec:perturb}, this result should still be taken with a grain of salt.

The power of the formalism introduced here, in the flat case akin to using a sledge-hammer to crack a nut, is shown to unfold at full strength in the subsequent subsection, where we obtain curvature induced corrections to the relation \eqref{flatres} perturbatively.

\subsubsection{Perturbing around flat space}\label{subsubsec:perturb}

On the base of the results obtained in the preceding section, we can start adding slight inhomogeneities to the problem.  To this aim, we first discuss the perturbative expansion of required quantities derived from the metric. Then, we deduce the consequences for the eigenvalue problem, and spend some more time on geodesic coordinates to highlight their relevance for the task.

\subsubsection*{Curvature-related quantities}

Consider a generic perturbative curvature effect on a three-dimensional Riemannian manifold, \ie for a spatial metric splitting
\begin{align}
\D s^2=\left[g^{(0)}_{ij}+ g^{(1)}_{ij}+ g^{(2)}_{ij}\right]\D x^i \D x^j ,
\label{splitting}
\end{align}
where $g^{(0)}_{ij}$ denotes the unperturbed, in the present case flat, metric in some set of coordinates $x^i$, $i=1, 2, 3$ (bear in mind that the sub- and superscripts $(0),$ $(1)$ and $(2)$ denote the perturbation order, and should not be understood as covariant or contravariant indices). Assuming the hierarchy $|g^{(2)}_{ij}|\ll |g^{(1)}_{ij}|\ll|g^{(0)}_{ij}|,$ the perturbation is treated at second order throughout this section. Correspondingly, the inverse metric can be approximated as
\begin{align}
\left[g^{(0)}_{ij}+ g^{(1)}_{ij}+ g^{(2)}_{ij}\right]^{-1}		
&\simeq g^{ij}_{(0)}-  g^{ik}_{(0)}g^{jl}_{(0)}g^{(1)}_{kl}+\left[g^{ik}_{(0)}g^{ml}_{(0)}g^{(1)}_{kl}g^{(1)}_{mn}g^{nj}_{(0)}-g^{ik}_{(0)}g^{jl}_{(0)}g^{(2)}_{kl} \right]\\ \nonumber
																			&=g^{ij}_{(0)}- g_{(1)}^{ij}+\left[g_{(1)}^{ik} g_k^{(1)j}- g_{(2)}^{ij}\right] ,
\end{align}
where the last line was just given for notational reasons, \ie to show that the unperturbed metric may be used to rise and lower indices. Furthermore, the determinant of the metric reads
\begin{align}
g\simeq g^{(0)}\left(1+g^{(1)}_{ij}g_{(0)}^{ij}\right).\label{detmetpert}
\end{align}
This perturbative expansion carries over to all curvature related quantities. In particular, the Laplace-Beltrami operator, defined in Eq. \eqref{squaredmom}, can generally be rewritten as
\begin{align}
\Delta=g^{ij}\left(\partial_i\partial_j-\Gamma^{l}_{ij}\partial_l\right),
\end{align}
an expression which is clearly affected by the perturbation. Subjecting the background manifold to the metric splitting \eqref{splitting}, the Christoffel symbols become (up to \nth{2} order)
\begin{align}
\Gamma_{ij}^k\simeq\prescript{(0)}{}{\Gamma}_{ij}^k+\prescript{(1)}{}{\Gamma}_{ij}^k+ \prescript{(2)}{}{\Gamma}_{ij}^k,
\end{align}
with the unperturbed symbol $\prescript{(0)}{}{ \Gamma}^{k}_{ij}$ and the higher-order expressions
\begin{align}
\prescript{(1)}{}{\Gamma_{ij}^k}=\frac{1}{2}	&\left[g_{(0)}^{kl}\left(\partial_i g^{(1)}_{jl}+\partial_j g^{(1)}_{il}-\partial_l g^{(1)}_{ij}\right)-g_{(1)}^{kl}\left(\partial_i g^{(0)}_{jl}+\partial_j g^{(0)}_{il}-\partial_l g^{(0)}_{ij}\right)\right] ,\\
\prescript{(2)}{}{\Gamma_{ij}^k}=\frac{1}{2}	&\Big[	g_{(0)}^{kl}\left(\partial_i g^{(2)}_{jl}+\partial_j g^{(2)}_{il}-\partial_l g^{(2)}_{ij}\right)+\left(g_{(1)}^{km}g^{(1)l}_{m}-g_{(2)}^{kl}\right)\left(\partial_i g^{(0)}_{jl}+\partial_j g^{(0)}_{il}-\partial_l g^{(0)}_{ij}\right)\nonumber\\
																		&-g_{(1)}^{kl}\left(\partial_i g^{(1)}_{jl}+\partial_j g^{(1)}_{il}-\partial_l g^{(1)}_{ij}\right)\Big] .
\end{align}
Accordingly, the Laplace-Beltrami operator inherits an order-by-order modification 
\begin{align}
\Delta\simeq\Delta^{(0)}+ \Delta^{(1)}+ \Delta^{(2)} ,\label{lapopexpans}
\end{align}
with the unperturbed operator $\Delta^{(0)}$ and the respective higher-order corrections
\begin{align}
\Delta^{(1)}	=	&-g_{(1)}^{ij}\partial_i\partial_j-\left (g_{(0)}^{ij}\prescript{(1)}{}{\Gamma_{ij}^k}-g_{(1)}^{ij}\prescript{(0)}{}{\Gamma_{ij}^k}\right)\partial_k ,\\
\Delta^{(2)}	=	&\left(g_{(1)}^{ik}g^{(1)j}_{k}-g_{(2)}^{ij}\right)\partial_i\partial_j\nonumber\\
&-\Big[g_{(0)}^{ij}\prescript{(2)}{}{\Gamma_{ij}^k}+\left(g_{(1)}^{ik}g^{(1)j}_{k}-g_{(2)}^{ij}\right)\prescript{(0)}{}{\Gamma_{ij}^k}-g_{(1)}^{ij}\prescript{(1)}{}{\Gamma_{ij}^k}\Big]\partial_k.\label{pertlap}
\end{align}
This clearly has an effect on the eigenvalue problem \eqref{evp}, which is the task of the subsequent section.

\subsubsection*{Eigenvalue problem and uncertainty relation}

Having put the Laplace-Beltrami operator into place, the eigenvalue problem \eqref{evp1} can be separated order by order expanding the eigenvalues and -functions as
\begin{align}
\lambda	&=\lambda^{(0)}+ \lambda^{(1)}+\lambda^{(2)} ,\\
\psi			&=\psi^{(0)}+ \psi^{(1)}+\psi^{(2)}.
\end{align} 
In comparison to standard time-independent nonsingular perturbation theory in quan\- tum mechanics, there is an additional subtlety arising at this point, though. According to Eq. \eqref{detmetpert}, perturbing the metric goes hand in hand with a perturbation of its determinant, which is part of the measure $\D\mu,$ with respect to which we define the scalar product of the investigated Hilbert space in accordance with section \ref{subsec:curved}. This translates to a perturbatively expanded measure $\D\mu\simeq\D\mu^{(0)}+\D\mu^{(1)},$ which, in turn, leads to a perturbation of the scalar product in the position representation $\braket{,}\simeq\braket{,}_0+\braket{,}_1.$ Thus, the structure of the Hilbert space itself, as it is expressed by the position representation, is altered by the perturbation.

The unperturbed operator $\Delta^{(0)}$ is self-adjoint  with respect to $\D\mu_0,$ \ie it is only almost (meaning up to higher-order corrections) self-adjoint with respect to $\D\mu.$ Hence, the solutions of the unperturbed eigenvalue problem
\begin{align}
\left(\Delta^{(0)}+\lambda_n^{(0)}\right)\psi_n^{(0)}=0
\end{align}
only furnish an almost orthonormal basis of $\hil.$ How to deal with this unusual kind of perturbation is explained in detail in appendix \ref{app_pert}, where this method is applied to an abstract operator $\op,$ in this case corresponding to $\op=-\Delta.$ In the main text we content ourselves with giving the perturbative corrections to the eigenvalues. At first order, we obtain
\begin{equation}
\lambda_n^{(1)}	=-\int\D\mu_0\psi_n^{(0)\dagger}\Delta^{(1)}\psi_n^{(0)},\label{eig1}
\end{equation}
which resembles ordinary perturbation theory.  In particular, the expectation value is evaluated with respect to the unperturbed measure. There is a significant change, though, to second order. The corresponding correction reads
\begin{align}
\lambda_n^{(2)}=	&-\Braket*{\psi_n^{(0)}}{\Delta^{(2)}\psi_n^{(0)}}_0-\sum_{m\neq n}\frac{\Braket*{\psi_m^{(0)}}{\Delta^{(1)}\psi_n^{(0)}}_0\Braket*{\psi_n^{(0)}}{\Delta^{(1)}\psi_m^{(0)}}_0}{\lambda_m^{(0)}-\lambda_n^{(0)}}\nonumber\\
							&-\Braket*{\psi_n^{(0)}}{\Delta^{(1)}\psi_n^{(0)}}_1+\sum_{m}\Braket*{\psi_m^{(0)}}{\Delta^{(1)}\psi_n^{(0)}}_0\Braket*{\psi_n^{(0)}}{\psi_m^{(0)}}_1,\label{eig2}
\end{align}
where, recall,  the subscripts to the Dirac brakets yield the perturbation-order of the scalar product.

Based on these results, we can determine the uncertainty relation perturbatively. Applying Eq. \eqref{vanishingmom} and the result in flat space \eqref{lowestunc}, which, as has been alluded to above, carries over to the slightly curved case, to Eqs. \eqref{stddevmomgen} and \eqref{uncgen}, we obtain
\begin{equation}
\sigma_\pi\rho\geq\sigma_\pi\left(\psi_{100}\right)\rho	\simeq\hbar\sqrt{\lambda^{(0)}_{100}+\lambda^{(1)}_{100}+\lambda^{(2)}_{100}}.%\simeq\pi\hbar\left[1+\frac{\lambda_{100}^{(1)}}{2\pi^2}+\frac{4\pi^2\lambda_{100}^{(2)}-\left(\lambda_{100}^{(1)}\right)^2}{8\pi^4}\right].
\end{equation}
Thus, the ensuing task consists in calculating the corrections to the eigenvalue of the ground state of the Laplace-Beltrami operator. 

However, there is another subtlety in the execution of this exercise we have tacitly glossed over until now. The eigenvalue problem \eqref{evp} consists not only of the differential equation \eqref{evp1} - the boundary conditions have to be taken into account as well. Depending on the choice of coordinates, they can depend on the perturbation order as well as all the other ingredients. In particular, this happens for ordinary spherical coordinates as those used in section \ref{subsubsec:flat}, creating inhomogeneities that mix up the order of perturbation. Fortunately, there is one (and only one) family of coordinate systems that can come to the rescue!

\subsubsection*{Geodesic coordinates}

As has been touched upon in section \ref{subsubsec:geoball}, geodesic coordinates $\sigma^i=(\sigma,\chi,\gamma)$ are defined such that they mimic spherical coordinates for geodesic balls, with the geodesic distance $\sigma,$ whose exterior derivative acts as a surface normal of geodesic spheres, playing the role of the radial coordinate. The boundary condition then simply reads
\begin{equation}
    \psi|_{\sigma=\rho}=0.
\end{equation}
It is, thus, completely independent of the order of perturbation. But where did we hide the corrections? 

Recall the definitions of geodesic coordinates \eqref{def_geo_dist} and \eqref{ortho}. Indeed, we camouflaged them into the coordinates themselves. Linearising those equations according to the perturbative expansion of the metric, we can write
\begin{align}
\sigma^i\simeq\sigma_0^i+\sigma_1^i+\sigma_2^i,\label{geod_coord}
\end{align}
with the geodesic coordinates to $\text{n}^{\text{th}}$ order $\sigma^i_n.$ Similarly, we perturbatively construct partial derivatives with respect to the employed coordinates as
\begin{align}
\frac{\partial}{\partial\sigma^i}\simeq	&\frac{\partial}{\partial\sigma_0^i}- \frac{\partial \sigma_1^j}{\partial \sigma_0^i}\frac{\partial}{\partial \sigma_0^j}+\left(\frac{\partial \sigma_1^j}{\partial \sigma_0^i}\frac{\partial \sigma_1^k}{\partial \sigma_0^j}\frac{\partial}{\partial \sigma_0^k}-\frac{\partial \sigma_2^j}{\partial \sigma_0^i}\frac{\partial }{\partial \sigma_0^j}\right),\label{derivpert}
\end{align}
which works up to second order because, expanded in this way, the relation
\begin{align}
\frac{\partial\sigma^j}{\partial\sigma^i}	&\simeq\delta^j_i
\end{align}
is satisfied.

As a starting point for the perturbative treatment, we have to express the given metric in terms of geodesic coordinates. Then,  Eqs. \eqref{def_geo_dist} and \eqref{ortho} become simply
\begin{align}
g_{\sigma i}=\delta_i^\sigma,\label{geod_met_char}
\end{align}
while the remaining three independent components of the metric contain all the information about the underlying space. If as in the present case the background is flat, the unperturbed coordinates $\sigma_0^i=(\sigma_0,\chi_0,\gamma_0)$ furnish a spherical coordinate system constructed around $p_0.$

Hence, we can solve the problem employing geodesic coordinates and expressing them in terms of ordinary spherical coordinates. In which way, though, does the perturbative splitting \eqref{splitting} precisely arise in the first place?

\subsubsection{Small position uncertainties and Riemann normal coordinates\label{subsubsec:RNC}}

We can understand the expansion \eqref{splitting} to be constructed on the base of the assumption of small geodesic balls respective to background curvature length scales such as the Ricci and Kretschmann ($K\equiv R_{ikjl}R^{ikjl}$) scalars. This can be achieved explicitly by invoking Riemann normal coordinates $x^a$ \cite{Riemann84,Brewin09}. We thus demand the extent of the geodesic balls to be restricted such that the entire domain is situated in a sufficiently small neighbourhood of $p_0$ for the metric to be approximated as
\begin{align}
g_{ij}\simeq \frac{\partial x^a}{\partial \sigma^i}\frac{\partial x^b}{\partial \sigma^j}\left(\delta_{ab}-\frac{1}{3}R_{acbd}|_{p_0}x^{c}x^{d}\right),\label{RNCmet}
\end{align}
with the Riemann tensor $R_{acbd}.$ Thus, to zeroth approximation space is flat and there is no first order correction.

Correspondingly, the inverse metric reads
\begin{align}
g^{ij}		&\simeq \frac{\partial \sigma^i}{\partial x^a}\frac{\partial \sigma^j}{\partial x^b}\left(\delta^{ab}+\frac{1}{3}R^{a~b}_{~c~d}|_{p_0}x^{c}x^{d}\right),
\end{align}
where the flat space metric $\delta_{ab}$ is used to rise and lower indices. As the first- and third-order corrections to the metric vanish the second and fourth order can be treated analogously to first- and second-order corrections.

First, it is shown that Riemann normal coordinates are "geodesic Cartesian coordinates" in the sense that spherical coordinates constructed around the center of the balls $p_0$ equal geodesic coordinates $\sigma_0^i=\sigma^i$ (\cf Eq. \eqref{geod_coord}), \ie
\begin{subequations}
\label{rncgeodesic}
\begin{align}
    x^1&=\sigma\sin\chi\cos\phi,\\
    x^2&=\sigma\sin\chi\sin\phi,\\
    x^3&=\sigma\cos\chi.
\end{align}
\end{subequations}
This evidently amounts to an enormous simplification in solving the general problem analytically. To proof it, it suffices to derive Eq. \eqref{geod_met_char} from Eqs. \eqref{RNCmet} and \eqref{rncgeodesic}. The former of the two equations implies that
\begin{align}
g_{\sigma i}\simeq \frac{\partial x^a}{\partial \sigma}\frac{\partial x^b}{\partial \sigma^i}\left(\delta_{ab}-\frac{1}{3}R_{acbd}|_{p_0}x^{c}x^{d}\right).\label{stepgeodsph}
\end{align}
Note that we can associate $x^a=\sigma l^a$ with the unit radial vector
\begin{align}
l^a\equiv(\sin\chi \cos\gamma,\sin\chi \sin\gamma,\cos\chi),\label{unradvec}
\end{align}
and similarly, according to Eq. \eqref{rncgeodesic}, $\partial x^a/\partial\sigma=l^a.$ Plugging those into Eq. \eqref{stepgeodsph}, we obtain
\begin{align}
g_{\sigma i}	&\simeq l^a \frac{\partial x^b}{\partial \sigma^i}\left(\delta_{ab}-\frac{\sigma_0^2}{3}R_{acbd}|_{p_0}l^{c}l^{d}\right),
\end{align}
which by the symmetries of the Riemann tensor and the orthonormality of spherical coordinates becomes
\begin{align}
g_{\sigma i}	&=l_a \frac{\partial x^a}{\partial \sigma^i}\\
				&=\delta_i^{\sigma},
\end{align}
thereby recovering Eq. \eqref{geod_met_char} as promised. In particular, the same reasoning applies to all higher-order corrections, proving that Riemann normal coordinates are in fact geodesic Cartesian coordinates. Thus, an additional change of coordinates is rendered unnecessary and we can immediately continue with the formalism developed above.

The second-order correction to the Christoffel symbols reads \cite{Brewin09}
\begin{equation}
    \prescript{(2)}{}{\Gamma_{ab}^c}=\frac{1}{3}\left(R_{bda}^{~~~c}+R_{adb}^{~~~c}\right)|_{p_0}x^d,
\end{equation}
where we used the fact that the metric in the point $p_0$ truly equals the identity $\delta_{ab}.$ Hence, we can easily derive the curvature's influence on the Laplace-Beltrami operator
\begin{equation}
    \Delta^{(2)}=\frac{1}{3}\left(R^{a~b}_{~c~d}|_{p_0}x^cx^d\partial_a-2R_a^b|_{p_0}x^a\right)\partial_b.\label{secondordlap}
\end{equation}
As, in accordance with Eq. \eqref{eig1}, this operator is acting solely on the wave function $\psi_{100},$ which is a function of the coordinate $\sigma$ alone, the only required quantity reads
\begin{equation}
\Delta^{(2)}\psi_{100}(\sigma)	=-\frac{\sigma}{3}R_{cd}|_{p_0}l^cl^d\partial_\sigma\psi_{100}.
\end{equation}
Taking into account Eq. \eqref{eig1}, this translates into the second-order correction to the eigenvalue
\begin{equation}
\lambda_{100}^{(2)}	=\frac{1}{3}\int_0^\rho \sigma^3\psi_{100}^{(0)\dagger}\partial_\sigma\psi_{100}^{(0)}\int_{S^2}R_{ab}|_{p_0}l^al^b\D\Omega,
\end{equation}
with the solid angle $\D\Omega=\sin\chi\D\chi\D\gamma.$ The two integrals can be evaluated independently to yield
\begin{align}
\int_0^\rho \sigma^3\psi_{100}^{(0)\dagger}\partial_\sigma\psi_{100}^{(0)}	&=-\frac{3}{8\pi},\\
\int_{S^2}R_{ab}|_{p_0}l^al^b\D\Omega	&=\frac{4\pi}{3}R|_{p_0}.
\end{align}
Plugging these back in, we obtain the correction
\begin{align}
\lambda_{100}^{(2)}=-\frac{R|_{p_0}}{6},\label{curvres2}
\end{align}
which leads to the uncertainty relation to second order
\begin{align}
\sigma_\pi\rho	&\gtrsim \pi\hbar\sqrt{1-\frac{R|_{p_0}}{6\pi^2} \rho^2}\\
						&\simeq \pi\hbar\left(1-\frac{R|_{p_0}}{12\pi^2} \rho^2 \right).\label{res}
\end{align}
A higher-order treatment of the problem is possible, yet tedious. An account to fourth order is given in appendix \ref{HOCORR}. In short, the third-order correction vanishes and plugging in Eq. \eqref{res4app}, the final formula for the uncertainty approximated at fourth order reads
\begin{align}
\sigma_\pi\rho	\gtrsim	& \pi\hbar\sqrt{1-\frac{R\mid_{p_0}}{6\pi^2}\rho^2- \frac{\eta}{\pi^2}\left(\Psi_2^{~2}+\frac{\Delta R}{15}\right)\Bigg|_{p_0} \rho^4},
\label{res4}
\end{align}
with the zeroth-order Cartan invariant $\Psi_2,$ which in three dimensions satisfies \cite{Musoke16}
\begin{align}
\Psi_2^{~2}=\frac{3R^{ab}R_{ab}-R^2}{72}
\end{align}
and the numerical constant
\begin{align}
\eta=\frac{2\pi^2-3}{8\pi^2}.
\end{align} 
Hence, the uncertainty relation may also be modified on manifolds with vanishing Ricci scalar such as the one described by the induced metric on hypersurfaces of constant time in the static Schwarzschild patch. 

Up until now the investigations in the present section derive from curved spaces alone. Correspondingly, as of yet the reasoning has been based on the canonical commutation relations \eqref{HeisAlg}. Yet, modified commutation relations may be applied to the given problem as well. This idea is investigated in the subsequent section.

\subsection{Asymptotic Generalized Extended Uncertainty Principle\label{subsec:AGEUP}}

For reasons of simplicity, we contend ourselves with the version of the modified algebra yielding commutative coordinates \eqref{GUPalgdDsimp}. As explained in section \ref{subsubsec:GUPEUPdD}, we can then express the momentum operator in terms of an auxiliary quantity $\hat{\pi}_i,$ which satisfies the canonical commutation relations \eqref{HeisAlg}, \ie it represents the normal quantum mechanical momentum operator \eqref{curvedspacemom}. Following this ansatz, the GUP-deformed momentum operator $\hat{\Pi}_i$ reads
\begin{equation}
\hat{\Pi}_i=\hat{\pi}_i\left(1+\frac{\beta l_p^2}{\hbar^2}\hat{\pi}^2\right),
\end{equation}
leading to coordinate-independent scalar operator from geometric calculus
\begin{equation}
\hatslashed{\Pi}=\hatslashed{\pi}\left(1+\frac{\beta l_p^2}{\hbar^2}\hat{\pi}^2\right).
\end{equation}
All the results obtained in this section were written in terms of eigenvalues of $\hat{\pi}^2.$ As $\hatslashed{\pi}$ and $\hatslashed{\Pi}$ commute, we can simply translate the already obtained results in terms of the observable $\hatslashed{\pi}$ into equivalent quantities related to the modified operator $\hatslashed{\Pi}.$ 

\emph{Per definitionem}, we have
\begin{align}
\sigma_\Pi\equiv\sqrt{\braket*{\hatslashed{\Pi}^2}-\braket*{\hatslashed{\Pi}}^2}=\sqrt{\braket*{\hat{\Pi^2}}-\delta^{ab}\braket*{\hat{\Pi}_a}\braket*{\hat{\Pi}_b}}.
\end{align}
For all wave-functions $\psi_n$ solving the eigenvalue problem \eqref{evp} we can write
\begin{align}
\braket*{\hatslashed{\Pi}}	&=\Braket*{\psi_n}{\hatslashed{\Pi}\psi_n}\\
					&=\Braket*{\psi_n}{\hatslashed{\pi}\left(1+\frac{\beta l_p^2}{\hbar^2}\hat{\pi}^2\right)\psi_n}\\
					&=\Braket*{\psi_n}{\hatslashed{\pi}\left(1+\beta l_p^2\lambda^2\right)\psi_n}\\
					&=\braket*{\hatslashed{\pi}}\left(1+\beta l_p^2\lambda^2\right)\\
					&=0,
\end{align}
where we used Eqs. \eqref{evp} and \eqref{vanishingmom}. If the correction induced from quantum gravity is small, the ground state $\psi_{100}$ continues to saturate the uncertainty relation by the same argument as in the slightly curved case (see section \ref{subsubsec:perturb}). 
On the other hand, we have
\begin{align}
\braket*{\hat{\Pi}^2}	&\simeq\Braket*{\psi}{\hat{\pi}^2\left(1+2\frac{\beta l_p^2}{\hbar^2}\hat{\pi}^2\right)\psi}\\
							&\simeq\hbar^2\lambda\left(1+2\beta l_p^2\lambda\right).
\end{align}
This information suffices to obtain the standard deviation of the modified momentum operator leading to the uncertainty relation
\begin{align}
\sigma_\Pi\rho	&\geq\hbar\rho\sqrt{\lambda\left(1+2\beta l_p^2\lambda\right)}\\
						&\gtrsim\hbar\sqrt{\rho^2\lambda}\left(1+\beta l_p^2\lambda\right).
\end{align}
Plugging in the eigenvalue in the general case to lowest nonvanishing order \eqref{curvres2} leads to the asymptotic form of both the EUP and the GUP, \ie to the AGEUP
\begin{align}
\sigma_\Pi\rho\gtrsim	&\pi\hbar\sqrt{1-\frac{R|_{p_0}\rho^2}{6\pi^2}}\left[1+\pi^2\beta\frac{l_p^2}{\rho^2}\left(1-\frac{R|_{p_0}\rho^2}{6\pi^2}\right)\right]\\
				\simeq			&\pi\hbar\sqrt{1-\frac{R|_{p_0}\rho^2}{6\pi^2}}\left(1+\beta\frac{l_p^2\sigma_\pi^2}{\hbar^2}\right)\\
				\simeq 		&\pi\hbar\left(1-\frac{R|_{p_0}\rho^2}{12\pi^2}+\beta\frac{l_p^2\sigma_\pi^2}{\hbar^2}\right)\label{derived_GEUP}
\end{align}
for small $R|_{p_0}\rho^2$ and $l_p^2/\rho^2.$ In the special case of flat space this recovers the usual GUP
\begin{align}
\sigma_\Pi\rho	\gtrsim\pi\hbar\left(1+\beta\frac{l_p^2\sigma_\Pi^2}{\hbar^2}\right).\label{compdomgup}
\end{align}
In every other case it leads to a GEUP in the given space once the EUP is known.

\subsection{Summary}
\label{subsec:3DEUPDis} 

We presented a formalism which allows for an asymptotic derivation of the EUP on arbitrary three-dimensional Riemannian manifolds, as given by our formula \eqref{res4}.  Interestingly, the leading (second-order) curvature-induced correction is proportional to the Ricci scalar, while the fourth-order correction is proportional to the zeroth-order Cartan invariant $\Psi_2$ and the curved space Laplacian of the Ricci scalar, all evaluated at the expectation value of the position operator. 

Finally, the formalism was extended phenomenologically to combine the result with the GUP by virtue of deformed commutators. Thus, we presented an asymptotic form of the GEUP in the low energy limit given by the formula \eqref{derived_GEUP}, which is our main achievement.

These results show that ordinary quantum mechanics entails an extended uncertainty relation if we include the curvature of space. What do we make out of this result? As mentioned at the beginning of this section, the motivation for GUPs and EUPs is quite vague inasmuch as it leaves open which kind of uncertainty relation is involved and which mathematical mechanism underlies the modification. Especially, the EUP in and of itself should just be an effect implied by the semiclassical combination of gravity and quantum mechanics alone as in quantum field theory in curved spacetime. Sure enough, this is exactly what we have obtained in its nonrelativistic limit - quantum mechanics on curved backgrounds. What does this imply for the GUP, though? Following the reasoning on Born reciprocity in section \ref{subsubsec:momunc}, an analogous derivation can be carried out in curved momentum space, yielding precisely such a relation. Furthermore, Eq. \eqref{compdomgup} implies the same on the base of modified commutators. This clearly suggests a link between the two approaches.  

\section{Generalisation to curved spacetime}\label{sec:4DEUP}

The reasoning of the preceding chapter must always be understood to be derived from and approximate to an underlying relativistic theory. Curved spaces do not arise out of nowhere, they can be understood as spacelike slices of spacetime, evolving according to the field equations of general relativity. How exactly this can be incorporated is the matter of the present section.

In that regard, we first (\cf section \ref{sec: effham}) have to deal with the Hamiltonian dynamics of a generally relativistic particle. Taking its non-relativistic limit, we define the physical momentum operator in section \ref{subsec:Setup},  which, as we discover, may contain a term akin to a magnetic one-form.  From there, we use section \ref{subsec:expsol4D} to follow a derivation,  which is analogous to the one performed in the previous section to provide an uncertainty relation in accordance the spacetime picture. The results are summarized in section \ref{subsec:sum4D}. This material is taken out of Ref.  \cite{Petruzziello21}.

\subsection{Derivation of the effective Hamiltonian\label{sec: effham}}

This subsection serves as a derivation of the effective dynamics of a non-relativistic particle on a curved four-dimensional background. To this aim, the action of a massive relativistic particle, subject to a curved geometry, can be written as
\begin{align}
S	&=-m\int \D s\\
	&=-m\int \sqrt{-g_{\mu\nu}(x)\dot{x}^\mu\dot{x}^\nu} \D \tau,\label{relaction}
\end{align}
with the background metric $g_{\mu\nu},$ the four-velocity $\dot{x}^\mu=\D x^\mu/\D\tau$ ($\tau$ denotes the affine parameter along the curve) and the mass of the particle $m.$ Correspondingly, the Lagrangian, giving rise to its dynamics, reads
\begin{equation}
L=-m\sqrt{-g_{\mu\nu}(x)\dot{x}^\mu \dot{x}^\nu}.\label{lagrangian2}
\end{equation}
Note that the action \eqref{relaction} is invariant under temporal reparameterizations $\tau'=\tau'(\tau)$ for any sufficiently well-behaved function $\tau'.$

After some algebra, the Lagrangian can be recast as
\begin{align}
L=  &-m\Big[\left(-g_{00}+g_{0i}g_{0j}h^{ij}\right)(\dot{x}^0)^2-\left(\dot{x}^0g_{0k}h^{ik}+\dot{x}^i\right)\left(\dot{x}^0g_{0l}h^{jl}+\dot{x}^j\right)g_{ij}\Big]^{1/2},
\end{align}
with $h^{ik}g_{kj}\equiv\delta^i_j.$ Thus, $h^{ij}$ is the inverse of the induced metric on hypersurfaces of constant $x^0.$ The Lagrangian can be further simplified by introducing the background field quantities
\begin{align}
    N=&\sqrt{-g_{00}+g_{0i}g_{0j}h^{ij}},\\
    N^i=&g_{0j}h^{ij},
\end{align}
which are readily identified as the lapse function and the shift vector in the 3+1 formalism \cite{Arnowitt59,Arnowitt60,Arnowitt08}. According to this approach to curved Lorentzian manifolds, any metric can be expressed as \cite{Gourgoulhon12}
\begin{equation}
    \D s^2=-N^2\left(\D x^0\right)^2+h_{ij}(N^i\D x^0+\D x^i)(N^j\D x^0+\D x^j),
\end{equation}
with $h_{ij}\equiv g_{ij}.$ Thence, the Lagrangian reads
\begin{equation}
    L=-m\sqrt{N^2\left(\dot{x}^0\right)^2-\left(\dot{x}^0N^i+\dot{x}^i\right)\left(\dot{x}^0N^j+\dot{x}^j\right)h_{ij}},
\end{equation}
which under the assumption that $\dot{x}^0>0$ (\ie that coordinate time is progressing in the same direction as the particle's proper time) can be written as
\begin{align}
     L=&-mN\dot{x}^0\sqrt{1-\frac{\left(\dot{x}^0N^i+\dot{x}^i\right)\left(\dot{x}^0N^j+\dot{x}^j\right)h_{ij}}{N^2\left(\dot{x}^0\right)^2}}\\
      \equiv&-mN\dot{x}^0\sqrt{1-\vel^2},\label{LR1}
\end{align}
where the last equality defines $\vel,$ the curved spacetime analogue of the velocity in units of the speed of light in special relativity. In terms of the conjugate momenta 
\begin{equation}
    P_\mu=mg_{\mu\nu}\dot{x}^\nu/\sqrt{-g_{\mu\nu}\dot{x}^\mu\dot{x}^\nu},
\end{equation}
we find that $\vel/(1-\vel)=h^{ij}P_iP_j/m^2$, implying that the non-relativistic limit corresponds to $\vel\ll 1,$ as expected. Therefore, we can expand Eq. \eqref{LR1} to obtain the effective nonrelativistic Lagrangian
\begin{equation}
    L_{NR}=\frac{m}{2\dot{x}^0}\left(\dot{x}^0N^i+\dot{x}^i\right)\left(\dot{x}^0N^j+\dot{x}^j\right)G_{ij}-mN\dot{x}^0,\label{LNR1}
\end{equation}
with the effective 3-metric $G_{ij}=h_{ij}/N.$ For the purpose of the subsequent sections, this metric is used to lower and raise indices and as the background for differential geometric quantities.

The effective nonrelativistic action $S_{NR}=\int L_{NR}\D\tau$ still harbours the time reparameterization invariance alluded to above. For simplicity, we fix the gauge by choosing $x^0=\tau$ to obtain the nonrelativistic Lagrangian
\begin{equation}
    L_{NR}=\frac{m}{2}\left(N^i+\dot{x}^i\right)\left(N^j+\dot{x}^j\right)G_{ij}-mN.\label{nonrellag}
\end{equation}
A closer look at this function tells us that it is of the form
\begin{equation}
    L_{NR}=\frac{m}{2}\dot{x}^i\dot{x}^jG_{ij}+m\dot{x}^iA_i-m\phi,
\end{equation}
with $A_i=N^jG_{ij}$ and $\phi=N-N^iN^jG_{ij}/2.$ This is clearly reminiscent of the Lagrangian describing a charged nonrelativistic particle minimally coupled to an electromagnetic gauge one-form $A_\mu=(\phi,A_i),$ where the mass $m$ plays the r\^ole of the charge.

On the other hand, the Lagrangian is additionally invariant under the gauge transformation $A\rightarrow A_i+\partial_i f,$ $\phi\rightarrow\phi-\dot{f},$ $G^{ij}\rightarrow G^{ij}$ for any scalar function $f(x^i,t),$ while the canonical momenta
\begin{equation}
    \pi_i=\frac{\partial L_{NR}}{\partial \dot{x}^i}=mG_{ij}\left(\dot{x}^j+N^j\right)
\end{equation}
are not, rendering them unobservable. Therefore, we define the gauge-invariant physical momenta as
\begin{equation}
    p_i\equiv\pi_i-mN^jG_{ij},\label{phys}
\end{equation}
in terms of which the Hamiltonian reads
\begin{equation}
    H_{NR}=\frac{1}{2m}p_ip_jG^{ij}+m\phi.\label{effhamiltonian}
\end{equation}
Having found the effective Hamiltonian, we are now able to give a quantum mechanical description of nonrelativistic particles in curved spacetime.

\subsection{Setup\label{subsec:Setup}}

Aiming towards an uncertainty relation derived from curved spacetime, we have to find the quantum mechanical counterpart of the theory outlined in the previous section. Therefore, we need an interpretation of the mathematics of spacetime we used there. Furthermore, we need to define the physical momentum operator and its standard deviation to be well-prepared for the explicit calculation. 

\subsubsection{Spacetime picture}

First, observe that the quantity $\vel,$ defined in Eq. \eqref{LR1}, is not a spacetime scalar. Thus, when cutting off the expansion \eqref{LNR1} at any order, we break the four-dimensional diffeomorphism invariance of the problem, introducing a preferred frame. Taking the nonrelativistic limit, this was expected. However, it is not necessarily the rest-frame of the particle we are dealing with. Instead, it should be understood as the rest frame of the lab containing the device used to perform a measurement on the system. 

The time-parameter appearing in the Schrödinger equation should then be defined in accordance with this slicing while the Hilbert space is equivalent to the one introduced in section \ref{subsubsec:posunc}, featuring an analogous background metric induced measure (\cf section \ref{subsec:curved}). 

However, note that the dynamics of the massive particle on these spacelike hypersurfaces are not governed solely by the respective induced metric. The lapse function, admittedly a slicing-dependent quantity, enters as well.  As the succession of spacelike hypersurfaces and the kinematics of the quantum theory are unambiguously described, we can now go on to define the operator describing the physical momentum.

\subsubsection{Physical momentum operator and its standard deviation}\label{subsubsec:physmomop}

As we pointed out above, the effective Hamiltonian \eqref{effhamiltonian} entails a certain amount of gauge freedom,  necessitating the definition of the gauge invariant, \ie physical, momenta $p_i$ in accordance with Eq. \eqref{phys}. If we want the uncertainty relation to be associated with measurable quantities, it has to contain the extra term introduced there. Accordingly, we define the momentum operator
\begin{equation}
\hat{p}_i\psi	=\left(\hat{\pi}_i-m\hat{A}_i\right)\psi\equiv-\left[i\hbar \left(\partial_i+\frac{1}{2}\Gamma^j_{ij}\right)+mN^jG_{ij}\right]\psi,
\end{equation}
where the canonical momentum operator $\hat{\pi}_i$ and the effective gravitomagnetic one-form $\hat{A}_i$ are defined according to the second equality. As this operator is, again, a vector, we can provide its scalar version applying geometric calculus as in Eq. \eqref{newmomop}
\begin{equation}
    \slashed{\hat{p}}\psi	\equiv\left(\hatslashed{\pi}-m\hatslashed{A}\right)\psi.
\end{equation}
Thus, the aim of the present section lies in compute its standard deviation
\begin{equation}
\sigma_p\equiv \sqrt{\braket{\hatslashed{p}^2}-\braket{\hatslashed{p}}^2}=\sqrt{\braket{\hat{p}^2}-\braket{\hatslashed{p}}^2}.\label{stddevphysmom}
\end{equation}
The square of the momentum operator, required for its determination, is of the form
\begin{equation}
\hat{p}^2=\hat{\pi}^2-2m\hat{p}_{mix}^2+m^2\hat{G}_{ij}\hat{N}^i\hat{N}^j,
\end{equation}
where $\hat{\pi}^2\psi=-\hbar^2\Delta\psi$ and $\hat{p}^2_{mix}$ mixes canonical momenta and the shift vector. Under the assumptions that it is Hermitian and leads to the correct classical limit, the latter operator reads
\begin{equation}
\hat{p}^2_{mix}=	\frac{1}{2}\left\{\hatslashed{\pi},\hatslashed{N}\right\}=\frac{1}{2}\left\{\hat{\pi}_i,\hat{N}^i\right\}.
\end{equation}
After some elementary algebra, it can be seen that it acts on wave functions as
\begin{equation}
\hat{p}^2_{mix}\psi=-\frac{i\hbar}{2}\left[\nabla_i\left(N^i\right)+2N^i\partial_i\right]\psi,\label{pmix}
\end{equation}
with the covariant derivative $\nabla_i$ with respect to the canonical connection of $G_{ij}.$ Note that the first term in \eqref{pmix} is anti-Hermitian.  Since $\hat{p}^2_{mix},$ on the other hand, is Hermitian, the anti-Hermitian part of the second term is cancelled by the first one. Thus, we can rewrite $\hat{p}^2_{mix}$ as 
\begin{align}
    \hat{p}^2_{mix}\psi=\left(-i\hbar N^i\partial_i\right)_H\psi=\left(\hat{N}^i\hat{\pi}_i\right)_H,\label{pmix2}
\end{align}
where the subscript $H$ denotes the Hermitian part.

Summing up the outcome of this subsection, the relevant operators act as
\begin{align}
\slashed{\hat{p}}\psi		=&\gamma^i\hbar\left[-i\left(\partial_i+\frac{1}{2}\Gamma^j_{ij}\right)-\frac{G_{ij}N^j}{\lambdabar_C}\right]\psi,\label{newmom4D}\\
\hat{p}^2\psi	=&\hbar^2\left[-\Delta+\frac{2}{\lambdabar_C}\left(iN^i\partial_i\right)_H+\frac{N^iN^jG_{ij}}{\lambdabar_C^2}\right]\psi,\label{squmom}
\end{align} 
with the reduced Compton wavelength $\lambdabar_C=\hbar/m.$

Having discovered the position space representation of the linear and squared momentum operators in \eqref{newmom4D} and \eqref{squmom}, respectively, we can, again, define the momentum uncertainty as in Eq. \eqref{defmomstddev}. The remainder of the present section is centered around the evaluation of this quantity.

\subsection{Explicit solution}\label{subsec:expsol4D}

Following the approach in section \ref{subsubsec:RNC}, we calculate the uncertainty relation perturbatively. In particular, we now expand $G_{ab}$ in Riemann normal coordinates  $x^a$ to second order
\begin{equation}
    G_{ab}\simeq \delta_{ab}-\frac{1}{3}R_{acbd}|_{p_0}x^cx^d.
\end{equation} 
This implies the perturbation of the eigenvalue problem \eqref{evp} treated at length in section \ref{subsubsec:perturb}. However, there are additional degrees of freedom to be considered. By analogy with the metric being expanded in Riemann normal coordinates, the shift vector should be treated perturbatively too, \ie
\begin{align}
    N^a\simeq&N^a_{(0)}+N^a_{(1)}+N^a_{(2)}\\
    =&N^a|_{p_0}+\nabla_bN^a|_{p_0}x^b+\nabla_b\nabla_cN^a|_{p_0}x^bx^c.\label{shiftexp}
\end{align}
This means that, in principle, the shift vector could yield zeroth and first order corrections. In contrast to section \ref{subsubsec:RNC}, we treat the perturbation at second order.

Hence, the standard deviation of the momentum operator, possibly having extra contributions at first order, now reads
\begin{equation}
\sigma_p\simeq\sqrt{\var{p}^{(0)}+\var{p}^{(1)}+\var{p}^{(2)}},\label{expansionstddevmom4D}
\end{equation}
where we introduced the variance $\sigma_p^2.$ Again, the task consists in evaluating this quantity order by order.

\subsubsection{Flat space}\label{subsubsec:flat4D}

By analogy with Eqs. \eqref{defmomstddev} and \eqref{expansionstddevmom4D}, we introduce the variances of the conjugate momentum operator $\hatslashed{\pi}$ and the shift vector $\hatslashed{N},$ denoted $\sigma_\pi^2$ and $\sigma_N^2,$ respectively. At zeroth order, the momentum uncertainty can then be expressed as
\begin{equation}
    \sigma_p^{(0)}=\sqrt{\var{\pi}^{(0)}+\frac{2\hbar}{\lambdabar_C}N^a|_{p_0}\braket*{-\left(\hat{\pi}_a\right)_H+\hat{\pi}_a}^{(0)}+\frac{\hbar^2}{\lambdabar^2_C}\var{N}^{(0)}}\label{0thorderunc}.
\end{equation}
As the momentum operator $\pi_a$ is Hermitian, the term in the brackets vanishes. Moreover, it is a simple exercise to show that a similar cancellation occurs to the variance of the shift vector leaving us with
\begin{equation}
    \sigma_p^{(0)}=\sigma_\pi^{(0)}.
\end{equation}
Hence, it has no influence at zeroth order. These considerations hold independently of the state with respect to which the uncertainty is calculated. Thus, the reasoning employed in section \ref{subsubsec:flat} applies equivalently at this point. In particular, the state of smallest momentum uncertainty continues to be the ground state of the Laplace-Beltrami operator $\psi^{(0)}_{100},$ defined in Eq. \eqref{flatsol}, implying the eigenvalue $\lambda_{100}^{(0)}=\hbar^2\pi^2/\rho^2$ (\cf Eq. \eqref{flateig}). 

This has direct nonperturbative consequences for the problem. First, according to Eq. \eqref{pmix2}, we can generally write
\begin{equation}
\braket*{\hat{p}_{mix}^2}=-\hbar\text{Im}\int\D\mu \psi^* N^a\partial_a\psi.
\end{equation}
As the eigenstates of the Laplace-Beltrami operator $\psi_n$ are real (see section \ref{subsubsec:flat}), the integrand appearing in this case is purely real and so is the integral. Thus, evaluated with respect to those states, in particular the ground state, the expectation value of $\hat{p}^2_{mix}$ vanishes. Furthermore, terms mixing expectation values of the shift vector and the momentum vanish identically due to Eq. \eqref{vanishingmom}. Then, the variance of the momentum operator in the ground state equals
\begin{equation}
    \sigma_p^2\left(\psi_{100}\right)=\sigma_{\pi}^2\left(\psi_{100}\right)+\frac{\hbar^2}{\lambdabar_C^2}\sigma_{N}^2\left(\psi_{100}\right)
\end{equation}
nonperturbatively.

Moreover, the unperturbed uncertainty relation is clearly unaltered with respect to section \ref{subsubsec:flat} (\cf Eq. \eqref{flatres})
\begin{equation} 
    \sigma_p^{(0)}\rho\geq\pi\hbar.\label{flatunc4D}
\end{equation}
As the state of lowest momentum uncertainty is identified, it is time to evaluate the curvature- induced corrections

\subsubsection{Perturbing around flat space}\label{subsubsec:perturb4D}

Contributions to $\var{\pi}$ from the nontrivial effective metric appear at second order and the first-order correction to $\var{N}$ is subject to similar cancellations as in Eq. \eqref{0thorderunc}. Hence, this contribution to the variance of the physical momentum operator vanishes
\begin{equation}
    \var{p}^{(1)}(\psi_{100})=0,
\end{equation}
from which we deduce that the shift vector corrects the uncertainty relation at the same order as the background curvature.

Besides, we already derived the correction to the standard deviation induced by spatial curvature in Eq. \eqref{eig1}, obtaining
\begin{equation}
    \var{\pi}^{(2)}(\psi_{100})=-\frac{1}{6}R|_{p_0}.
\end{equation}
Have in mind that the Ricci scalar is deduced from the effective metric $G_{ab}$ here. Thus, we are left with the second-order correction to the variance of the shift vector, which, after cancellations, reads
\begin{align}
    \var{N}^{(2)}=\nabla_a N^i\nabla_b N^j G_{ij}|_{p_0}\left(\braket{\hat{x}^a \hat{x}^b}-\braket*{\hat{x}^a}\braket{\hat{x}^b}\right)^{(0)}.
\end{align}
When evaluated with respect to the ground state, the second term in the bracket, being an integral over an odd function, vanishes, while the first yields
\begin{equation}
    \Braket*{\psi_{100}}{\hat{x}^a\hat{x}^b\psi_{100}}^{(0)}=\frac{\rho^2}{18}\left(2-\frac{3}{\pi^2}\right)\delta^{jk}.\label{doublexint}
\end{equation}
Finally, lowering and raising indices with the effective metric $G_{ab},$ the second-order correction to the variance of the physical momentum operator equals
\begin{equation}
    \var{p}^{(2)}(\psi_{100})=-\frac{R|_{p_0}}{6}+\xi\frac{\rho^2}{2\lambdabar_C^2}\nabla_jN_i\nabla^jN^i|_{p_0},
\end{equation}
where we introduced the mathematical constant $\xi=(2-3/\pi^2)/9.$

Observe that, though it seems to be of fourth order at first glance due to the factor $\rho^2,$ the second term is actually quadratic because the expansion done here is performed in terms of $\rho\sqrt{\mathcal{R}}$ where $\mathcal{R}$ denotes any curvature invariant with dimensions of squared inverse length. 

In fact, the nonrelativistic limit implies $p^2/m^2\ll 1.$ By virtue of Eq. \eqref{flateig}, we know that the ground state of the Laplacian obeys $\braket*{\hat{p}^2}\sim(\hbar/\rho)^{2}$ (actually this holds for all its eigenstates). Adding the definition of the reduced Compton wave length, we thus obtain $\rho^2/\lambdabar^2_C\sim m^2/\braket*{\hat{p}^2}\gg 1.$ Depending on the magnitude of the shift vector, which is usually small, in the nonrelativistic context, this term can become dominant.

\subsubsection{Result}

Gathering all the results from the previous sections and introducing the Compton wavelength $\lambda_C\equiv2\pi\lambdabar_C$, we obtain the uncertainty relation
\begin{align}
    \sigma_p\rho\gtrsim &\pi\hbar\left[1-\frac{\rho^2R|_{p_0}}{12\pi^2}+\xi\frac{\rho^4}{\lambda_C^2}\nabla_jN_i\nabla^jN^i|_{p_0}\right].\label{finalunc}
\end{align}
In short, given a four-metric $g_{\mu\nu}$ and an observer defining a foliation of spacetime, the uncertainty relation \eqref{finalunc} can be computed asymptotically by evaluating the Ricci scalar derived from $G_{ab}$ and the corresponding covariant derivative of the shift vector at $p_0.$ 

The above expression can be applied to any spacetime metric written in 3+1 form, and is heavily influenced by its intrinsic characteristics. Therefore, the resulting effect is directly dependent on the background spacetime and the foliation given by the observer, \ie the measurement apparatus.

\subsection{Summary}\label{subsec:sum4D}

The present section served to include the spacetime picture into the approach to gravitationally induced uncertainty relations pursued in section \ref{sec:3DEUP}. In that vein, the four-dimensional background metric is split in accordance with the ADM decomposition, \ie into lapse function, shift vector and induced metric on hypersurfaces of constant coordinate time. Correspondingly, the dynamics of nonrelativistic particles are governed by an effective background metric, which does not equal the induced metric on hypersurfaces, the shift vector and lapse function, which, together, minimally couple to the mass of the particle by analogy with the electromagnetic one-form.

Similarly to section \eqref{sec:3DEUP}, the investigated Hilbert space consists of wave functions confined to a geodesic ball by applying Dirichlet boundary conditions. Within this domain, placed on a flat background, we find the state which saturates the uncertainty relation, the ground state of the Laplacian. Afterwards, we gradually add in curvature following the Riemann normal coordinate expansion.

As a result, the uncertainty relation in flat space is corrected by terms depending on the norm of the gradient of the shift vector, evaluated at the center of the ball, in addition to the contribution stemming from the Ricci scalar of the effective metric. Thus, this expression can also be used to describe rotating backgrounds.

However, the calculations performed in the present section were situated well in the nonrelativistic realm. One might wonder how to obtain a general relativistic uncertainty relation. This will be the matter of the subsequent section.

\section{Relativistic generalization}\label{sec:relgen}

The approach introduced in the preceding section was manifestly nonrelativistic. Therefore, a generally covariant formulation of the problem is difficult to conceive. In this section we try to take the first step towards making this relation relativistically invariant by leaving out the nonrelativistic limit taken in going from Eq. \eqref{LR1} to Eq. \eqref{LNR1} (\cf section \ref{subsec:droplim}). This leads to ordering ambiguities in the quantum mechanical description, which are dealt with in section \ref{subsubsec:opord} and fortunately cancel in the result. Furthermore, we follow a derivation similar to those in the previous sections, introducing the Riemann Normal coordinate expansion in section \ref{subsec:RNCrel} and performing the explicit calculation in section \ref{subsec:explicitsolrel}. Section \ref{subsec:covrel} provides an outline on how to extend the framework to incorporate general covariance. Finally, section \ref{subsec:discrel} is intended to summarize the results. The material for this section is extracted from Ref. \cite{Wagner21b}.

\subsection{Dropping the relativistic limit}\label{subsec:droplim}

We start at the relativistic Lagrangian provided in Eq. \eqref{LR1}. The corresponding action continues to be invariant under time reparameterizations. Therefore, we can, again, pick $x^0=\tau.$ Applying this choice, the canonical momenta read
\begin{equation}
    \pi_i=m\frac{G_{ij}(\dot{x}^j+N^i)}{\sqrt{1-\vel^2}}.\label{relmom1}
\end{equation}
This relation has to be inverted to be able to express the velocities in terms of the momenta, which can be achieved by squaring the relation \eqref{relmom1}
\begin{equation}
    \pi^2\equiv\pi_i\pi_jG^{ij}=m^2N\frac{\vel^2}{1-\vel^2}.\label{momeps}
\end{equation}
Hence, we can rewrite the magnitude of the relativistic velocity $\vel$ in terms of the canonical momenta as
\begin{equation}
    \vel^2=\frac{\pi^2}{m^2N+\pi^2}.\label{velmomsq}
\end{equation}
Correspondingly, we indeed find an inversion of Eq. \eqref{relmom1} in
\begin{equation}
    \dot{x}^i=\frac{\sqrt{1-\vel^2}}{m}G^{ij}\pi_j-N^i.
\end{equation}
In the nonrelativistic limit the physical momentum is defined as $p_i=mG_{ij}\dot{x}^j.$ In a generally relativistic context, though, we have to redefine the physical momentum, multiplying the equivalent of the $\gamma$-factor in special relativity
\begin{equation}
    p_i\equiv\frac{mG_{ij}\dot{x}^j}{\sqrt{1-\vel^2}}=\pi_i-\sqrt{1+\frac{\pi^2}{Nm^2}}mG_{ij}N^j\label{relclassmom}
\end{equation}
to account for standard relativistic effects. Note that due to Eq. \eqref{momeps} the extra factor appearing here becomes trivial in the nonrelativistic limit $\vel\ll 1$ or, by Eq. \eqref{velmomsq}, $\pi^2/Nm^2\ll 1.$ For small canonical momenta with respect to the particle's mass, this clearly recovers Eq. \eqref{phys}. In the ultrarelativistic limit, \ie $\vel\simeq 1$ or $\pi^2/Nm^2\gg 1$, on the other hand, it results in
\begin{equation}
    \left.p_i\right|_{\vel\simeq 1}\equiv \left.mG_{ij}\dot{x}^j\right|_{\vel\simeq 1}=\pi_i-\sqrt{\pi^2}G_{ij}N^i/\sqrt{N}, \label{ultrelmom}
\end{equation}
which, being independent of the mass, also applies to massless particles.

Before we dive into the quantum mechanical calculations, we need to settle an additional ambiguity, which, in principle, always arises when turning commuting into noncommuting variables.

\subsection{Operator ordering ambiguities}\label{subsubsec:opord}

Quantization, provided we understand it as such in the first place, is not an injective map. In fact, given any classical function there is an infinite number of possible quantum operators corresponding to it. Consider, for example, the squared position in one dimension, $x^2,$ which could be derived as the classical limit of any operator of the symmetric form 
\begin{equation}
    F(\hat{p})\hat{x}F^{-2}(\hat{p})\hat{x}F(\hat{p})=\hat{x}^2+\hbar^2\left[\left(\frac{F'}{F}\right)^2+\frac{F''}{F}\right],\label{opordex}
\end{equation}
for a general real function $F.$ In the end, it is up to experiment to decide on the correct definition even though there may be theoretical reasons to prefer one ordering over another. For instance, the Laplace-Beltrami operator, a quantization of the classical function $g^{ij}(x)p_ip_j,$ apart from being backed by experiment, has the added advantage of being invariant under spatial diffeomorphisms as expected from the squared magnitude of the physical momentum. In the more primitive case of Eq. \eqref{opordex}, however, there is just no reason to expect anything else than $\hat{x}^2$ to be the quantization of the squared position. If no other principles can be found to guide the choice of ordering, it is, thus, intuitive to refrain from adding more ingredients. Furthermore, as they are supposed to be observables, the resulting operators have to be symmetric.

In comparison to the nonrelativistic version \eqref{phys}, the relativistic momentum \eqref{relclassmom} mixes positions and canonical momenta. Thus, it is not of the primitive form featured in Eq. \eqref{opordex}. Not specifying the exact prescription, its quantum mechanical counterpart can be written as
\begin{equation}
     \slashed{\hat{p}}=\hatslashed{\pi}-m\left(\sqrt{1+\frac{\hat{\pi}^2}{\hat{N}m^2}}\hatslashed{N}\right)_\Ord,\label{nonpertrelmomop}
\end{equation}
where the subscript $\Ord$ stands for any symmetric ordering without addition of extra operators. Then, the inequality, which is at the heart of the present work, has to be derived from the standard deviation of the physical momentum operator provided in Eq. \eqref{stddevphysmom}. The rest of the treatment is analogous to the one introduced in section \ref{subsec:expsol4D}. 

For the purpose of calculating this quantity, it is solely required to show that any symmetric term containing a power of the squared conjugate momentum and a position coordinate, satisfies
\begin{align}
    \frac{1}{2}\left(\hat{\pi}^{2J}\hat{x}^a\hat{\pi}^{2(\mathcal{N}-J)}+\hat{\pi}^{2(\mathcal{N}-J)}\hat{x}^a\hat{\pi}^{2J}\right)=&\frac{1}{2}\left\{\hat{x}^a,\hat{\pi}^{2\mathcal{N}}\right\}+\left[\hat{\pi}^{2(\mathcal{N}-J)},\left[\hat{x}^a,\hat{\pi}^{2J}\right]\right]\\
    =&\frac{1}{2}\left\{\hat{x}^a,\hat{\pi}^{2\mathcal{N}}\right\},
\end{align}
with $J\in (0,1,\dots,\mathcal{N})$ and that similarly, once two coordinates are included
\begin{align}
    \frac{1}{2}&\left(\hat{\pi}^{2J}\hat{x}^a\hat{\pi}^{2K}\hat{x}^b\hat{\pi}^{2[\mathcal{N}-(J+K)]}+\hat{\pi}^{2[\mathcal{N}-(J+K)]}\hat{x}^b\hat{\pi}^{2K}\hat{x}^a\hat{\pi}^{2J}\right)\nonumber\\
    =&\frac{1}{2}\left\{\hat{x}^a\hat{x}^b,\hat{\pi}^{2\mathcal{N}}\right\}+i\hbar\left\{J\left[\hat{\pi}^{2[\mathcal{N}-(J+K)]}\hat{x}^b,\hat{\pi}^a\hat{\pi}^{2(J+K-1)}\right]\right.\nonumber\\
    &\left.-\left(J+K\right)\left[\hat{x}^a,\hat{\pi}^b\hat{\pi}^{2(\mathcal{N}-1)}\right]\right\}\\
    =&\frac{1}{2}\left\{\hat{x}^a\hat{x}^b,\hat{\pi}^{2\mathcal{N}}\right\}+\hbar^2\left\{J\hat{\pi}^{2(\mathcal{N}-1)}\delta^{ab}\right.+2\left[(J+K)(\mathcal{N}-1)\right.\nonumber\\
    &\left.\left.-J(J+K-1)\right]\hat{\pi}^{2(\mathcal{N}-2)}\hat{\pi}^a\hat{\pi}^b\right\},
\end{align}
with $J,K\in (0,1,\dots,\mathcal{N})$ and $J+K\in (0,1,\dots,\mathcal{N}).$ This implies that every function $f$ which is analytic on $\reals^+$ and, therefore, can be expanded nonsingularly for all elements in the spectrum of $\hat{\pi}^2$ will satisfy
\begin{align}
    \left[f(\hat{\pi}^2)x^a\right]_\Ord=&\frac{1}{2}\left\{\hat{x}^a,f\left(\hat{\pi}\right)\right\},\label{ordering1}\\
        \left[f(\hat{\pi}^2)\hat{x}^a\hat{x}^b\right]_\Ord=&\frac{1}{2}\left\{\hat{x}^a\hat{x}^b,f\left(\hat{\pi}\right)\right\}+\hbar^2\left[\mathcal{G}\left(\hat{\pi}^2\right)\delta^{ab}+\tilde{\mathcal{G}}\left(\hat{\pi}^2\right)\hat{\pi}^a\hat{\pi}^b\right],\label{ordering2}
\end{align}
where the subscript $\Ord$ symbolizes a general symmetric ordering without adding extra operators, and we introduced the two additional, not specified, but equally analytic functions $\mathcal{G}$ and $\tilde{\mathcal{G}}.$ Similar results hold for symmetric orderings of the forms $[f(\hat{\pi}^2)\{x^a\pi_b\}g(\hat{\pi}^2)]_\Ord$ and $[f(\hat{\pi}^2)\{x^ax^b\pi_c\pi_d\}g(\hat{\pi}^2)]_\Ord,$ where the curly brackets indicate that the ordering in their interior is fixed. These identities suffice to show that the resulting uncertainty relation is independent of employed ordering.

As position- and momentum-dependent operators appear within one square root in the expression \eqref{nonpertrelmomop}, the ordering has to be enforced at the perturbative level, which, fortunately, is exactly what is required for the purpose of this section. 

\subsection{Riemann normal coordinates}\label{subsec:RNCrel}

Assuming small position uncertainties, the geometry of the relevant neighbourhood of the underlying three-dimensional manifold may be approximated by describing the effective spatial metric in terms of Riemann normal coordinates $x^a,$ defined around the point $p_0$ as in Eq. \eqref{RNCmet} and the shift vector as in Eq. \eqref{shiftexp}. Furthermore, the lapse function may be expanded as
\begin{align}
    N\simeq&N|_{p_0}+\nabla_bN|_{p_0}x^b+\nabla_b\nabla_cN|_{p_0}x^bx^c.\label{lapseexp}
\end{align}
Being a quantity derived from the metric, the canonical momentum operator is expanded as $\hatslashed{\pi}\simeq\hatslashed{\pi}_{(0)}+\hatslashed{\pi}_{(2)}.$ This implies that, applying Eqs. \eqref{ordering1} and \eqref{ordering2} and the relation $[\hatslashed{\pi},\hat{x}^a]=[\hatslashed{\pi}_{(0)},\hat{x}^a],$ the physical momentum operator satisfies order by order
\begin{subequations}\label{physmomopexp}
\begin{align}
    \slashed{\hat{p}}_{(0)}=&\hatslashed{\pi}_{(0)}-m\slashed{N}|_{p_0}\sqrt{1+\hat{\Pi}^2},\label{p0}\\
    \slashed{\hat{p}}_{(1)}=&\frac{m}{2}\left(\frac{1}{2}\left.\slashed{N}\nabla_a\ln{N}\right|_{p_0}\left\{\hat{x}^a,\frac{\hat{\Pi}^2}{\sqrt{1+\hat{\Pi}^2}}\right\}-\nabla_a\slashed{N}|_{p_0}\left\{\hat{x}^a,\sqrt{1+\hat{\Pi}^2}\right\}\right),\label{p1}\\
    \slashed{\hat{p}}_{(2)}=&\hatslashed{\pi}_{(2)}+\frac{m}{4}\left[\left.\left(\nabla_a\slashed{N}\nabla_{b}\ln{N}+\slashed{N}\frac{\nabla_a\nabla_{b}N}{2N}\right)\right|_{p_0}\left\{\hat{x}^a\hat{x}^b,\frac{\hat{\Pi}^2}{\sqrt{1+\hat{\Pi}^2}}\right\}\right.\nonumber\\
    &-\left.\slashed{N}\frac{\nabla_{a}N\nabla_{b}N}{4N^2}\right|_{p_0}\left\{\hat{x}^a\hat{x}^b,\frac{\hat{\Pi}^2\left(4+3\hat{\Pi}^2\right)}{\left(1+\hat{\Pi}^2\right)^{3/2}}\right\}-\nabla_a\nabla_b\slashed{N}|_{p_0}\left\{\hat{x}^a\hat{x}^b,\sqrt{1+\hat{\Pi}^2}\right\}\nonumber\\
    &-\left.\frac{\slashed{N}}{2Nm^2}\right|_{p_0}\left\{\hat{\pi}^2_{(2)},\frac{1}{\sqrt{1+\hat{\Pi}^2}}\right\}\Bigg]+\slashed{\mathcal{G}}_{ab}\left(\hat{\pi}^2_{(0)}\right)\delta^{ab}+\tilde{\slashed{\mathcal{G}}}_{ab}\left(\hat{\pi}^2_{(0)}\right)\hat{\pi}_{(0)}^a\hat{\pi}_{(0)}^b,\label{p2}
\end{align}
\end{subequations}
where we introduced the ordering-dependent tensor- and vector-valued functions $\slashed{\mathcal{G}}_{ab}$ and $\tilde{\slashed{\mathcal{G}}}_{ab},$ which are analytic on $\reals^+,$ and the operator $\hat{\Pi}^2\equiv \hat{\pi}_{(0)}^2/N|_{p_0}m^2.$ Its expectation value
\begin{equation}
    \braket*{\hat{\Pi}^2}=\frac{\vel^2_{(0)}}{1-\vel^2_{(0)}}
\end{equation}
measures the degree of relativity of the given state at lowest order. Functions of $\hat{\Pi}^2$ can be expanded in the eigenstates of the Laplacian
\begin{equation}
    f\left(\hat{\Pi}^2\right)\equiv\sum_{n,l,m}f\left(\frac{\hbar^2\lambda^{(0)}_{nl}}{m^2N|_{p_0}}\right)\ket*{\psi_{nlm}^{(0)}}\bra*{\psi_{nlm}^{(0)}},
\end{equation}
where the states are represented as in Eq. \eqref{flatsol} and the eigenvalues are provided in Eq. \eqref{flateig}.  

In the nonrelativistic limit, \ie $\hat{\Pi}^2\to 0,$ the expansion of the physical momentum operator \eqref{physmomopexp} clearly recovers the expressions provided in section \ref{subsec:expsol4D} as expected. Having thus obtained an expansion of the momentum operator around a point on our background manifold, it is time to tackle the main goal of this section.

\subsection{Explicit solution}\label{subsec:explicitsolrel}

In this subsection, we explicitly derive the uncertainty relation for a general curved background. The result is first obtained analytically on flat space to be further generalized to small perturbations around it as indicated by the expansion in the preceding section.

\subsubsection{Flat space}\label{subsubsec:flatrel}

\begin{figure}[h!]
    \centering
    \includegraphics[width=\linewidth]{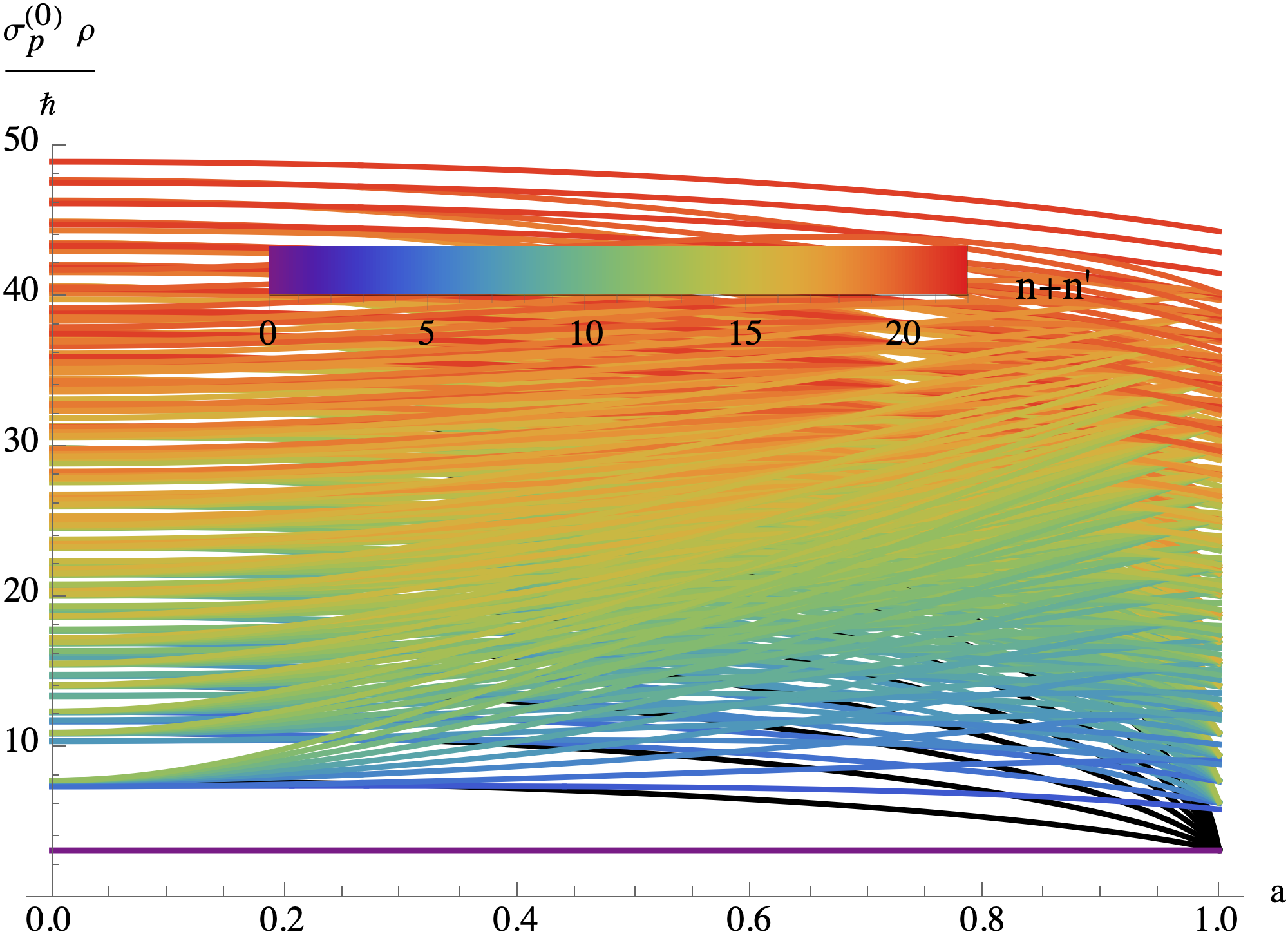}
    \caption{Lower bounds on the ultrarelativistic momentum uncertainty of all linear combinations of two eigenstates of the Laplacian \eqref{flatsol} with principal quantum numbers $(n,n')\leq (10,11)$ to lowest nonvanishing order as functions of the parameter $a\in[0,1],$ characterizing the relative weight, and in units of $\hbar/\rho$. Black curves correspond to linear combinations including the ground state \eqref{unperteig100}, while the colour measures the sum of the principal quantum numbers $n$ and $n'$ for the others. The eigenvalue of the ground state is represented by the violet line.}
    \label{fig:smallestuncrel}
\end{figure}

As for nonrelativistic particles, we begin with the uncertainty relation in flat space. In this case, the linear and squared momentum operators are given by Eq. \eqref{p0} and as
\begin{equation}
    \hat{p}^2_{(0)}=\hat{\pi}^2_{(0)}-2m\slashed{N}|_{p_0}\left\{\hatslashed{\pi}_{(0)},\sqrt{1+\hat{\Pi}^2}\right\}+m^2N^aN_a|_{p_0}\left(1+\hat{\Pi}^2\right).
\end{equation}
Thus, the variance of the momentum operator $\sigma_p^2$ can be expressed as
\begin{equation}
    \var{p}^{(0)}=\var{\pi}^{(0)}+\var{p}^{(0)}_{\text{rel}},\label{varsplitflat}
\end{equation}
where the global minimum of $\var{\pi}^{(0)},$ stemming from the ground state of the Laplacian $\psi_{100},$ was found in section \ref{subsubsec:flat} and we introduced the relativistic correction
\begin{align}
    \var{p}^{(0)}_{\text{rel}}=&2m\slashed{N}|_{p_0}\left(\braket*{\hatslashed{\pi}^{(0)}}\left\langle\sqrt{1+\hat{\Pi}^2}\right\rangle-\braket*{\hatslashed{\pi}^{(0)}\sqrt{1+\hat{\Pi}^2}}\right)\nonumber\\
    &+m^2\left.N^aN_a\right|_{p_0}\left[\braket*{1+\hat{\Pi}^2}-\braket*{\sqrt{1+\hat{\Pi}^2}}^2\right].
\end{align}
Clearly, the first two terms can decrease the uncertainty when $\braket{\hatslashed{\pi}_{(0)}}\neq 0,$ \ie for superpositions of eigenstates of the Laplacian with relative phase, the kind which was treated in section \ref{subsubsec:flat}. Expressed as a linear combination $\Psi$ (\cf Eq. \eqref{genstaten}) of the eigenstates of the Laplacian and applying Eqs. \eqref{evp1} and \eqref{vanmomrealwave}, it becomes
\begin{align}
    \var{p}^{(0)}_{\text{rel}}=&2m\slashed{N}|_{p_0}\sum_{n\neq  n'}\text{Re}\left(a^*_{n'}a_n\Braket*{\psi^{(0)}_{n'}}{\hatslashed{\pi}_{(0)}\psi^{(0)}_n}\right)\nonumber\\
    &\times\left(\sum_{n''} |a_{n''}|^2\sqrt{1+\frac{\lambdabar_C^2\lambda_{n''}}{N|_{p_0}}}- \sqrt{1+\frac{\lambdabar_C^2\lambda_n}{N|_{p_0}}}\right)\\
    \geq&m^2\left. N^aN_a\right|_{p_0}\left[1+\sum_n\left|a_n\right|^2\left(\frac{\lambdabar_C^2\lambda_n}{N|_{p_0}}-\sum_{n'}\left|a_{n'}\right|^2\sqrt{1+\frac{\lambdabar_C^2\lambda_{n}}{N|_{p_0}}}\sqrt{1+\frac{\lambdabar_C^2\lambda_{n'}}{N|_{p_0}}}\right)\right]\nonumber\\
    &-2m\left|\left|\slashed{N}\right|\right|_{p_0}\sum_{n\neq n'}|a_{n}||a_{n'}|\left|\left|\text{MaxRe}\left(e^{i\Delta\phi_{nl,n'l'}}\Braket*{\psi^{(0)}_{n'}}{\hatslashed{\pi}_{(0)}\psi^{(0)}_n}\right)\right|\right|\nonumber\\
    &\times\left|\sum_{n''} \left(|a_{n''}|^2-\delta_{n''n}\right)\sqrt{1+\frac{\lambdabar_C^2\lambda_{n''}}{N|_{p_0}}}\right|,
\end{align}
where $\lambdabar_C$ stands for the reduced Compton wave length, MaxRe indicates a choice of relative phase $\Delta\phi_{nln'l'}$ between the coefficients $a_n,$ $a_{n'}$ such that the real part of the resulting quantity is maximized and we introduced the norm 
\begin{equation}
\left|\left|\slashed{N}\right|\right|=\sqrt{\slashed{N}^2}=\sqrt{N^aN_a}.
\end{equation}
In the nonrelativistic limit the corrections multiply the factor $\sum_{n''}(|a_{n''}|^2-\delta_{n''n})=0,$ and the contribution vanishes as expected. After some straight-forward yet tedious algebra, displayed in appendix \ref{app_maxre}, and recovering all quantum numbers in three dimensions, we can estimate
\begin{align}
    \var{p}^{(0)}_{\text{rel}}\geq&\left. N^aN_a\right|_{p_0}\left[1+\sum_n\left|a_n\right|^2\left(\frac{\lambdabar_C^2j_{l,n}^2}{\rho^2N|_{p_0}}-\sum_{n'}\left|a_{n'}\right|^2\sqrt{1+\frac{\lambdabar_C^2j_{l,n}^2}{\rho^2N|_{p_0}}}\sqrt{1+\frac{\lambdabar_C^2j_{l',n'}^2}{\rho^2 N|_{p_0}}}\right)\right]\nonumber\\ &-\frac{2\hbar^2}{\rho\lambdabar_C}\sqrt{N^aN^bG_{ab}}\big|_{p_0}\sum_{n,l\neq n',l'}|a_{n}||a_{n'}|\frac{j_{l,n}j_{l',n'}}{|j^2_{l,n}-j^2_{l',n'}|}\nonumber\\
    &\times\left|\sum_{n'',l''} \left(|a_{n'',l''}|^2-\delta_{n''n}\delta_{l''l}\right)\sqrt{1+\frac{\lambdabar_C^2j^2_{l'',n''}}{\rho^2N|_{p_0}}}\right|.
\end{align}
As the nonrelativistic case has been treated already in section \ref{subsubsec:flat}, a possible change in the state of smallest uncertainty should be expected to result in the ultrarelativistic limit, \ie for states for which $\lambdabar_Cj_{l'',n''}/N|_{p_0}\rho^2\gg 1.$ Then, the relativistic correction becomes approximately
\begin{align}
    \frac{\rho^2}{\hbar^2}\left.\var{p}^{(0)}_{\text{rel}}\right|_{\braket{\hat{\Pi}}\gg 1}\gtrsim&\left. \frac{N^aN_a}{N}\right|_{p_0}\sum_n\left|a_n\right|^2\left(j_{l,n}^2-\sum_{n'}\left|a_{n'}\right|^2j_{l,n}j_{l',n'}\right)\nonumber\\
    &-2\sqrt{\frac{N_aN^a}{N}}\Bigg|_{p_0}\sum_{n,l\neq n',l'}|a_{n}||a_{n'}|\frac{j_{l,n}j_{l',n'}}{|j^2_{l,n}-j^2_{l',n'}|}\nonumber\\
    &\times \left|\sum_{n'',l''} \left(|a_{n'',l''}|^2-\delta_{nn''}\delta_{ll''}\right)j_{l'',n''}\right|.
\end{align}
This implies, that the sum of both relevant contributions to the variance \eqref{varsplitflat} satisfies at the ultrarelativistic level
\begin{align}
    \frac{\rho^2}{\hbar^2}\left.\var{p}^{(0)}\right|_{\braket{\hat{\Pi}}\gg 1}\gtrsim&\sum_n\left|a_n\right|^2\left[j_{l,n}^2\left(1+\left. \frac{N^aN_a}{N}\right|_{p_0}\right)-\left. \frac{N^aN_a}{N}\right|_{p_0}\sum_{n'}\left|a_{n'}\right|^2j_{l,n}j_{l',n'}\right] \nonumber\\
    &-2\sqrt{\frac{N_aN^a}{N}}\Bigg|_{p_0}\sum_{n,l\neq n',l'}|a_{n}||a_{n'}|\frac{j_{l,n}j_{l',n'}}{|j^2_{l,n}-j^2_{l',n'}|}\nonumber\\
    &\times \left|\sum_{n'',l''} \left(|a_{n'',l''}|^2-\delta_{nn''}\delta_{ll''}\right)j_{l'',n''}\right|.
\end{align}
As the transition amplitude $\Braket{\psi_{nlm}}{\hatslashed{\pi}_{(0)}\psi_{n'l'm'}}$ is only nonvanishing if $\Delta l=|l'-l|=1,$ which is shown in appendix \ref{app_maxre}, the effect of linearly combining more than two eigenstates of the Laplacian cannot be stronger than just adding two of them. Thus, we can consider only the former without loss of generality. Then, we can define the relative weight $a\equiv|a_{n,l}|=\sqrt{1-|a_{n',l'}|^2},$ leading to the relation
\begin{align}
    \left.\var{p}^{(0)}\right|_{\braket{\hat{\Pi}}\gg 1}\geq&\frac{\hbar^2}{\rho^2}\left[ A(a)-B(a)\left.\sqrt{\frac{N^aN_a}{N}}\right|_{p_0}+C(a)\left.\frac{N^aN_a}{N}\right|_{p_0}\right],
\end{align}
where, denoting the quantum numbers of the two states as $n,l$ and $n',l'$ by a slight abuse of notation, we introduced the functions of the parameter
\begin{align}
    A(a)=&a^2j_{l,n}^2+(1-a^2)j_{l',n'}^2,\\
    B(a)=&2a\sqrt{1-a^2}\left|1-2a^2\right|
    \frac{j_{l',n'}j_{l,n}}{|j^2_{l',n'}-j^2_{l,n}|}\left|j_{l,n}-j_{l',n'}\right|,\\
    C(a)=&a^2(1-a^2)\left(j_{l,n}-j_{l',n'}\right)^2.
\end{align}
As a function of the shift vector and the lapse function, the uncertainty clearly has a global minimum at $\sqrt{N^aN_a/N}|_{p_0}=\frac{B}{2C}.$ Thus, we can estimate
\begin{align}
    \left.\var{p}^{(0)}\right|_{\braket{\hat{\Pi}}\gg 1}\geq&\frac{\hbar^2}{\rho^2}\left[ A-\frac{B^2}{4C}\right]\\
    =&\frac{\hbar^2}{\rho^2}\left[a^2j_{l,n}^2+(1-a^2)j_{l',n'}^2-\left(1-2a^2\right)^2\frac{j_{l',n'}j_{l,n}}{|j^2_{l',n'}-j^2_{l,n}|}\right].
\end{align}
The resulting uncertainties as functions of the parameter $a$ are plotted for all eigenfunctions of the Laplacian \eqref{flatsol} characterized by quantum numbers $n\leq 10,n'\leq 11$ in Fig. \ref{fig:smallestuncrel}.
Not a single one of those states has a smaller uncertainty than the ground state of the Laplacian $\psi_{100}^{(0)},$ defined in Eq. \eqref{unperteig100}. Those mixing with the ground state as 
\begin{equation}
    \Psi=ae^{i\Delta\phi}\psi_{100}^{(0)}+\sqrt{1-a^2}\psi_{nlm}^{(0)},
\end{equation}
coloured black, for example, only reach their minimum value at $a=1.$ Furthermore, the difference only grows with increasing $n+n'$ as can be inferred from the colour of the other graphs. 

To put it in a nutshell, the ground state of the Laplacian $\psi_{100}^{(0)}$ remains the state of smallest uncertainty, \ie we have found
\begin{equation}
    \Psi_0^{(0)}=\psi_{100}^{(0)},
\end{equation}
and we recover the inequality \eqref{flatres} in the relativistic setting. This result will be modified by gradually adding in curvature.

\subsubsection{Corrections}\label{subsubsec:pertrel}

As perturbative corrections are comparably small by definition, the fact that the ground state of the Laplacian uniquely saturates the uncertainty relation carries over to the slightly curved setting, in general meaning that
\begin{equation}
    \Psi_0=\psi_{100}.
\end{equation}
This implies that Eq. \eqref{vanmomrealwave}, \ie the vanishing of the expectation value of the conjugate momentum operator, continues to hold to all orders in the expansion. Furthermore, the integration measure in flat space, with respect to which we compute corrections to expectation values (\cf appendix \ref{app_pert}), is even in the radial coordinate. Hence, all expectation values of operators which are odd in the sum of the numbers of positions and unperturbed momenta vanish. Therefore, the derivation of the correction
\begin{equation}
    \var{p}^{(1)}(\psi_{100})=\Braket*{\psi_{100}^{(0)}}{\hat{p}^2_{(1)}\psi_{100}^{(0)}}_0-2\Braket*{\psi_{100}^{(0)}}{\hat{p}_{(1)}\psi_{100}^{(0)}}_0\Braket*{\psi_{100}^{(0)}}{\hat{p}_{(0)}\psi_{100}^{(0)}}_0
\end{equation}
simplifies significantly.

In fact, at this order all possibly arising corrections to $\braket{\hatslashed{p}}$ are even in the momenta and odd in the coordinates, implying that $\braket{\hatslashed{p}_{(1)}}=0.$ Most of the contributions to $\braket{\hat{p}^2}$ vanish for the same reason. The remaining terms assume the form
\begin{equation}
    \Braket*{\psi_{100}^{(0)}}{\hat{p}_{(1)}^2\psi_{100}^{(0)}}_0\propto \Braket*{\psi_{100}^{(0)}}{\left\{\hat{x}^a\hat{\pi}_b\right\}\psi_{100}^{(0)}}_0.
\end{equation}
This can be shown to equal zero, applying the canonical commutation relations \eqref{HeisAlg} and computing explicitly that
\begin{equation}
    \Braket*{\psi_{100}^{(0)}}{\hat{x}^a\hat{\pi}_b\psi_{100}^{(0)}}_0=\frac{i\hbar}{2}\delta^a_b.\label{xpiint}
\end{equation}
To put it in a nutshell, there are no first-order corrections to the uncertainty relation. In order to obtain curvature-induced contributions, it is necessary to treat the system at higher order.

The quadratic modification of the variance of the momentum operator
\begin{equation}
    \var{p}^{(2)}=\braket*{\left(\hatslashed{p}^{(1)}\right)^2}_0-\braket*{\hatslashed{p}^{(1)}}^2_0+\braket*{\left\{\hatslashed{p}^{(2)},\hatslashed{p}^{(0)}\right\}}_0-2\braket*{\hatslashed{p}^{(2)}}_0\braket*{\hatslashed{p}^{(0)}}_0,
\end{equation}
for example, yields meaningful terms.  Still, it simplifies considerably taking into account generic cancellations. The second term was shown to vanish when treating the calculations at first order. Furthermore, Eq. \eqref{p0} implies that the third and the fourth terms largely cancel for all eigenstates of the Laplacian, leaving us with
\begin{equation}
    \var{p}^{(2)}\left(\psi_{nlm}\right)=\braket*{\left(\hatslashed{p}^{(1)}\right)^2}_0+\braket*{\left\{\hatslashed{p}^{(2)},\hatslashed{\pi}^{(0)}\right\}}_0+\braket*{\left[\hatslashed{p}^{(0)}-\hatslashed{\pi}^{(0)},\hatslashed{p}^{(2)}\right]}_0.
\end{equation}
All contributions to $\hatslashed{p}^{(2)},$ as provided in Eq. \eqref{p2}, except for $\hatslashed{\pi}^{(2)}$ are even in $\hat{\pi}_a$ and $\hat{x}^b$ while $\hatslashed{\pi}^{(0)}$ is evidently odd. Thus, when evaluated respective to the ground state of the Laplacian, this sum experiences a further simplification to read
\begin{equation}
    \var{p}^{(2)}\left(\psi_{100}\right)=\braket*{\hat{\pi}^2_{(2)}}_0+\braket*{\left(\hatslashed{p}^{(1)}\right)^2}_0+\braket*{\left[\hatslashed{p}^{(0)}-\hatslashed{\pi}^{(0)},\hatslashed{p}^{(2)}\right]}_0.\label{var2ndaftcanc}
\end{equation}
The first term appearing at the right-hand side just equals $\var{\pi}^{(2)}$ as derived in section \ref{subsubsec:perturb}, yielding
\begin{equation}
    \braket*{\pi_{(2)}^2}=-\frac{R|_{p_0}}{6},
\end{equation}
with the Ricci scalar $R$ derived from the effective metric $G_{ab},$ while the correction obtained in section \ref{subsubsec:perturb4D} is hidden in the second term. Making use of Eq. \eqref{p1}, this expectation value is of the form
\begin{align}
    \braket*{\left(\hatslashed{p}^{(1)}\right)^2}_0=&\frac{m^2}{4}\braket*{\left\{\hat{x}^a,\slashed{F}_a\left(\hat{\Pi}^2\right)\right\}^2}_0\\
    =&m^2\left\{F_{ac}F_b^{~c}\left(\Pi^2\right)\braket*{x^ax^b}_0\right.\nonumber\\
    &\left.+\rho^2\frac{\Pi^2}{\pi^2}\left[\left(F_{ac}F^{ac}\right)'-\frac{\rho^2}{\hbar^2}\frac{\Pi^2}{\pi^2}\left(2F_{ac}F_{bc}''+F_{ac}'F_{bc}'\right)\braket*{\hat{\pi}^a\hat{\pi}^b}\right]\right\},\label{p1sqexprel}
\end{align}
where $\Pi^2=\hbar^2\pi^2/\rho^2m^2N|_{p_0}=\pi^2\lambdabar_C^2/\rho^2N|_{p_0},$ and we introduced the dimensionful, tensor-valued function
\begin{equation}
F_{ac}\left(\hat{\Pi}^2\right)=\frac{1}{2}N_c\nabla_a\ln{N}|_{p_0}\frac{\hat{\Pi}^2}{\sqrt{1+\hat{\Pi}^2}}-\nabla_aN_c|_{p_0}\sqrt{1+\hat{\Pi}^2}.
\end{equation}
The required expectation values can be evaluated explicitly, yielding Eq. \eqref{doublexint} and
\begin{equation}
    \Braket*{\psi_{100}^{(0)}}{\hat{\pi}^a\hat{\pi}^b\psi_{100}^{(0)}}_0=\frac{\hbar^2\pi^2}{3\rho^2}\delta^{ab}.\label{doublepiint}
\end{equation}
Plugging these explicit results back into Eq. \eqref{p1sqexprel}, we obtain
\begin{equation}
    \braket*{\left(\hatslashed{p}^{(1)}\right)^2}_0=m^2\rho^2\left\{\frac{\xi}{2} F_{ac}F^{ac}+\frac{\Pi^2}{\pi^2}\left[\frac{1}{2}\left(F_{ac}F^{ac}\right)'-\frac{\Pi^2}{3}\left(2F_{ac}''F^{ac}+F_{ac}'F^{\prime ac}\right)\right]\right\}.
\end{equation}
Fortunately, the third term of Eq. \eqref{var2ndaftcanc} turns out to be such that the dependence on the ordering in the second-order correction to the momentum operator \eqref{p2} exactly cancels. Resultingly, this contribution can be expressed as
\begin{align}
    \braket*{\left[\hatslashed{p}^{(0)}-\hatslashed{\pi}^{(0)},\hatslashed{p}^{(2)}\right]}_0=&\Braket*{\psi_{100}^{(0)}}{\hat{\pi}^a\hat{\pi}^b\psi_{100}^{(0)}}_0\left[\frac{m^2\rho^4}{\hbar^2\pi^4}\frac{\Pi^4}{\sqrt{1+\Pi^2}}\left(\frac{\tilde{F}_{ab}\left(\Pi^2\right)}{1+\Pi^2}-\tilde{F}_{ab}'\left(\Pi^2\right)\right)\right.\nonumber\\
    &\left.+\frac{\rho^2}{6\pi^2}\frac{\Pi^2}{1+\Pi^2}\left.\frac{N^cN_cR_{ab}}{N}\right|_{p_0}\right]\\
    =&\frac{\rho^2m^2}{3\pi^4}\frac{\Pi^2}{\sqrt{1+\Pi^2}}\left(\frac{\tilde{F}_a^a}{1+\Pi^2}-\tilde{F}_a^{\prime a}\right)\nonumber\\
    &+\frac{\hbar^2}{36}\frac{\Pi^2}{1+\Pi^2}\left.\frac{N^cN_cR}{N}\right|_{p_0},
\end{align}
where we introduced the dimensionful, tensor-valued function
\begin{align}
    \tilde{F}_{ab}=&\frac{ N_c}{2}\left[\frac{1}{2N}\nabla_a\left(N^c\nabla_{b}N\right)\frac{\hat{\Pi}^2}{\sqrt{1+\hat{\Pi}^2}}-\nabla_a\nabla_bN^c\sqrt{1+\hat{\Pi}^2}\right.\nonumber\\
    &\left.\left.-\frac{1}{8}N^c\nabla_{a}\ln{N}\nabla_{b}\ln{N}\frac{\hat{\Pi}^2\left(4+3\hat{\Pi}^2\right)}{\left(1+\hat{\Pi}^2\right)^{3/2}}\right]\right|_{p_0}.
\end{align}
Plugging all those terms back into Eq. \eqref{var2ndaftcanc}, the correction to the variance of the momentum operator reads
\begin{align}
    \var{p}^{(2)}=&\frac{\hbar^2}{2}\left[\left.\frac{ R}{9}\left(\frac{N_aN^a}{N}\frac{\Pi^2}{1+\Pi^2}-3\right)\right|_{p_0}+\mathcal{F}_1\left(\Pi^2\right)\left.\frac{N_bN^b}{N}\nabla_{a}\ln{N}\nabla^{a}\ln{N}\right|_{p_0}\right.\nonumber\\
    &-\mathcal{F}_2\left(\Pi^2\right)\left.\frac{N^b\nabla_{a}N_b}{N}\nabla^{a}\ln{N}\right|_{p_0}+
    \mathcal{F}_3\left(\Pi^2\right)\left.\frac{\nabla_{a}N_b\nabla^aN^b}{N}\right|_{p_0}\nonumber\\
    &\left.-\mathcal{F}_4\left(\Pi^2\right)\left.\frac{N^b\Delta N_b}{N}\right|_{p_0}+\mathcal{F}_5\left(\Pi^2\right)\left.\frac{N_bN^b}{N^2}\Delta N\right|_{p_0}\right],
\end{align}
where we introduced the functions of the degree of relativity
\begin{align}
    \mathcal{F}_1=&\frac{\pi^2}{4}\xi\frac{\Pi^2}{1+\Pi^2}+\frac{\Pi^2\left(-4+5\Pi^2+2\Pi^4\right) }{12\left(1+\Pi^2\right)^3},\\
    \mathcal{F}_2=&\pi^2\xi+1-\frac{\Pi^2}{3\left(1+\Pi^2\right)^2},\\
    \mathcal{F}_3=&\pi^2\xi\frac{1+\Pi^2}{\Pi^2}+1+\frac{1}{6}\frac{\Pi^2}{1+\Pi^2},\\
    \mathcal{F}_4=&\frac{1}{6}\frac{\Pi^2}{1+\Pi^2},\\
    \mathcal{F}_5=&\frac{1}{12}\frac{\Pi^2\left(4+\Pi^2\right)}{\left(1+\Pi^2\right)^2}.
\end{align}
The final result of this section thus reads
\begin{align}
    \sigma_p\rho\geq& \pi\hbar\left\{1+\frac{\rho^2}{4\pi^2}\left[\left.\frac{ R}{9}\left(\frac{N_aN^a}{N}\frac{\Pi^2}{1+\Pi^2}-3\right)\right|_{p_0}+\mathcal{F}_1\left(\Pi^2\right)\left.\nabla_{a}\ln{N}\nabla^{a}\ln{N}\frac{N_bN^b}{N}\right|_{p_0}\right.\right.\nonumber\\
    &-\mathcal{F}_2\left(\Pi^2\right)\left.\nabla^{a}\ln{N}\frac{\nabla_{a}N_bN^b}{N}\right|_{p_0}+\mathcal{F}_3\left(\Pi^2\right)\left.\frac{\nabla_{a}N_b\nabla^aN^b}{N}\right|_{p_0}-\mathcal{F}_4\left(\Pi^2\right)\left.\frac{N^b\Delta N_b}{N}\right|_{p_0}\nonumber\\
    &\left.\left.+\mathcal{F}_5\left(\Pi^2\right)\left.\frac{N_bN^b}{N^2}\Delta N\right|_{p_0}\right]\right\}.\label{relres}
\end{align}
Undoubtedly, this is a quite involved expression. Therefore, it is instructive to consider its asymptotic behaviour. On the one hand, for $\Pi\ll 1,$ implying the nonrelativistic limit,  we recover the relation derived in section \ref{subsubsec:perturb4D}
\begin{equation}
    \sigma_p\rho\geq\pi\hbar\left[1-\rho^2\left.\left(\frac{R}{12\pi^2}-\xi\frac{\rho^2}{\lambda_C^2}\nabla_aN_b\nabla^aN^b\right)\right|_{p_0}\right]\label{nonrelunc}
\end{equation}
as expected. Ultrarelativistic particles satisfying $\Pi\gg 1$, on the other hand, obey the uncertainty relation
\begin{align}
    \sigma_p\rho\geq\pi\hbar&\left\{1+\frac{\rho^2}{4\pi^2}\left[\frac{ R}{9}\left(\frac{N_aN^a}{N}-3\right)+\tilde{\xi}\nabla_a\left(N_b/\sqrt{N}\right)\nabla^a\left(N^b/\sqrt{N}\right)\right.\right.\nonumber\\
    &\left.\left.\left.-\frac{N_a}{6\sqrt{N}}\Delta\left(N^a/\sqrt{N}\right)\right]\right|_{p_0}\right\},
\end{align}
with the mathematical constant $\tilde{\xi}=\xi+7/6\pi^2\sim\Ord (10^{-1}),$ a result, which reflects the form of the ultrarelativistic momentum \eqref{ultrelmom}. In particular, it is independent of $\Pi.$ Hence, there is no divergence at high energies. Instead, the relation asymptotes towards a constant value. Therefore, it also applies to massless particles. On the other hand, the nonrelativistic, shift-dependent correction in Eq. \eqref{nonrelunc} scales linearly with the mass of the particle. Thus, the gravitational influence is strongest on very massive particles as expected. Furthermore, as mentioned above, all relativistic corrections are dependent on the value of the shift vector. If the latter vanishes, the corrections only depend on the scalar curvature of the effective spatial geometry. 

This result can be applied to any particle, irrespective of mass and velocity, and any background. Have we thus obtained a covariant uncertainty relation?

\subsection{A real covariant uncertainty relation}\label{subsec:covrel}

\begin{figure}[h!]
    \centering
    \includegraphics[width=\linewidth]{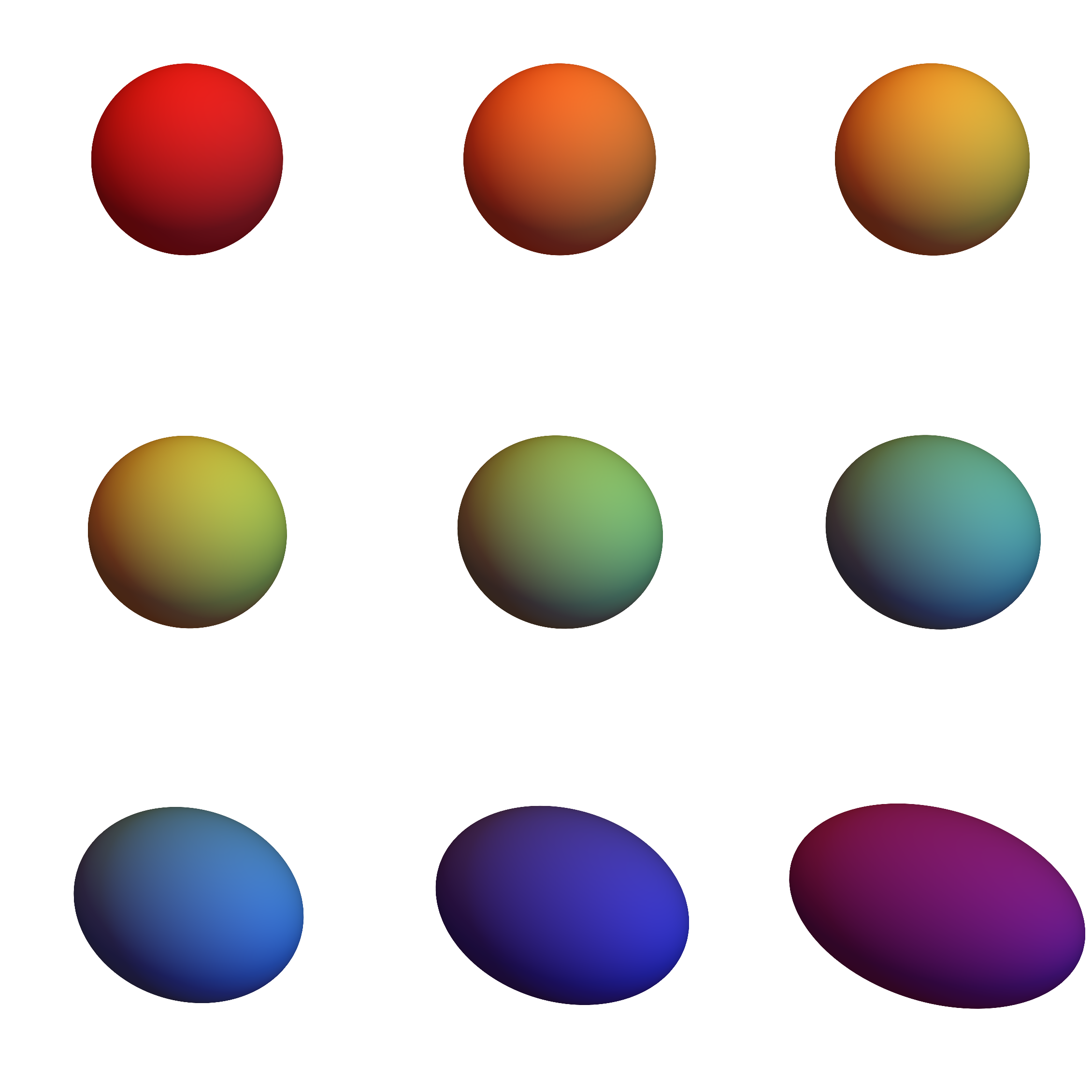}
    \caption{A geodesic ball on a flat background, deformed by Lorentz transformations. The velocity parameter corresponding to the transformation, in units of the speed of light, is colour-coded, increasing in steps of $.1$ from red ($0$) to violet ($.8$).}
    \label{fig:lorentzgeoball}
\end{figure}

The derivation we followed throughout this section was based on a given division of the underlying spacetime manifold $M$ into a time direction $\reals$ and spacelike hypersurfaces $\Sigma,$ on which the uncertainty relation is determined. To be more precise, this was indicated by the choice of domain. The investigated particle is confined to a geodesic ball at a certain time, which is clearly a slicing-dependent statement. Changing the time coordinate, \eg by a global Lorenz transformation in flat spacetime (see Fig. \ref{fig:lorentzgeoball}), would lead to a deformation of the ball and change the problem entirely. Clearly, the core of this peculiarity lies in the original objective to obtain an uncertainty principle relating positions and momenta. This approach relates quantities, which are intrinsically noninvariant. 

\begin{figure}[h!]
    \centering
    \includegraphics[width=\linewidth]{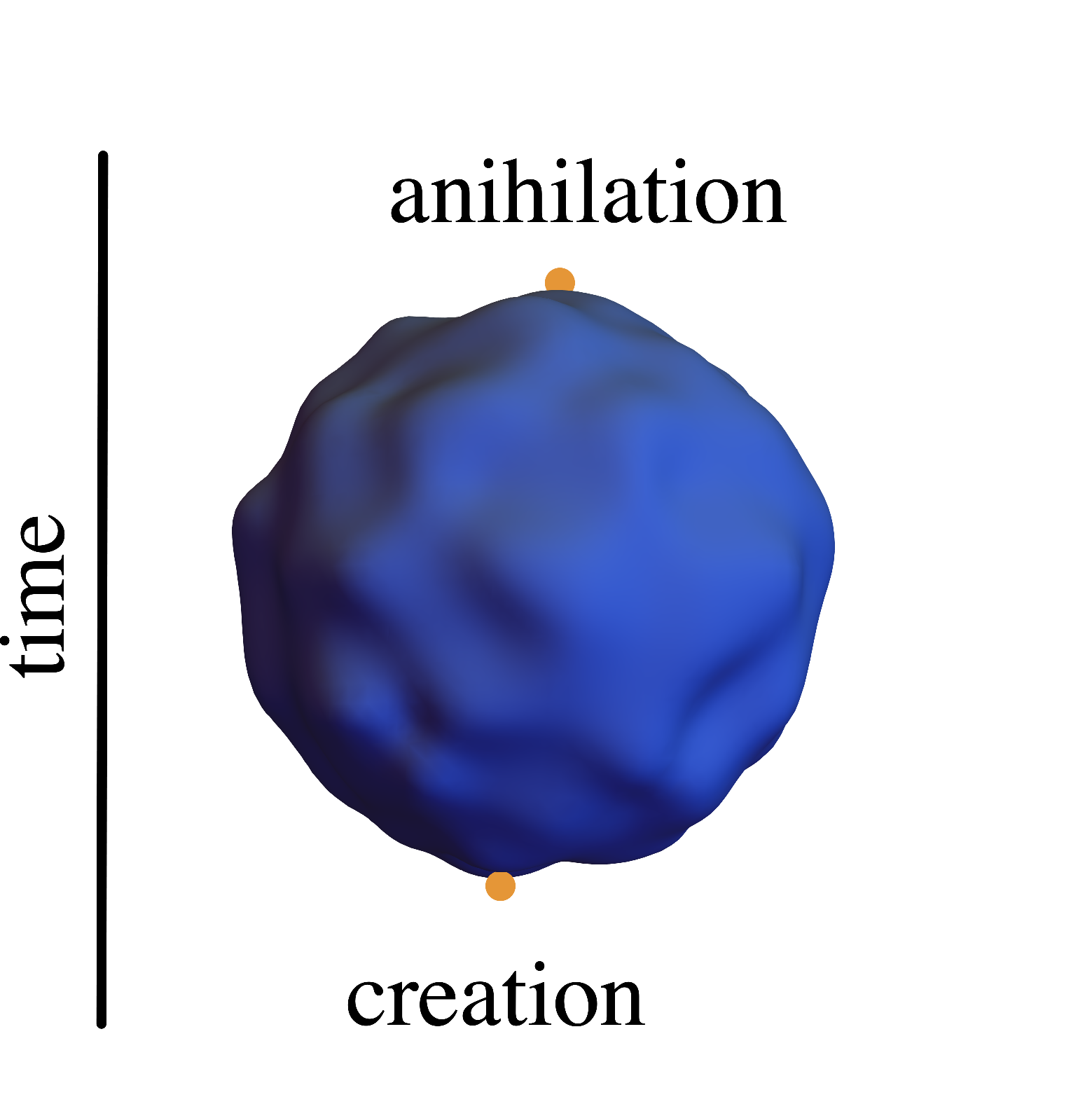}
    \caption{Schematic visualisation of the interpretation of a random compact domain in three-dimensional spacetime and its meaning as the place of a quantum fluctuation surged by the uncertainty relation.}
    \label{fig:spacetimedomain}
\end{figure}

In principle, the relevant quantity to study to stay in the covariant realm would be the standard deviation of the Dirac operator $\hatslashed{D}=-i\hbar\gamma^\mu\partial_\mu.$ By analogy, this requires a domain which is compact not only in space but also in time, thus treating both entities equally. Hence, it seems necessary to consider it to be expanding into the future from an initial hypersurface to afterwards recollapse into another hypersurface, basically like the creation and subsequent annihilation of an excitation by the uncertainty. A cartoonish visualisation of this process is displayed in Fig. \ref{fig:spacetimedomain}.

As a result, this yields an uncertainty relation between the spacetime volume of the domain (related to the proper time measured along a fixed curve inside of it) and the particle's mass due to Einstein's relation $p^2=m^2$. Figuratively speaking, the ensuing inequality would predict the combined mass of a virtual excitation that could be created by the uncertainty alone when the spacetime is restricted to a certain volume. More precisely, we are unable to resolve excitations of a certain combined mass inside of a given finite spacetime volume. This will be the subject of future research.

\subsection{Summary}\label{subsec:discrel}

Starting at the dynamics of relativistic particles in curved spacetime, we have derived an uncertainty relation between the positions and momenta on hypersurfaces of constant time in accordance with the 3+1-formalism. This was achieved by, first, finding the relevant relativistic physical momentum operator and then obtaining its standard deviation on a compact domain, \ie a geodesic ball. Thus, this result can be understood as a relativistic generalization of the reasoning in sections \ref{sec:3DEUP} and \ref{sec:4DEUP}.

To second order, the resulting uncertainty relation is, again, proportional to the Ricci scalar of the effective spatial metric as well as a couple of terms, which depend on the gradients of the shift vector and the lapse function. In particular, all relativistic corrections to the nonrelativistic result in section \ref{sec:4DEUP} are proportional to the shift vector and/ or its gradients. Thus, they all vanish in the absence of meaningful nondiagonal entries in the original spacetime metric. Interestingly, the ultrarelativistic limit yields a constant correction and does not diverge. Thus, the relation may, in principle, also be used to describe massless particles. Yet, this inequality is clearly not covariant -- it presupposes a given slicing of spacetime. Therefore, we indicate an avenue towards obtaining locally Lorentz invariant results.

The last three sections were aimed at reaching Eq. \eqref{relres} by further and further generalizing the original approach invented by Schürmann \cite{Schuermann18}. However, it is instructive to examine the resulting effect by virtue of some examples.

\section{Applications}\label{sec:applications}

The investigations presented in the preceding sections lead to modifications to the uncertainty relation induced by gravity in an ever more general setting. Finally, this culminated in an equality characterizing relativistic particles in a general curved background spacetime \eqref{relres}. But what does this, admittedly, convoluted expression imply in special cases of interest?

This section will shed some light on the uncertainty relations accelerated particles as well as particles subject to spherically and axially symmetric spacetimes will experience. It consists of combined and considerably extended material from Refs. \cite{Dabrowski19,Dabrowski20,Petruzziello21,Wagner21b}.

\subsection{Rindler horizon}

As has been known for a long time, accelerated observers on Minkowski spacetime follow hyperbolic trajectories \cite{Born09,Einstein12}. This scheme was further generalized to curved spacetime by Wolfgang Rindler \cite{Rindler60}.

In this regard, assume as given a general background metric, expanded in terms of spatial Riemann normal coordinates $(X,Y,Z)$ constructed around a point $p_0$ as in Eq. \eqref{RNCmet} and a time coordinate $T$ such that the uncertainty relation is evaluated at $T=0.$  After the transformation
\begin{equation}
    t=\frac{1}{\alpha}\arctanh\left(\frac{T}{X+\frac{1}{2\alpha}}\right),\hspace{.8cm}x=\sqrt{\left(X+\frac{1}{2\alpha}\right)^2-T^2}\hspace{.8cm}y=Y,\hspace{.8cm}z=Z,
\end{equation}
the same geometry is described from the point of view of an observer at the point $p_0$ moving with an acceleration $\alpha$ in the $x$-direction. On the spacelike hypersurface at $T=0$, this implies that the shift vector and lapse function transform as
\begin{equation}
    \left.\left(N,N^i\right)\right|_{T=0}\rightarrow \left(\frac{1}{2}+\alpha x\right)\left.\left(N,N^i\right)\right|_{T=0},
\end{equation}
while the induced metric is unmodified.

For reasons of simplicity, we assume that the original frame is chosen in such a way that the shift vector vanishes. Then, the influence of the acceleration on the uncertainty relation is solely due to a conformal factor in the effective metric
\begin{equation}
    G_{ab}\rightarrow \frac{2}{1+2\alpha x}G_{ab}\equiv\Omega^2G_{ab}.
\end{equation}
Correspondingly, the Ricci scalar of the effective metric transforms as \cite{Wald84}
\begin{equation}
    R\rightarrow \Omega^{-2}\left[R-4\Delta\ln\Omega-2G^{ab}\nabla_a\ln\Omega\nabla_b\ln\Omega\right].
\end{equation}
Therefore, the uncertainty relation obeyed by a possibly relativistic, accelerated particle at $p_0,$ \ie at $x=1/2\alpha,$ reads
\begin{equation}
    \sigma_p\rho\geq\pi\hbar\left[1+\frac{\rho^2}{12\pi^2}\left(\frac{5\alpha^2}{2}-R|_{p_0}\right)\right],
\end{equation}
which in a flat background clearly becomes
\begin{equation}
    \sigma_p\rho\geq\pi\hbar\left(1+\frac{5}{24\pi^2}\alpha^2\rho^2\right).\label{rinduncflat}
\end{equation}
Thus, the acceleration increases the uncertainty irrespective of background. In flat space, for example, the correction mimics a globally minimal momentum uncertainty $\sigma_{p}\geq \sqrt{5/6}\hbar\alpha\propto T_U,$ with the Unruh temperature $T_U$. This has an intuitive interpretation as the maximal possible wavelength of a particle due to a thermal bath in the background.

\subsection{Spherically symmetric solutions} 
\label{subsec:sphapp}

Spherically symmetric metrics are naturally expressed in such a way that the shift vector vanishes. Correspondingly, the relativistic uncertainty relation \eqref{relres} depends only on the Ricci scalar of the effective metric. In the following, we first find a general solution for these kinds of backgrounds to afterwards apply it to Schwarzschild black holes and the static patch of the de Sitter spacetime.

\subsubsection{General result}\label{subsubsec:genressphapp}

Written in terms of the coordinates $r^i=(r,\theta,\phi)$ in Lema\^itre-Tolman-Bondi form \cite{Lemaitre33,Tolman34,Bondi47}, every spherically symmetric line element can be expressed as
\begin{align}
\D s^2=\frac{\D r^2}{1+2E(r)}+r^2\left(\D\theta^2+\sin^2\theta\D\phi^2\right).\label{effLTB}
\end{align}
The shape of the function $E(r)$ indicates that we deal with spatially dependent curvature. This ansatz was used in many cosmological attempts of modelling dark energy caused by spherically symmetric voids of matter in the universe \cite{GarciaBellido09,February09}, leading to a dipolar distribution of objects in the sky \cite{Schwarz16,Migkas20}. 

We denote the distance of the geodesic ball from the center of symmetry by $r_0.$ Due to the symmetry of the background spacetime, this completely characterizes the position of the center of the ball $p_0$ because the curvature function and its derivatives satisfy $E^{(n)}_0\equiv E^{(n)}|_{p_0}=E^{(n)}(r_0),$ where the subscript in this case denotes the $\text{n}^{\text{th}}$ derivative with respect to $r.$ Then, the leading-order contribution reads
\begin{align}
\sigma_\pi\rho	&\geq \pi\hbar\sqrt{1+2\frac{E_0+r_0 E'_0}{3\pi^2r_0^2} \rho^2}\\
						&\simeq \pi\hbar\left(1+\frac{E_0+r_0E'_0}{3\pi^2r_0^2} \rho^2 \right).\label{sphleadcont1}
\end{align}
Plugging in \eg $E_0=(-1/2)Kr_0^2,$ \ie assuming constant curvature $K$, this equation recovers the result that was already obtained nonperturbatively in ref. \cite{Schuermann18} and stated in Eq. \eqref{Kgeom}. Additionally, the fourth-order contribution, displayed in Eq. \eqref{res4app}, vanishes, thereby validating the formalism - a perturbative expansion of a polynomial trivially equals itself, confirming the result cited above. 

In general, we have to derive the function $E$ from the lapse function and the induced metric of the chosen background.

\subsubsection{Schwarzschild static patch}
\label{subsubsec:Schwarzsphapp}

For the static patch of the Schwarzschild geometry, \ie an eternal black hole or the exterior of a massive planet as it is seen by the observer at spacelike infinity, the effective three-metric reads
\begin{align}
    \D s^2=\frac{1}{\sqrt{1+2\phi_{\text{GR}}}}\left(\frac{\D r^2}{1+2\phi_{\text{GR}}}+r^2\D\Omega^2\right),
\end{align}
where we introduced the gravitational potential of a point particle of mass $M$ as $\phi_{\text{GR}}\equiv -GM/r.$ To rewrite this expression in the form \eqref{effLTB}, it is necessary to find a new radial coordinate $R,$ satisfying
\begin{equation}
    R=\frac{r}{\sqrt[4]{1+2\phi_{\text{GR}}}},
\end{equation}
which is a meaningful transformation as long as the gravitational potential satisfies $\phi_{\text{GR}}>-\frac{5}{2},$ \ie until just outside the horizon. Then, the characteristic function of the spherical geometry reads to second order in the gravitational potential \begin{equation}
    E(R)\simeq \frac{3}{2}\phi_{\text{GR}}+\frac{5}{8}\phi_{\text{GR}}^2.
\end{equation}
This implies that the relativistic uncertainty relation becomes at the same level of approximation
\begin{equation}
    \sigma_p\rho\gtrsim\pi\hbar\left(1+\frac{5}{24\pi^2}\frac{\left.\phi_{\text{GR}}^2\right|_{r_0}\rho^2}{r_0^2}\right)=\pi\hbar\left[1+\frac{5}{24\pi^2}\left(\left.\vec{\nabla}\phi_{\text{GR}}\right|_{r_0}\right)^2\rho^2\right]\label{schwarzresapp},
\end{equation}
which comparing to Eq. \eqref{rinduncflat} is an exceedingly complicated way of writing that, to lowest order in the gravitational potential, its gradient acts like a force or $\vec{\alpha}=\vec{\nabla}\phi_{\text{GR}}.$

A more precise calculation, omitting the expansion in the gravitational potential, shows that the full correction to the uncertainty relation equals
\begin{equation}
    \sigma_p\rho\geq\pi\hbar\left(1+\frac{5}{24\pi^2}\frac{\left.\phi_{\text{GR}}^2\right|_{r_0}}{\sqrt{1+2\left.\phi_{\text{GR}}\right|_{r_0}}}\frac{\rho^2}{r_0^2}\right)\label{schwarzres}.
\end{equation}

\subsubsection{de Sitter static patch}
\label{subsubsec:dSsphapp}

The effective three-metric on a spacelike hypersurface of the static patch of the de Sitter geometry, which manifestly describes an unchanging cosmological horizon, reads
\begin{align}
    \D s^2=\frac{1}{\sqrt{1-\frac{r^2}{r_H^2}}}\left(\frac{\D r^2}{1-\frac{r^2}{r_H^2}}+r^2\D\Omega^2\right).
\end{align}
This geometry is of particular interest because the EUP, as it is usually applied phenomenologically, is motivated from the maximum length introduced by this very horizon. Therefore, which influence is dominant in earth-bound experiments is an important question to answer. 

By analogy with the previous case, the metric first has to be described in terms of the new coordinate
\begin{equation}
    R=\frac{r}{\sqrt{1-\frac{r^2}{r_H^2}}},
\end{equation}
a transformation, which is valid within the entire patch. Correspondingly, for small $R/r_H$ the required function can be approximated as
\begin{equation}
    E(R)\simeq\frac{R^4}{8r_H^4}.
\end{equation}
This leads to the approximated result 
\begin{equation}
    \sigma_p\rho\gtrsim\pi\hbar\left(1+\frac{5}{24\pi^2}\frac{r_0^2\rho^2}{r_H^4}\right),
\end{equation}
which, interestingly is of exactly the same form as Eqs. \eqref{rinduncflat} and \eqref{schwarzresapp}. This fact renders a direct comparison to the influence of massive bodies straight-forward. At the linear level, the two contributions will simply add in describing the uncertainty of a particle moving in the Schwarzschild-de Sitter geometry, yielding
\begin{equation}
    \sigma_p\rho\gtrsim\pi\hbar\left[1+\frac{5\rho^2}{24\pi^2r_0^2}\left(\frac{r_0^4}{r_H^4}+\phi_{\text{GR}}|_{r_0}^2\right)\right].\label{dSresapp}
\end{equation}
Introducing the Schwarzschild radius $r_S=2GM,$ the effect of the massive body in the center of symmetry dominates as long as
\begin{equation}
    r_0<r_{\text{lim}}\equiv\sqrt[3]{r_Sr_H^2}.
\end{equation}
For the earth, which corresponds to a Schwarzschild radius $r_S\sim 1\millimeter,$ and the cosmological horizon of a size of around $r_H\sim 10^{26}\meter,$ the limiting distance from the center of symmetry amounts to about $r_{\text{lim}}\sim 10^{16}\meter\sim 10\hspace{1.5pt}\text{ly},$ \ie far beyond Alpha Centauri. We can thus safely neglect the EUP induced by the cosmological horizon and rather concentrate on the local deviations.

For completeness, the uncertainty relation induced by the de Sitter horizon without further assumptions on its size becomes by almost complete analogy with Eq. \eqref{schwarzresapp}
\begin{equation}
    \sigma_p\rho\geq\pi\hbar\left(1+\frac{5}{24\pi^2}\frac{\frac{r_0^4}{r_H^4}}{\sqrt{1-\frac{r_0^2}{r_H^2}}}\frac{\rho^2}{r_0^2}\right)\label{dSres}.
\end{equation}

All the applications treated up until now where simple, inasmuch as they did not include rotation as indicated by a nonvanishing shift vector. It is time to start manoeuvring more difficult waters.

\subsection{Rotating geometries}

Running the risk of entering comparably complicated terrain, this subsection is aimed at obtaining the uncertainty relation for spacetimes with nonvanishing shift vector. In that vein, we start our analysis with the G\"odel universe. Subsequently, we focus our attention on the weak-field approximation of the Kerr-geometry, \ie the metric describing a rotating source in the context of general relativity. Finally, we investigate spacetimes stemming from rotating compact objects in the framework of higher-order theories of gravity with the purpose of pinpointing the main differences with respect to the standard scenario.

\subsubsection{G\"odel universe}

The G\"odel solution \cite{Goedel49,Kajari04,Buser13} is a homogeneous and anisotropic spacetime arising from Einstein's field equations for a perfect fluid with nonvanishing angular momentum. It essentially describes a rotating universe in which closed timelike curves are allowed, thus, in principle, permitting time travel. The line element associated with such a curved background, written in cylindrical coordinates $(t,r^i)=(t,r,\phi,z),$ reads \cite{Goedel49,Kajari04,Buser13}
    \begin{align}\nonumber
\D s^2  =&-\D t^2-\frac{2r^2}{\sqrt{2}a}\D t \D\phi+\frac{\D r^2}{\left(1+\frac{r^2}{4a^2}\right)}+r^2\left(1-\frac{r^2}{4a^2}\right)\D\phi^2+\D z^2,\label{eqgodel}
\end{align}
with the constant parameter $a>0$ which has units of length and quantifies the angular momentum of matter. Have in mind that this slicing only covers the region $r<2a.$

An observer orbiting circularly around the $z$-axis (\ie co-rotating with the Gödel universe) experiences the flow of proper time according to the time coordinate $t.$ The effective lapse function, three-metric and shift vector from the point of view of this observer read
\begin{align}
N                                       =&\sqrt{1+\frac{2}{4\frac{a^2}{r^2}-1}},\\
G_{ij}\D r^i\D r^j                      =&\frac{1}{N}\left[\frac{\D r^2}{1+\frac{r^2}{4a^2}}+\left(1-\frac{r^2}{4a^2}\right)r^2\D\phi^2+\D z^2\right],\\
N^i\frac{\partial}{\partial r^i}        =&-\frac{a}{\sqrt{2}\left(a^2-\frac{r^2}{4}\right)}\partial_\phi.
\end{align}
As $r<2a$ in this slicing, and the prefactors of terms containing higher powers of $r_0/2a$ (where $r_0=r|_{p_0}$) in the resulting uncertainty relation get ever smaller, we only display the leading order for the sake of brevity. Correspondingly, the observer defined above measures the uncertainty relation
\begin{equation}
    \sigma_p\rho\gtrsim \pi\hbar\left\{1+\frac{\rho^2}{4\pi^2a^2}\left[\mathcal{F}_3\left(\Pi|_{N=1}\right)-1\right]\right\},
\end{equation}
thus continuously increasing the uncertainty. In the nonrelativistic and ultrarelativistic regimes, this inequality becomes
\begin{align}
    \sigma_p\rho|_{\Pi\ll1}\gtrsim & \pi\hbar\left[1-\frac{\rho^2}{4\pi^2a^2}\left(1-\frac{\pi^2\xi}{\Pi^2}\right)\right],\\
    \sigma_p\rho_{\Pi\gg1}\gtrsim & \pi\hbar\left[1-\frac{\rho^2}{4\pi^2a^2}\left(1-\pi^2\tilde{\xi}\right)\right],
\end{align}
implying that the correction is minimal for relativistic, \ie light, particles. Clearly, this should be expected from a gravitational effect.

\subsubsection{Rotating source in general relativity\label{subsec:lense}}

The phenomenon of frame-dragging was discovered only few years after the final settlement of GR. As a matter of fact, in 1918 Lense and Thirring found the weak-field limit for the spacetime generated by a rotating body \cite{Lense18}. The main prediction of this solution is the existence of a precession of the orbits drawn by a test body, a feature that is completely absent in Newtonian mechanics. To get to this conclusion, they argued that in isotropic spherical coordinates $(t,r^i)=(t,r,\theta,\varphi)$ the metric tensor originating from a rotating source takes the form \cite{Lense18}
\begin{align}\label{eqlense}
\D s^2= &-(1+2\phi_{\text{GR}})\D t^2+(1-2\phi_{\text{GR}})\left(\D r^2+r^2\D\Omega^2\right)+4\phi_{\text{GR}} a_J\sin^2\theta \D\varphi \D t,
\end{align}
where $a_J=J/M$ is the rotational parameter with $J$ denoting the angular momentum of the source. 

Note that the time coordinate $t$ used here corresponds to the time measured by a static observer at infinite distance from the gravitating body in the center. In turn, the uncertainty is calculated as it would be measured by this observer. In general, though, this provides a good approximation of the slicing carved out by the dynamical rest frame of the particle itself as long as it does not get too close to the horizon. Therefore, corrections to the results are expected to be of higher order. Analogous considerations apply to the extended models of gravity which are treated as corrections to Eq. \eqref{eqlense} below.

The static observer at infinity experiences the lapse function, shift vector and effective metric 
\begin{align}
N                                           \simeq&1-\phi_{\text{GR}},\\
N^i\frac{\partial}{\partial r^i}            \simeq&-2a_J\frac{\phi_{\text{GR}}}{r^2},\\
    G_{ef}\D r^e\D r^f                      \simeq&\left(1+3\phi_{\text{GR}}\right)\left(\D r^2+r^2\D\Omega^2\right).
\end{align}
Unfortunately, this reasoning leads to an uncertainty which is at least quadratic in the gravitational potential, while the Lense-Thirring solution corresponds to a first-order expansion. In light of this, we generalize the discussion by approximating the Kerr metric in Boyer-Lindquist coordinates
\begin{align}
    \D s^2=&-\left(1+2\tilde{\phi}_{\text{GR}}\frac{\tilde{r}^2}{\Xi^2}\right)\D t^2+4\tilde{\phi}_{\text{GR}}a_J\sin^2\theta\frac{\tilde{r}^2}{\Xi^2}\D t\D \varphi\nonumber\\
    &+\frac{\Xi^2}{\Sigma}\D \tilde{r}^2+\Xi^2\D\theta^2+\tilde{r}^2\left(1+\frac{a_J^2}{\tilde{r}^2}-2\tilde{\phi}_{\text{GR}}\sin^2\theta\frac{a_J^2}{\Xi^2}\right)\sin^2\theta\D\varphi^2,
\end{align}
with $\tilde{\phi}_{\text{GR}}=-GM/\tilde{r}$, $\Xi=\tilde{r}\sqrt{1+\cos^2\theta a_J^2/\tilde{r}^2}$ and $\Sigma=\tilde{r}^2(1+\tilde{\phi}_{\text{GR}}+a_J^2/\tilde{r}^2),$ to fourth order in $\tilde{\phi}_{\text{GR}}$ and $a_J/\tilde{r}$ simultaneously. Bear in mind that the Schwarzschild-like $\tilde{r}$ relates to the radial coordinate introduced with the Lense-Thirring metric as $\tilde{r}=r(1-\phi_{\text{GR}}/2)^2.$ As the resulting uncertainty relation is 3-diffeomorphism invariant, we nevertheless provide it in terms of $r$ for the sake of future convenience.

Consequently, the static observer at infinity measures the uncertainty relation
\begin{align}
    \sigma_P\rho\geq &\pi\hbar\left\{1+\frac{\left.\phi^2_{\text{GR}}\right|_{r_0}\rho^2}{48\pi^2 r_0^2}\left[10+30\left.\phi_{\text{GR}}\right|_{r_0}+55\left.\phi^2_{\text{GR}}\right|_{r_0}\right.\right.\nonumber\\
    &\left.\left.-\frac{a_J^2}{r_0^2}\left(469-217\cos 2\theta-96\mathcal{F}_3|_{N=1}\left(7-3\cos 2\theta\right)\right)\right]\right\}.\label{KerrRes}\\
    \equiv&\pi\hbar\left(1+\lambda_{LT}\right),
\end{align}
where the last line defines the (generalized) Lense-Thirring correction $\lambda_{LT}.$ Note that this result recovers Eq. \eqref{schwarzresapp} when restricted to quadratic order in Newton's potential as expected.

\begin{figure}[h!]
\begin{minipage}{.49\linewidth}
\centering
        \includegraphics[width=\linewidth]{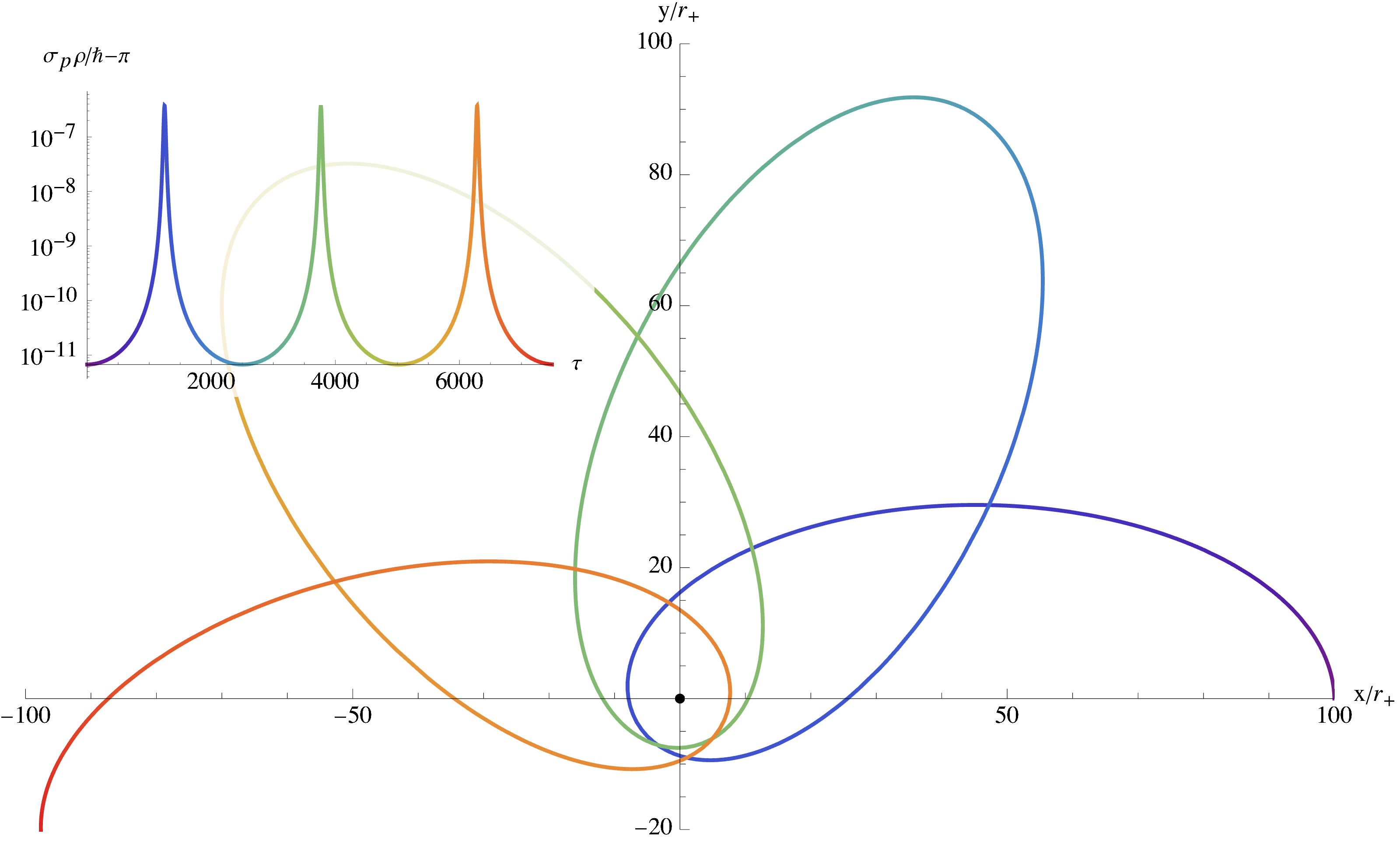}
\end{minipage}
\begin{minipage}{.49\linewidth}
\centering
        \includegraphics[width=.95\linewidth]{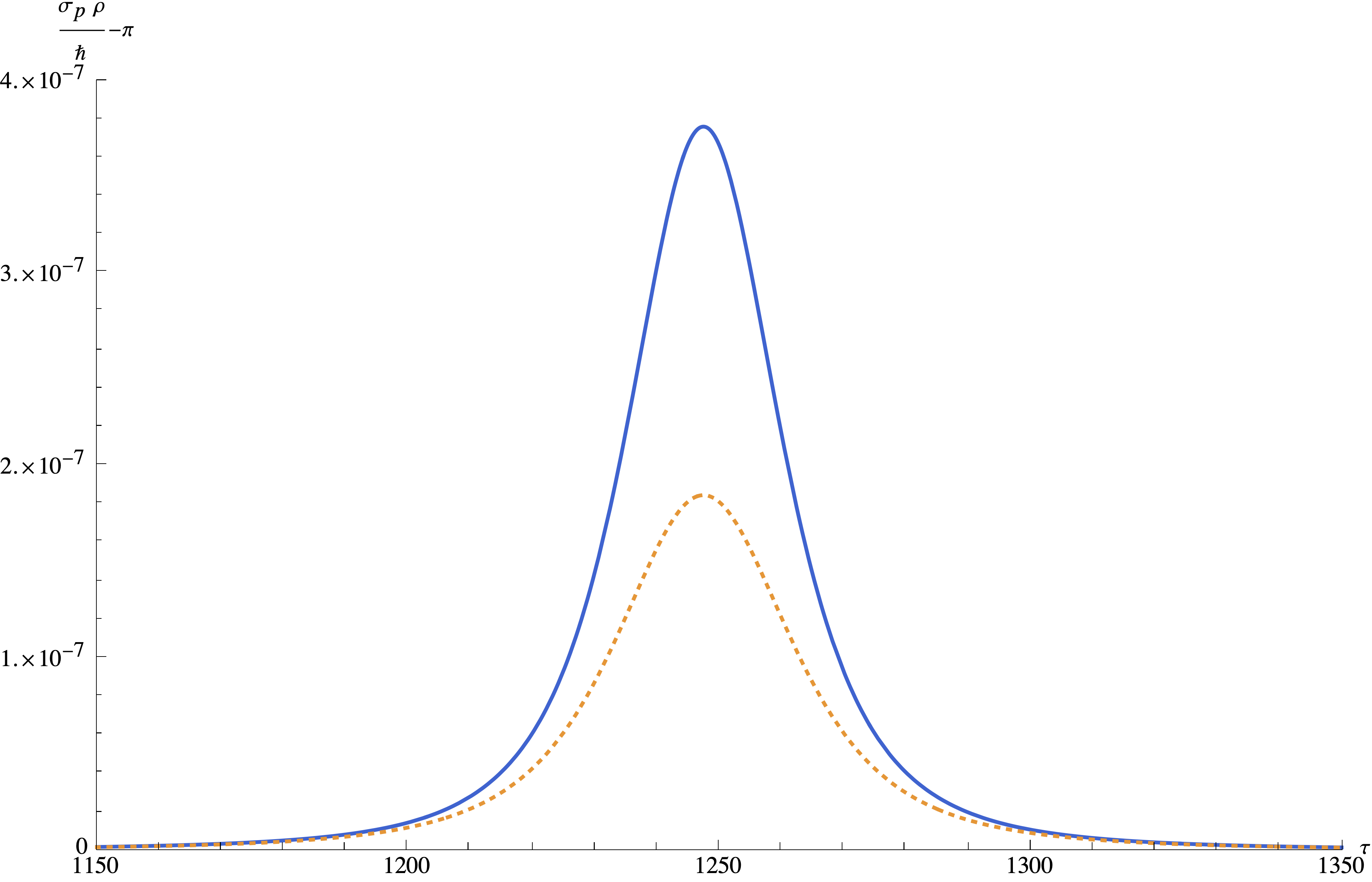}
\end{minipage}
\caption{The left plot shows the trajectory followed by a massive particle in the equatorial ($x$-$y$) plane of a fast black hole, rotating as $a_J/GM=0.5,$ and with an outer horizon of radius $r_+,$ symbolized by the black disk in the center. Its starting point lies on the $x$ axis at a distance $100r_+$ from the source, with initial velocity $u(\tau =0)\simeq(1.010,-0.0010,0.0000,0.0001)$ in Boyer-Lindquist coordinates. The color, ranging from violet to red, indicates an increase in the affine parameter $\tau.$ Inset in the top left corner is a plot displaying all corrections to the uncertainty relation in flat space in units of $\hbar$ logarithmically, experienced by a particle of mass $m=\hbar /r_+,$ with position uncertainty $\rho=10^{-1}r_+$ along this orbit as a function of proper time. On the right-hand side, the fully relativistic (blue) and nonrelativistic (orange, dotted) corrections are compared, allowing a closer look at the first peak of the uncertainty relation.}
    \label{fig:relnonrelcomp}
\end{figure}

A sample orbit in the equatorial plane of the Kerr metric approximated as indicated above is given on the left-hand-side of Fig. \ref{fig:relnonrelcomp}, where the color of the curve changes with increasing proper time. The ensuing correction to the uncertainty relation is displayed in the inset plot in the top-left corner as a function of said affine parameter. The peaks correspond to the times of closest approach to the outer horizon along the trajectory. Thus, the influence is strongest, when the curvature is large as expected. The nonrelativistic and relativistic expressions are compared graphically at the first peak in the plot to the right of Fig. \ref{fig:relnonrelcomp}. The new corrections thus lead to an increase of the effect by a factor of two for the given choice of parameters.

\subsubsection{Fourth-order gravity}

Fourth-order gravity, introduced by Stelle \cite{Stelle77}, represents one of the first attempts to cure the quantization problems of the gravitational interaction. In particular, it was pointed out \cite{Stelle77} that the introduction of higher-derivative terms in the Einstein-Hilbert action can make the model renormalizable. To be more precise, according to the prescription in \cite{Stelle77,Stelle78}, the gravitational action from which to build up quantum gravity should be given by
\begin{equation}\label{actionstelle}
S=\frac{1}{16\pi^2G^2}\int d^4x\sqrt{-g}\left(R+\frac{R^2}{2\hbar^2m_s^2}-\frac{1}{2\hbar^2m_t^2}R_{\mu\nu}R^{\mu\nu}\right),
\end{equation}
where $m_s$ and $m_t$ are dimensionful constants measured in units of mass. However, the drawback of this model consists in the appearance of ghost-like degrees of freedom which undermine the unitarity of the underlying quantum field theory. Such a circumstance is a typical feature of local higher-order derivative gravity \cite{Asorey97}.

For the current model, a Lense-Thirring-like solution has been recently obtained when analyzing the light bending due to these theories \cite{Buoninfante20c}. In isotropic spherical coordinates, the aforementioned solution is
\begin{align}\label{geq}
\D s^2= &-(1+2\phi)\D t^2+(1-2\psi)\left(\D r^2+r^2\D\Omega^2\right)+2\xi\sin^2\theta \D\varphi \D t\,,
\end{align}
where the gravitational potentials are given by
\begin{align}
\phi=& \,\phi_{\text{GR}}\left(1+\frac{1}{3}e^{-m_0r/\hbar}-\frac{4}{3}e^{-m_2r/\hbar}\right)\label{eqstelle1},\\[2mm]
\psi=& \,\phi_{\text{GR}}\left(1-\frac{1}{3}e^{-m_0r/\hbar}-\frac{2}{3}e^{-m_2r/\hbar}\right)\label{eqstelle2},\\[2mm]
\xi=& \,2\phi_{\text{GR}}a_J\left[1-(1+m_2r/\hbar)e^{-m_2r/\hbar}\right],\label{eqstelle3}
\end{align}
with $m_0=2m_t/\sqrt{12m_t^2/m_s^2-1}$ and $m_2=\sqrt{2}m_t$ being the masses of the spin-0 and spin-2 massive modes, respectively.

As the influence stemming from the higher-derivative terms ought to be small in comparison to the general relativistic effect, results should be given as corrections to the Lense-Thirring outcome. Written this way, the uncertainty relation reads for small gravitational potentials
\begin{align}
    \sigma_p\rho\geq&\pi\hbar\Bigg[1+\lambda_{LT}-\frac{\phi_{\text{GR}}|_{p_0}}{18\pi^2}\left(\frac{\rho^2}{\lambda_{m_0}^2}e^{-\frac{m_0 r_0}{\hbar}}+\frac{8\rho^2}{\lambda_{m_2}^2}e^{-\frac{m_2 r_0}{\hbar}}\right)\Bigg],
\end{align}
where $\lambda_{m_0}$ and $\lambda_{m_2}$ denote the Compton wavelengths corresponding to the respective massive gravitational modes.

As straightforwardly recognizable in the previous equation, for the current example we do not have to resort to a higher-order expansion of the Kerr-like solution in the context of the examined extended model of gravity as the leading-order correction is already linear in $\phi_{\text{GR}}.$ Thus, it clearly stems from the Ricci scalar of the effective metric, which is why it is equal for relativistic and nonrelativistic particles. This feature is shared by the upcoming analysis.

\subsubsection{Infinite-derivative gravity}

Starting from the above scenario, and recalling that the reasoning in \cite{Asorey97} prevents any local higher-order derivative gravity from being free from ghost fields, it is clear that locality must be given up on to arrive at a quantum gravitational model which is simultaneously renormalizable and unitary. However, nonlocality should be manifest only in the currently unexplored UV regime since all the available data acquired from gravitational experiments comply with the local behaviour of gravity. Along this direction, it is possible to encounter the so-called infinite-derivative gravity theory, which precisely possesses the characteristics listed above. As the name suggests, the usual Einstein-Hilbert action is now accompanied by nonlocal functions of the curvature invariants; in the simplest form, the nonlocal gravitational action reads \cite{Biswas12,Buoninfante18}
\begin{align}
S=&\frac{1}{16\pi^2G^2}\int \D^4x\sqrt{-g}\left(R+R\frac{1-e^{-{\hbar^2\Box}/{\kappa^2}}}{4\Box}R\right.-\left.R_{\mu\nu}\frac{1-e^{-{\hbar^2\Box}/{\kappa^2}}}{2\Box}R^{\mu\nu}\right),\label{actionidg}
\end{align}
where $\kappa$ is the energy scale at which the nonlocal aspects of gravity are expected to be prominent.

As for the previous example, a Lense-Thirring-like solution can be analytically computed in this framework. Formally, the shape of the metric tensor is the same as the one exhibited in \eqref{geq}, with the difference that the gravitational potentials are instead represented by
\begin{align}
\phi=&\psi= \,\phi_{\text{GR}}\rm\left(\frac{\kappa\,r}{2\hbar}\right),\label{eqidg1}\\[2mm]
\xi=& \,2\phi_{\text{GR}}a_J\left[\rm{Erf}\left(\frac{\kappa\,r}{2\hbar}\right)-\frac{\kappa\,r}{\sqrt{\pi}\hbar}e^{-\kappa^2r^2/4\hbar^2}\right],\label{eqidg2}
\end{align}
where $\rm{Erf}(x)$ denotes the error function.

Again expressed as corrections to the Lense-Thirring result, the uncertainty relation for small gravitational potentials becomes
\begin{equation}
    \sigma_p\rho\geq\pi\hbar\left(1+\lambda_{LT}-\frac{\phi_{\text{GR}}|_{p_0}}{4\pi^{\frac{5}{2}}}\rho^2r_0\frac{\kappa^3}{\hbar^3}e^{-\kappa^2 r_0^2/4\hbar}\right).
\end{equation}

\subsection{Summary}

The uncertainty relation \eqref{relres}, as nontrivial as it is, makes an immediate understanding difficult. Therefore, this section was used to display its behaviour in several important cases. 

In particular, assuming a vanishing shift vector, an accelerated particle experiences an additional increase in uncertainty proportional to the squared acceleration with respect to the corresponding inertial frame. This reflects the fact, that the dynamics are not governed by the induced metric on the spacelike hypersurfaces but the effective metric, additionally containing the lapse function. Accordingly, there is an effect through acceleration even when the background is Minkowskian.

Furthermore, quantum mechanical objects moving on a Schwarzschild geometry undergo approximately the same uncertainty-growth as they approach the source of the gravitational field. Comparison of this effect to the problem in Rindler coordinates quantitatively recovers the acceleration induced by a conservative gravitational potential. The investigated static patch of the de Sitter spacetime yields an effect at fourth order in the distance from the horizon. Therefore, a back-of-the-envelope estimate clearly shows that its influence can be safely neglected on the surface of the earth when comparing it to the effect induced by the planet itself, thus implying that the EUP, as it is often applied phenomenologically, may have to be questioned.

Additionally, we have computed the form of the uncertainty relation for the G\"odel universe, the Lense-Thirring solution and its extension in the framework of fourth-order and infinite-derivative gravity. In all cases, the corrections increase the uncertainty. In the explicit scenario of a relatively close orbit to a rotating black hole \`a la general relativity, the relativistic corrections significantly increased its value. Remarkably, whilst the leading-order contribution goes like $\phi_{\text{GR}}^2$ in the Lense-Thirring scenario, for the extended models we observe a proportionality to $\phi_{\text{GR}}$. Therefore, there may be a regime in which the two terms are comparably important, thus leading to a simultaneous ``coexistence'' of the two quantities. A similar occurrence has also been addressed in different contexts, as for the case of the Casimir effect \cite{Buoninfante19b}.  

\section{Modified commutators as manifestation of curved spaces}
\label{sec:modcomorcurspa}

The last four sections very much created a link between deformed Heisenberg algebras and curved spaces, showing that the latter, when looked at operationally, lead to similarly modified uncertainty relations. In this section we take the opposite route. In particular, we derive a theory of curved momentum space from the kind of algebra underlying the minimum length paradigm, \ie GUPs.

Note that this kind of connection had already been studied superficially in Ref. \cite{Chang10}. Furthermore, it is on display in the context of doubly special relativity \cite{Amelino-Camelia02b,Kowalski-Glikman02b,Amelino-Camelia02c,Amelino-Camelia02c,Amelino-Camelia10}, which may be interpreted as a theory defined on de Sitter-momentum space \cite{Kowalski-Glikman02c,Kowalski-Glikman03}. Moreover, it has been corroborated further from the geometric point of view \cite{Carmona21,Relancio21}. Those results provided a strong motivation to search for an equivalence between GUP-deformed quantum mechanics and quantum mechanics on curved momentum space. 

The aim of the present section lies in promoting the said link to an exact duality by introducing a novel set of conjugate variables $\hat{X}^i$ and $\hat{P}_i$ satisfying the $d$-dimensional Heisenberg algebra. Those new coordinates in the cotangent bundle can be used to describe the investigated kinds of modifications in $d$ dimensions canonically. As for the transformation often applied in case of the GUP on commutative space \cite{Brau06,Bosso18b}, this naturally leads to a modification of the single-particle Hamiltonian. The thus arising dynamics constitute motion on a nontrivial momentum space. For the quadratic GUP the curvature tensor is proportional to the coordinate noncommutativity. However, a commutative space does not imply that the corresponding background is trivial. On the contrary, the resulting basis in momentum space is nonlinearly related to the one underlying the Euclidean metric.

Therefore, it is possible to import bounds on the curvature of momentum space and the deviation from the canonical basis from the literature on noncommutative geometry and GUPs with commuting coordinates, respectively. We thus obtain a distinct interpretation for the already existing phenomenology. Furthermore, the new set of phase space variables allows for a rather simple treatment of noncommutative space in quantum mechanics, mapping it onto a theory which is analogous to quantum mechanics on curved manifolds as described in section \ref{sec:measures}. Note that an instance of this duality was obtained along a complementary road in Ref. \cite{Singh21}.

This section, containing work published in Ref. \cite{Wagner21a}, is organized as follows. Subsections \ref{subsec:curvedmom} and \ref{subsec:GUPQM} briefly recall the required information on curved momentum space and GUP-like deformations in quantum mechanics, respectively. The equivalence of those two theories is established in subsection \ref{subsec:equivalence} providing the map connecting them. Subsequently, the newly appearing geometrical observables are constrained in subsection \ref{subsec:constraints}. Finally, subsection \ref{subsec:Conclusion} is intended as summary and conclusion of the results.

\subsection{Curved momentum space}\label{subsec:curvedmom}

In order to understand curved momentum space, a short introduction to the geometry of generalized Hamilton spaces is indispensable. On the base of this reasoning and under the assumption that the metric bears no position-dependence, it is straightforward to construct the corresponding quantum theory.

\subsubsection{Geometry}\label{subsubsec:geo}

The theory of curved momentum spaces is derivative from the geometry of generalized Hamilton spaces \cite{Miron01,Miron12}, which is gradually seeing more application to physics, in particular in the context of the phenomenology of quantum gravity \cite{Barcaroli15,Carmona19,Relancio20a}. The starting point for this investigation is a metric which not only depends on the position but also the momentum of the investigated object
\begin{equation}
    g^{ij}=g^{ij}(x,p),
\end{equation}
where we, for the moment, assume that $\pi_i=p_i.$ To study the corresponding geometry, it is necessary to find a nonlinear connection $N_{ij}$ which governs the division of the cotangent bundle into horizontal ("position") and vertical ("momentum") space. This choice is highly nontrivial, though it can be simplified in a special case. Define the Cartan tensor of the background space as
\begin{equation}
    C^{kij}\equiv \frac{1}{2}\mpar^kg^{ij},\label{cartan}
\end{equation}
where the partial derivative with respect to momenta is denoted as $\mpar^i=\partial/\partial p_i.$ If this tensor turns out to be totally symmetric, the metric can be derived from the Hamiltonian of a free particle of mass $m$
\begin{equation}
    H=\frac{1}{2m}p_ip_jg^{ij},
\end{equation}
according to the relation
\begin{equation}
    g^{ij}=m\mpar^i\mpar^jH.
\end{equation}
Furthermore, a canonical nonlinear connection can be found as
\begin{equation}
    N_{ij}=\frac{1}{4}\left(\left\{g_{ij},H\right\}-g_{ik}\mpar^k\partial_jH-g_{jk}\mpar^k\partial_iH\right),
\end{equation}
where the symbols $\{,\}$ denote the Poisson bracket. Once the nonlinear connection is known, it is possible to derive the covariant derivatives in position and momentum space and the curvature tensors. 

Assuming that the metric is solely a function of the momenta
\begin{equation}
    g^{ij}=g^{ij}(p),\label{mommet}
\end{equation}
the nonlinear connection immediately vanishes, making the problem particularly simple. Correspondingly, the covariant derivative in position space is just the partial derivative. Motion in momentum space, on the other hand, is described by a Levi Civita-like connection related to the Cartan tensor
\begin{equation}
    C^{ij}_k=-\frac{1}{2}g_{kl}\left(\mpar^{i}g^{jl}+\mpar^{j}g^{il}-\mpar^{l}g^{ij}\right).
\end{equation}
Defining covariant differentiation in momentum space denoted by the symbol $\dot{\nabla}$ in the usual way, this makes it possible to construct a scalar from the Cartan tensor 
\begin{equation}
    C\equiv g_{(jk}g_{il)}\dot{\nabla}^lC^{ijk},\label{cartsca}
\end{equation}
where the parenthesis in the indices implies total symmetrization. If the Cartan tensor is totally symmetric, this quantity is uniquely defined and measures the departure from Riemannian geometry. Moreover, the curvature tensor in position space vanishes while its counterpart in momentum space $S_k^{~ilj}$ takes the familiar form
\begin{equation}
    S_k^{~ilj}=\mpar^jC_k^{il}-\mpar^lC_k^{ij}+C_k^{ml}C_m^{ij}-C_k^{mj}C_m^{il},\label{MomRiem}
\end{equation}
which is clearly reminiscent of the Riemann tensor. Therefore, the Hamilton geometry derived from a purely momentum-dependent metric is simply of Riemannian type. We can further define the Ricci scalar as usual
\begin{equation}
    S\equiv g_{ij}S_k^{~ikj}.\label{MomRic}
\end{equation}

Unfortunately, the metric, which will be treated below, does not generally yield a totally symmetric Cartan tensor \eqref{cartan}. Thus, we are dealing with a generalized Hamilton space. In this case, the nonlinear connection must be provided beforehand. By analogy with the simpler case, we choose the nonlinear connection to vanish because the metric harbours no position dependence. Then, the same reasoning follows. 

A note of caution might be in order, though. Have in mind, that the metric still constitutes a tensor and thus transforms as such. It can only be independent of the position if the system is described in Cartesian coordinates. Otherwise, several issues arise which complicate the process of quantization enormously. Fortunately, this set of coordinates suffices for the purpose of the present section. 

\subsubsection{Quantum mechanics}\label{subsubsec:curvqm}

Given a metric \eqref{mommet} and a vanishing nonlinear connection, it is possible to construct the line element in momentum space
\begin{equation}
    \D \tilde{s}^2=g^{ij}(p)\D p_i\D p_j.
\end{equation}
First and foremost, this implies that the dynamics of a single particle derive from a Hamiltonian operator of the form
\begin{equation}
    \hat{H}=\frac{1}{2m}\hp_i\hp_jg^{ij}\left(\hp\right)+V\left(\hx^i\right).\label{hamcurvmomspagen}
\end{equation}
On a rather cautionary note, bear in mind that the kinetic term could equivalently be chosen proportional to the squared geodesic distance from the origin in momentum space, \ie $\tilde{\sigma}^2(\hp)$ \cite{Amelino-Camelia11,Relancio21}, which satisfies a differential equation analogous to Eq. \eqref{def_geo_dist} with initial condition $\tilde{\sigma}(0)=0.$ As a result, the Hamiltonian would be covariant with respect to momentum diffeomorphisms in addition to the required covariance in position space. However, as was stressed in Sec. \ref{subsec:introcurvmom}, it is unclear whether this principle actually applies to our world \cite{Amelino-Camelia19,Gubitosi21}. Most importantly, either choice of kinetic part yields the same main results in Secs. \ref{subsec:applquadGUP} and \ref{subsec:applicationcot}. Indeed, in normal coordinates (\cf Sec. \ref{subsubsec:RNC}) both coincide.

How to quantize a single-particle theory governed by the Hamiltonian \eqref{hamcurvmomspagen} was already explained in section \ref{sec:measures}. Correspondingly, it acts on wave functions as
\begin{equation}
    \hat{H}\psi=\left[\frac{1}{2m}g^{ij}(p)p_ip_j+V\left(i\hbar\dot{\nabla}^i\right)\right]\psi.\label{curvham}
\end{equation}
Furthermore, the geodesic distance $\sigma$, the only possible position-dependent scalar appearing in the Hamiltonian, can be computed solving the differential equation 
\begin{equation}
    g^{ij}\partial_i\sigma^2\partial_j\sigma^2=4\sigma^2.
\end{equation}
In the given case, this procedure results in the expression
\begin{equation}
    \sigma^2=g_{ij}(p)\left(x-x_0\right)^i\left(x-x_0\right)^j,
\end{equation}
where $x_0^i$ denote the coordinates of the point, with respect to which the distance is calculated. For reasons of simplicity, choose it to coincide with the origin, \ie $x_0^i=0.$ In accordance with Eq. \eqref{xsqmomrep} and defining the determinant of the metric as $\det g^{ij}\equiv g$, the squared geodesic distance may be represented as the Laplace-Beltrami operator in momentum space
\begin{equation}
    \hat{\sigma}^2\psi=-\hbar^2\frac{1}{\sqrt{g}}\dot{\partial}^i\left(\sqrt{g}g_{ij}\dot{\partial}^j\psi\right),\label{actgeoddist}
\end{equation}
which is clearly Hermitian with respect to the measure $\D\mu=\sqrt{g}\D^dp.$ 

Evidently, this description bears much resemblance to quantum mechanics on a spatially curved manifold. Keep in mind, though, that this picture does not hold under general coordinate transformations.

\subsection{GUP-deformed quantum mechanics}\label{subsec:GUPQM}

A more in-depth treatment of deformed single-particle quantum mechanics was already provided in section \ref{sec:GUPsEUPs}. The required parts are repeated here for convenience. In contrast to the theory described in the previous section, quantum mechanics with a minimum length is derived from a deformed algebra of observables
\begin{subequations}
\label{GUPalg}
\begin{align}
    [\hat{x}^a,\hat{x}^b]=&i\hbar\theta^{ab}(\hat{x},\hat{p}),\\
    [\hat{p}_a,\hat{p}_b]=&0,\\
    [\hat{x}^a,\hat{p}_b]=&i\hbar f^a_b(\hat{p}),\label{posmomgupalg}
\end{align}
\end{subequations}
where we introduced the tensor-valued functions $\theta^{ab}(\hat{x},\hat{p})$ and $f^a_b(\hp),$ which are not independent. Instead, they are constrained by the Jacobi identity
\begin{equation}
    \left[\theta^{ab},\hat{p}_c\right]=2\left[f^{[a}_c,\hx^{b]}\right].\label{JacobiId}
\end{equation}

As described in section \ref{subsec:theo}, the usual way to go at this point consists in finding a representation for this algebra in momentum space. For example, the position operator may read \cite{Kempf94}
\begin{equation}
    \hat{x}^a\psi=i\hbar f^a_b(p)\dot{\partial}^b\psi\label{GUPposop}.
\end{equation}
Within this representation, the Jacobi identity \eqref{JacobiId} can be solved yielding
\begin{equation}
    \theta^{ab}=2f^{[a}_c\dot{\partial}^{|c|}f^{b]}_d\left(f^{-1}\right)^d_ex^e\propto \hat{J}^{ba}.\label{noncom}
\end{equation}
At first glance, the theory of GUPs and the theory of curved momentum space differ substantially. How, then, can they be reconciled with each other?

\subsection{Equivalence of the modifications}\label{subsec:equivalence}

The algebra \eqref{GUPalg} indicates that the kinematical description in the GUP-approach is based on unusual coordinates in phase space. In particular, they are not of Darboux-form, which would imply the canonical commutation relations \eqref{HeisAlg} to be satisfied. The Darboux theorem \cite{Darboux82}, however, states that symplectic manifolds, like phase space, have vanishing curvature. Thus, provided the necessary transformation is found, every system can be expressed in terms of Darboux coordinates. The task of this section entails finding new operators 
\begin{equation}
    \hat{x}^a\rightarrow \hat{X}^i\left(\hat{x},\hat{p}\right),\hspace{1cm} \hat{p}_a\rightarrow\hat{P}_i\left(\hat{p}\right)\label{pstrans}
\end{equation}
such that $\hat{X}^i$ and $\hat{P}_i$ satisfy the Heisenberg algebra \eqref{HeisAlg}. A similar approach albeit with different realization and goals was followed in Ref. \cite{Galan07} in the context of DSR.

\subsubsection{Transformation}

Let us, in particular, assume that the transformation takes the shape
\begin{subequations}\label{trans}
\begin{align}
    \hx^a=&\left(e^{-1}\right)^a_i(\hP)\hX^i\label{xtrans},\\
    \hp_a=&e_a^i(\hP)\hP_i,\label{ptrans}
\end{align}
\end{subequations}
where the coordinates transform according to the operator ordering imposed by geometric calculus (\cf section \ref{subsec:vecop}), applied to momentum space. Note that other operator orderings would yield equivalent theories \cite{Bosso20c,Bosso21}, which, however, would not manifestly unveil the nontriviality of momentum space. 

This transformation immediately implies that the Hamiltonian describing the dynamics of a nonrelativistic particle may be reexpressed as
\begin{equation}
    \hat{H}=\frac{1}{2m}\hP_i\hP_je^i_ae^j_b\delta^{ab}+V\left[\left(e^{-1}\right)^a_i\hX^i\right].\label{hamiltaftertrafo}
\end{equation}
Moreover, the geodesic distance in the original flat background transforms in a similar way to the kinetic energy
\begin{equation}
\hat{\sigma}^2=\delta_{ab}\hx^a\hx^b=\delta_{ab}\left(e^{-1}\right)^a_i\hX^i\left(e^{-1}\right)^b_j\hX^j.
\end{equation} 
Thus, the matrix $e^i_a$ may be understood as \emph{vielbein}. Then, we may construct the metric and its inverse as
\begin{align}
    g^{ij}=&\delta^{ab}e^i_ae^j_b\label{metfromviel},\\
    g_{ij}=&\delta_{ab}\left(e^{-1}\right)_i^a\left(e^{-1}\right)_j^b.
\end{align}
 For this structure to be consistent, the measure has to read
\begin{equation}
\D\mu=\det \left(e^i_a\right)\D^dp,
\end{equation} 
\ie represent the volume form derived from the metric (\cf sections \ref{subsec:curved} and \ref{subsec:curvedmom}). Correspondingly, the Hamiltonian acts in momentum space as
\begin{align}
    \hat{H}\psi(P)=& \frac{P_i P_jg^{ij}}{2m}\psi(P)
    + V\left[i\hbar\left(e^{-1}\right)^a_i\dot{\nabla}^i\right]\psi(P),\label{HamAfterTrafo}
\end{align}
while the geodesic distance exactly obeys Eq. \eqref{actgeoddist}.

Under the assumption that the transformed phase space coordinates obey the Heisenberg algebra, the commutator of positions and momenta \eqref{GUPalg} implies the Jacobian
\begin{equation}
    \dot{\partial}^aP_j=\left(f^{-1}\right)_b^a\left(e^{-1}\right)_j^b,\label{Jacobian}
\end{equation}
which may be rewritten as a condition on the \emph{vielbein}
\begin{equation}
    f^{[a}_d\left[\dot{\partial}^{|d|}\left(e^{-1}\right)^{b]}_je^j_c-\dot{\partial}^{|d|}f^{b]}_d\left(f^{-1}\right)^d_c\right]=0.\label{conscondgen1}
\end{equation}
Then, the tensor measuring the spatial noncommutativity reads after some algebra
\begin{equation}
	\theta^{ab}=2f_c^{[a}\dot{\partial}^{|c|}f^{b]}_d\left(f^{-1}\right)^d_ex^e.\label{conscondgen2}
\end{equation}
Fortunately, this relation, derived from the assumptions that the new phase space coordinates obey the Heisenberg algebra and that the original variables satisfy the commutation relations \eqref{GUPalg} and the Jacobi identity \eqref{JacobiId}, reproduces the condition on the noncommutativity of space in the original representation \eqref{noncom}. Thus, the transformation introduced in this section can always be performed.

To put it in a nutshell, it is possible to describe the dynamics implied by any set of deformed commutators of the form \eqref{GUPalg} by Darboux coordinates, defined in Eqs. \eqref{xtrans} and \eqref{ptrans}, if the matrix characterizing the transition satisfies the consistency condition \eqref{conscondgen1}, and the noncommutativity of the spatial coordinates is of the form \eqref{conscondgen2}. The background, the system is moving on, will then necessarily be nontrivial.

Note, though, that this is how the metric can be determined in terms of the original momenta $\hp_a.$ In principle, as can be seen from the relation
\begin{equation}
    e^i_a\left(\hp_b\right)=e^i_a\left[e^j_b\left(\hp_c\right)\hP_j\right]=\dots,\label{infreg}
\end{equation}
trying to express the result in terms of the transformed momenta $\hP_i$ leads to an infinite regress. Yet, this problem can be circumvented by solving it iteratively as in perturbation theory. Before  we get to this point, though, it is instructive to show how the consistency conditions turn out when $\theta^{ab},$ $f^a_b$ and $e^i_a$ are expressed in terms of scalar functions.

\subsubsection{Conditions on scalars}\label{subsubsec:scacond}

As may be deduced from the Jacobi identity \eqref{JacobiId}, the spatial noncommutativity depends on the original phase space variables as
\begin{equation}
	 \theta^{ab}=\theta\left(\hp^2\right)\hat{J}^{ba},
\end{equation}
where the newly introduced dimensionful scalar $\theta$ measures the noncommutativity of space. Furthermore, expressed in a way similar to Refs. \cite{Chang10,Chang11,Chang16}, the quantity $f^a_b,$ being a tensor, assumes the form
\begin{equation}
    f^a_b=A\left(\hp^2\right)\delta^a_b+B\left(\hp^2\right)\frac{\hp^a\hp_b}{\hat{p}^2},\label{introAB}
\end{equation}
where we introduced the dimensionless scalars $A$ and $B.$ Note that they have to satisfy the conditions $A(0)=1$ and $B(0)=0$ for the given phase space variables to reduce to ordinary canonical conjugates in the low-energy limit. Both scalars are related to the function $\theta$ according to Eq. \eqref{conscondgen2}
\begin{equation}
    \theta=2\left(\log A\right)'(A+B)-\frac{B}{\hp^2},\label{noncomcondsca}
\end{equation}
where the prime denotes derivation with respect to $\hp^2.$ 

Furthermore, providing the \emph{vielbein} in the most general form compatible with the GUP
\begin{equation}
    e^i_a=C\left(\hp^2\right)\delta^i_a+D\left(\hp^2\right)\frac{\hp^i\hp_a}{\hp^2},\label{introCD}
\end{equation}
Eq. \eqref{conscondgen1} suffices to determine the newly introduced dimensionless scalar functions $C$ and $D,$ implying the relation
\begin{equation}
    \frac{D}{C}=\left[\theta+2\left(\log C\right)'(A+B)\right]\hp^2=A-1,\label{conscondgen2res}
\end{equation}
which, assuming that the background reduces to flat space in the low-energy limit, \ie $C(0)=1$ and $D(0)=0,$ can be solved to yield
\begin{align}
    C=&\exp\left(\frac{1}{2}\int_0^{\hp^2}\frac{A-1-\theta}{A+B}(q)\D q\right),\label{Cres}\\
    D=&(A-1)C\label{Dres}.
\end{align}

Have in mind, though, that the expression for the \emph{vielbein} \eqref{introCD} needs to be translated to a description in terms of the canonical momenta in accordance with Eq. \eqref{infreg}. The metric can then be obtained from Eq. \eqref{metfromviel} as
\begin{equation}
    g^{ij}=C^2\delta^{ij}+\left(2CD+D^2\right)\hP^i\hP^j.
\end{equation}
In short, we can understand the GUP as dual description to a quantum theory on nontrivial momentum space. Additionally, the newly found set of phase space variables allows for applications in its own right.

\subsubsection{Note on canonical variables}\label{subsubsec:canvar}

Classically, the dynamics of any system are governed by the action describing it. Alternatively, in quantum theory it suffices to provide a Hamiltonian and an algebra relating the dynamical variables. In the Heisenberg picture the evolution of the system may then be obtained according to the Heisenberg equations. To provide the corresponding Schrödinger equation and the action of a system, however, it is compulsory to find canonically a set of conjugate variables, obeying the Heisenberg algebra \eqref{HeisAlg}. By construction, this is the case considering the phase space coordinates introduced in the preceding subsection  \eqref{trans}. Furthermore, it is evident that the Heisenberg equations of motion in terms of both sets provided in this section are equivalent. Thus, the action of the system, subject to a GUP including spatial noncommutativity, reads
\begin{align}
    S=&\int\D t\left[\dot{X}^iP_i-H(X,P)\right].
\end{align}
Up until now, this kind of result had only been obtained in the case of a commutative space \cite{Brau06,Bosso18b} which is related to the one provided in the present section by a canonical transformation.

\subsubsection{Iterative approach}\label{subsubsec:itap}

For all intents and purposes, it suffices to solve Eqs. \eqref{noncomcondsca} and \eqref{conscondgen2res} iteratively. Assume as given the coefficients of a power series expansion of $A$ and $B$
\begin{equation}
    A=\sum_nA_n\left(\frac{l\hp}{\hbar}\right)^{2n}, \hspace{.5cm} B=\sum_nB_n\left(\frac{l\hp}{\hbar}\right)^{2n},
\end{equation}
with some length scale $l,$ and where $B_0=0$ to avoid divergences. Similarly, describe the scalars $\theta,$ $C$ and $D$ using power series
\begin{align}
    \theta=&\frac{1}{\hp^2}\sum_n\theta_{n-1}\left(\frac{l\hp}{\hbar}\right)^{2n},\\ 
    C=&\sum_nC_n\left(\frac{l\hp}{\hbar}\right)^{2n},\\ 
    D=&\sum_nD_n\left(\frac{l\hp}{\hbar}\right)^{2n},
\end{align}
where now $D_0=\theta_{-1}=0.$ Then, Eq. \eqref{noncomcondsca} becomes at $N^{\text{th}}$ order
\begin{equation}
    \sum_{n=0}^NA_{N-n}\left[2(N-n)\left(A_n+B_n\right)-B_n-\theta_{n}\right]=0,
\end{equation}
determining the coefficients $\theta_n$ order by order. Moreover, the Eqs. \eqref{conscondgen2res} uniquely specify the dependence of the coefficients $C_n$ and $D_n$ on $A_n$ and $B_n$ in an analogous fashion
\begin{align}
    D_N=&\sum_{n=0}^NC_{N-n}\left[\theta_{n}+2(N-n)\left(A_{n}+B_{n}\right)\right]\\
    =&\sum_{n=0}^NC_{N-n}A_n-C_N.
\end{align}

In short, the coefficients of the power series expansions describing the functions $C$ and $D$ are related to the ones representing the given scalars $A$ and $B$ such that there is no ambiguity. This opens up the possibility for a perturbative treatment.

\subsubsection{Application to the quadratic GUP}\label{subsec:applquadGUP}

As mentioned above, under the assumption that the GUP recovers Heisenberg's relation in the low-energy limit, the unperturbed scalars have to satisfy $A_0=1$ and $B_0=\theta_{-1}=0.$ Furthermore, denote $A_1=\beta,$ $B_1=\beta'$ and choose the Planck length to describe the scale to compare to ($l=l_p$) in accordance with the literature \cite{Das09,Tawfik14,Hossenfelder12}. Accordingly, we find
\begin{align}
    \theta_0=&0,&\theta_1=&2\beta-\beta',\\%&\theta_2=&4\gamma-\gamma'+2\beta\beta'\\
    C_0=&1,&C_1=&\frac{\beta'-\beta}{2},\\%&C_2=&\frac{3\beta^2-6\beta\beta'-(\beta')^2-6\gamma+\gamma'}{8}\\
    D_0=&0,&D_1=&\beta.%&D_2=&\frac{\beta'-\beta}{2}\beta+\gamma
\end{align}
At second order, the contribution stemming from the iterative appearance of the \emph{vielbein} \eqref{infreg} is trivial. Thus, the metric reads
\begin{equation}
    g^{ij}=\delta^{ij}+h^{ij},
\end{equation}
where the correction to the Euclidean part results as
\begin{equation}
    h^{ij}=\left(\beta'-\beta\right)\left(\frac{l_p\hP}{\hbar}\right)^2\delta^{ij}+2\beta\left(\frac{l_p}{\hbar}\right)^2\hP^i\hP^j.\label{metpert}
\end{equation}
Hence, we can derive the Cartan tensor from it, yielding
\begin{equation}
    C^{ijk}=2\left(\frac{l_P}{\hbar}\right)^2\left[\left(\beta'-\beta\right)\hP^i\delta^{jk}+2\beta \hP^{(j}\delta^{k)i}\right]\label{GUPcartan}.
\end{equation}
The Cartan tensor is totally symmetric if and only if $\beta'=2\beta,$ \ie $\theta\simeq 0,$ implying a commutative background. Then, the scalar \eqref{cartsca} derived from it reads in the low-energy limit
\begin{equation}
    \left.C\right|_{\hP=0}=2d(d+2)\beta\left(\frac{l_p}{\hbar}\right)^2.\label{Cquad}
\end{equation}
Otherwise, this metric does not belong to the class of Hamilton spaces as claimed in section \ref{subsubsec:curvqm}. Nevertheless assuming a vanishing nonlinear connection as was argued in the same section, the curvature tensor in momentum space \eqref{MomRiem} can be determined. In the low-energy limit it reads
\begin{equation}
    \left.S^{ikjl}\right|_{\hP=0}=2\theta_1\left(\frac{l_p}{\hbar}\right)^2\left(\delta^{ij}\delta^{kl}-\delta^{il}\delta^{kj}\right).\label{GUPRiem}
\end{equation}
Given this result, it is possible to compute the Ricci scalar in accordance with Eq. \eqref{MomRic}
\begin{equation}
    \left.S\right|_{\hP=0}=2d(d-1)\theta_1\left(\frac{l_p}{\hbar}\right)^2.\label{Squad}
\end{equation}
Thus, at first order the curvature of momentum space, provided the system is represented canonically, measures the noncommutativity of space described in terms of the original coordinates. This is why, the Cartan tensor is totally symmetric in the case of a GUP with a commutative background. Note, though, that, despite the background being flat, the momentum basis in terms of which the system is hence described is not the usual one. As the symplectic structure is not invariant under nonlinear transformations of momenta, the resulting theory is not equivalent to ordinary quantum mechanics, notwithstanding the flat background. This effect is measured by the quantity $C$ \eqref{Cquad}.

In short, quadratically deformed Heisenberg algebras may be understood as a normal-frame description of a momentum space harbouring essentially Planckian curvature if space is noncommutative. Thus, we can import much information from the phenomenology of GUPs to this arena.

\subsection{Constraints from existing literature}\label{subsec:constraints}

The preceding subsection served to point out a correspondence between models of the quadratic GUP and quantum mechanics on a non-Euclidean momentum space. This connection implies that bounds on the noncommutativity of space $\theta_1$ immediately carry over to the curvature tensor in momentum space, in accordance with equation \eqref{GUPRiem}. Some of these, mostly extracted from Ref. \cite{Hinchliffe02}, are displayed in table \ref{tab:labgupS}. The dominating constraint on the curvature scalar \eqref{Squad} stems from the electron dipole moment, yielding
\begin{equation}
    \left.S\right|_{p=0}<10^{27}m_p^{-2}.
\end{equation}
Note that spatially noncommutative geometry may lead to direct violations of Lorentz-invariance \cite{Anisimov01}, which would push this bound into the Planckian regime. However, depending on the relativistic generalization of the model, the symmetry might only be deformed, implying much weaker constraints.

\begin{table}[h!]
    \centering
\begin{tabular}{ c c || c } 
 Experiment & Ref.  & Upper bound on $Sm_p^2$\\
 \hline
 electron dipole moment & \cite{Hinchliffe01} & $10^{27}$\\
 lamb shift & \cite{Chaichian00a,Chaichian02} & $10^{29}$\\
 $^9$Be decay & \cite{Carroll01} & $10^{29}$\\ 
 composite quarks/ leptons & \cite{Ghoderao18,Tanabashi18} & $10^{29}$\\
 M\o ller scattering & \cite{Hewett00} & $10^{31}$\\
 muon $g-2$ & \cite{Kersting01} & $10^{31}$\\
 hydrogen spectrum & \cite{Gnatenko14,Akhoury03} & $10^{33}$\\
 $^{133}$Cs decay & \cite{Carroll01} & $10^{35}$\\
 star energy loss & \cite{Schupp02} & $10^{35}$\\
 Pauli oscillator & \cite{Heddar21} & $10^{41}$\\ 
 Aharonov-Bohm & \cite{Chaichian00b} & $10^{43}$
\end{tabular}
\caption{Upper bounds on the low-energy-limit of the scalar curvature in momentum space as in Eq. \eqref{Squad}, given in units of $l_p^2/\hbar^2= m_p^{-2}.$ \label{tab:labgupS}}
\end{table}

Furthermore, in the case of a commutative background space ($\theta_1=0$), bounds on the parameter $\beta$ can be translated as limits to the deviation from the usual momentum basis embodied by the scalar $C$ \eqref{Cquad}. A selection of bounds extracted from tables \ref{tab:labgup} and \ref{tab:gravigup} and obtained in the said way is on display in table \ref{tab:labgupC}. Recall, however, the issues with experiments involving harmonic oscillators \cite{Bawaj14,Bushev19}, optomechanical setups \cite{Khodadi17} and the Equivalence principle \cite{Ghosh13} explained in section \ref{subsec:constraints}. As there are reservations towards the direct adoption of results from multiparticle states to the mechanics of single particles due to the already mentioned inverse soccer ball problem (see \eg Ref. \cite{Amelino-Camelia13}), those should be taken with a grain of salt. The strongest constraint, excluding macroscopic experiments, is derived from corrections to state transitions in the hydrogen atom \cite{Bouaziz10,AntonacciOakes13}, implying that
\begin{equation}
    \left.C\right|_{p=0}<10^{25}m_P^{-2}.\label{CConst}
\end{equation}

Summarizing, both the curvature of momentum space as well as the deviation from the canonical momentum basis in the flat case are constrained experimentally from bounds on the noncommutativity of space and on the $\beta$-parameter of the commutative quadratic GUP, respectively.

\begin{table}[h!]
    \centering
\begin{tabular}{ c c || c } 
 Experiment & Ref.  & Upper bound on $Cm_p^2$\\
 \hline
 harmonic oscillators & \cite{Bushev19,Bawaj14} & $10^6$\\ 
 equivalence principle & \cite{Ghosh13} & $10^{19}$\\ 
 hydrogen spectrum & \cite{Bouaziz10,AntonacciOakes13} & $10^{25}$\\
 quantum noise & \cite{Girdhar20} & $10^{27}$\\
 tunnelling microscope & \cite{Das08} & $10^{32}$\\
 muon $g-2$  & \cite{Das11} & $10^{32}$\\
  gravitational bar detectors & \cite{Marin13,Marin14} & $10^{32}$\\ 
 lamb shift & \cite{Das08,Ali11b} & $10^{35}$\\ 
 $^{87}$Rb interferometry & \cite{Gao16,Khodadi18a} & $10^{39}$
\end{tabular}
\caption{Upper bounds on the deviation from the canonical basis in momentum space $C$ as in Eq. \eqref{Cquad} given in units of $l_p^2/\hbar^2= m_p^{-2}.$ \label{tab:labgupC}}
\end{table}

\subsection{Summary}\label{subsec:Conclusion}

The preceding sections suggested a deep connection between GUPs and EUPs on the one hand and non-Euclidean momentum and position spaces on the other hand, respectively. In this section we have further strengthened this connection presenting a noncanonical transformation, which provides a direct map from theories involving GUPs to quantum mechanics on curved momentum space.

In that vein, we have first reviewed quantum mechanics on a background described by a purely momentum-dependent metric, which had already been given an account of in section \ref{sec:measures}. Running the risk of being redundant, we have further displayed the kind of general changes to the canonical commutation relations, which are usually associated to GUPs including noncommutativity of the position coordinates, a subject explained more deeply in section \ref{sec:GUPsEUPs}. Bringing those two lines of thought together, we have found an explicit dual description of this type of deformation in terms of a nontrivial momentum space. In other words, every GUP entailing a certain set of non-Darboux coordinates yields its counterpart in a specific set of canonically conjugated phase space variables. The resulting dynamics strongly indicate the presence of a nontrivial momentum space.

In particular, in the case of the quadratic GUP, the curvature tensor in momentum space is proportional to the spatial noncommutativity. However, the dual description of a commutative space does not imply a trivial background because the corresponding basis in momentum space is curvilinear. As nonlinear basis transformations in momentum space are not canonical, the resulting theory is inequivalent to ordinary quantum mechanics. The deviation from Riemannian geometry, induced by this unusual basis, can then be measured by a scalar derived from the Cartan tensor.

This has allowed us to import constraints on the curvature of momentum space from bounds on the noncommutativity of space, yielding for the Ricci scalar in momentum space $S|_{p=0}<10^{27}m_p^{-2}.$ Moreover, the literature on GUPs with commutative space has been helpful in constraining the deviation from Riemannian geometry when the curvature is vanishing, implying $C_{p=0}<10^{17}m_p^2.$

Evidently, the reasoning applied in the present section has been general enough to be applied to EUPs in an analogous fashion. Correspondingly, those can be mapped to theories of quantum mechanics on curved position space thus establishing exactly the connection hinted at in the preceding sections. 

To make a long story short, the interplay of EUPs and non-Euclidean momentum space as well as EUPs and curved position space yields a rich phenomenology, that justifies further investigation. In particular, there is a strong motivation to find a formulation of quantum mechanics on generalized Hamilton spaces away from Cartesian coordinates such that the metric may depend on positions and momenta. 

\section{Quantum Mechanics on the curved cotangent bundle}\label{sec:qmhamilton}

The previous section established a direct connection between modified uncertainty relations and curved spaces, showing that the former yield a description of the latter in non-Darboux coordinates. However, these results could only be expressed in terms of Cartesian coordinates -- the metric would not have been solely momentum-dependent otherwise. A simple transition to spherical coordinates, for example,  implies that it acquires a position dependence. While DeWitt's approach \cite{DeWitt52} introduced in section \ref{subsec:curved} proves successful in the regime of curved position space in the position representation or curved momentum space in the momentum representation, it cannot be applied once the metric is of the form
\begin{equation}
    g^{ij}=g^{ij}(\hat{x},\hat{p}).\label{met}
\end{equation}
Furthermore, extrapolating from the results on deformed Heisenberg algebras, it is natural to expect that GEUPs, which, recall, combine both position- and momentum-dependent modifications, in fact constitute an effective description of quantum mechanics on a generally nontrivial cotangent bundle. Thus, there is a strong motivation for the construction of a quantum mechanical formalism which can be applied to backgrounds described by metrics depending on both positions and momenta
\begin{align}
    \D s^2=g_{ij}\left(x,p\right)\D x^i\D x^j, &&\D \tilde{s}^2=g^{ij}\left(x,p\right)\D p_i\D p_j,\label{lengthel}
\end{align}
where the second equality defines the metric in momentum space as the inverse of the one in position space as usual. This immediately implies that the Hilbert space measure in both momentum and in position space inherits an analogous dependence. The present section delineates an avenue towards the quantum description of such a theory in the position representation. 

In particular, we aim at a consistent formulation of the quantum mechanics of a single particle on a background of the form \eqref{met}. In that vein, we promote the measure of the Hilbert space, \ie the volume form, to a, by assumption, Hermitian operator. Furthermore, by analogy with geometric quantization \cite{Woodhouse97}, we split it into two pieces to merge them symmetrically with each wave function in the scalar product, thus creating wave densities. Then, assuming the scalar product of their eigenvectors consists of plane waves, the position and momentum operators possess a particularly simple position representations. Furthermore, we find representations for the squared momentum, \ie the Hamiltonian of the free particle, and, equivalently, the geodesic distance from the origin. Correspondingly, we define perturbation theory in this context and finally deal with central potentials, particularly the isotropic harmonic oscillator and the hydrogenic atom, on a background harbouring curvature in position as well as momentum space, described by an expansion akin to Riemann normal coordinates. In this context, we find that the isotropic harmonic oscillator, given a suitable prescription of operator ordering, retains its reciprocal nature, thereby resulting in an exact example of Born reciprocity on the curved cotangent bundle. Specifically, this behaviour is indicated by a symmetry akin to T-duality in string theory \cite{Sathiapalan86,Alvarez94}. Thus, combining the concepts of Born reciprocity, preserved by the Hamiltonian of the harmonic oscillator, and the generally curved cotangent bundle, i. e. nonvanishing Ricci tensors in position and momentum space, we obtain the kind of UV-IR mixing which is generally expected from quantum gravity \cite{Minwalla99}.

This section, containing work published in Ref. \cite{Wagner21c}, is organized as follows. First, in section \ref{subsec:curvcot} we summarize the mathematics required for a consistent description of the curved cotangent bundle. Based on this background, we introduce the main formalism as a generalization of DeWitt's ideas in section \ref{subsec:formalism}. Section \ref{subsec:applicationcot} is devoted to phenomenology following a perturbative application of the approach to central potentials. Finally, we wrap up and conclude in section \ref{subsec:sumcot}.

\subsection{The curved cotangent bundle\label{subsec:curvcot}}

As mentioned in the previous section, the language of Lagrange and Hamilton geometries is explained extensively in Refs. \cite{Miron01,Miron12}. Therefore, this introduction is solely intended to briefly summarize the ingredients required for the purpose of the present section. 

The main goal of the said mathematical program lies in the description of spaces within which proper length measurements may be momentum-dependent in accordance with Eq. \eqref{lengthel}, while retaining covariance with respect to coordinate transformations of the form
\begin{equation}
    x^i\rightarrow X^i(x),\hspace{1cm}p_i\rightarrow P_i=\frac{\partial x^j}{\partial X^i}p_j.\label{gencoordtrans}
\end{equation}
If the cotangent bundle, however, exhibits curvature -- even if solely position space is curved --, the canonical variables do not provide a clear partition into position and momentum space. In particular, a nonlinear connection $N_{ij}(x,p)$ is required as a bookkeeping device dividing the cotangent bundle into the horizontal (position) and vertical (momentum) distributions. Under reasonable conditions, the nonlinear connection can be derived from the metric in a canonical way. For example, in the Riemannian case, \ie for purely positional curvature, its canonical version reads
\begin{equation}
    N_{ij}=p_k\Gamma^k_{ij}(x),\label{levicivnonlin}
\end{equation}
with the Christoffel symbols of the Levi-Civita connection $\Gamma^k_{ij}.$

Given a nonlinear connection, we can define the orthogonal vector fields on phase space
\begin{equation}
    \frac{\delta}{\delta x^i}=\frac{\partial}{\partial x^i}+N_{ij}\frac{\partial}{\partial p_j},\hspace{1cm}\frac{\partial}{\partial p_i},\label{modder}
\end{equation}
both of which manifestly transform covariantly under the coordinate transformation \eqref{gencoordtrans}
\begin{equation}
    \frac{\delta}{\delta x^i}=\frac{\partial X^j}{\partial x_i}\frac{\delta}{\delta X^j},\hspace{1cm}\frac{\partial}{\partial p_i}=\frac{\partial x^i}{\partial X^j}\frac{\partial}{\partial P_j}.
\end{equation}
Most importantly, these two vector fields are related by a map $F(\delta_i)=-g_{ij}\dot{\partial}^j,$ $F(\dot{\partial}^i)=g^{ij}\delta_j,$ with $\delta_i\equiv\delta/\delta x^i,$ $\dot{\partial}^i\equiv\partial/\partial p_i.$ Thus, the present system admits an almost complex structure ($F^2=-\id$), which is precisely the mathematical premise underlying quantum mechanics -- in physicist's terms: Born reciprocity.

Crucially, the positional partial derivative is modified, yielding a curvature tensor on the cotangent bundle
\begin{equation}
    \left[\frac{\delta}{\delta x^i},\frac{\delta}{\delta x^j}\right]\equiv R_{kij}\frac{\partial}{\partial p_k},
\end{equation}
which for Riemannian backgrounds assumes the form
\begin{equation}
    R_{kij}=p_mR^{m}_{~kij}(x),
\end{equation}
with the Riemann curvature tensor $R^{m}_{~kij}.$ Furthermore, by analogy with Riemannian geometry the antisymmetric part of the nonlinear connection yields the torsion on the cotangent bundle which is assumed to vanish for the remainder of the present section.

An analogous reasoning leads to the basis of one-forms
\begin{equation}
    \D x^i,\hspace{1cm}\delta p_i=\D p_i-N_{ji}\D x^j,\label{fundforms}
\end{equation}
which, too, transform as expected
\begin{equation}
    \D x^i=\frac{\partial x^i}{\partial X^j}\D X^j,\hspace{1cm}\delta p_i=\frac{\partial X^j}{\partial x^i}\delta p_j.
\end{equation}
Here, the symbol $\delta$ is to be understood as an exterior covariant derivative. Correspondingly, in $d$ dimensions we can construct the covariant integration measures
\begin{equation}
    \sqrt{g}\D^dx,\hspace{1cm} \frac{1}{\sqrt{g}}\delta^d p,\label{modint}
\end{equation}
with the determinant of the covariant metric $g=\det g_{ij}.$

From Eqs. \eqref{modder} and \eqref{modint}, it is clear, that it is much easier to maintain a position representation involving only position-dependent wave functions $\psi(x)$ (such that $\delta_i\psi=\partial_i\psi$) than to deal with the momentum representation. Therefore, unless $N_{ij}=0,$ \eg for purely momentum-dependent metrics, we will mainly restrict ourselves to the the former case. %The inclusion of momentum-dependence will likely require a doubling of the phase space variables as it is done in Born geometries \cite{Freidel13,Freidel14,Freidel15,Freidel17,Freidel18} and double field theory \cite{Siegel93a,Siegel93b,Hull09}.

Incidentally, this simplification in position space is also why it is possible to deal with quantum mechanics on Riemannian backgrounds \`a la deWitt as described in section \ref{subsec:curved}, \ie without reverting to the nonlinear connection.

\subsection{Promoting the volume element to an operator}\label{subsec:formalism}

Clearly, once it depends on positions and momenta, the determinant of the metric being part of the volume form in the scalar product has to be promoted to an operator $\sqrt{\hat{g}}\D^d x,$ which is assumed to be Hermitian and whose eigenvalues are, by definition, positive. In order to achieve this in a symmetric way, denote its square root as $\hat{\mu}(\hat{x},\hat{p})\equiv\sqrt[4]{\hat{g}}.$ Then, the position operator can be defined as (\cf Eq. \eqref{curvposop})
\begin{equation}
\hat{x}^i\equiv\int \D^d x x^i\ket{\hat{\mu}x}\bra{\hat{\mu}x}.\label{posop}
\end{equation}
Thus, its eigenstates are of the form $\ket{\hat{\mu}x}=\hat{\mu}\ket{x}$ such that $\hat{x}^i\ket{\hat{\mu}x}=x^i\ket{\hat{\mu}x}.$ As for deWitt's approach, these eigenstates furnish an orthonormal basis, hence satisfying
\begin{align}
\int \D^d x\ket{\hat{\mu}x}\bra{\hat{\mu}x}=&\mathds{1},\label{completex}\\
\Braket{\hat{\mu}x}{\hat{\mu}x'}=	&\delta^d(x-x'),\label{orthox}
\end{align}
which generalize Eqs. \eqref{curvposid} and \eqref{curvpostransamp}. Then, every state in the Hilbert space can be expanded in the given basis, yielding
\begin{equation}
    \ket{\psi}=\int\D^d x\Braket{\hat{\mu}x}{\psi}\ket{\hat{\mu} x}.
\end{equation}
Defining the position space wave function as $\psi\equiv\Braket{x}{\psi},$ we can express amplitudes in terms of the wave density $\Psi\equiv\Braket{\hat{\mu}x}{\psi}=\hat{\mu}(\psi).$ Note that we revert to the term "wave density" because the quantities $\Phi,\Psi\dots$ and the equivalents in momentum space $\tilde{\Phi},\tilde{\Psi}\dots,$ defined below, change under coordinate transformations not as scalar functions but as scalar densities of weight $1/2$ and $-1/2,$ respectively.  

This leads to the corresponding scalar product
\begin{align}
\Braket{\psi}{\phi}=	&\int\D^d x\Braket{\psi}{\hat{\mu}x}\Braket{\hat{\mu}x}{\phi}\\
			\equiv	&\int\D^d x \Psi^*\Phi.\label{scapro}
\end{align}
Note that if $\hat{\mu}=\hat{\mu}(\hat{x}),$ the measure commutes with the position operator, thus rendering both of them simultaneously diagonalisable.  Then, $\ket{x}$ can be understood as eigenstate of $\hat{\mu},$ which implies $\ket{\hat{\mu}x}=\mu(x)\ket{x},$ recovering the description \`a la deWitt (\cf Eqs. \eqref{curvposop}, \eqref{curvposid} and \eqref{curvpostransamp}). As can be inferred from  Eq. \eqref{scapro}, this procedure provides a map from quantum mechanics on the curved cotangent bundle into ordinary flat space-quantum mechanics, given the actions of operators on wave densities like $\Phi$ and $\Psi.$ 

\subsubsection{A note on the momentum representation}

As the framework developed in this section is intended to be invariant under coordinate transformations, the discussion in momentum space is highly nontrivial. In particular, the measure $\D^d p/\sqrt{g}$ does not reflect this principle. Instead, equation \eqref{fundforms} indicates that the unique covariantly transforming basis of one-forms in momentum space reads
\begin{equation}
    \omega_i\equiv \delta p_i. 
\end{equation}
However, this set of forms is not closed, \ie $\D \omega_i\neq 0,$ unless $R_{ijk}=\dot{\partial}^kN_{ij}=0,$ which is already violated by a position-dependent metric (\cf Eq. \eqref{levicivnonlin}). Thus, they cannot be exact, a necessary condition for them to be derivable from some new kind of new phase space coordinate. Therefore, an application of the reasoning introduced above to momentum space is not immediate. Instead, it seems plausible that a full treatment will likely require a doubling of the phase space coordinates as in Born geometries \cite{Freidel13,Freidel18} or double field theory \cite{Siegel93a,Siegel93b,Hull09}. This line of research will be dealt with in future work, while we presently restrict ourselves mainly to the position representation.

There is an exception to this rule, however -- if the metric turns out to be purely momentum-dependent, the nonlinear connection can be chosen to vanish \cite{Miron01,Wagner21a,Relancio21}, which leads to a situation in momentum space akin to the problem solved by deWitt \cite{DeWitt52}. Yet, it can also be formulated in the new language which is advocated for in the present work. By analogy with Eq. \eqref{posop}, we can then define the momentum operator as
\begin{equation}
\hat{p}_i=\int \D^d p p_i\ket{\hat{\mu}^{-1}p}\bra{\hat{\mu}^{-1}p},\label{momop}
\end{equation}
whose eigenstates $\ket{\hat{\mu}^{-1}p}$ obey the relations
\begin{align}
\int \D^d p\ket{\hat{\mu}^{-1}p}\bra{\hat{\mu}^{-1}p}=						&\mathds{1},\label{completep}\\
\Braket{\hat{\mu}^{-1}p}{\hat{\mu}^{-1}p'}=	&\delta^d(p-p').\label{orthop}
\end{align}
Accordingly, this provides us with the momentum space representation
\begin{equation}
    \ket{\psi}=\int\D^d p\braket{\hat{\mu}^{-1}p}{\psi}\ket{\hat{\mu}^{-1} p}\equiv\int\D^d p\hat{\mu}^{-1}(\tilde{\psi})(p)\ket{\hat{\mu}p}
\end{equation}
and, defining $\tilde{\psi}(p)\equiv\braket{p|\psi}$ and $\tilde{\Psi}(p)\equiv\braket{\hat{\mu}^{-1}p|\psi}=\hat{\mu}^{-1}(\tilde{\psi}),$ with the scalar product
\begin{align}
\Braket{\psi}{\phi}=	&\int\D^d p\Braket{\psi}{\hat{\mu}^{-1}p}\Braket{\hat{\mu}^{-1}p}{\phi}\\
						 \equiv	&\int\D^d p \tilde{\Psi}^*\tilde{\Phi}.
\end{align}
Before this formalism can be applied to examples, though, it remains to be shown that this construction is well-defined. 

\subsubsection{Representations of conjugate operators}

The phase space basis displayed in Eq. \eqref{modder} indicates that the momentum operator ought to be represented in position space as
\begin{equation}
    \Braket{\hat{\mu} x}{\hat{p}_i\psi}=-i\delta_i\Psi(x)=-i\partial_i\Psi(x).
\end{equation}
In particular, the corresponding commutator with the position operator reads
\begin{equation}
    \left[x^i,-i\hbar\delta_j\right]\Psi=i\hbar\left(\delta_j^i\Psi-\dot{\partial}^iN_{jk}\dot{\partial}^k\Psi\right)=i\hbar\delta ^i_j\Psi.
\end{equation}
Thus, the position and momentum operators continue to satisfy the Heisenberg algebra \eqref{HeisAlg}. Furthermore, a general amplitude featuring the momentum operator reads
\begin{equation}
    \Braket{\psi}{\hat{p}_i\phi}=\int\D^d x\Psi^*\left(-i\hbar\partial_i\Phi\right),
\end{equation}
which is trivially symmetric and in the case of a purely position-dependent measure just recovers the effect of Eq. \eqref{curvedspacemom}. 

Analogously, if the metric is purely momentum-dependent, we obtain for the momentum representation of the position operator
\begin{equation}
        \Braket{\hat{\mu}^{-1}p}{\hat{x}^i\psi}=\hat{x}^i\tilde{\Psi}=i\hbar\dot{\partial}^i\tilde{\Psi},
\end{equation}
in accordance with Eq. \eqref{modder}. Then, the Fourier transform can be generalized such that it reads
\begin{align}
    \tilde{\Psi}=&\frac{1}{\sqrt{2\pi\hbar}}\int\D^d x \Psi e^{-i\frac{p_i x^i}{\hbar}},\\
   \Psi=&\frac{1}{\sqrt{2\pi\hbar}}\int\D^d p \tilde{\Psi}e^{i\frac{p_ix^i}{\hbar}}.
\end{align}
Thus, it is at least possible to find a complete picture of quantum mechanics on curved momentum space, dual to the GUP-formalism.

It is evident that the representations of the position and momentum operators defined in this section are indeed Hermitian (due to the symmetric choices in Eqs. \eqref{posop} and \eqref{momop}) and satisfy the canonical commutation relations \eqref{HeisAlg}. Thus, $\hat{\mu},$ itself a real function of $\hat{x}^i$ and $\hat{p}_j,$ can be made Hermitian by suitable symmetrization and is well-defined modulo operator ordering ambiguities. We conclude that the procedure outlined above is consistent.

\subsubsection{Free particle}

The Hamiltonian of a free particle of mass $M$ subject to a general metric $g^{ij}(x,p)$ continues to be of the form $H_{\text{fp}}=g^{ij}p_ip_j/2M.$ Recall that the quantum operator reflecting the Hamiltonian in the position representation is aptly chosen to be the Laplace-Beltrami operator \eqref{Laplace-Beltrami} if the metric is solely dependent on the coordinates, \ie the particle is living in curved space only. This implies the general amplitude
\begin{equation}
    \Braket*{\psi}{\hat{H}_{\text{fp}}\phi}=-\frac{h^2}{
2M}\int \D^d x\psi^*\partial_i\left(\sqrt{g}g^{ij}\partial_j\phi\right).\label{kinenamp1}
\end{equation}
In the more general case treated here, the said amplitude can be naturally generalized to the form
\begin{equation}
    \Braket*{\hat{\mu}x}{\hat{H}_{\text{fp}}\psi}=-\frac{\hbar^2}{4M}\hat{\mu}^{-1}\partial_i\left[\{\hat{\mu}^2,\hat{g}^{ij}\}\partial_j\left(\hat{\mu}^{-1}\Psi\right)\right],
\end{equation}
where the metric in its dependence on positions and momenta has been promoted to a self-adjoint operator, which does not necessarily commute with the square root of its determinant $\hat{\mu}^2.$ In other words, we obtain
\begin{equation}
    \hat{H}_{\text{fp}}\Psi=\frac{1}{4M}\hat{\mu}^{-1}\hat{p}_i\{\hat{\mu}^2,\hat{g}^{ij}\}\hat{p}_j\hat{\mu}^{-1}\Psi,
\end{equation}
which, as a symmetric product of symmetric operators, is symmetric. Furthermore, the expression $\hat{H}_{\text{fp}}\Psi$ transforms like a scalar density $\hat{\mu}$ multiplying a scalar, \ie a scalar density of weight $1/2,$ as required.

A general amplitude featuring the Hamiltonian of a free particle can thus be expressed as
\begin{align}
    \Braket{\psi}{\hat{H}_{\text{fp}}\phi}=&-\frac{h^2}{4M}\int \D^d x\Psi^*\hat{\mu}^{-1}\partial_i\left[\{\hat{\mu}^2,\hat{g}^{ij}\}\partial_j\left(\hat{\mu}^{-1}\Phi\right)\right]\\
    =&-\frac{h^2}{4M}\int \D^d x\psi^*\partial_i\left[\{\hat{\mu}^2,\hat{g}^{ij}\}\partial_j\phi\right],
\end{align}
which in the case of a purely position-dependent metric exactly recovers \eqref{kinenamp1}.

Having derived the action of the basic operators, these can be used to apply the framework at hand to classic textbook problems.

\subsection{Phenomenology}\label{subsec:applicationcot}

Classical and quantum gravity corrections to quantum mechanical experiments are usually negligibly small. For instance, the relative correction of the earth's gravitational field to the energy spectrum of the hydrogen atom is of the order $10^{-38}$ \cite{Zhao07}, while the corresponding magnitude stemming from a Planckian GUP is expected to lie around $10^{-46}$ (\cf table \ref{tab:factorest}). Thus, it suffices to treat the corresponding effects perturbatively, \ie we can assume the background to be flat at leading order. If solely position space is curved, this expansion around a point $x_0$ is, again, given in its simplest form in terms of Riemann Normal coordinates $x^i,$ yielding corrections of second order depending on the Riemann tensor $R_{ikjl}$ evaluated at $x_0.$ When both position and momentum space are curved, the corresponding curvature tensors (denoted $R_{ikjl}$ in position space and $S^{ikjl}$ in momentum space) can still be defined even though they may both be position- and momentum-dependent. Assuming that there is no mixing of positions and momenta to second order, we then propose an analogous expansion at the point $Y_0=(x_0,p_0)$ in the cotangent bundle. As an aside, the low energy limit usually taken when considering quantum gravity effects would put $p_0$ into the origin of momentum space. 

Correspondingly, we can write the metric as
\begin{equation}
    g_{ij}\simeq\delta_{ij}-\frac{1}{3}\left(R_{ikjl}|_{Y_0}x^kx^l-S_{i~j}^{~k~l}|_{Y_0}p_kp_l\right),
\end{equation}
where indices are raised and lowered using the flat metric $\delta_{ij}.$ The sign difference between the correction terms compensates for the fact that $g_{ij}$ denotes the inverse of the metric in momentum space. Note that this difference disappears when considering the metric as an operator, \eg in the position space representation.

Therefore, the operator $\hat{\mu}$ has to be expanded too, yielding
\begin{equation}
    \hat{\mu}\Psi\simeq\left[1-\frac{1}{12}\left(R_{kl}|_{Y_0}x^kx^l+\hbar^2S^{kl}|_{Y_0}\partial_k\partial_l\right)\right]\Psi,
\end{equation}
with the Ricci tensors in position and momentum space $R_{ij}$ and $S^{ij},$ respectively. This results in an expansion of the free particle Hamiltonian
\begin{align}
    \hat{H}_{\text{fp}}\Psi\simeq& \left(-\frac{\hbar^2}{2M}\Delta_0+\hat{H}^{(2)}_{\text{fp}}\right)\Psi,\\
    \hat{H}^{(2)}_{\text{fp}}\Psi     =&\frac{\hbar^2}{6M}\Big[R^j_i|_{Y_0}x^i\partial_j-R^{i~j~}_{~k~l}|_{Y_0}x^kx^l\partial_i\partial_j\Big]\Psi,\label{hfp}
\end{align}
with the Laplacian in flat space $\Delta_0,$ and where we used the symmetries of the curvature tensors and absorbed a constant tern into a redefinition of the energy -- neglecting gravitational backreaction, only energy differences can be detected.

\subsubsection{Perturbation theory}

How to deal with nonsingular perturbation theory, which additionally exerts an influence on the scalar product measure, is explained in detail in appendix \ref{app_pert}. Fortunately, written in terms of general wave densities $\Psi,$ this reduces to an instance of ordinary perturbation theory. Under the assumption that $\braket{\Psi|\Psi}=\braket{\Psi^{(0)}|\Psi^{(0)}}=1$ we then obtain
\begin{equation}
    \int\D^dx\text{Re}\left(\Psi^{*(2)}\Psi^{(0)}\right)=0.\label{wavedenspert}
\end{equation}
Say a hermitian operator $\op$ with discrete eigenvalues $\lambda_n$ and eigenstates $\Psi_n$ ($n$ can stand for several quantum numbers) is corrected by second-order contributions to the metric as $\op\simeq\op^{(0)}+\op^{(2)}.$ Then, Eq. \eqref{wavedenspert} implies for the corrections $\lambda^{(2)}_n$ to its eigenvalues $\lambda^{(0)}$ that
\begin{align}
    \lambda_n^{(2)}=\int\D^dx\Psi_n^{*(0)}\op^{(2)}\Psi_n^{(0)}.\label{eigvalpert}
\end{align}
In the given setup $\Psi^{(0)}=\psi^{(0)},$ which is why we can write
\begin{align}
    \lambda_n^{(2)}=\int\D^dx\psi_n^{*(0)}\op^{(2)}\psi_n^{(0)}.
\end{align}
Note, though, that this equality is restricted to normal coordinates. 

\subsubsection{Central potentials in three dimensions}

Central potentials are usually expressed in terms of the geodesic distance from the origin. If only space is curved and in Riemann normal coordinates, this distance just reads $\sigma=r=\sqrt{x^ix^j\delta_{ij}}.$ This changes, though, once momentum space is allowed to become nontrivial as well. In general, the geodesic distance from the origin, which in the given coordinate system coincides with $x_0$, again, satisfies the differential equation 
\begin{equation}
    g^{ij}\partial_i\sigma\partial_j\sigma=1\label{geoddistdiff}
\end{equation} 
and the initial condition $\sigma|_{x_0}=0.$ In the perturbative case, we then expand $\sigma\simeq\sigma^{(0)}+\sigma^{(2)},$ with $\sigma^{(0)}=r.$ Furthermore, linearising Eq. \eqref{geoddistdiff}, we obtain the correction
\begin{align}
    \sigma^{(2)}=\frac{x^ix^j}{6r}S_{i~j}^{~k~l}p_kp_l .
\end{align}
Thus, we shall expand any central potential as
\begin{equation}
    V(\sigma)=V(r)+\frac{V'(r)}{6}\frac{x^ix^j}{r}S_{i~j}^{~k~l}|_{Y_0}p_kp_l, \label{genpot}
\end{equation}
which, depending on the background curvature, may cease to be isotropic. 

As this potential is both position and momentum-dependent, promoting it to a hermitian operator is an ambiguous task. The ensuing operator ordering ambiguities have to be treated with care on a case-by-case basis.

%Taking into account all possible linear orderings, we obtain the correction to the potential operator
%\begin{align}
%    \hat{V}^{(2)}\Psi=&-\frac{\hbar^2}{12}\Big[V'\frac{x^ix^j}{r}S_{i~j}^{~k~l}\partial_k\partial_l+\frac{V'}{r}S^j_i x^i\partial_j\nonumber\\
%    &+\alpha S\frac{ V'}{r}+S_{ij}\frac{x^ix^j}{r^2}\left(\frac{\beta V'}{r}-\gamma V''\right)\Big]\Psi
%\end{align}
%where the parameters $\alpha,\beta,\gamma\in [0,1]$ parametrize the ambiguity which take the values $1/4,$ $1/3$ and $1/3$ respectively if all orderings are weighted equally. 

\subsubsection*{Isotropic harmonic oscillator}

According to Eq. \eqref{genpot}, the potential of the harmonic oscillator reads
\begin{equation}
    V(\sigma)\simeq\frac{1}{2}M\omega^2\left(r^2+\frac{1}{3}x^ix^jS_{i~j}^{~k~l}|_{Y_0}p_kp_l\right),
\end{equation}
with the oscillation frequency $\omega.$ Taking into account all possible operator orderings combining the four noncommuting contributions $\hat{x}^i,\hat{x}^j,\hat{p}_k,\hat{p}_l$ and after application of simple algebra, we obtain the correction to the corresponding operator
\begin{align}
    \hat{V}_{(2)}=&\frac{1}{6}M\omega^2\left(\hat{x}^i\hat{x}^jS_{i~j}^{~k~l}|_{Y_0}\hat{p}_k\hat{p}_l+i\hbar S^j_i|_{Y_0}\hat{x}^i\hat{p}_j\right),\label{harmoscpot}
\end{align}
where the dependence on the operator ordering is constant and can be removed by another redefinition of the energy. A close comparison of Eqs. \eqref{hfp} and \eqref{harmoscpot} makes apparent that the implications of curvature in both position and momentum space are analogous and Born reciprocity, a crucial feature of the harmonic oscillator, is restored in adapted "natural units" ($M=\omega=1$).

The eigenstates and -values of the unperturbed Hamiltonian describing the harmonic oscillator $\hat{H}_{(0)}=\hat{H}_{\text{fp}}^{(0)}+\hat{V}_{(0)}$ read in the position representation and in spherical coordinates \cite{Messiah99}
\begin{align}
    \psi_{nlm}^{(0)}=&N_{nl}r^le^{-\frac{M\omega}{2\hbar}r^2}L_{\frac{n-l}{2}}^{l+1/2}\left(\frac{M\omega}{\hbar}r^2\right)Y^l_{m}(\theta,\phi),\\
    N_{nl}=&\sqrt{\sqrt{\frac{M^3\omega^3}{\pi\hbar^3}}\frac{2^{\frac{n+3l}{2}+3}\left(\frac{n-l}{2}\right)!\left(\frac{M\omega}{2\hbar}\right)^l}{\left(2n+l+1\right)!!}},\\
    E^{(0)}_n=&\braket{\psi_{nlm}^{(0)}|\hat{H}_{(0)}\psi_{nlm}^{(0)}}=\hbar\omega\left(n+\frac{3}{2}\right),
\end{align}
with the generalized Laguerre polynomials $L_n^\alpha (x),$ the spherical harmonics $Y_{lm}(\theta ,\phi),$ and where we introduced the quantum numbers $n\geq 0$ (radial), $l=n\Mod{2},n\Mod{2}+2,\dots n$ and $m=-l,-l+1,\dots,l$ (angular). Correspondingly, the curvature in both position and momentum space induces corrections, which can be calculated according to Eq. \eqref{eigvalpert} as
\begin{equation}
    E^{(2)}_{nlm}=\int\D^3x\Psi_{nlm}^{(0)}\hat{H}^{(2)}\Psi_{nlm}^{(0)}.
\end{equation}
These integrals can be solved analytically. The resulting contribution exactly follows the pattern
\begin{align}
    E^{(2)}_{nlm}=\frac{\hbar\omega}{6}&\left[\left(l(l+1)-3m^2\right)\left(\frac{M\hbar\omega}{2}S_{zz}+\frac{\hbar^2R_{zz}}{2M\hbar\omega}\right)\right.\nonumber\\
    &\left.\left.+m^2\left(\frac{M\hbar\omega}{2}S+\frac{\hbar^2R}{2M\hbar\omega}\right)\right]\right|_{Y_0},\label{Ecorrharm}
\end{align}
with the Ricci scalars in position and momentum space $R$ and $S$, respectively. This has been checked for all quantum numbers up to $n=10.$ 

In general, quantum harmonic oscillators are composite objects. As a reaction to experimental results in this context \cite{Pikovski11,Bawaj14}, it has been pointed out \cite{Amelino-Camelia13,Kumar19} that quantum gravity effects as embodied by the GUP do not scale with powers of the Planck mass but its product with the number of fundamental constituents -- an effect dubbed inverse soccer ball problem in section \ref{sec:GUPsEUPs}. The corresponding relative corrections at Planckian-per-constituent momentum space-curvature $S\sim (Nm_p)^{-2},$ with the effective number of constituents (elementary particles) contained in the oscillator $N,$ read for table top experiments
\begin{equation}
    \delta E_{nlm}\equiv\frac{E^{(2)}_{nlm}}{E^{(0)}_{nlm}}\sim\frac{M\hbar\omega}{N^2m_p^2}+\frac{E_{\text{surf}}^2}{M\hbar\omega},\label{harmoscres}
\end{equation}
with the energy scale corresponding to the spatial curvature acting on objects on the surface of the earth $E_{\text{surf}}=\sqrt{R}|_{\text{surf0}}\hbar\sim 10^{-19}\text{eV}.$ Essentially, those effects are important in reciprocal regimes -- classical gravity modifies processes at small energies, \ie large distances and oscillation periods, while quantum gravity acts at high energies or small distances as expected. In particular, in the regime of high frequency and replacing $N=M/\bar{m}_c,$ with the average mass of the constituents $\bar{m}_c$, the corrections are of the form
\begin{equation}
    \delta E_{nlm}\sim\frac{\hbar\omega}{M}\frac{\bar{m}_c^2}{m_p^2}.
\end{equation}
This implies that it is most effective to use probes with high average constituent mass. Furthermore, the ratio $\hbar\omega/M$ favours microscopic oscillators, \ie such containing a small number of constituents. For the applications considered in the literature, \eg in Ref. \cite{Pikovski11}, the relative corrections are minute once the number of constituents is taken into account in the way indicated here, which corresponds to the parameter $\alpha=1$ in Ref. \cite{Kumar19}. For example, the setup in Ref. \cite{Bawaj14} leads to relative corrections of $\delta E_{nlm}\sim 10^{-70}.$

More importantly, Eq. \eqref{harmoscres} implies that it is impossible to distinguish position from momentum space-curvature with a single harmonic oscillator of mass $M$ and frequency $\omega .$ More precisely, the corrections \eqref{harmoscres} are invariant under the transformation
\begin{equation}
    M\hbar\omega S_{ij}\longleftrightarrow\frac{\hbar^2R_{ij}}{M\hbar\omega},
\end{equation}
which is clearly reminiscent of T-duality in string theory \cite{Sathiapalan86}, connecting the low and high energy behaviour. This is not the first time T-duality has been encountered in the context of the curved cotangent bundle. In fact, it is manifest and as such one of the defining features of metastring theory and its underlying Born geometry \cite{Freidel13,Freidel14,Freidel15,Freidel17,Freidel18}. Thus, according to Eq. \eqref{harmoscres}, the harmonic oscillator maintains its Born reciprocal property on arbitrary nontrivial backgrounds.

In a nutshell, under the assumption of a reciprocal Hamiltonian like the isotropic harmonic oscillator, we exactly obtain the behaviour, Born was striving for, when trying to merge quantum theory with general relativity through the curved cotangent bundle.

\subsubsection*{Coulomb potential}

The coulomb potential, describing a hydrogenic atom, is corrected in a way similar to the isotropic harmonic oscillator. Again using Eq. \eqref{genpot}, it reads to second order
\begin{equation}
    V(\sigma)\simeq\frac{Z\alpha\hbar}{r}\left(1-\frac{1}{6}S_{i~j}^{~k~l}|_{Y_0}\frac{x^ix^j}{r^2}p_kp_l\right),
\end{equation}
with the fine-structure constant $\alpha\simeq 1/137$ and the number of elementary charges in the origin $Z.$ The corresponding quantum operator can be found unambiguously under the assumption that a perturbative treatment is indeed possible: The corrections induced by terms, which are dependent on the ordering, scale as $\hat{r}^{-3}.$ Thus, the corresponding expectation values with respect to the unperturbed eigenstates of the Hamiltonian diverge. These contributions to the potential are clearly too singular. Hence, omitting them is equivalent to renormalizing the problem. In short, the quantum operator representing the corrections to the potential may be unambiguously expressed as
\begin{equation}
    \hat{V}_{(2)}=-\frac{Z\alpha\hbar}{6\hat{r}^3}\left(S_{i~j}^{~k~l}|_{Y_0}\hat{x}^i\hat{x}^j\hat{p}_k\hat{p}_l+i\hbar S_i^j|_{Y_0}\hat{x}^i\hat{p}_j\right).
\end{equation}
The unperturbed version of this problem leads to the eigenstates and -values \cite{Messiah99}
\begin{align}
    \psi_{nlm}^{(0)}=&N_{nl}\left(\frac{2nr}{a_0^*}\right)^le^{-\frac{nr}{a_0^*}}L_{n-l-1}^{2l+1}\left(\frac{2nr}{a_0^*}\right)Y^l_m(\theta,\phi),\\
    N_{nl}=&\sqrt{\left(\frac{2}{na_0^*}\right)^3\frac{(n-l-1)!}{2n(n+l)!}},\\
    E^{(0)}_n=&-\frac{Z^2\alpha^4M}{2n^2},
\end{align}
with the reduced Bohr radius $a_0^*=\hbar/Z\alpha M$ and the quantum numbers $n>0,$ $l=0,1,\dots,n$ and $m=-l,-l=1,\dots,l.$

By analogy with the treatment of the harmonic oscillator, the following pattern, found for the corrections to the energy eigenvalues, has been verified for all quantum numbers until $n=10.$ On the one hand, we obtain
\begin{equation}
    E_{n00}^{(2)}=0.
\end{equation}
On the other hand, if $l\neq 0,$ the resulting corrections read
\begin{align}
    E^{(2)}_{nlm}=&\frac{M}{6}\left[\left(l(l+1)-3m^2\right)\left(C_{ln}M^2S_{zz}+\frac{\hbar^2R_{zz}}{2M^2}\right)\right.\nonumber\\
    &\left.\left.+m^2\left(C_{ln}M^2S+\frac{\hbar^2R}{2M^2}\right)\right]\right|_{Y_0},\label{coulres}
\end{align}
where we defined the function of the quantum numbers
\begin{equation}
    C_{nl}=-\frac{Z^2\alpha^4}{n^3l(l+1)(2l+1)}.
\end{equation}
Again, the curvature of space acts most strongly on large length scales (small $M$) while the curvature in momentum space is most effective for small distances (large $M$). Neglecting the contribution of the spatial curvature, the effect is largest for $n=2,$ $l=1$ and $m=0.$ At Planckian curvature $S_{ij}\sim m_p^2,$ \ie assuming that $M$ reflects the mass of a elementary particle, the relative correction to the energy is of the order
\begin{equation}
    \delta E_{nlm}\sim\frac{M^2}{m_p^2}.
\end{equation}
For the hydrogen atom, where $M$ denotes the mass of the electron, this corresponds to a value $\delta E_{nlm}\sim 10^{-44}$ as had been reported before in the context of the GUP \cite{Bouaziz10,AntonacciOakes13,Brau99}. This result can be improved upon by applying it to more massive charged particles such as the $W^{\pm},$ yielding $\delta E_{nlm}\sim 10^{-35}.$  

Over all, the ensuing corrections \eqref{coulres} bear much resemblance to those appearing in the context of the harmonic oscillator \eqref{Ecorrharm}. For fixed $n$ and $l,$ there continues to be an invariance under the transformation
\begin{equation}
    C_{nl}M^2S_{ij}\longleftrightarrow \frac{\hbar^2R_{ij}}{2M^2},
\end{equation}
which, considering the inverse scaling with the squared mass, again has a taste of T-duality. However, taking into account states of distinct $l$ and/ or $n,$ this symmetry is broken in general -- a hydrogenic atom could clearly be used as a means of discriminating between curvature in position and momentum space. In particular, the resulting changes induced by momentum space-curvature strongly decay towards higher values of $n$ and $l$ and vanish for $l=0.$ Clearly, it is the Hamiltonian that explicitly breaks Born reciprocity, thereby also breaking the T-duality-like invariance.

\subsection{Summary}\label{subsec:sumcot}

The duality between GUP-deformed quantum mechanics and nontrivial momentum space found in the preceding section necessitate the creation of a formalism capable of describing quantum mechanics on background metrics, which are simultaneously position- and momentum-dependent. In particular, an approach alike is required to investigate the position representation of quantum mechanics on a nontrivial momentum space. Furthermore, such a formalism would set the stage to find position and momentum representations of the GEUP, a task which had proven illusive thus far.

The present section marks a first step towards such a description, taking a rather intuitive route. In particular, the Hilbert space measure, derived from the metric \'a la DeWitt \cite{DeWitt52}, has been promoted to an operator. It has been subsequently split into two pieces and merged symmetrically with the wave functions entering the scalar product, thus yielding wave densities as in the geometric quantization program \cite{Woodhouse97}. Resultingly, under the assumption that they comply with the geometrical modifications to derivatives derived from the nonlinear connection on the curved cotangent bundle, we have defined the momentum operator in the position representation by analogy with its Euclidean counterpart. Furthermore, we have found a representation of the Hamiltonian of a free particle and, analogously, the geodesic distance to the origin.

This has made it possible to investigate a metric, which was defined by an expansion similar to  Riemann normal coordinates in position and momentum space. After a short discussion of perturbation theory in this context, we have applied the formalism to two central potentials, the harmonic oscillator and the hydrogenic atom. As a result, we have analytically obtained corrections to the eigenvalues of the Hamiltonian in terms of the Ricci scalars and tensors in position and momentum space. Interestingly, given the right choice of ordering, the isotropic harmonic oscillator retains its symmetry between positions and momenta, thus, in principle, making it impossible to distinguish between curvature in position and momentum space. Resultingly, we have achieved an instantiation of Born reciprocity in quantum mechanics on the curved cotangent bundle - exactly as intended by Born \cite{Born38,Born49}. This property is accompanied by a T-duality-like behaviour of the ensuing relative corrections to the energies of stationary states. Thus, the findings presented in this section corroborate the relation between T-duality, Born reciprocity and the curved cotangent bundle manifest in metastring theory and Born geometries \cite{Freidel13,Freidel14,Freidel15,Freidel17,Freidel18}.

These encouraging results will make it possible to investigate the position representation of theories of curved momentum space dual to noncommutative geometries and GUPs, \ie instances of quantum spaces. Furthermore, it would be interesting to see, whether it is possible to describe theories of GEUP-deformed quantum mechanics in this language. These ideas will be the subject of future research.

\section{Conclusion\label{sec:conc}}

This thesis has been aimed at investigating the relation between modified Heisenberg algebras, \ie GUPs and EUPs, and non-Euclidean background manifolds, \ie curved position and momentum space, in quantum mechanics. Along these lines, we have obtained a number important results, which merit being enumerated in the present section -- the most important ones,  in the author's view, are displayed in boldface.
\begin{itemize}
\item {\bfseries A relativistic EUP can be derived from semiclassical gravity alone without prior modification of the canonical commutation relations. }
\end{itemize}
Assuming that wave functions in the studied Hilbert space are constrained to a geodesic ball, we have found a global lower bound on the standard deviation of the momentum operator, suitably formulated in curved space, as a function of the radius of the said ball, thus yielding the desired uncertainty relation. This inequality has been successively generalized to curved spacetime and relativistic particles making use of the ADM-decomposition. Thus, we have explicitly related semiclassical Einstein-type gravity to the EUP.  As a result, we have quantified corrections to the relation in flat space, which depend on the curvature scalar of the effective spatial metric, the lapse function and the shift vector as well as their covariant derivatives. Interestingly, all relativistic contributions depend on the shift vector, \ie for static backgrounds in the common slicings the full relation survives the nonrelativistic limit. Furthermore, they do not diverge in the limit of vanishing mass. As a result, they equally apply to massless particles. 
\begin{itemize}
\item Event horizons have analogous curvature-induced contributions.
\end{itemize}
We have studied the previously derived, admittedly involved, EUP for a number of important physical applications. In particular, the corresponding expressions for accelerated particles as well as such subject to the gravitational influence of massive bodies as described by the Schwarzschild metric and those surrounded by a cosmological horizon have turned out almost identical. 
\begin{itemize}
\item Phenomenologically, astrophysical corrections to the uncertainty relation are dominant over cosmological ones.
\end{itemize}
We have further estimated that the EUP induced by the cosmological horizon is negligible in comparison to the contribution from the earth's gravitational field up to distances far beyond the next star system, Alpha Centauri. This is clearly of importance for phenomenological applications, contributions to which stemming from astrophysical sources had been consistently ignored before. 
\begin{itemize}
\item Relativistic corrections may be dominant for light particles in rotating spacetimes.
\end{itemize}
We have further investigated metrics containing nonvanishing space-time components, \ie the type usually describing rotating backgrounds. After dealing with the G\"odel universe, we have calculated the contribution of rotating sources in general relativity as well as Stelle- and infinite derivative gravity. In the former case, we have displayed the evolution of an object in an eccentric orbit around a fast rotating black hole and the resulting deviation from the uncertainty relation in flat space. Comparing the nonrelativistic to the relativistic contributions for light particles we have shown that the latter cannot be neglected in the regime of strong curvature.
\begin{itemize}
\item {\bfseries There is an explicit duality between theories exhibiting GUPs and quantum dynamics on non-Euclidean momentum space. }
\end{itemize}
Motivated by the previous results, we have followed the inverse approach to relate the GUP and curved momentum space. Specifically, we have found new conjugate variables, \ie such which obey the canonical commutation relations, for GUP-deformed quantum systems including noncommutative geometries. In this new, dual formulation, we have identified the quantum dynamics of a particle moving on nontrivial momentum space. In particular, we have explicitly obtained the momentum-dependent metric corresponding to an arbitrary GUP. We have, thus, established a duality between nontrivial momentum space and GUP-deformed quantum mechanics. Hence, we have found that the physics of general GEUPs is characterized by the curved cotangent bundle.
\begin{itemize}
\item {\bfseries Coordinate noncommutativity implies momentum space curvature.}
\end{itemize}
Applying this formalism to the most commonly invoked quadratic GUP, we have shown that the resulting curvature tensor is precisely proportional to the noncommutativity of the original coordinates. However, as the Hamiltonian is not invariant under diffeomorphisms in momentum space, commutative coordinates do not imply an Euclidean background. In this case, the GUP is reflected in a different momentum space basis, whose deviation from the trivial counterpart is measured by an additional scalar derived from the Cartan tensor. Finally, we have constrained this quantity as well as the curvature in momentum space on the basis of already existing bounds in the literature on the commutative GUP and noncommutative geometry, respectively.
\begin{itemize}
\item Quantum mechanics can be consistently defined on the curved cotangent bundle.
\end{itemize} 
To the knowledge of the author, there had not been a consistent formulation of quantum mechanics in these kinds of backgrounds, \ie when the metric is a function of both positions and momenta, even though the preceding results had clearly indicated a need for such a description, \eg when dealing with the position representation of the GUP,  going beyond Cartesian coordinates or describing GEUPs. Therefore, we have made a first step towards identifying the position representation on arbitrary backgrounds. In particular, we have promoted the Hilbert space scalar product measure to an operator and merged it with the wave functions. As a result, the position and momentum space representations of the momentum as well as the position operator have turned out to be particularly simple. The same has held for the free particle Hamiltonian and the geodesic distance operator. Assuming that the scalar product of position and momentum eigenstates continue to be plane waves, we have manifested the consistency of the approach.
\begin{itemize}
\item The harmonic oscillator allows for a Born reciprocal description also in the context of curved spaces.
\end{itemize} 
Expanding a general metric in Riemann normal coordinate-like variables in position and momentum space, implying curvature in both, we have applied the resulting formalism to the hydrogenic atom as well as the harmonic oscillator, and found the corresponding corrections to the spectrum of the Hamiltonian. Interestingly, after suitable choice of operator ordering, the resulting contributions to the harmonic oscillator, the only inherently Born reciprocal system usually used in nonrelativistic mechanics, are such that it is impossible to distinguish between curvature in position and momentum space. Hence, we have discovered an instantiation of exact Born reciprocity on the curved cotangent bundle which presents itself as a T-duality-like structure of the said corrections.

In the introduction we posed the central question -- can we establish curved momentum space as the overarching principle connecting all areas of quantum gravity phenomenology related to the minimum length? We have come a long way in showing that this concept is indeed underlying the missing puzzle piece, the GUP. Similar considerations hold for the EUP and curved spacetime. Therefore, not unexpectedly, our answer is in the affirmative.  As this thesis corroborates, if we accept the concept of minimum length, in the low-energy regime the tale of quantum gravity is written in the language of curved momentum space.

\addsec{Appendices}
%\addcontentsline{toc}{section}{Appendices}
\begin{appendix}
%\fancyhead[L]{\nouppercase{\rightmark}} 
\renewcommand{\thesubsection}{\Alph{subsection}}
\renewcommand{\theequation}{\thesubsection.\arabic{equation}}
%\addappheadtotoc
%\titlecontents{section}
%  [3.8em]
%  {}{\contentslabel{2.3em}}
%  {\hspace*{-2.3em}}
%  {\titlerule*[0.7em]{.}\contentspage}
%\begingroup
%\renewcommand{\cleardoublepage}{}
%\renewcommand{\clearpage}{}
\subsection{Evaluation of $\text{MaxRe}\left[e^{i\phi}\Braket*{\psi_{n'l'm'}}{\hatslashed{\pi}\psi_{nlm}}\right]^2$ in flat space}\label{app_maxre}
%\endgroup

In order to be able to evaluate the expectation value of the momentum operator with respect to a general state $\Psi$ written in the basis of the Laplacian (see Eq. \eqref{genstaten}), we need to compute the transition amplitudes $\Braket*{\psi_{n'}}{\hatslashed{\pi}\psi_n}.$ In particular, confined to geodesic balls of radius $\rho$ and on a flat three-dimensional background they read
\begin{equation}
    \Braket*{\psi_{n'l'm'}}{\hatslashed{\pi}\psi_{nlm}}=\int_{B_\rho}\D\mu\psi_{n'l'm'}^*\hatslashed{\pi}\psi_{nlm},
\end{equation}
where the functions $\psi_{nlm}$ were defined in Eq. \eqref{flatsol}. 

According to Eq. \eqref{vanishingmom}, those amplitudes evidently vanish if $n'=n,$ $l'=l$ and $m'=m.$ To be more precise, this result can be extended to cases where $m'\neq m.$ Having in mind that non-vanishing $\Delta m\equiv m'-m$ leads to a phase difference $\psi_{nlm}=\exp{i \Delta m\gamma}\psi_{nlm'},$ the only possible change in the transition amplitude has to stem from derivatives with respect to the coordinate $\gamma.$ Due to the proportionality
\begin{equation}
    \partial_\gamma\psi_{nlm}\propto \psi_{nlm},
\end{equation}
we infer that the relevant integrals, \ie the ones which could prevent the transition amplitude from vanishing, share the behaviour
\begin{equation}
    \int_0^{2\pi} e^{-i\Delta m \gamma}\sin\gamma\D\gamma =\int_0^{2\pi} e^{-i\Delta m\gamma}\cos\gamma\D\gamma=0,
\end{equation}
where the last equality holds irrespective of the value of $\Delta m.$ Thus, varying $\Delta m$ does not change the transition amplitude, yielding
\begin{equation}
    \Braket*{\psi_{nlm'}}{\hatslashed{\pi}\psi_{nlm}}=0.
\end{equation}

As the eigenvalues of the Laplacian are functions of the quantum numbers $l$ and $m$ (\cf Eq. \eqref{flateig}), the remaining transition amplitudes feature states with distinct eigenvalues. The evaluation of those can be simplified considering a different amplitude
\begin{align}
    \Braket*{\psi_{n'l'm'}}{\hatslashed{\pi}^3\psi_{nlm}}=&-\hbar^2\lambda_{nl}\braket*{\hatslashed{\pi}}\\
    =&-\hbar^2\lambda_{n'l'}\braket*{\hatslashed{\pi}}-\hbar^3\int_{B_\rho}\D\mu\partial^j\left[\left(-i\slashed{\partial}\psi_{n'l'm'}\right)^*\partial_j \psi_{nlm}\right],\label{transappint}
\end{align}
where the boundary condition \eqref{evp2} has been applied. The last term, being a total derivative, can be turned into a surface integral by Stokes' theorem such that we can rewrite Eq. \eqref{transappint} as
\begin{align}
    \braket*{\hatslashed{\pi}}=&\frac{\hbar}{\lambda_{nl}-\lambda_{n'l'}}\int_{\partial B_{\rho}}\D\tilde{\mu}\left(-i\slashed{\partial}\psi_{n'l'm'}\right)^*n^j\partial_j\psi_{nlm}\\
    =&\frac{\sigma^2\hbar}{\lambda_{nl}-\lambda_{n'l'}}\int_{S^2}\D\Omega\left(-i\slashed{\partial}\psi_{n'l'm'}\right)^*\partial_\sigma\psi_{nlm}\Big|_{\sigma=\rho},\label{surfint}
\end{align}
with the determinant of the induced metric on the surface of the geodesic ball, which in spherical coordinates is proportional the volume element of the two-sphere $S^2$ (of radius $\sigma$) $\D\tilde{\mu}=\sigma^2\D\Omega=\sigma^2\sin\chi\D\chi\D\gamma,$ and the outward normal $n^i=\delta^i_\sigma.$ Writing the basis states decomposed in terms of their radial and angular parts, 
\begin{equation}
R_{nl}(\sigma)=\sqrt{\frac{2}{\rho^3j^2_{l+1}(j_{l,n})}}j_{l}\left(j_{l,n}\frac{\sigma}{\rho}\right)    
\end{equation}
and $Y^l_m(\chi,\gamma)$ respectively, Eq. \eqref{surfint} can be reexpressed as
\begin{equation}
    \Braket*{\psi_{n'l'm'}}{\hatslashed{\pi}\psi_{nlm}}=-\frac{i\rho^2\hbar}{\lambda_{nl}-\lambda_{n'l'}}\partial_{\sigma}R_{n'l'}\partial_{\sigma}R_{nl}\big|_{\sigma=\rho}\int\D\Omega \gamma_\sigma\left(Y^{l'}_{m'}\right)^*Y^l_m,
\end{equation}
where $\gamma_\sigma=\gamma_al^a$ in the sense of section \ref{subsubsec:momunc} denotes the unit radial vector ($l^a$ being defined in Eq. \eqref{unradvec}) and we used the boundary condition \eqref{evp2}, yielding $R_{nl}|_{\sigma=\rho}=0.$ Without loss of generality, we can choose $l\geq l'$ because the inverse case can be obtained from this one by complex conjugation. Then the remaining integral can be calculated explicitly, resulting in the expression
\begin{align}
    \int\D\Omega\gamma_\sigma\left(Y^{l'}_{m'}\right)^*Y^l_m=&\delta_{l'}^{l+1}\left[s^{-1}_{l'm'}\delta_{m'}^{m+1}\left( -\gamma_x+i\gamma_y\right)\right.\nonumber\\
    &\left.+s^{1}_{l'm'}\delta_{m'}^{m-1}\left( \gamma_x+i\gamma_y\right)+is^0_{l'm'}\delta^m_{m'}\gamma_z\right],
\end{align}
where we introduced the unit vectors in $x,$ $y$ and $z$ directions denoted $\gamma_x,$ $\gamma_y$ and $\gamma_z,$ respectively, and the sequences $s^{\Delta m}_{lm}$
\begin{align}
    s^0_{lm}=&\sqrt{\frac{(l+1)^2-m^2}{4(l+1)^2-1}},\\
    s^{\pm 1}_{lm}=&\frac{1}{2}\sqrt{\frac{(1+l\mp m)(2+l\mp m)}{3+4l(2+l)}}.
\end{align}
Formulated explicitly, a general transition amplitude featuring the momentum operator reads
\begin{align}
    \Braket*{\psi_{n'l'm'}}{\hatslashed{\pi}\psi_{nlm}}=&\frac{\rho^2\hbar}{\lambda_{n'l'}-\lambda_{nl}}\partial_{\sigma}R_{n'l'}\partial_{\sigma}R_{nl}\big|_{\sigma=\rho}\Big\{\nonumber\\
    &\times\delta_{l'}^{l+1}\left[s^1_{l'm'}\delta_{m'}^{m+1}\left(i \gamma_x+\gamma_y\right)
    +s^{-1}_{l'm'}\delta_{m'}^{m-1}\left(-i \gamma_x+\gamma_y\right)+is^0_{l'm'}\delta^m_{m'}\gamma_z\right]\nonumber\\
    &+\delta_{l'+1}^{l}\left[s^1_{lm}\delta_{m'+1}^{m}\left(-i \gamma_x+\gamma_y\right)+s^{-1}_{lm}\delta_{m'-1}^{m}\left(i \gamma_x+\gamma_y\right)-is^0_{l'm'}\delta^m_{m'}\gamma_z\right]\Big\}.
\end{align}
Furthermore, the square of its maximal real part including a relative phase $\phi$ becomes
\begin{align}
    \text{MaxRe}\left[e^{i\phi}\Braket*{\psi_{n'l'm'}}{\hatslashed{\pi}\psi_{nlm}}\right]^2=&\left(\frac{\rho^2\hbar}{\lambda_{n'l'}-\lambda_{nl}}\left.\partial_{\sigma}R_{n'l'}\partial_{\sigma}R_{nl}\right|_{\sigma=\rho}\right)^2\nonumber\\
    &\times\delta_{l'}^{l+1}\left[\left(s^1_{l'm'}\right)^2\delta^{m+1}_{m'}+\left(s^{-1}_{l'm'}\right)^2\delta^{m-1}_{m'}+\left(s^0_{l'm'}\right)^2\delta^m_{m'}\right]\nonumber\\
    &\times\delta_{l'+1}^{l}\left[\left(s^1_{lm}\right)^2\delta^{m}_{m'+1}+\left(s^{-1}_{lm}\right)^2\delta^{m}_{m'-1}+\left(s^0_{lm}\right)^2\delta^m_{m'}\right].\label{transnormmaxreal}
\end{align}
Introducing the sign function $\text{sgn}(x),$ which equals one for $x>0$ and negative one for $x<0,$ we can readily evaluate
\begin{equation}
    \partial_{\sigma}R_{nl}|_{\sigma=\rho}=\text{sgn}\left[j_{l+1}\left(j_{l,n}\right)\right]\sqrt{2\lambda_{nl}/\rho^3},
\end{equation}
and estimate $0<s^{\pm 1}_{lm},s^0_{lm}< 1/2$. This implies for Eq. \eqref{transnormmaxreal} that
\begin{align}
    \text{MaxRe}\left[e^{i\phi}\Braket*{\psi_{n'l'm'}}{\hatslashed{\pi}\psi_{nlm}}\right]^2\leq & \frac{\hbar^2}{\rho^2}\frac{\lambda_{nl}\lambda_{n'l'}}{\left(\lambda_{n'l'}-\lambda_{nl}\right)^2}\left[\delta^{l'+1}_{l}\left(\delta^{m+1}_{m'}+\delta^{m-1}_{m'}+\delta^m_{m'}\right)\right.\nonumber\\
    &\left.+\delta^{l'}_{l+1}\left(\delta^{m}_{m'+1}+\delta^{m}_{m'-1}+\delta^m_{m'}\right)\right].\label{res_app_maxre}
\end{align}
\setcounter{equation}{0}

\subsection{Measure changing nonsingular perturbation theory to \nth{2} order\label{app_pert}}

Consider an operator $\op$ with domain $D$ acting on the Hilbert space $\hil=L^2(D\subseteq{\rm I\!R^3,}\D\mu),$ which is self-adjoint with respect to the measure $\D\mu.$ Then, it satisfies the eigenvalue equation
\begin{align}
\op \psi_n	&=\lambda_n\psi_n~\text{in~} D,\label{ev}\\
\psi_n			&=0~~~\text{on~}\partial D,
\end{align}
with its eigenvalues $\lambda_n$ and eigenvectors $\psi_n$ ($n$ may stand for multiple quantum numbers), which are assumed to be normalized such that
\begin{align}
\Braket*{\psi_n}{\psi_n}\equiv\int_D\D\mu\psi_n^\dagger\psi_n=1.\label{norm}
\end{align}
Further assume that the operator can be expanded perturbatively as
\begin{align}
\op			&=\sum_{N=0}^\infty\epsilon^N\op^{(N)},
\end{align}
with $\epsilon\ll1$ and where all operators $\op^{(N)}$ are $\op^{(0)}$-bounded. The latter is denotes the unperturbed operator whose eigenvalues and -vectors are assumed to be known. This is the starting point for perturbation theory in quantum mechanics.

A perturbative treatment of the eigenvalue problem \eqref{ev} can be complicated by the fact that the unperturbed operator might not be self-adjoint with respect to the measure $\mu.$ Instead, it may be self-adjoint with respect to a measure $\D\mu_{0}.$ In usual introductions to perturbation theory $\D\mu=\D\mu_0$ is assumed.

Yet, this is not generally the case. In principle, the measures may be related as
\begin{align}
\D\mu(x)=\D\mu_0(x) a^2(x),\label{measure}
\end{align}
with a positive non-vanishing function of the coordinates $a^2(x).$ A simplifying assumption that has to be made here requires that the function $a^2(x)$ may be expanded perturbatively as
\begin{align}
a^2(x)=1+\sum_{N=1}^{\infty} a_{N}^{2}(x).\label{measass}
\end{align}
It is questionable whether this problem is solvable or even mathematically well defined without this additional assumption. Anyway, it certainly applies to the physical example studied in section \ref{subsubsec:perturb}. Expanding $a^2(x)$ as in Eq. \eqref{measass} is equivalent to introducing a perturbed scalar product as
\begin{align}
\Braket*{\psi}{\phi}=\sum_{N=0}^{\infty}\Braket*{\psi}{\phi}_N.\label{prodexp}
\end{align}

Additionally, we assume the spectrum of $\op^{(0)}$ to be discrete, \ie we consider non-singular perturbation theory. The formalism could be straightforwardly generalised to the singular case, which is,  however, not needed at this point.  By virtue of Eq. \eqref{measass} the unperturbed operator is almost self-adjoint, meaning that the non-self-adjointness is of higher order, \ie
\begin{align}
\Braket*{\psi}{\op^{(0)}\phi}_0=\Braket*{\op^{(0)}\psi}{\phi}_0.
\end{align}
In fact, the eigenvectors of the unperturbed operator $\psi_n^{(0)}$ can be normalized (without loss of generality) as $\Braket*{\psi^{(0)}_n}{\psi^{(0)}_n}_0=1.$ Thus, they span an orthonormal basis of the Hilbert space corresponding to the unperturbed problem $\hil_0=L^2(D\subseteq{\rm I\!R^3,}\D\mu_0),$ while the eigenvectors of the full operator $\psi_n$ span an orthonormal basis of $\hil.$ Hence, we can expand
\begin{align}
\psi^{(i)}_n	&=\sum_{m}c^{(i)}_{nm}\psi_m\\
					&=\sum_{m}c^{(1)}_{nm}\psi_m^{(0)}+\Ord(|g{(1)}|),
\end{align}
with the  complex $i^{\text{th}}$-order coefficients $c^{(i)}_{nm}.$
In particular, we approximate
\begin{align}
\psi^{(1)}_n	&\simeq \sum_{m}c^{(1)}_{nm}\psi^{(0)}_m.\label{1stperteigvec}
\end{align}
Furthermore, from Eq. \eqref{norm} and the normalization of $\psi^{(0)}_n$, we deduce
\begin{align}
\Braket*{\psi^{(1)}_n}{\psi^{(0)}_n}_0+\Braket*{\psi^{(0)}_n}{\psi^{(1)}_n}_0+\Braket*{\psi^{(0)}_n}{\psi^{(0)}_n}_1	&=0,\label{norm1st}\\
\Braket*{\psi^{(2)}_n}{\psi^{(0)}_n}_0+\Braket*{\psi^{(0)}_n}{\psi^{(2)}_n}_0+\Braket*{\psi^{(0)}_n}{\psi^{(0)}_n}_2~~~&\nonumber\\+\Braket*{\psi^{(1)}_n}{\psi^{(1)}_n}_0+\Braket*{\psi^{(1)}_n}{\psi^{(0)}_n}_1+\Braket*{\psi^{(0)}_n}{\psi^{(1)}_n}_1&=0\label{norm2nd}.
\end{align}
The first of these relations Eq. \eqref{norm1st} translates to
\begin{align}
\text{Re}\left(c_{nn}^{(1)}\right)=-\frac{1}{2}\Braket*{\psi_n^{(0)}}{\psi_n^{(0)}}_1,
\end{align}
which yields without loss of generality (discarding a global phase)
\begin{align}
c_{nn}^{(1)}=-\frac{1}{2}\Braket*{\psi_n^{(0)}}{\psi_n^{(0)}}_1.\label{coeffnn}
\end{align}
At this point, we see the first influence of the non-self-adjointness of $\op^{(0)}:$ The first order eigenvectors cannot be taken to be orthogonal to the unperturbed ones because they amongst themselves are only orthogonal to \nth{0} order, \ie
\begin{align}
\Braket*{\psi_n^{(0)}|\psi_m^{(0)}}=\delta_{nm}+\Ord(|g^{(1)}|).\label{almorth}
\end{align}
Perturbing Eq. \eqref{ev} yields order by order
\begin{align}
\mathcal{O}(1)&
\begin{cases}
\left(\op^{(0)}-\lambda_n^{(0)}\right)	&\left.\psi_n^{(0)}=0\right|_{D},\\
																		&\left.\psi_n^{(0)}=0\right|_{\partial D},
\end{cases}\label{eig0th}\\
\mathcal{O}(\epsilon)&
\begin{cases}
\left(\op^{(0)}-\lambda^{(0)}_n\right)	&\left.\psi_n^{(1)}=-\left(\op^{(1)}-\lambda_n^{(1)}\right)\psi_n^{(0)}\right|_{D},\\
																		&\left.\psi_n^{(1)}=0\right|_{\partial D},
\end{cases}\label{eig1st} \\
\mathcal{O}(\epsilon^2)&
\begin{cases}
\left(\op^{(0)}-\lambda^{(0)}_n\right)	&\left.\psi_n^{(2)}=-\left(\op^{(2)}-\lambda^{(2)}_n\right)\psi_n^{(0)}-\left(\op^{(1)}-\lambda_n^{(1)}\right)\psi_n^{(1)}\right|_{D}, \\
																		&\left.\psi_n^{(2)}=0\right|_{\partial D}.
\end{cases}  \label{eig2nd}
\end{align}
These equations will now be contracted with $\bra{\psi_m}$ for some $m.$ The case $m=n$ yields the corrections to the eigenvalues while the coefficients $c^{(1)}_{nm}$ can be determined from the opposite case. Note that said contraction shifts the order of perturbation of some terms due to the expansion not only  of $\psi_m$ but also of the scalar product. 

The zeroth-order part corresponds to solving the unperturbed eigenvalue problem because
\begin{align}
\lambda_n^{(0)}=\Braket*{\psi^{(0)}_n}{\op^{(0)}\psi_n^{(0)}}_0
\end{align}
takes place entirely within $\hil_0.$ Furthermore, the first-order corrections to the eigenvalues read (applying Eq. \eqref{eig0th})
\begin{align}
\lambda_n^{(1)}=	&\Braket*{\psi_n^{(1)}}{\op^{(0)}\psi_n^{(0)}}_0+\Braket*{\psi_n^{(0)}}{\op^{(0)}\psi_n^{(1)}}_0+\Braket*{\psi_n^{(0)}}{\op^{(0)}\psi_n^{(0)}}_1+\Braket*{\psi_n^{(0)}}{\op^{(1)}\psi_n^{(0)}}_0\\
						=	&\lambda^{(0)}\Big(\Braket*{\psi_n^{(1)}}{\psi_n^{(0)}}_0+\Braket*{\psi_n^{(0)}}{\psi_n^{(1)}}_0+\Braket*{\psi_n^{(0)}}{\psi_n^{(0)}}_1\Big)+\Braket*{\psi_n^{(0)}}{\op^{(1)}\psi_n^{(0)}}_0,
\end{align}
which with Eq. \eqref{norm1st} becomes
\begin{align}
\lambda_n^{(1)}=\Braket*{\psi_n^{(0)}}{\op^{(1)}\psi_n^{(0)}}_0.
\end{align}
Hence, $\op^{(0)},$ not being self-adjoint, does not significantly change the first-order contribution to the eigenvalues. However, the expectation value is taken with respect to the measure $\D\mu_0.$ Contracting the eigenvalue problem with $\bra{\psi_m},$ while assuming $n\neq m,$ yields,  after some algebra and applying Eq. \eqref{eig1st},
\begin{align}
c_{nm}^{(1)}=\frac{\Braket*{\psi_m^{(0)}}{\op^{(1)}\psi_n^{(0)}}_0}{\lambda_n^{(0)}-\lambda_m^{(0)}}~\text{for}~n\neq m.\label{coeffnm}
\end{align}
Again, this is equivalent to the case where $\op^{(0)}$ is self-adjoint and the amplitude is calculated with respect to the unperturbed measure. Thus, to first order the only clearly visible change is in the coefficient $c_{nn}^{(1)}$ according to Eq. \eqref{coeffnn}.

Analogously, the second-order correction to the eigenvalues reads after application of Eqs. \eqref{norm2nd} and \eqref{eig0th}
\begin{align}
\lambda_n^{(2)}=	&\Braket*{\psi_n^{(0)}}{\op^{(2)}\psi_n^{(0)}}_0+\Braket*{\psi_n^{(0)}}{\op^{(1)}\psi_n^{(0)}}_1+\Braket*{\psi_n^{(1)}}{\op^{(1)}\psi_n^{(0)}}_0+\Braket*{\psi_n^{(0)}}{\op^{(1)}\psi_n^{(1)}}_0\nonumber\\
							&+\Braket*{\psi_n^{(1)}}{\left(\op^{(0)}-\lambda_n^{(0)}\right)\psi_n^{(1)}}_0+\Braket*{\psi_n^{(0)}}{\left(\op^{(0)}-\lambda_n^{(0)}\right)\psi_n^{(1)}}_1.
\end{align}
Finally, expanding $\psi_n^{(1)}$ as in Eq. \eqref{1stperteigvec} and applying Eqs. \eqref{coeffnn}, \eqref{almorth} and \eqref{coeffnm}, this equation can be expressed in terms of already known quantities
\begin{align}
\lambda_n^{(2)}=	&\Braket*{\psi_n^{(0)}}{\op^{(2)}\psi_n^{(0)}}_0+\sum_{m\neq n}\frac{\Braket*{\psi_m^{(0)}}{\op^{(1)}\psi_n^{(0)}}_0\Braket*{\psi_n^{(0)}}{\op^{(1)}\psi_m^{(0)}}_0}{\lambda_n^{(0)}-\lambda_m^{(0)}}+\Braket*{\psi_n^{(0)}}{\op^{(1)}\psi_n^{(0)}}_1\nonumber\\
&-\sum_{m}\Braket*{\psi_m^{(0)}}{\op^{(1)}\psi_n^{(0)}}_0\Braket*{\psi_n^{(0)}}{\psi_m^{(0)}}_1.\label{gen_eig2}
\end{align}
The first two terms of this equation appear in usual perturbation theory as well, which is indicated by the fact that they are evaluated with respect to the unperturbed measure, while the last two are entirely new,  a fact that can be inferred from the higher-order scalar products they are featuring. Thus, the change in the measure has a strong effect on the second-order eigenvalues and cannot be neglected.
\setcounter{equation}{0}

\subsection{Higher-order corrections to the asymptotic uncertainty relation}
\label{HOCORR}

In this appendix we expand upon the perturbative derivation of the asymptotic uncertainty relation in section \ref{subsubsec:perturb} to get to fourth order, thereby explaining the origin of the mathematical constant $\xi.$ In principle, the required math is the same as in the text although the calculations get rather lengthy and can in their entirety only be performed numerically. The Riemann normal coordinate expansion approximates the metric to fourth order as \cite{Brewin09}
\begin{align}
g_{ij}\simeq &e^a_ie^b_j\Big[\delta_{ab}-\frac{1}{3}R_{acbd}|_{p_0}x^{c}x^{d}-\frac{1}{6}\nabla_e R_{acbd}|_{p_0}x^{c}x^{d}x^e\nonumber\\
&+\left(\frac{2}{45}R_{acdg}R^{~~~g}_{bef}|_{p_0}-\frac{1}{20}\nabla_e\nabla_f R_{acbd}|_{p_0}\right)x^{c}x^{d}x^ex^f\Big].\label{geod_met4}
\end{align}
Furthermore, the determinant of the metric equals to second order
\begin{align}
\sqrt{g}\simeq 1-\frac{1}{3}R_{ab}|_{p_0}x^ax^b.
\end{align}
As was touched upon in section \ref{subsubsec:RNC}, the symmetries of the Riemann tensor imply that spherical coordinates constructed around $p_0$ are automatically geodesic coordinates. In other words, Eq. \eqref{geod_met4} already satisfies Eq. \eqref{geod_met_char}. Accordingly, the higher-order contributions to the part of the Laplace-Beltrami-operator acting on the state saturating the uncertainty relation $\psi^{(0)}_{100}(\sigma)$ read
\begin{align}
\Delta^{(3)}\psi_{100}^{(0)}	&=-\frac{\sigma^2}{4}\nabla_c\psi_{100}^{(0)} R_{ab}|_{p_0}l^al^bl^c\partial_\sigma,\\
\Delta^{(4)}\psi_{100}^{(0)}	&=-\frac{\sigma^3}{5}\mathcal{R}_{abcd}|_{p_0}l^al^bl^cl^d\partial_\sigma\psi_{100}^{(0)},
\end{align}
where we defined for simplicity
\begin{align}
\mathcal{R}_{abcd}\equiv\frac{1}{9}R^{e~f}_{~a~b}R_{ecfd}+\frac{1}{2}\nabla_c\nabla_dR_{ab}.
\end{align}
As there is no first-order contribution to the metric, the third-order correction can be treated as if it was of first order,  while the fourth-order contribution can be understood to be of second order in the sense of appendix \ref{app_pert}. Hence, the third-order correction reads
\begin{align}
\lambda_{100}^{(3)}	&=\Braket*{\psi_{100}^{(0)}}{\hat{p}^2_{(3)}\psi_{100}^{(0)}}_0/\hbar^2\\
						&=\int_0^\rho\D\sigma\frac{\sigma^4}{4}\psi_{100}^{(0)}\partial_\sigma\psi_{100}^{(0)} \int_{S^2}\D\Omega \nabla_{c}R_{ab}|_{p_0} l^al^bl^c\\
						&=0,
\end{align}
where the last equality stems from the fact that the angular integral vanishes. Thus, the role of the next-to-leading order is played by the fourth one. According to Eq. \eqref{gen_eig2}, the corresponding eigenvalue reads
\begin{align}
\lambda_{100}^{(4)}=&\hbar^{-2}\Bigg(\Braket*{\psi_{100}^{(0)}}{\hat{p}^2_{(4)}\psi_{100}^{(0)}}_0-\hbar^{-2}\sum_{nlm\neq 100}\frac{\Braket*{\psi_{nlm}^{(0)}}{\hat{p}^2_{(2)}\psi_{100}^{(0)}}_0\Braket*{\psi_{100}^{(0)}}{\hat{p}^2_{(2)}\psi_{nlm}^{(0)}}_0}{\lambda_{nlm}^{(0)}-\lambda_{100}^{(0)}},\nonumber\\
							&+\Braket*{\psi_{100}^{(0)}}{\hat{p}^2_{(2)}\psi_{100}^{(0)}}_2-\sum_{nlm}\Braket*{\psi_{nlm}^{(0)}}{\hat{p}^2_{(2)}\psi_{100}^{(0)}}_0\Braket*{\psi_{100}^{(0)}}{\psi_{nlm}^{(0)}}_2\Bigg).\label{ev4}
\end{align} 
In order to be able to compute the values of the terms which are not summed over, we need the following integrals
\begin{align}
\int_0^\rho\D\sigma\sigma^5\psi_{100}^{(0)}\partial_\sigma\psi_{100}^{(0)} 	&=\frac{5(3-2\pi^2)}{48\pi^3}\\
\int_{S^2}\D\Omega l^al^bl^cl^d																	&=\frac{4\pi}{15}\left(\delta^{ab}\delta^{cd}+\delta^{ac}\delta^{bd}+\delta^{ad}\delta^{bc}\right).
\end{align}
Furthermore, in three dimensions the Riemann tensor can be expressed in terms of the Ricci tensor and the metric (note that at $p_0$ we have $g_{ab}=\delta_{ab}$) as
\begin{align}
R_{abcd}|_{p_0}=	&\delta_{ac}R_{bd}|_{p_0}-\delta_{ad}R_{bc}|_{p_0}-\delta_{bc}R_{ad}|_{p_0}+\delta_{bd}R_{ac}|_{p_0}-\frac{R|_{p_0}}{2}\left(\delta_{ac}\delta_{bd}-\delta_{ad}\delta_{bc}\right),
\end{align}
and by the contracted Bianchi identity
\begin{align}
\nabla_c\nabla_dR_{ab}|_{p_0}\left(\delta^{ab}\delta^{cd}+\delta^{ac}\delta^{bd}+\delta^{ad}\delta^{bc}\right)=3\Delta R|_{p_0}.
\end{align}
Combining these, we obtain
\begin{align}
\Braket*{\psi_{100}^{(0)}}{\Delta^{(4)}\psi_{100}^{(0)}}_0=	&\rho^2\frac{3-2\pi^2}{120\pi^2}\Bigg(-\frac{R^2|_{p_0}}{9}+\frac{14}{27}R^{ab}R_{ab}|_{p_0}+\Delta R|_{p_0}\Bigg),\\
\Braket*{\psi_{100}^{(0)}}{\Delta^{(2)}\psi_{100}^{(0)}}_2=	&\rho^2\frac{3-2\pi^2}{648\pi^2}\left(R^2|_{p_0}+2R^{ab}R_{ab}|_{p_0}\right).																			
\end{align}

Concerning the sums in Eq. \eqref{ev4}, both $\Braket*{\psi_{nlm}^{(0)}}{\Delta^{(2)}\psi_{100}^{(0)}}_0$ and $\Braket*{\psi_{100}^{(0)}}{\psi_{nlm}^{(0)}}_2$ only give non-vanishing contributions if $l=m=0$ or $l=2,$ $m=-2,-1,0,1,2.$ The exact limits of these sums are not known (the summands are intricate terms containing spherical Bessel functions evaluated at special zeros of other Bessel functions). Yet, they show fast convergence and can thus be obtained numerically. Consider the quantity
\begin{align}
S_{lm}=\sum_{n=2}^{\infty}-\hbar^{-2}\left(-\frac{\Braket*{\psi_{nlm}^{(0)}}{\hat{p}^2_{(2)}\psi_{100}^{(0)}}_0\Braket*{\psi_{100}^{(0)}}{\hat{p}^2_{(2)}\psi_{nlm}^{(0)}}_0}{\hbar^{2}\left(\lambda_{nlm}^{(0)}+\lambda_{100}^{(0)}\right)}+\Braket*{\psi_{nlm}^{(0)}}{\hat{p}^2_{(2)}\psi_{100}^{(0)}}_0\Braket*{\psi_{100}^{(0)}}{\psi_{nlm}^{(0)}}_2\right).
\end{align}
Using this notation, we can express the fourth-order contribution to the eigenvalue as
\begin{align}
\lambda^{(4)}_{100}=&\hbar^{-2}\left(\Braket*{\psi_{100}^{(0)}}{\hat{p}^2_{(4)}\psi_{100}^{(0)}}_0+\Braket*{\psi_{100}^{(0)}}{\hat{p}^2_{(2)}\psi_{100}^{(0)}}_2\right)+\lambda_{100}^{(2)}\Braket*{\psi_{100}^{(0)}}{\psi_{100}^{(0)}}_2+S_{00}+\sum_{m=-2}^2S_{2m}.\label{ev42}
\end{align}
Computing the remaining terms, we obtain
\begin{align}
\lambda_{100}^{(2)}\Braket*{\psi_{100}^{(0)}}{\psi_{100}^{(0)}}_2	&=\frac{3-2\pi^2}{648\pi^2}R^2|_{p_0},\\
S_{00}																							&=\frac{3-2\pi^2}{1944\pi^2}R^2|_{p_0},\\
S_{20}																							&= -\frac{S_{2c}}{3} (T_{20}^{ab}R_{ab}|_{p_0})^2,\\
S_{21}+S_{2-1}																				&= -S_{2c}|T_{21}^{ab}R_{ab}|_{p_0}|^2,\\
S_{22}+S_{2-2}																				&= -S_{2c}|T_{20}^{ab}R_{ab}|_{p_0}|^2,
\end{align}
where we introduced the matrices
\begin{align}
T_{20}^{ab}	&=\text{Diag}(1,1,-2),\label{T20}\\
T_{21}^{ab}	&=
\begin{bmatrix}
0	&	0	&	1\\
0	&	0	&	i\\
1	&	i	&	0\\
\end{bmatrix},\label{T21}\\
T_{22}^{ab}	&=
\begin{bmatrix}
1	&	i	&	0\\
i	&	-1	&	0\\
0	&	0	&	0\\
\end{bmatrix},\label{T22}
\end{align}
and the newly introduced numerical constant $S_{2c}\simeq 6*10^{-3}.$ As the matrices \eqref{T20}, \eqref{T21} and \eqref{T22} are not tensors, the contributions $S_{20},$ $S_{21}$ and $S_{22}$ are not scalars. The reason for this unusual behaviour lies in the fact that $\Delta^{(2)}$ and $\braket*{,}_2$ are by themselves not scalar quantities. However, summed up they yield
\begin{align}
\sum_{m=-2}^2S_{2m}=S_{2c}\left(\frac{R^2}{6}-\frac{R^{ab}R_{ab}}{2}\right),
\end{align}
which is a scalar as expected. Finally, plugging all terms into Eq. \eqref{ev42}, we obtain the fourth-order contribution to the eigenvalue
\begin{align}
\lambda_{100}^{(4)}=-\rho^2\left.\left[\eta_1\left(3R_{ab}R^{ab}-R^2\right)+\eta_2\Delta R\right]\right|_{p_0},
\end{align}
where we introduced the numerical constants 
\begin{align}
\eta_1	&\simeq 3.0*10^{-3},\\
\eta_2	&=\frac{2\pi^2-3}{120\pi^2}.
\end{align}
Hence, we can approximately say that
\begin{align}
\eta\equiv 72\eta_1\simeq 15\eta_2.
\end{align}
Furthermore, the result can be expressed in terms of the zeroth-order Cartan invariant $\Psi_2,$ which in three dimensions reads \cite{Musoke16}
\begin{align}
\Psi_2^{~2}=\frac{3R^{ab}R_{ab}-R^2}{72},
\end{align}
thus yielding
\begin{align}
\lambda_{100}^{(4)}\simeq -\rho^2\eta\left(\Psi_2^{~2}+\frac{\Delta R}{15}\right)\Bigg|_{p_0},
\label{res4app}
\end{align}
with the numerical constant
\begin{align}
\eta=\frac{2\pi^2-3}{8\pi^2}.
\end{align} 
This concludes our treatment of the extended uncertainty relation to fourth order.
\end{appendix}

\sloppy

\printbibliography[title=Bibliography]

%\pagestyle{empty}

%\begin{otherlanguage}{polish}
%	\input{Main_and_Chapters/streszczenie}
%\end{otherlanguage}

%\newpage

%\cleardoublepage

%\input{Main_and_Chapters/abstract2}

\end{document}